%% file: toprev.tex
\documentclass{ws-ijmpa}

\usepackage{graphicx}
\usepackage{dcolumn}
\usepackage{bm}
\usepackage{rotating}

\begin{document}

\newcommand{\dzero}     {D\O}
\newcommand{\ttbar}     {\mbox{$t\bar{t}$}}
\newcommand{\bbbar}     {\mbox{$b\bar{b}$}}
\newcommand{\ccbar}     {\mbox{$c\bar{c}$}}
\newcommand{\ppbar}     {\mbox{$p\bar{p}$}}
\newcommand{\herwig}    {\sc{herwig}}
\newcommand{\pythia}    {\sc{pythia}}
\newcommand{\vecbos}    {\sc{vecbos}}
\newcommand{\alpgen}    {\sc{alpgen}}
\newcommand{\qq}        {\sc{qq}}
\newcommand{\evtgen}    {\sc{evtgen}}
\newcommand{\tauola}    {\sc{tauola}}
\newcommand{\geant}     {\sc{geant}}
\newcommand{\met}       {\mbox{$\not\!\!E_T$}}
\newcommand{\mex}       {\mbox{$\not\!\!E_x$}}
\newcommand{\mey}       {\mbox{$\not\!\!E_y$}}
\newcommand{\metcal}    {\mbox{$\not\!\!E_{Tcal}$}}
\newcommand{\rar}       {\rightarrow}
\newcommand{\eps}       {\epsilon}

\markboth{Kehoe, Narain, Kumar}
{Review of Top Quark Physics Results}

%
\catchline{}{}{}{}{}
%

\title{Review of Top Quark Physics Results}

\author{R. Kehoe}

\address{Physics Department, Southern Methodist University, Dallas, Texas\\
Dallas, TX 75275, USA\\
kehoe@physics.smu.edu}

\author{M. Narain}
\address{Physics Department, Brown University\\
Providence, RI, 02912 USA\\
narain@hep.brown.edu}

\author{A. Kumar}
\address{Physics Department, State University of New York at Buffalo\\
Buffalo, New York 14260, USA\\
ashishk@fnal.gov}

\maketitle

\begin{history}
\received{14 December 2007}
\published{12 February 2008}
\end{history}

\begin{abstract}

As the heaviest known fundamental particle, the top quark has taken a central
role in the study of fundamental interactions.  
Production of top quarks
in pairs provides an important probe of
strong interactions.  The top quark mass is a key fundamental parameter
which places a valuable constraint on the Higgs boson mass and electroweak
symmetry breaking. Observations of the relative rates and kinematics of 
top quark final states constrain potential new physics.  In many cases,
the tests available with study of the top quark are both critical and
unique.  Large increases in data samples from the Fermilab Tevatron have
been coupled with major improvements in experimental techniques to produce
many new precision measurements of the top quark.
The first direct evidence for electroweak production of top quarks
has been obtained, with a resulting direct determination of $V_{tb}$.
Several of the properties of the top quark have been measured.
Progress has also been made in obtaining improved limits on
potential anomalous production and decay mechanisms.
This review presents an overview of
recent theoretical and experimental developments in this field.  We also
provide a brief discussion of the implications for further efforts.

\end{abstract}

\ccode{PACS numbers: 11.25.Hf, 123.1K}

\newpage

\section{Introduction}
\input{intro/intro}

{\section{Production and Decay of the Top Quark}
\label{sec:prodDecay}}
\input{theory/theory}

\section{Experimental Facilities and Reconstruction}
\input{apparatus/experiment}

{\section{Top Quark Pair Production Cross Section Measurements}
\label{sec:topPairs}}
\input{ttbar_csec/ttcsec}

{\section{Single Top Quark Searches}~\label{secSTsearch}}
\input{single_top/SingleTop}

{\section{Measurements of the Top Quark Mass}
\label{sec:mTop}}
\input{mtop/mtop.tex}

{\section{Properties of the Top Quark}
\label{sec:Properties}}
\input{proper/properties.tex}

{\section{Searches for Non-Standard Top Quark Production and Decay}
\label{sec:nonStandard}}
\input{bsm/bsm.tex}

{\section{Prospects for Future Measurements}
\label{sec:Conclu}}
\input{prospects/conclu.tex}

\end{document}

%% file: intro/intro.tex
The top quark has played a key role in particle physics for well over
a decade. Its discovery\cite{d0Obs,cdfObs} marks a triumph of the modern 
theoretical
framework of $SU(3)\times SU(2)\times U(1)$ interactions.  It also hallmarked
a new generation of experimental approaches that will be central to
the intensifying search for the Higgs boson and new physics. 
Subsequent analyses have measured the mass of the top quark with remarkable 
precision. What is more, this particle is uniquely placed
to probe strong interactions, electroweak interactions, and the Higgs
mechanism itself.  Many challenges confront the 
experimentalist, particularly because of the diversity and complexity 
of the final state.  

We review this exciting
new field with the underlying goal of illustrating not only the measurements
themselves, but also the techniques being developed to extract them.
In the remainder of this section, we synopsize the role of the top quark
within our understanding of fundamental interactions and particles.  In
subsequent sections we discuss production and decay expectations, 
as well as the experiments that study top quark events.  Measurements of
strong 
and electroweak production are reviewed in Section 
\ref{sec:topPairs} and \ref{secSTsearch}, respectively.  Section
\ref{sec:mTop} treats precision measurement of the top quark mass.
This is followed in Section \ref{sec:Properties} with coverage of other
properties of this particle.  Lastly, we finish in Section 
\ref{sec:nonStandard} with searches for non-standard physics within
the top quark sector and we conclude with some considerations for future efforts.
There are excellent reviews of this subject which primarily cover 
results obtained at $\sqrt{s} = 1.8$ TeV from 1992-1996
\cite{bhatRev,dhimanRev}, termed Run I of the 
Fermilab Tevatron.  Informative reviews with a more phenomenological 
emphasis are also available\cite{dawsonRev,wagnerRev}.
We concentrate in this review on results obtained at
$\sqrt{s} = 1.96$ TeV and much higher integrated luminosity in the
Tevatron's Run II which started in 2001.  We also make a point of including
the later Run I results that may not have been completely discussed before.
To provide completeness and also a better indication of the direction of
the field, we include several preliminary results from CDF and \dzero\ when
published results are not yet available.  As these results may change before
publication, we indicate them in the text.

\subsection{Electroweak and strong interactions}

We now view the weak and electromagnetic interactions as different
manifestations of the same underlying electroweak interaction. The
crucial progress in this realization came with the development of
a viable $SU(2)\times U(1)$ model\cite{salam-weinberg}. This
interaction is mediated by four gauge bosons. The $W^{\pm}$ 
bosons mediate the charge changing weak currents for left-handed fermions.  
The $Z$ boson and the photon propagate the weak and electromagnetic neutral
currents. The symmetry of the theory's Lagrangian is broken in the ground
state of the system by introducing the Higgs field\cite{higgs} 
which has a non-zero vacuum expectation value. This then provides mass to the 
$W^\pm$ and $Z$ bosons.
This model, originally designed around first and second generation leptons, was
extended first to include two generations of quarks, and then a full
third generation of fundamental fermions. The first generation consists of 
the electron ($e$) and its neutrino ($\nu_e$) and the up ($u$) and down 
($d$) quarks. The second 
generation includes the muon ($\mu$), muon neutrino ($\nu_\mu$), 
charm ($c$) quark, and strange ($s$) 
quark and the third generation consists of the tau ($\tau$), tau neutrino
($\nu_\tau$), top ($t$) quark, 
and bottom ($b$) quark. In each generation, the left-handed charged lepton 
and neutrino form a doublet under $SU(2)$ and so do the left-handed quarks. 
The quantum number associated with the $SU(2)$ symmetry is called weak 
isospin and the left-handed fermions have a third component of
weak isospin, $t_{3L} = \pm \frac{1}{2}$. 
Right-handed fermions are isosinglets.  The Higgs couples to these
fundamental fermions $f$ with Yukawa couplings, $y_f$. 
Each fermion propagates through the Higgs field
and, by virtue of this coupling, acquires mass. The value of
$y_f$ is therefore related to the mass of the fermion, $m_f$, by 
$y_f = \sqrt{2} m_f/v$, where $v$ is the vacuum expectation value
of the Higgs field.
The value of $y_f$ is different for each fermion and is not
predicted by the model. We can only ascertain its value by a measurement
of each fermion's mass.

Strong interactions were given a coherent theoretical description
by the non-abelian $SU(3)$ theory of quantum chromodynamics (QCD)\cite{qcd}
in conjunction with the quark model of hadrons\cite{quarkmod}. Eight
massless gauge bosons (gluons) propagate the interaction and 
carry a color charge. Quarks are the
underlying constituents of hadrons. Their quantum numbers explain 
the isospin and baryon number properties of hadrons
in nature. The strength of the strong interaction, governed by $\alpha_{s}$,
is large, making perturbative calculations difficult. The attraction
between quarks exhibits asymptotic freedom, meaning that at large
energies or very small distances, the interaction strength declines.
The converse means that isolated quarks are prohibited from being
extracted
from hadrons. This `confinement' means that bare colored particles
are not observable in nature. Two colored partons receding from each
other will produce a series of colorless hadrons in the final state.  This
process is called hadronization.  These hadrons are collimated
into jets moving roughly in the directions of the original partons.
The process is understood phenomenologically by several 
models, eg. Ref. \cite{fieldFeyn-stringFrag}. In
the string fragmentation model, a narrow tube or `string' connects
two colored partons. As they recede, particles are created along the
string which give rise to jets and energy flow along lines of color
in an event.  The physics of hadronization provides great
challenges for precise measurements that involve jets, like those of the
top quark.

\subsection{Top quark and the flavor spectrum}

The existence of the top quark is firmly placed in our picture of
fundamental interactions. In our current understanding, all matter
is made of fundamental fermion fields in two categories. Quarks
are sensitive to the strong interaction; leptons are not. The electroweak
eigenstates of quarks are a mix of the mass eigenstates, described by the 
CKM matrix\cite{CKM}. As a result, 
generations that contain only a single quark lead to the expectation of 
flavor changing neutral current interactions. The absence of such 
interactions in experimental observations motivated
the successful prediction of the charm quark\cite{GIM}.

The discovery of the third generation tau lepton ($\tau$)\cite{tauDiscover} 
and bottom quark ($b$)\cite{bDiscover} led to the search for the top
quark ($t$) as the $SU(2)$ partner of the bottom quark. Experimentally, 
several measurements showed that the bottom quark was a member of 
an $SU(2)$ doublet. Several early models\cite{georgi}$^{-}$\cite{gursey}
allowed for the absence of a sixth quark.
If the bottom quark were an $SU(2)$ singlet,
then flavor changing neutral current interactions would result in the 
$B$ meson system. The limit on the branching fraction for 
$B\rightarrow \mu^+\mu^- < 10^{-3}$, however, put a stringent
limit on these processes and ruled out the singlet hypothesis\cite{fcnclimits}.
Measurements of $B^0$-$\bar{B}^0$ mixing\cite{BmesonMix} 
are sensitive to
the magnitude of the CKM matrix elements $V_{td}$ and $V_{ts}$. These 
were observed to be non-zero, indicating the existence of the top quark, 
although this could also be explained without the presence of a top 
quark\cite{topless}.  Measurements of the third component of the 
weak isospin of the $b-$quark, $t_{3L}^b$, provided strong 
additional evidence.   The $b-$quark coupling
to the $Z$ boson is dependent on
$t_{3L}^{b}+\frac{1}{3}\sin^{2}\theta_{w}$, where $\theta_w$ is
the weak mixing angle.
Measurements of the $Z\rightarrow\bbbar$\cite{Zwidth} rate, as well
as the $\bbbar$ charge asymmetry in $e^+e^-$ 
collisions\cite{T3b,LEPT3b} both probed this coupling.
Both measurements indicated $t_{3L}^b = -\frac{1}{2}$\cite{T3b}, necessitating
a $t_{3L} = +\frac{1}{2}$ partner to the $b$ quark. 
A model was developed\cite{topless} 
to explain the $Z$ width measurement without the top quark, but this could 
not account for the charge asymmetry measurement.

Theoretical arguments also favored the existence of a top quark
once the $\tau$ lepton was identified.  In the $SU(2)\times U(1)$ electroweak 
model, triangle diagrams with V and A couplings give rise to
anomalies unless the total charge in a given generation sums
to zero. With three color degrees of freedom per quark, anomalies
are avoided in the first two generations.
The existence of the $\tau$ (and its surmised neutrino)
indicated the presence of a third generation quark doublet. Once 
the $b$ quark was found, another quark of charge = $+\frac{2}{3}$ was 
required.

As a result, a search for the top quark was pursued from the late 1970's
onward (e.g. Refs. \refcite{petraLim}$-$\refcite{d0limit131}). 
As accelerator energies
continued to rise, the lower limits on the top quark mass increased.
A fuller description of this period of research can be found in Ref. 
\refcite{bhatRev}.
By the mid-1990's, it was established that the top quark was very
heavy\cite{d0limit131}.  
This had the result of focusing the search strategies of
the CDF and \dzero\ collaborations and led to the discovery of top-antitop 
quark pair production in 1995 at the Tevatron\cite{d0Obs,cdfObs}.

\subsection{Top quark properties}

A massive top quark has several properties which make it quite interesting
as a probe of known strong and electroweak physics, as well as a 
sensitive window to potential
new physics.  Perturbative QCD calculations of pair production 
can be carried out with significant precision.  
Electroweak production of single top quarks gives direct sensitivity
to $|V_{tb}|$.  The value of $m_t$
places an important constraint on the mass of the Higgs boson.
It has been speculated that the resulting top quark 
Yukawa coupling, $y_t \approx 1$, 
could point to new dynamics beyond the standard model\cite{topBSM}.  
The exceedingly short predicted lifetime of the top quark of $O(10^{-25}s)$ is an 
order of magnitude smaller than hadronization timescales and this
permits a glimpse of the properties of a bare quark.  The decay modes
of the top quark may harbor evidence of new physics if alternative models
are correct. Measuring the charge of the top quark is important to establish
that the top quark is as expected.  A direct determination of the
$|V_{tb}|$ matrix element probes for potential new physics.

%% file: theory/theory.tex
For the next several years, the top quark will continue to be produced
solely in hadron collisions, specifically involving protons.  In such
collisions, the top quark can be produced strongly in pairs or
electroweakly alone.  In general, the proton can be considered to harbor
three `valence' quarks ($uud$) which dictate its quantum numbers.
These valence quarks typically carry much of the momentum of the proton.
There are also virtual or `sea' quarks and gluons in the proton
which carry less momentum individually.  When a proton and antiproton
collide, a hard interaction 
occurs between one of the constituents (`partons') of the proton with
a parton of the antiproton.  Soft interactions involving
the remainder of the hadron constituents produce many low energy
particles which are largely uncorrelated with the hard 
collision.  Because of the large mass of the top quark, production
usually involves the higher momentum valence quarks at the Tevatron.

\subsection{Strong pair production}
\input{theory/pairprod.tex}

\subsection{Electroweak production of single top quarks}\label{EWKtop}
\input{theory/singletop_theory.tex}

\subsection{Top quark decays and final states}
\input{theory/decay.tex}

\subsection{Modeling top quark signal and background}
\input{theory/modeling.tex}

%% file: theory/pairprod.tex
\begin{figure*}[!h!tbp]
\begin{center}
\epsfig{figure=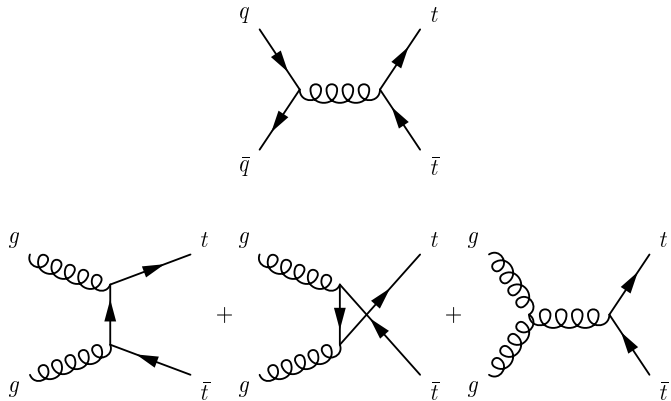,width=0.7\textwidth}
\end{center}
\vspace*{8pt}
\caption{LO Feynman diagrams for quark-antiquark annihilation ($q\bar{q}$) and
gluon fusion ($gg$).
}
\label{fig:feyndiag} 
\end{figure*}

Production of a top-antitop quark pair (\ttbar) occurs dominantly via strong
processes.  
At leading order (LO), valence quarks supply the primary production
probability at the Tevatron via quark-antiquark annihilation ($q\bar{q}$).  
Approximately 15\% of the cross section comes from gluon ($gg$)
fusion. Fig. \ref{fig:feyndiag} shows these diagrams.

In the center-of-mass frame in which the proton and antiproton are rapidly 
moving,
the hard interactions between constituent partons are fast relative to the time
for partons to interact.  As a result, the hadronic collision can be
factorized\cite{factoriz}
into a parton collision weighted by `parton distribution functions' 
($pdf$'s), $F_i(x_i)$ which express the probability for parton $i$ to carry
momentum fraction, $x_i$, of its parent hadron.
These $pdf$'s are properties of specific hadrons and are independent of the
specific hard scatter interaction at parton level.
They encompass non-perturbative soft processes.
As a result, they are extracted from the examination of inelastic interactions
involving hadrons.  The pair production cross section is then calculated as
\begin{equation}
\sigma (\ppbar\rightarrow \ttbar +X) = \Sigma_{i,j} \int dx_i dx_j
    \times F^p_i (x_i, \mu_f) F^{\bar{p}}_j(x_j, \mu_f) \hat{\sigma}_{ij}(x_i, x_j, m_t^2, \mu_f^2)
\end{equation}
where the sum runs over gluons and light quarks in the 
colliding proton and antiproton, and $\hat{\sigma}_{ij}$ is the perturbative 
cross section for collisions of partons $i$ and $j$.  
The factorization scale, $\mu_F$, defines the splitting of perturbative
and non-perturbative elements.  It is common to treat this scale in common
with the scale appropriate to the renormalization of the perturbative 
cross section, $\mu_R$, since both parameters are arbitrary.  This common
scale, $\mu$, is usually taken to be equal to the top quark mass ($m_t$). 
An exact
calculation would not depend on these scales.  Finite order calculations
have a sensitivity that must be assessed as an uncertainty in the theoretical
calculation, usually by bounding the predictions with $\mu=m_t/2$ and 
$\mu=2m_t$ calculations.

Initial LO cross section calculations were performed in Ref.~\refcite{LO}. 
Next-to-leading order (NLO) calculations\cite{NLO,NLOb} accounted 
for associated quark production and gluon bremstrahlung,
and virtual contributions to the LO processes. Because top quark production
at the Tevatron occurs near threshold, a large uncertainty in these
calculations results from initial state soft gluon radiation.
In order to fully account for this, several groups included
resummation of dominant soft logarithms 
to all orders in perturbation theory for hadronic cross sections.
Results from resummed
Drell-Yan production\cite{sterman,catani} were 
generalized to handle the color elements in the initial
heavy quark calculations\cite{HQresum}$^-$\cite{BCMN}.
While the corrections are small for \ttbar\ production,  
the dependence of the cross section
calculation on the choice of scale is reduced.
This work has been improved with more
recent $pdf's$ and more accurate uncertainties in the determination
of the pair production cross section\cite{cacciari}.  The results
of this calculation for Tevatron energies are given in Table~\ref{paircsec}.
A total uncertainty of 10\% to 15\% is obtained for the NLO cross
section\cite{cacciari}.
Parton distribution functions, particularly
for gluons at high $x$, provide the main source of theoretical uncertainty.
The modest increase in collision energy from 1.8 TeV to 1.96 TeV 
causes a 30\% increase in the production
cross section.  This is because of the substantial gain in valence
quark phase space from the parton distribution functions.

\begin{table}[ht]
\begin{center}
\tbl{Calculated \ttbar\ cross sections from Refs.~\protect \refcite{cacciari,kidonakis}
assuming $m_t = 175 \rm GeV/c^2$. The former is a complete
NLO calculation where the uncertainty is primarily from the choice in $pdf$.
The latter provides an NLO calculation with additional higher order terms and
an uncertainty taken from the choice in kinematic scheme.  
The rightmost column provides the 
change in calculated cross section ($\Delta\sigma$) appropriate to a change in 
$m_t$ ~\protect\cite{cacciari}.}
{\begin{tabular}{lcccc}\toprule
& $\sqrt{s}$ (TeV) & Kidonakis, et al.\protect\cite{kidonakis} & Cacciari, et al.\protect\cite{cacciari} & $\Delta\sigma$ (pb)\\
&&&& ($m_t = 170, 180$ GeV)\protect\cite{cacciari} \\
\colrule
& 1.8            & $5.24\pm 0.31$ pb & $5.19^{+0.52}_{-0.68}$ pb & +0.91, -0.76 \\
& 1.96           & $6.77\pm 0.42$ pb & $6.70^{+0.71}_{-0.88}$ pb & +1.13, -0.95 \\
\botrule
\end{tabular}
\label{paircsec}}
\end{center}
\end{table}

Resummed calculations have also been 
performed including additional higher order terms\cite{kinchoice,kidonakis}.  
Both Refs. \refcite{cacciari} and \refcite{kidonakis} are based on
\refcite{NLO,NLOb} with resummation at least to next-to-leading
logarithms.  A difference between them is 
that the former resums soft gluons to all orders, while the latter ignores
small terms beyond next-to-next-to leading order.
Because top quark production is at threshold and both calculations are not
complete beyond NLO, there
is an ambiguity in the soft-gluon resummation calculation.  
We do not attempt a detailed review of this topic here, but in short the
situation is the following.  These calculations are
performed in terms of a choice of kinematic parameters,
and different choices will give slightly different
parton level calculations\cite{kinchoice}.  
To address this, Ref. \refcite{kidonakis} has expanded the ersummed cross 
section to next-to-next-to leading order.  The resulting corrections are 
small for choices of scale $=m_t$, but the observed dependence of
the cross section on scale and kinematics is further reduced.  The results
of this calculation for $\sqrt{s}=1.8$ and $1.96$ TeV are given in
Table~\ref{paircsec}.  The remaining difference from choice of kinematics
provides the uncertainty estimate \cite{kidonakis}.

Because of its short lifetime, the \ttbar\ pair is not
expected to form a bound meson.  However, non-standard production mechanisms
have been proposed which do form a \ttbar\ resonance, eg. 
Ref. \refcite{toptechnicolor2}.  Testing such production
mechanisms will help determine if the top quark fills a special role
among fundamental fermions.  Such analyses will be discussed in
Section \ref{sec:resonance}.

%% file: theory/singletop_theory.tex

\begin{figure}[!h!tbp]
\begin{center}
\epsfig{figure=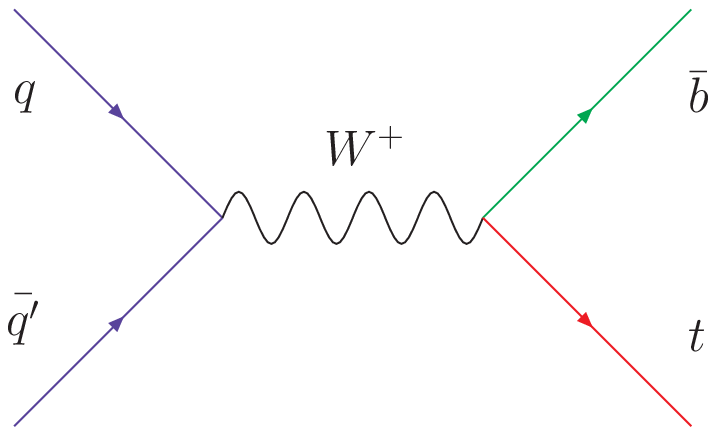,width=0.30\textwidth}
\epsfig{figure=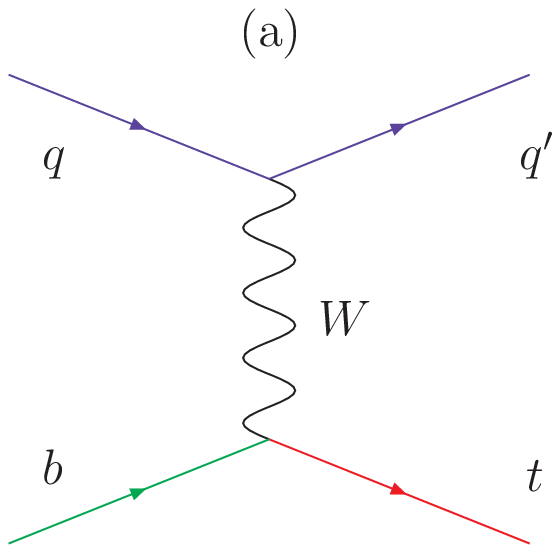,width=0.25\textwidth}
\epsfig{figure=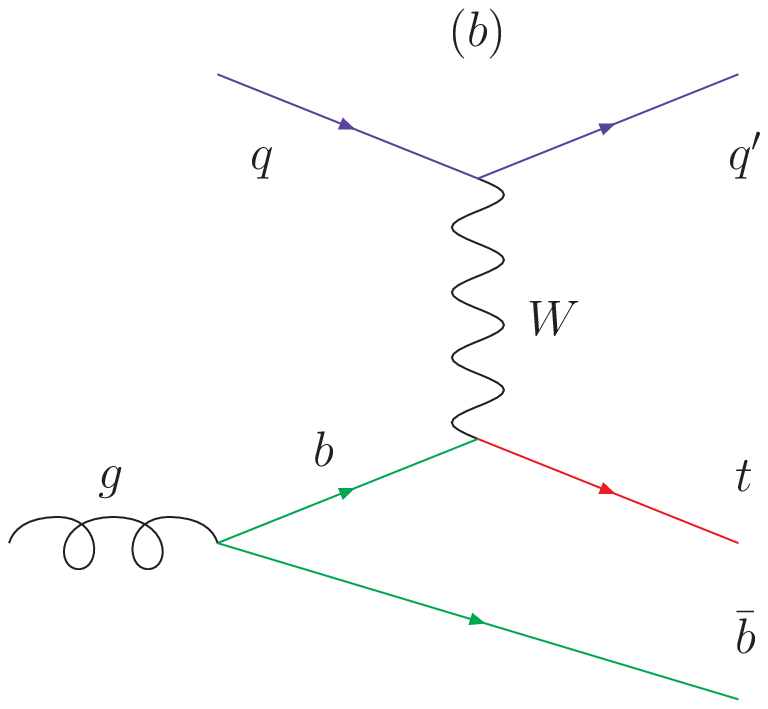,width=0.25\textwidth}
\end{center}
\caption{Feynman diagram for leading order
$s$-channel single top quark production (left) and $t$-channel single top
quark production (middle and right). The middle diagram is the leading order and 
right one is  the $O(\alpha_s)$ $W$-gluon fusion diagram .}
\label{fig:STfeynprod}
\end{figure}

In addition to the pair production of top quarks via the strong interaction 
at the Tevatron, they can also be produced singly in electroweak
interactions.  There are three modes of single top quark production, which
differ in the virtuality, $Q^2$, of the participating $W$ boson, where $Q^2$ is
the negative square of the $W$ boson four momentum $q$. Two of the three
modes are labeled by the corresponding $Mandelstam$ variables $t$ and $s$
involved in the transition matrix elements. The three dominant modes
of single top quark production are listed below in descending order of
their expected production cross section:

\begin{itemize}
\item $t$-channel: In the process $p\overline p\rightarrow t q \overline
  b+X$,  the $W$ boson is spacelike ($-Q^2 =   q^2=t<0$). The predicted cross
  section for this channel, for the Tevatron center of mass
  energy $\sqrt s=1.96$ TeV, computed at   NLO is   
 1.98$^{+0.23}_{-0.18}$ pb\cite{ST-xsec-sullivan}. 
Inclusion of NNLO and NNNLO threshold soft-gluon  corrections leads to
a cross section of  2.30$\pm$0.14 pb\cite{ST-xsec-kidonakis}.
  Two Feynman diagrams, as shown in Fig.~\ref{fig:STfeynprod}, 
   contribute to this 
   channel. In the leading order diagram,
  the $b$ quark is from the sea of quarks in the proton or anti-proton which 
  couples with the virtual $W$ to produce a top quark. In
  the next-to-leading-order diagram, the anti-bottom ($\bar{b}$) quark comes from the
  splitting of the gluon into a $b\overline b$ pair. The $b$ quark
  couples to the virtual $W$ boson and produces the top quark. Thus this
  channel is also  known as  $W$-gluon fusion. The $\overline b$ quark in the final
  state has low transverse momentum ($p_T$) and is at high $\eta$. The $t$-channel
  denotes the processes with the following quarks in the final state: 
  $t q \overline b, \overline t \overline q,
  t\overline q \overline t q$. We denote this channel as $tqb$.

\item $s$-channel:  In the process  $p\overline p\rightarrow t\overline b
  +X$, the $W$ boson is timelike ($-Q^2 =   q^2=s\ge (m_t+m_b)^2>0$). For
  this channel the predicted cross section at NLO for $\sqrt s=1.96$ TeV is
0.88$^{+0.08}_{-0.07}$ pb\cite{ST-xsec-sullivan}. Corrections 
at NNLO and NNNLO leads to a predicted  cross section of  
1.08$\pm$0.08 pb\cite{ST-xsec-kidonakis}.
  The two initial-state quarks annihilate
  into a virtual $W$ boson which decays into a top quark and a bottom quark,
  see  Fig.~\ref{fig:STfeynprod}. The $s$-channel includes both  
   $t\overline b$ and    $\overline  t b$ and is also
  referred to as $W^*$ production. We will refer to this channel as $tb$.
 
\item associated single top quark production: In the process $p\overline
  p\rightarrow t W$, $Q^2=m_W^2$, an on-shell  $W$ boson is produced together
  with a top quark. The  production cross section for this channel is
  predicted to be very small, about 0.28$\pm$0.06 pb at NLO plus
  NNLO and NNNLO threshold soft-gluon  corrections\cite{ST-xsec-kidonakis}. 

\end{itemize}
There are also other diagrams for  single top quark production  which involve
$Wt\overline s$ or $W t \overline d$ vertices, but they are highly suppressed due
to the small CKM matrix elements. Their contribution to the cross section 
is expected to be less than a percent.

One of the primary interests in establishing the cross section for
single top quark production is that its production cross section 
 is directly sensitive to the transition width of $t\rightarrow Wb$, 
and consequently to $|V_{tb}|$. 
The partial decay width of the top quark, $\Gamma(t\rightarrow Wb)$, can be 
measured,  and hence it will be possible to measure the  top quark lifetime.
At the Tevatron, single top quark events have the same final state signature 
as the production of a Higgs boson in association with the $W$ boson, thus
understanding single top quark productions is an important prerequisite for 
Higgs searches  at the Tevatron.  Single top quarks are also one of the
irreducible backgrounds to  associated Higgs production. 
Studies of single top quark events could lead to insights into 
the existence of non standard couplings (e.g. right-handed couplings and
polarization) in the top quark decays to $W$ boson and $b$-quark. 
Single top quark production is very sensitive to 
new physics. The $s$ and $t$-channels are sensitive to different new
physics and hence can be used to distinguish between various exotic 
models (Top Flavor, 4th generation or those which produce flavor changing
neutral currents etc).

%% file: theory/decay.tex
With three generations of quarks, the unitarity of the 
CKM matrix\cite{ckm_measurement} in conjunction with the 
measured values of $V_{ub}$ and $V_{cb}$
means the decay probability of $t\rightarrow Wb$
is virtually 100\%. The W decay is divided approximately equally
between each of the three lepton-neutrino ($l\nu$) pairs and each of the three
colored doublets of up-down quarks and of charm-strange quarks. The
quarks will hadronize to produce jets and heavy flavor mesons from 
$b$ quarks will decay.  Typically, the $b-$jet and $W$
decay products will have large momenta in the plane transverse to the 
beam (`transverse momentum', or `$p_T$') because of the high value of $m_t$.

Events from top quark pair production consist of two $W$ bosons, and
two $b$-quarks: $t\overline t\rightarrow W^+b W^-\overline b$.
Both W's may decay leptonically, one may decay leptonically and the
other hadronically, or both may decay to hadrons. 
The first case, termed `dilepton', gives a final state of
$\ell^+\nu\ell^-\overline\nu b \overline b$ and manifests itself in the
detector as two high $p_T$ leptons, significant unobserved
$p_T$ from the $\nu$s (`\met'), plus two $b$-quark
jets.  The second `$\ell +$jets' scenario provides 
$\ell^\pm\nu q \overline q' b \overline b$ and is observed as
one high $p_T$ lepton, large \met, and four jets including two
$b$-jets.  The last decay chain is the `all-jets' channel and
it results in a
$q_1 q_2 \overline q_3 \overline q_4 b \overline b$ final state
appearing as six jets including two from $b$-quarks. If the leptons
are restricted to $e$ and $\mu$, the usual case, then
the total branching fractions are approximately $5\%$, $30\%$ and
$45\%$ for dilepton, $\ell +$jets and all-jets channels, respectively.
Quark color degrees of freedom are primarily responsible for the 
substantially larger branching fraction of \ttbar\ to the $\ell+$jets
and all-jets final states.

Final states arising from single top quark production consist of 
a top quark, a bottom quark and a light quark ($t$-channel or $tqb$), or 
a top quark and a $b$ quark ($s$-channel or $tb$). 
To successfully identify single top quark production in the presence of
backgrounds, the $W\rightarrow e\nu (\mu\nu)$ decays are required.
Thus single top quark events contain an isolated hight $p_T$ lepton,
significant \met\ due to  
the neutrino, two $b$-jets (and a light quark jet in the t-channel
production).

In addition, for both the \ttbar\ and single top quark production events, 
the heavy flavor hadrons from $b$ quark
fragmentation will have a long lifetime.  This means that final state
particles will originate from a location which is some distance away 
from the event primary vertex.  Sometimes these $b$-jets will exhibit
a soft, non-isolated $e$ or $\mu$ from semileptonic $b$ decay.

%% file: theory/modeling.tex
Measurements of the top quark require a firmly understood simulation
of signal and background processes.  This understanding includes a
need for reliable estimates of event selection efficiency and the
kinematic and angular distributions in the final state.
Event generators are used to simulate the
physics of production, decay and hadronization.  Two kinds are employed.
Exact leading-order calculations such as in {\sc Alpgen} 1.2\cite{alpgen} 
result in the partonic final states at the origin of observed events.
{\sc Alpgen} in particular also includes the spins of the particles.
Showering Monte Carlos such as {\sc Pythia}~6.2\cite{pythia} or 
{\sc Herwig}~6.4\cite{herwig} simulate the hard scatter with 
leading order elements
and employ a phenomenological showering mechanism to account for
hadronization.  
An approximate perturbative QCD description of gluon radiation is implemented in
these simulations.  {\sc Pythia} also adds the 
underlying event from the proton antiproton interaction.
The two types of generator are often used sequentially so that
the hard and soft physics are dealt with as completely as possible.   
In more recent experimental analyses, a significant effort is 
being made to correctly account
for potential double-counting of soft QCD processes when
this is done.  A proposed mechanism to match observed jets with
partons is now employed in these cases\cite{matching1,matching2}.
An alternative approach is also available\cite{matchingCatani}.
Specialized simulators are also
employed to perform important calculations for $\tau$ decays ({\sc
Tauola}\cite{tauola}) or $b$ and $c$ quark decays  ({\sc qq}~9.1\cite{QQgen},
{\sc Evtgen}\cite{evtgen}).  These generators  are reviewed in detail in
Ref.~\refcite{MCgenRev}.  The specifics of their use are documented in 
some of the papers reviewed below (e.g. Ref.~\refcite{CDFMCgen}).

Various Monte Carlo generators were used to model single top quark events. 
\dzero\ uses  the {\sc singletop}
package\cite{STstmc}, based on the {\sc CompHep} Monte Carlo generator\cite{STcomphep}
to produce the parton  four-vectors of the single top quark signal
events. This package uses a leading order simulation for the $s$-channel and
a next-to-leading-order simulation  for the $t$-channel processes. Spin
information in both the production and  decay is included. The matrix element
event generator  {\sc MadGraph/MadEvent}\cite{STmadevent} is used by the CDF
experiment to simulate the signal processes.

Most of the Run II CDF analyses described in this paper use these generators
in the following way.  In general, the CTEQ5L structure functions are used.
Monte Carlo samples for \ttbar\ to  calculate acceptances and efficiencies,
and to understand kinematic shapes, are generated with both {\sc Pythia} and
{\sc Herwig}.   {\sc Alpgen} plus {\sc Herwig} is used with the parton
matching algorithm  to model the $W+$ heavy flavor backgrounds.  Diboson
backgrounds are also generally modeled with {\sc Alpgen}+{\sc Herwig}.
Single top quark backgrounds for $\ell +$jets channels, and $Z/\gamma *$
backgrounds for dilepton channels are studied using {\sc Pythia}.  In
general, $b$ and $c$ quark decays are handled by the QQ generator\cite{QQgen}.

The \dzero\ Run II simulation generally uses {\sc Alpgen} fed into {\sc Pythia}
for signal estimation.  $W+$jets and diboson production are also simulated
with {\sc Alpgen} and {\sc Pythia}.  $Z\rightarrow\tau\tau$ backgrounds are
studied using {\sc Pythia}.  In general, {\sc Tauola} and {\sc Evtgen} are
used for $\tau$ and $b/c$ hadron decays, respectively.
The $Q^2$ scale used by \dzero\ single top quark $s$-channel 
and $t\overline t$ samples is
$m_t^2$, while the $t$-channel is generated at $(m_t/2)^2$. CTEQ6M (CTEQ6.1M)
parton distribution functions are used for single top quark ($t\overline t$)
events.  The components of the $W+$jets
process: heavy flavor process $\bar{q}q^\prime\rightarrow Wg$ with
$g\rightarrow b\bar{b}$ or $g\rightarrow c\bar{c}$, and $gq\rightarrow Wc$)
are included in their relative proportions estimated using {\sc Alpgen} and
normalized to the data.

%% file: apparatus/experiment.tex
\subsection{Experiments}

The Tevatron provides sufficient energy to create
the extremely massive top quark. The two experiments at the Tevatron, \dzero\ 
and CDF, are therefore unique in their ability to study the top quark directly.
The accelerator complex at Fermilab accelerates protons through a linear accelerator and
three synchrotrons, the Booster, the Main Injector, and the superconducting
Tevatron ring\cite{tevatron}. The Main Injector was
built for Run II to increase the number of antiprotons that can be
injected into the Tevatron. 
Antiprotons are produced by extracting protons at intermediate energy onto a 
nickel target. They are selected from the secondaries that are
produced and then cooled and accelerated before they are injected into 
the proton 
acceleration chain. Proton and antiproton beams are accelerated to a final
energy of 980 GeV in the Tevatron to provide proton antiproton collisions at
1.96 TeV, up from 1.8 TeV in Run I. The use of antiprotons is important 
because it provides the ability to use valence quark annihilation, 
with available higher parton center of mass energies, for top quark production.
Instantaneous luminosities are high, so far reaching $2.6\times 10^{32}/cm^2/s$.
Typically around three proton-antiproton interactions occur with each 
beam crossing.

\begin{figure*}[!h!tbp]
\begin{center}
\epsfig{figure=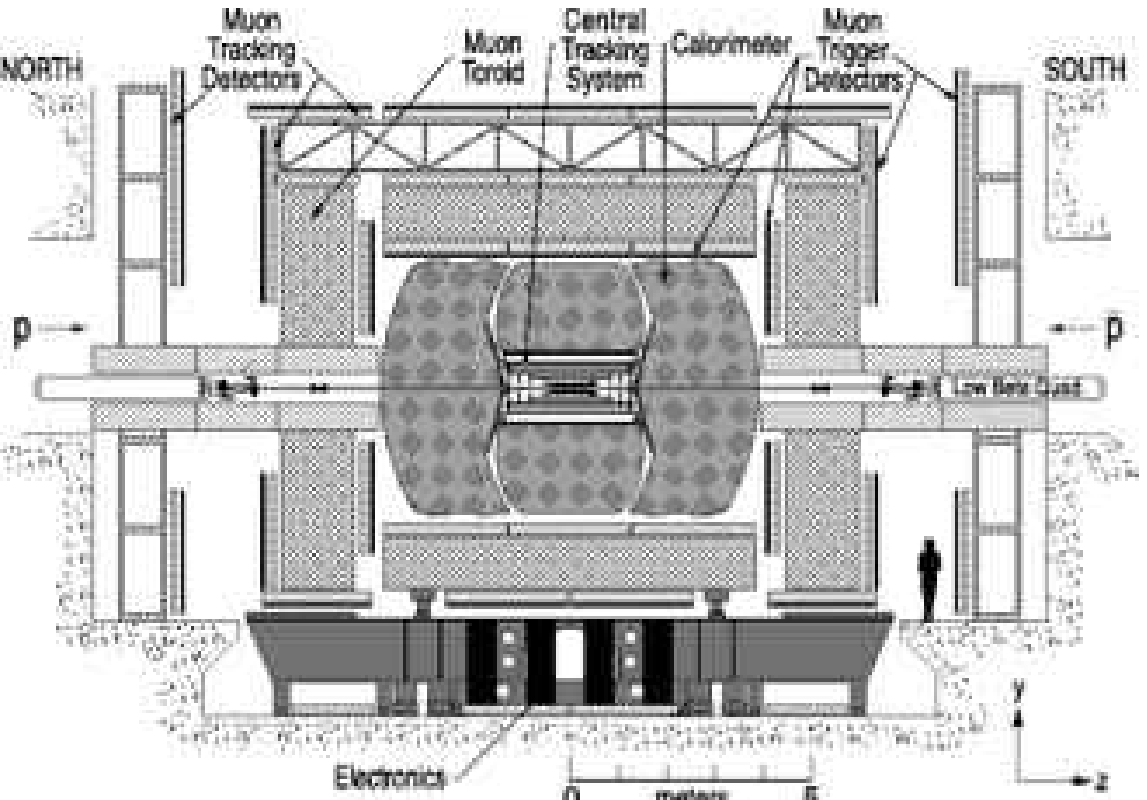,width=0.40\textwidth}
\epsfig{figure=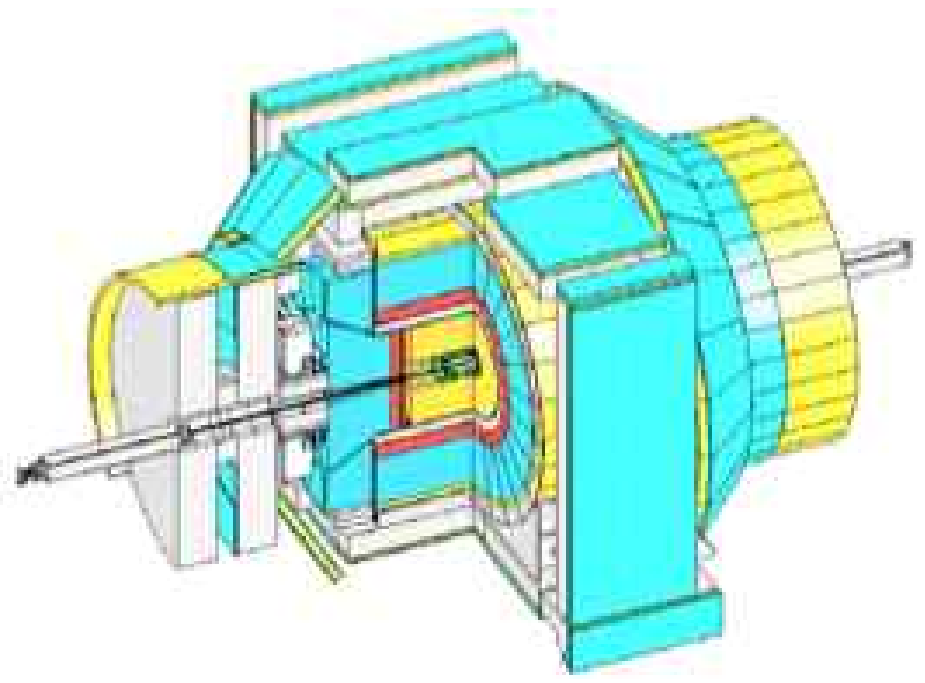,width=0.40\textwidth}
\end{center}
\vspace*{8pt}
\caption{The Run II \dzero\ detector (left) and CDF detector(right).  Each
detector has an inner tracking region surrounded by calorimeters and
outer muon spectrometers.
}
\label{fig:cdfd0det} 
\end{figure*}

Collisions occur in the two locations on the Tevatron Ring where 
the \dzero\ and CDF
experiments are located. Both detectors (see Fig. \ref{fig:cdfd0det}) 
have magnetic central tracking
regions which include silicon microstrip trackers for precise vertex
measurements.  Field strengths of 1.4 T and 2.0 T are achieved via
superconducting solenoids for CDF and \dzero, respectively. The 
higher value for \dzero\ 
partially compensates for a smaller tracking volume.  Primary charged
particle tracking is performed for CDF by a drift chamber which 
obtains 96 position measurements for each particle. Acceptance ranges to
$|\eta| < 2.0$ for silicon and $|\eta|<1.0$ for the drift chamber.  \dzero\ 
utilizes a scintillating fiber tracker composed of sixteen doublet 
fiber layers.  
In tandem with the silicon microstrip detector an acceptance of $|\eta| <
3.0$ is achieved.  All trackers replace Run I
equivalents\cite{d0run1,cdfrun1}, resulting in better $b$-tagging efficiency
for CDF\cite{cdfrun2}, and a new $b$-tagging and track momentum measurement
capability for \dzero \cite{d0run2}. 

Sampling calorimeters with large acceptance provide energy measurements for 
electrons, photons, and hadrons. \dzero\ utilizes uranium absorber bathed in
a liquid argon sampling medium to achieve near-compensating performance,
good energy resolution and low backgrounds for the surrounding muon
spectrometer\cite{d0run1}.  Coverage 
extends to $|\eta| < 4.2$. Scintillator-absorber layers provide these
measurements for CDF in the region $|\eta| < 3.6$\cite{cdfrun1}.  New
endplugs for CDF\cite{cdfrun2}, and electronics and hardware trigger upgrades
for \dzero \cite{d0run2} deliver improved performance for the high 
instantaneous luminosities achieved in Run II.

Drift tubes and scintillator arranged in layers outside of the calorimeter
are used for muon identification for both experiments\cite{d0run1,cdfrun1}.
In the \dzero\ case, these layers are distributed 
in front of and behind thick background-suppressing magnetized iron
toroids which provide a 1.8 T field for the full coverage to
$|\eta| < 2$. 
The CDF detector has coverage to $|\eta| < 1.0$.  The primary
enhancements from Run I are improved forward coverage for both
experiments\cite{cdfrun2,d0muon}.

Data are taken using a three-tiered trigger 
system composed of one hardware and two software levels.  Luminosity
is measured via gas Cerenkov or plastic scintillation counters mounted
in the forward regions of CDF and \dzero, respectively.  Analyses utilize
data taken from 2002 until Spring 2006.  By Spring of 2007, 
each experiment had accumulated about 2.5 fb$^{-1}$ of collision data.

{\subsection{Reconstruction }~\label{particleID}
\label{sec:perf}}
\input{apparatus/perform.tex}

%% file: apparatus/perform.tex
The effective identification of $e, \mu, \nu$, light quark jets, $b$ jets and $\tau$s is
crucial to the analyses reviewed in this paper.  One cannot
properly demonstrate the requisite understanding of signal and background
efficiency without a 
thorough understanding of the performance of the detector to measure
their energies or momenta, their efficiencies, and their rates to 
be incorrectly reconstructed.  This subsection describes the methods
of identification for these primary objects as well as the ways in which
their performance is determined and propagated into experimental models
of the processes being considered.

\paragraph{Electrons and Muons:}

Electrons are reconstructed as grouped or `clustered' electromagnetic 
calorimeter (EM) elements which
exhibit significant energy.  Cells are the fundamental elements used by
\dzero.  Three adjacent towers are used by CDF.  These clusters are associated 
with a well-reconstructed charged particle track.  A range of parameters
are used to identify good electrons, including the ratio of EM to hadronic
energy, isolation from more energy in a wider cone around the centroid, 
cluster energy divided by track momentum ($E/p$), and shower shape.  CDF
applies these cuts for `tight' electrons, and omits the isolation requirement
for `loose' electrons.  \dzero\  applies basic cuts on electron candidates and 
then assembles a likelihood discriminant which combines these and other 
calorimeter and tracking parameters.  
For \dzero, only isolation and EM-fraction cuts are applied for loose electrons.
Muons are identified from track signatures in the outer muon spectrometers of
each experiment, which are then matched to charged particle tracks.  In the
case of \dzero, isolation cuts are applied based on tracks and calorimeter energy.
Loose muons may pass a looser isolation selection.  

The efficiency,
momentum resolution and energy scale performance of $e$ and $\mu$ in the
detector are typically measured by examination of $Z\rightarrow ee,\mu\mu$ events.
By requiring one `tag' lepton to be well-identified, and employing other cuts to 
suppress non-$Z$ backgrounds, the other `probe' lepton can be measured in an unbiased way.
The reconstructed mass of the $Z$ boson is a direct reflection of the electron
energy scale.
Isolated leptons can be misidentified typically from jet activity.  
This can occur through fragmentation to
$\pi^0\rightarrow\gamma\gamma +$ track where the track is from a conversion
or charged hadron overlap.  Semileptonic decay of hadrons in the jet
can also rarely provide isolated leptons with significant $p_T$.  The probability
at which jets mimic leptons is termed the `fake rate'.  It 
is often extracted from multijet samples where $W$ and $Z/\gamma^*$ contributions
are suppressed.  For instance, \dzero\ uses such samples with leptons passing
very loose requirements.  It is important that signal trigger conditions on the 
faking lepton be reproduced.

These selections are applied to find leptons from the $W$ bosons in top quark decay.  
It is also 
important to be able to identify leptons from semileptonic
heavy flavor meson decay.  Such leptons can be used to flag a reconstructed
jet as having come from the fragmentation of a $b-$quark.  These
`soft leptons' do not have to satisfy the isolation requirements.
Estimation of the rate in data to tag $b-$jets with a soft lepton is usually
ascertained from a dijet sample where one jet is tagged and a non-isolated
lepton is sought in the other.  The rate for typical
jets, primarily light flavor jets, to produce a `mistag' is measured
from a multijet sample, such as $\gamma+jet$ events.

\paragraph{Jets, $\tau$s and \met:}

An iterative, fixed-cone algorithm\cite{ConeAlg} is applied to
electromagnetic and hadronic calorimeter elements to reconstruct jets.  
Because of the high jet multiplicity in top quark events, cone sizes used are
generally small: for \dzero\  (CDF) most analyses use $\Delta R =
\sqrt{\Delta\phi^2 + \Delta\eta^2} = 0.5 (0.4)$. 
Jet energies are corrected for various effects such as non-linearities in
single particle response, unregistered energy at detector element boundaries
and cracks, particle showers, and underlying event and multiple
interactions. The jet energy scale is ultimately measured using $\gamma+$
jets events where the photon in the electromagnetic calorimeter section
is calibrated via the $Z\rightarrow ee$ resonance.  The general approach
compares the momentum of the jet in the transverse plane
(`transverse momentum', or `$p_T$') to the reference provided by the photon 
$p_T$, although in practice several effects mean that a direct
$p_T$ comparison is not optimal\cite{D0runIJES}. 
A subset of jets can be further identified as
originating from the fragmentation of heavy quarks, particularly $b-$quarks,
if they encompass an associated, non-isolated low $p_T$ electron or muon from
semileptonic decay or if they match a secondary vertex.
Hadronic $\tau$ lepton decays are also identified as
jets.  In this case selections are applied based on shower profile in the
calorimeter, as well as requirements of one or three tracks in the jet cone.

Neutrinos escape the detectors undetected. Their presence can be inferred 
from a significant imbalance in the $p_T$ of all observed
particles in an event.  Recall that, since most final state particles are
essentially massless compared to the magnitudes of their momenta, 
$p \sim E$.  So the event-wide $p_T$ imbalance is generally 
termed missing $E_T$, or \met.  The \met\
is calculated as the negative of the vector sum of the $E_T$'s of individual
calorimeter elements.  This quantity is then corrected for the $p_T$'s of
any reconstructed $\mu$'s in the event, as well as the energy scale 
corrections for jets and electrons.

\paragraph{Vertices:}

The \ppbar\ collision point is called the event primary vertex. 
It is reconstructed with well-measured tracks that are consistent with
origination from the same point.  In many cases, more than one primary
vertex are reconstructed.  These may be due to extra \ppbar\
interactions resulting from the high operating instantaneous luminosity.
There will generally be only one vertex corresponding to the hard scatter
giving rise to a top quark.

The secondaries from the decay of long-lived $B$ hadrons produce charged
particle tracks that originate from a vertex that is displaced from the
primary vertex. Both CDF and \dzero\  employ algorithms that identify
(``tag'') jets as
originating from $b-$quarks by reconstructing such secondary vertices
within a particular jet (`secondary vertex tag'). 
Such secondary vertices are reconstructed from at
least two well-reconstructed tracks with $p_T>1.0$ GeV that match a jet.  
These tracks must have an impact parameter in the transverse $(x,y)$ plane
with a significance of $> 3 \sigma$ with respect to the primary vertex for CDF.
\dzero\  applies a $3.5 \sigma$ cut.  CDF adds a second, three-track vertex
with relaxed track $p_T$ and impact parameter cuts. The vertex reconstructed
from these tracks must be significantly displaced from the primary vertex in
the transverse plane, given the known tracking resolutions and it must be on
the same side of the primary vertex as the jet, to tag a jet as a $b$-jet.

Other tagging algorithms exist.  For instance, an `impact parameter tag' 
algorithm has been used by \dzero\ which 
is based on counting the number of tracks with  impact 
parameter significance above a certain value.  Both experiments
have also implemented `jet probability taggers'.  These use
the knowledge of track resolutions to calculate a probability that tracks in
jets originate from an event primary vertex and compute a jet probability 
based on the combined track probabilities.  The jets are tagged if
the jet probability has a low value, indicating an inconsistency 
with the hypothesis for that jet to originate from the primary vertex. 
\dzero\  and CDF also use neural networks to combine the various parameters
associated with a $b$-tag such as impact parameter, momentum and invariant
mass of all tracks associated with the vertex, vertex displacement, etc. into
a more powerful discriminant.

It is important to understand the efficiency for $b-$jets to be
tagged, and light quark and gluon jets to be mistagged.  Jets
from $c-$quarks and their relevant tagging rates must also be accounted for.
The efficiency for a particular $b-$tagging algorithm is generally measured
in heavy-flavor enriched inclusive jet samples in data.  Such an enrichment
can be achieved by lepton-tagging a jet for which the $b-$tag efficiency is
to be measured.   Monte Carlos are used to obtain an estimate of the residual
$c-$quark content and other selection  biases.
The mistag rate to falsely identify light quark or gluon jets
as $b-$jets is estimated from samples of inclusive jet events.  
The primary backgrounds produce displaced vertices that are
equally probable to be `behind' the primary vertex
(negative tags) from the point of view of the jet they correspond to, 
and `in front of' the primary vertex (positive 
tags).  However, heavy flavor will produce an asymmetric distribution
with more positive vertex displacements.  The negative tag rate 
is a useful first approximation to the mistag rate.  It must be 
corrected for heavy flavor in the jet samples, as well as for
long-lived particles in actual light
flavor jets that are not reflected in the negative tag rate.  
Final tag and mistag rates are parametrized as functions of $p_T$ 
and $\eta$.

\paragraph{Performance and parametrization of simulation:}

The final element tying the simulation of signal and background samples
to the data involves the simulation
of the detector.  For CDF and \dzero\ , {\sc Geant3}\cite{geant3} is 
used with a full detector simulation.  At this stage minimum bias Monte Carlo
events were added to the  hard scatter such that the Poisson distribution
mean matches the  average instantaneous luminosity of the data sample.

In general, the simulation does not replicate
exactly the detailed performance parameters of the detector.  The parameters
of interest are: $e$ and $\mu$ identification efficiency, momentum resolution
and energy/momentum scale; jet efficiency, energy resolution and energy scale;
\met\ resolution; and $b-$tag algorithm efficiency.  As a result, it is
necessary to quantify each element of performance in the simulation and in
the actual detector using isolatable control samples.  For leptons, these are
generally the $Z\rightarrow ee,\mu\mu$ samples.  For jets, an important
sample is the $\gamma +$jet sample.  $b-$tagging requires simulated
inclusive jet samples.  The simulation is then corrected so that
the performance for real data is replicated.  A few performance measures must
be measured only for use in data.  These are generally the lepton fake rates
and the $b-$tagging mistag rates.

%% file: ttbar_csec/ttcsec.tex
Measurement of the top quark pair (\ttbar) production cross section 
permits a unique
test of perturbative QCD predictions.  Unveiling discrepancies between
measured rates and expectations for various final state channels provides
a potential indication of new physics.  The selection of clean, well-understood
analysis channels facilitates the measurement of the top quark mass and
other properties.  In this section, we discuss the primary channels 
that have been employed for the cross section measurement.  These consider
variants of the dilepton, $\ell+$jets and all-jets channels.
Analyses of $\tau$ channels, while
they have been pursued, have not yet provided significant constraints on 
$\sigma_{t\bar{t}}$ and are omitted from this section.

Two primary aspects of these analyses are general in their impact.  The
event selection that is used for all top quark measurements has been
developed based on extensive studies of the expected properties of
\ttbar\ events as well as those from known backgrounds.  The lepton,
jet and \met\ cuts are derived from these studies, as are more sophisticated
parameters which consider angular correlations, scalar $E_T$ sums,
or event topology.  Given a particular event selection, the behavior
of signal and background events in terms of these variables must be 
understood well enough to keep uncertainties in the measurements under control.
It is particularly important that the background models be validated
wherever possible to optimize the analysis sensitivity.  
Achievement of an effective
selection whose performance can be confirmed not only ensures an optimal
cross section measurement, but it also serves as the foundation for 
measurements of the properties of the top quark.  We 
discuss these two issues in this section and then review the individual
measurements from \dzero\ and CDF.

{\subsection{Selection variables}
\label{sec:ttSelVars}}

Events containing a \ttbar\ pair have several distinctive characteristics
that are not reflected by the anticipated backgrounds.  In general, the total scale
of energy in the event, particular in jets, is quite large.  Decay products
are roughly isotropically distributed in top quark events.  These two
qualities follow largely from the high mass of the \ttbar\ pair.  A third
category comes from the observation that some backgrounds come from 
mismeasurement and this reflects itself in angular correlation between objects.
The discriminating variables used fall into the following categories:

\paragraph{Energy scale:} Due to the large mass of the top quark, the 
characteristic energy scale of the \ttbar\ event is significantly larger 
than that of the average QCD background event. This means that \ttbar\ events
generally have more energetic jets and larger multijet invariant masses.
This is especially true of the leading two jets from the
$b$ quarks which come directly from top quark decay.  This total energy
scale is most often expressed as a scalar sum of object $E_T (p_T)$s,
termed $H_T$.  Sometimes the 
discriminating power can be increased by also adding observed lepton
$p_T$s and perhaps also the \met.  Invariant mass parameters, such as for 
the leading two jets ($M_{min}^{1,2}$), the second leading two jets 
($M_{min}^{3,4}$) or all of the objects ($\ell$, jets, \met), also 
indicate the production of a high mass state.  

\paragraph{Event topology:} In general, backgrounds with jets originating from
gluon radiation provide steeply falling $p_T$ spectra for jets.
The manner in which color flows in such events tends to produce particles
that congregate in a planar geometry.  Additionally, QCD events usually 
have a more back-to-back spectrum because of their hard scatter origin.
The jets from \ttbar\ decay, on the other hand, are 
almost isotropically distributed.
These differences can be quantified using event-shape parameters, such 
as sphericity, $\mathcal{S}$, and aplanarity, $\mathcal{A}$, 
calculated from the 
normalized momentum tensor\cite{sphericity}. Top quark events
tend to have higher $\mathcal{A}$ and $\mathcal{S}$ than the background events.

\paragraph{Rapidity spectrum:} \ttbar\ events on average
are expected to have more jets of higher energy and with less boost in the 
beam direction, resulting in events with many central jets. The 
QCD background tends to have jets that are more forward-backward
in rapidity. One variable
considered is the centrality ($\mathcal{C}$), defined as 
$\mathcal{C}=\frac{\sum E_T}{\sqrt{\hat {s}}}$, where $\sqrt{\hat {s}}$ 
is the total energy in the event (or it may represent the scalar
sum of the momenta (energies) of jets $\Sigma E^{jets}$). Another variable is
$<\eta^2>$, the $p_T$-weighted mean square of the $\eta$ of the jets. 
The QCD multijet background is expected to have a broader distribution
in $<\eta^2>$ than the \ttbar\ signal. 

\paragraph{Angular Correlations:} Several of the backgrounds 
arise because the $E_T (p_T)$ of one or more objects in candidate 
events have been significantly mismeasured.  Often, this
mismeasurement will create the appearence of \met\ either in the
azimuthal direction of the offending object, or opposite it.  Jet
energy and $\mu$ momentum resolutions are most often the cause.  As a
result, cuts are placed on the angular correlation of objects:
$\Delta\phi(x,y_i)$ is the azimuthal angle between the transverse momentum
vector of objects $x$ and $y_i$.  Here, the subscript $i$ denotes
leading, second leading, etc. of object $y$.  Common uses are
$\Delta\phi(\met,\mu_1)$ or $\Delta\phi(\met,j)$ where $\mu_1$ is the leading
muon or $j$ is a jet.

\subsection{Background extraction from data}

Production of top quark pairs presents a unique set of final states.  
However, the high mass of the top
quark also means that the production cross section is quite small.  
Use of the variables discussed in Section~\ref{sec:ttSelVars} entail
a heavy reliance on measurements of jets and \met.  These are complex
observables defined by sophisticated algorithms and subtle physics
at a hadron collider.  Simulation of background processes must be
complemented by study of those processes in the data.  Often, the attempt
to validate the background model will result in corrections to the 
simulation that will improve agreement.  In other cases, the background
will be taken from the data alone or in combination with the simulation.
The simulation can still provide a valuable cross-check of the model, since 
it represents the best attempt to estimate the background from first
principles.  This in turn gives confidence in the signal model which can 
only be obtained from Monte Carlo.  
Common schemes to be elaborated in the following sections are:

\paragraph{Modeling Jet Production:}
The physics behind jets involves soft processes and 
the algorithms used to reconstruct jets have complex behavior.  For
all analyses, it is therefore important to demonstrate an understanding of
the background (and signal) at all jet multiplicities.  In 
low jet multiplicities, background will dominate.  Agreement 
between models and data provides a confirmation of the background
model in the signal rich bins.  Typically this will be reflected as
an estimate of the sum of signal and background (`$S+B$') vs. 
jet multiplicity, with the observed data in comparison.  
Another approach is to
consider the agreement of the modeled and observed jet $p_T$s.  
This has occasionally been used, for instance, to determine the 
background level from a fit to the leading jet $p_T$.

\paragraph{Lepton Backgrounds:}
Jets also impact analyses because they can mimic charged
leptons or evidence of $\nu$s.  In the former case, it is 
difficult to precisely model those jets that shower in the detector
to look like electrons, or that produce a lepton from semileptonic
decay which can pass lepton isolation and $p_T$ cuts.  The rate
for jets to fake leptons is obtained in data as mentioned in
Section~\ref{sec:perf}.  Two things are crucial in the use of these
rates.  First, samples must be identified in data 
which have negligible signal contribution.  Second, it is important
that the faking, which is a result of hadronization or decay,
be uncorrelated with the kinematics of the event.  
Typically, samples are selected with all kinematic selections and one
lepton identification relaxed.  In the case of \met, the rate
to satisfy certain \met\ cuts in a $\nu$-less process (termed
the `\met\ fake rate') is 
studied in $Z\rightarrow \ell\ell (\ell=e,\mu)$.

\paragraph{Jet Flavor Modeling:}
Identification of $b-$jets from evidence of 
displaced secondary vertices or non-isolated soft leptons plays 
an important role in selecting a sample enriched with top quark
events.  In all $b-$tag analyses, the rate for light quark or
gluon jets to accidentally satisfy the soft lepton or displaced
vertex criteria is called the `mistag rate' as described in Section
~\ref{sec:perf}.  The application of these mistag rates to
an untagged background sample can be somewhat more complicated
than for the leptons.  The reason is that jets arising from the
hadronization of $c-$quarks can also harbor high impact parameter
tracks.  Also, all backgrounds contain some relative fraction
of light quark, $c$ and $b$ jets.  In
$W+$jets production, most of the
associated jets arise from light quarks or gluons.  Occasionally,
gluon splitting will generate a $\bbbar$ or $\ccbar$ pair which
give rise to heavy flavor jets.  More rarely, sea 
$s-$quarks radiate a $W$ boson
giving a single $c-$quark and subsequent jet. 
It is common in this instance
to use Monte Carlo to provide some estimate of the relative
flavor content, and then to apply the mistag rates or tagging
efficiencies to the relevant background fractions.

\subsection{Dilepton final states~\label{sec:llcsec}}
\input{ttbar_csec/llcsec.tex}

\subsection{Single lepton channels~\label{sec:ljcsec}}
\input{ttbar_csec/ljetscsec.tex}

{\subsection{All-jets channel}
\label{sec:csec6j}}
\input{ttbar_csec/all-jets.tex}

\subsection{Combined cross section}

The most sensitive of each type of cross section measurement 
discussed in this section are given 
in Table~\ref{tab:csecs}.  For collisions with $\sqrt{s}=1.8$ TeV, the quoted
measurements are final results.  Work is still progressing
on the $\sqrt{s}=1.96$ TeV measurements.  
The input $m_t$ is 175 (178) GeV for many of 
the \dzero\ (CDF) measurements. 
Generally, the measured cross section will be higher if $m_t$ is
lower because the efficiency to select top decreases somewhat with $m_t$.
For the dilepton and $\ell+$jets channels, the slope of measured
cross section with difference in mass from the input value ($\Delta\sigma/\Delta m$)
is $\sim 0.1$.  The all-jets channels exhibit significantly more variation.
The range over which the $\Delta\sigma$ is
quoted at least includes $170-180$ GeV, and several measurements test from
160 to 190 GeV.
Currently, the total uncertainty is reaching
the 15\% level.  This should still be reduced as higher statistics are accumulated.
More events will also translate into somewhat lower systematic uncertainties, largely
because of the increase in samples dedicated to parametrizing detector performance
such as the jet energy scale.

\begin{table}[ht]
\begin{center}
\tbl{Measured cross sections for Tevatron experiments, including the function
for different top quark masses.  The channel, experiment, collider energy
and integrated luminosity are listed in the first four columns.  The measured
cross section is in column 5. Preliminary results are indicated by * in column 1.}
{\begin{tabular}{lcccc}\toprule
Channel & Expt. & $\sqrt{s}$ (TeV) & Lum. (pb$^{-1}$) & $\sigma_{\ttbar}$ (pb)  \\ 
\colrule
all-jets/$\mu-$tag \protect\cite{d0r1alljetscsec} & \dzero\ & 1.8 & 110 & $7.1\pm2.8(stat)\pm1.5(sys)$ \\
all-jets/$b-$tag \protect\cite{cdfr1alljetscsec} & CDF & 1.8 & 110 & $10.1\pm1.9(stat)^{+4.1}_{-3.1}(sys)$ \\
$\ell\ell,\ell+$jets, all-jets \protect\cite{d0r1ttcsec} & \dzero\ & 1.8 & 125 & $5.69\pm 1.21(stat)\pm 1.04$(sys) \\
$\ell\ell, \ell+$jets, all-jets \protect\cite{cdfr1ljsltcsec} & CDF & 1.8 & 110 &  $6.5^{+1.7}_{-1.4}$(stat+sys) \\
\hline
$\ell\ell, \ell+$track \protect\cite{cdfr2llcsec} & CDF & 1.96 & 197 & $7.0^{+2.4}_{-2.1}(stat)^{+1.6}_{-1.1} (sys) \pm 0.6$ (lum) \\
$\ell\ell, \ell+$track \protect\cite{d0425llcsec} & \dzero\ & 1.96 & 425  & $7.4\pm1.4(stat)\pm 1.0 (sys)$ \\
$\ell\ell$* \protect\cite{cdf1fbllcsec}           & CDF     & 1.96 & 1200 & $6.2\pm 1.1(stat) \pm 0.7(sys) \pm 0.4(lum)$ \\
$\ell+$track* \protect\cite{cdf1fbllcsec}         & CDF     & 1.96 & 1000 & $8.3\pm 1.3(stat) \pm 0.7(sys)\pm 0.5(lum)$ \\
$\ell\ell, \ell+$track* \protect\cite{d01fbllcsec} & \dzero\ & 1.96 & 1000 & $6.2^{+0.9}_{-0.8}(stat)^{+0.8}_{-0.7}(sys)\pm 0.4(lum)$ \\
\hline
$\ell+$jets/topo \protect\cite{cdfljtopo194} & CDF & 1.96 & 194 & $6.6\pm 1.1(stat) \pm 1.5 (sys)$  \\
$\ell+$jets/topo \protect\cite{d0lj425topo} & \dzero\ & 1.96 & 425 & $6.4^{+1.3}_{-1.2}(stat)\pm 0.7(sys) \pm 0.4(lum)$ \\
$\ell+$jets/topo* \protect\cite{d0lj900topo} & \dzero\ & 1.96 & 900 & $6.3^{+0.9}_{-0.8}(stat)\pm 0.7(sys) \pm 0.4(lum)$ \\
\hline
$\ell+$jets/$\mu-$tag \protect\cite{cdfr2ljmutagcsec} & CDF & 1.96 & 194 & $5.3\pm 3.3(stat) ^{+1.3}_{-1.0} (sys)$ \\
$\ell+$jets/$\mu-$tag* \protect\cite{cdf760mutagcsec} & CDF & 1.96 & 760 & $7.8\pm 1.7(stat) ^{+1.1}_{-1.0} (sys)$ \\
$\ell+$jets/$\mu$-tag* \protect\cite{d0r2ljsmt} & \dzero\ & 1.96 & 425 & $7.3^{+2.0}_{-1.8}(stat+sys)\pm 0.4 (lum)$ \\
\hline
$\ell+$jets/$b-$tag-kin \protect\cite{cdfr2btagkinfitcsec} & CDF     & 1.96 & 162 & $6.0\pm 1.6(stat.)\pm 1.2(sys)$ \\
$\ell+$jets/jprob \protect\cite{cdfr2jetprobcsec} & CDF     & 1.96 & 318 & $8.9\pm 1.0(stat)^{+1.1}_{-1.0}(sys)$  \\
$\ell+$jets/$b-$tag \protect\cite{cdfr2btag318csec} & CDF     & 1.96 & 318 & $8.7\pm 0.9(stat)^{+1.1}_{-0.9}(sys)$ \\
$\ell+$jets/$b$-tag \protect\cite{ljbtag425} & \dzero\ & 1.96 & 426 & $6.6\pm 0.9(\rm stat+sys)\pm 0.4(lum)$ \\
$\ell+$jets/$b-$tag* \protect\cite{cdf1fbbtagcsec} & CDF     & 1.96 & 1100 & $8.2\pm 0.5(stat)\pm 0.8(sys)\pm 0.5(lum)$ \\
$\ell+$jets/$b$-tag* \protect\cite{d0900ljbtag} & \dzero\ & 1.96 & 900 & $8.3^{+0.6}_{-0.5}(stat)^{+0.9}_{-1.0}(sys)\pm 0.5 (lum)$ \\
\hline
all-jets/$b-$tag \protect\cite{cdfr2alljetscsec} & CDF & 1.96 & 311 & $7.5 \pm 2.1(stat)^{+3.3}_{-2.2}(sys)^{+0.5}_{-0.4} (lum)$  \\
all-jets/$b-$tag \protect\cite{d0r2alljetscsec} & \dzero\ & 1.96 & 405 & $4.5^{+2.0}_{-1.9} (stat)^{+1.4}_{-1.1}(sys) \pm 0.3(lum)$ \\ 
\botrule
\end{tabular}
\label{tab:csecs}}
\end{center}
\end{table}

%% file: ttbar_csec/llcsec.tex
The signature of two isolated, high $p_T$ leptons 
and \met\ in association with two high $E_T$ jets is
a striking consequence of the \ttbar\ quark decay chain 
where both $W$ bosons decay leptonically.  The only processes 
exhibiting the leptonic signature are
diboson production, particularly $WW$ where both
$W$'s decay leptonically as in the \ttbar\ case, or $Z\rightarrow\tau\tau$
production where both $\tau$'s decay leptonically.  Diboson
cross sections are of the same order as that expected for \ttbar\ production.  
$Z\rightarrow\tau\tau$ production is much greater but the dilepton
branching fraction is small and the charged lepton and $\nu$ $p_T$'s
are soft.  In both cases, additional jets are only produced at
a rate of approximately $\alpha_s$ each and so jets become a primary 
discriminator of these backgrounds. 

Beyond these processes, it is possible to mimic the top quark signature 
through instrumental mismeasurement.  The chief background arising
this way comes from $Z/\gamma^*\rightarrow ee,\mu\mu +$ jets production.
Here the reconstructed \met\ is the result of tails in the 
jet or lepton energy or momentum resolutions.  An important background
also comes from $W\rightarrow e,\mu +$ jets production where one of the jets
fakes an isolated lepton.  QCD multijet production is a background
when both leptonic and \met\ mismeasurement occur.

In order to extract a measurement of the top quark production cross section
in the dilepton channels,
different techniques have been developed.  One of the most widely used
has been a selection where loose and tight versions
of the lepton identification are employed.  For `explicit' dileptons,
these distinctions
still refer to selections requiring convincing signals in the subdetectors
responsible for lepton identification, particularly the 
electromagnetic layers of the calorimeter and the outer muon spectrometers.  
However, these detectors 
have significant gaps and holes in their coverage, and lepton identification
and isolation requirements are inefficient.  The resulting loss
in acceptance seriously reduces the signal event yield in the dilepton
channel.  In the most extreme `implicit' dilepton strategies,
identification requirements are relaxed to mere requirements of an 
isolated, high $p_T$ track or even to no evidence of a
second lepton at all.  Such an approach can accept dilepton events
where one of the leptons is a $\tau$ which decays hadronically.
Event-wide kinematic variables and tagging of jets from $b-$quarks are 
valuable ways to isolate the dilepton final state, particularly when
the lepton requirements are relaxed and backgrounds are higher.

\subsubsection{Explicit channels}

Explicit dilepton analyses are defined as those in which 
both leptons are fully reconstructed as either $e$ or $\mu$.
$W\rightarrow\tau\rightarrow e,\mu$ decay modes cannot be distinguished from,
and so are grouped with,
the direct $W\rightarrow e$ or $W\rightarrow\mu$ channels.  Three channels
have been examined by both Tevatron experiments: $ee, e\mu$ and 
$\mu\mu$.

The \dzero\ experiment has used explicit channels to produce a cross section
measurement at $\sqrt{s} = 1.96$ TeV with 243 pb$^{-1}$ of data~\cite{d0r2llcsec}.
Events were triggered with two leptons in the
first level hardware trigger and one or two leptons in the higher
level software triggers.  In general, the Run II measurements derive
many techniques from the Run I analyses\cite{d0r1ttcsec} (see final combined
Run I cross section in Table~\ref{tab:csecs}).  
Although the differences are less than in Run I, momentum resolutions are
significantly different for $e$'s and $\mu$'s.  This means event selections 
across the three channels are somewhat varied.  Also, because $b$-jets
carry away a large momentum directly from top quark decay, 
\dzero\ uses relatively soft lepton $p_T$ cuts and stiff jet $E_T$ cuts.  
The primary background which arises
after these cuts is mainly $Z/\gamma*$+jets production.  A substantial
cut on \met\ reduces this background considerably, but it is still
the primary background in the like-flavored channels.  Suppression of
this background involves rejecting
events with dielectron or dimuon invariant mass ($M_{ee}$ or $M_{\mu\mu}$)
consistent with that measured
for the $Z$ boson $M_Z$\cite{LEPZmass}.  In the $ee$ channel, the excellent
electron energy resolution allows a narrow window to be drawn around
the reconstructed $M_{ee}$.  The cut on \met\ is elevated for events
within this window.  For the $\mu\mu$ channel, the muon momentum resolution
from the tracker degrades at higher $p_T$ and
has significant non-gaussian tails.  Instead of a window cut, a kinematic 
fit is employed which quantifies a $\chi^2$ test with respect to $M_Z$, 
$\chi^2_{M_Z}$.  
In Run I, this test used the muon and calorimeter 
measurements and knowledge of their resolutions.
The improved Run II detector has permitted a simpler test based on the muon
momentum measurements alone.

The selections were optimized to produce the 
smallest fractional statistical uncertainty on the cross section, taken to be
proportional to $S/\sqrt{S+B}$.  These optimizations used several variables, 
including event shape variables, dilepton mass and \met\ cuts.
The most important of these variables have been
the jet $E_T$ and $H_T$.  The final selection criteria are given in
Table~\ref{tab:llcscuts}.

Instrumental backgrounds consist of 
$Z\rightarrow ee,\mu\mu$ background events 
where the \met\ may be mismeasured, and $W+$jets and multijet events where
the jets fake leptons.  These are generally determined from data.  
Lepton backgrounds are estimated by folding fake rates into a signal-like
sample with one tight lepton.  For the 
$Z\rightarrow ee$ background, the rate for \met\ to pass the \dzero\ cuts
is assessed in a sample of $\gamma+2$ jets events that are 
kinematically similar to the $Z+2$ jets sample.  Agreement in the \met\
spectrum is observed for the $Z+$jets data sample confined to the region
80 GeV$<M_{ee}<100$ GeV and a high statistics 
$\gamma+$jets sample.   The $\gamma+2$ jets behavior also agrees well with 
a high statistics $Z(\rightarrow ee) +2$ jets Monte Carlo sample where 
all known effects on the \met\ resolution in data have been accounted for.
The ratio of events to pass divided by those that fail the \met\ cut
is extracted from the data and multiplied by the number of $ee$ events
failing the \met\ cut in data.  The $Z\rightarrow\mu\mu +$ jets background
is estimated by extracting the \met\ fake rate from fully
simulated {\sc Alpgen} Monte Carlo
and multiplying that by the number of observed $\mu\mu$ events in data.

Physics backgrounds are those in which the full signature from the
top quark pair is mimicked at the particle level.  These are
$Z\rightarrow\tau\tau\rightarrow ll+2\nu$s, 
$WW\rightarrow 2\ell + 2\nu$ and 
$WZ\rightarrow 3\ell + \nu$.  The former is the
dominant background for the $e\mu$ channel.  These backgrounds were
estimated from Monte Carlo corrected for efficiencies measured in data.

\begin{figure*}[!h!tbp]
\begin{center}
\epsfig{figure=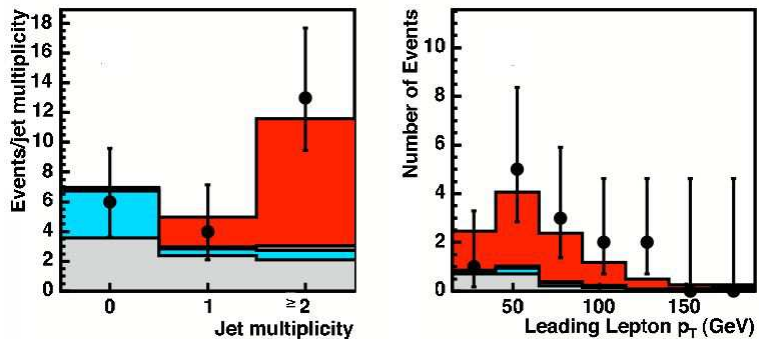,width=0.80\textwidth}
\end{center}
\vspace*{8pt}
\caption{Kinematic characteristics of dilepton events
in 243 pb$^{-1}$ of \dzero\ data~\protect\cite{d0r2llcsec}.  Event jet multiplicity (left) and
leading lepton $p_T$ (right) are shown.  Data are indicated by points and error
bars, and the sums of signal and background distributions are overlaid as solid histograms.}
\label{fig:d0ll230csec} 
\end{figure*}

The cross section for $\sqrt{s}=1.96$ TeV 
is taken by maximizing the likelihood among the three
channels that, given a value of the assumed cross section, the estimated 
backgrounds and top quark 
efficiencies can produce the number of observed events.  The cross 
section is $8.6^{+3.2}_{-2.7}(stat)\pm 1.1(sys)\pm 0.6(lum)$ pb.  
Fig. \ref{fig:d0ll230csec} indicates the jet multiplicity of the dilepton
sample with all other selections applied. The excess of events over
background in the 2-jet bin is consistent with the expected top quark 
contribution.
Fig. \ref{fig:d0ll230csec} also provides the $p_T$ distribution for the 
leading lepton
in the final event sample. Kinematic distributions of dilepton events
have been of some interest. The CDF collaboration has observed
some unexpected properties of their dilepton samples in Run I and
Run II\cite{llkin}. The \dzero\ distributions are consistent with standard
model effects. 
The dominant uncertainties for the \dzero\ measurement are roughly equal
from lepton identification and triggering, and from jet reconstruction
and energy scale.  

\begin{figure*}[!h!tbp]
\begin{center}
\epsfig{figure=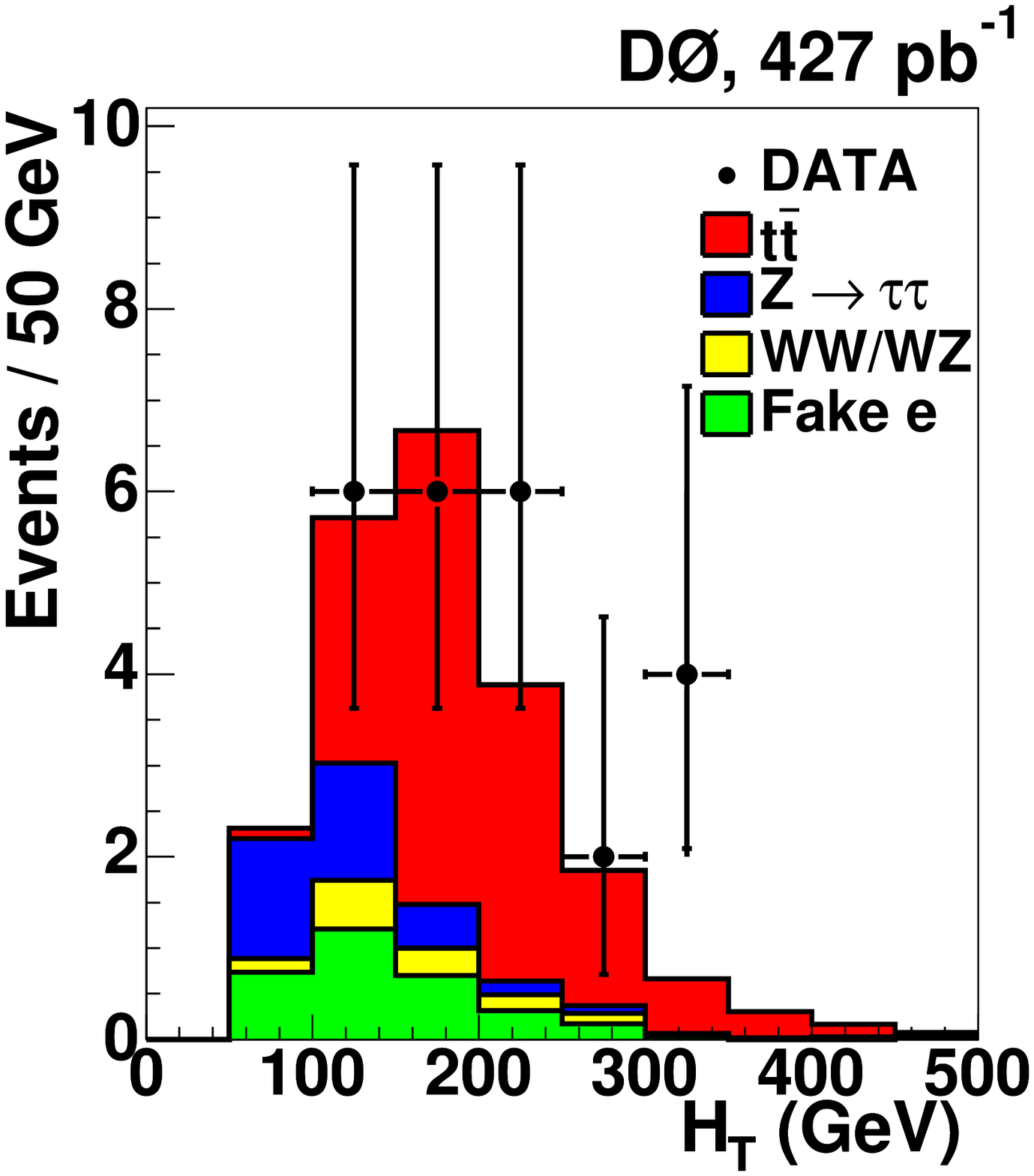,width=0.40\textwidth}
\epsfig{figure=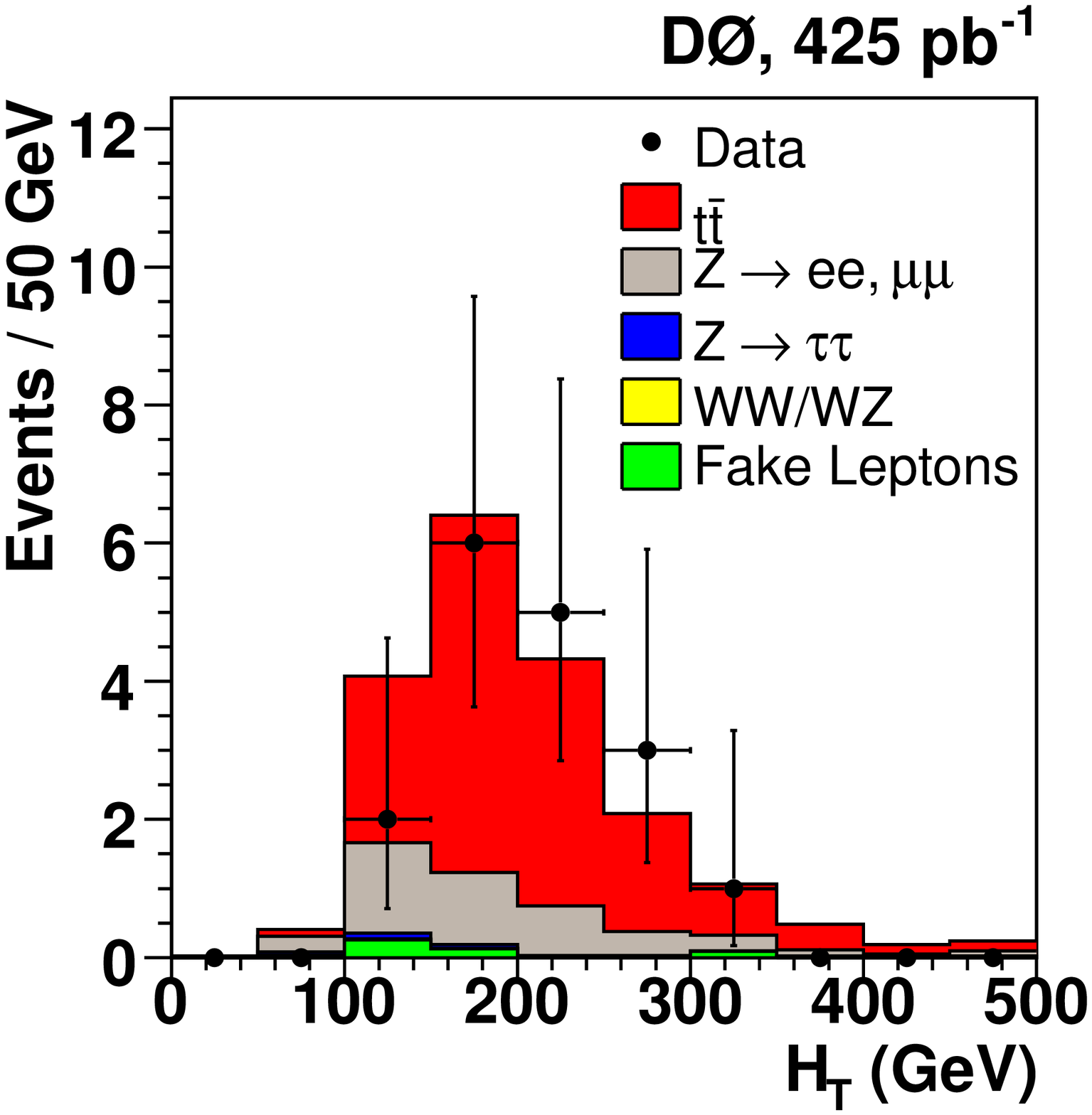,width=0.45\textwidth}
\end{center}
\vspace*{8pt}
\caption{Kinematic characteristics of dilepton and $\ell+$track events
in \dzero\ analysis of 425 pb$^{-1}$ sample~\protect\cite{d0425llcsec}.  
Event $H_T$ distributions are shown for the $e\mu$ channel (left) and
the combined $\ell+$track channel (right).  Data are indicated by points
with error bars, the estimated signal and background contributions are shown
via overlaid histograms.}
\label{fig:d0ll425csec} 
\end{figure*}

This basic analysis has been performed in 425 pb$^{-1}$
of data with somewhat modified selection on the $e\mu$ and $\mu\mu$ channels.
In the latter case, substantial improvement was made in the 
rejection of $Z$ bosons.  Figure \ref{fig:d0ll425csec} shows the
$H_T$ distribution for the $e\mu$ data, and for \ttbar\ and background
expectations.  Good agreement is observed and the statistical sample
is starting to permit a fairly distinct
\ttbar\ component to become more evident.
In combination with the $\ell+$track channels
described in the next section, a combined cross section measurement
of $7.4\pm 1.4(stat)\pm 1.0(sys)$ pb was obtained~\cite{d0425llcsec}.  A 
preliminary result in 1 fb$^{-1}$ from the dilepton and $\ell+$track
channels has
given $6.2\pm 0.9 (stat)^{+0.8}_{-0.7}(sys)\pm 0.4 (lum)$ pb 
\cite{d01fbllcsec}.  These results are given in Table \ref{tab:csecs}.

For Run II, CDF has analyzed an explicit dilepton 
data sample from $197 \pm 12$ pb$^{-1}$\cite{cdfr2llcsec}.
Triggering involved one high $p_T$ lepton, $e$ or $\mu$.
In general, the CDF analyses are characterized by a common approach across
channels which aids in background understanding and justification of
event selection.  The selection in Run II is similar to that used in Run I
\cite{cdfr1llcsec} (see combined measurement in Table \ref{tab:csecs}).
Top quark events were selected by requiring
two leptons with $p_T >$ 20 GeV, plus
two energetic jets with $p_T >$ 15 GeV and large \met ($>$25 GeV).  
One of the two leptons must satisfy a tight selection, while the other 
may pass looser cuts.
For both the $ee$ and $\mu\mu$ channels, 
events are rejected if the dilepton invariant mass, $M_{ll}$, 
is near the $Z$ boson mass, $M_Z$, to suppress Z backgrounds.  The 
basic kinematic selection is given in Table~\ref{tab:llcscuts}.
An opposite sign requirement for the
two leptons is applied.  
Events were removed when $\Delta\phi(\met,j)$ is small and when
$\Delta\phi(\met,\ell) < 20\deg$.  CDF did not require that explicit
and implicit channels have orthogonal event selection.
 
\begin{figure*}[!h!tbp]
\begin{center}
\epsfig{figure=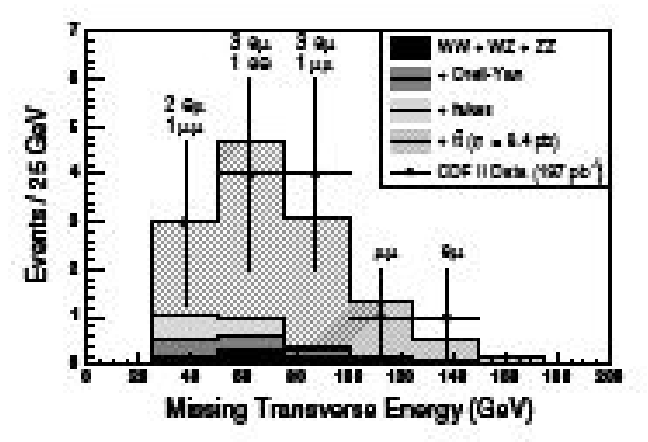,width=0.47\textwidth}
\epsfig{figure=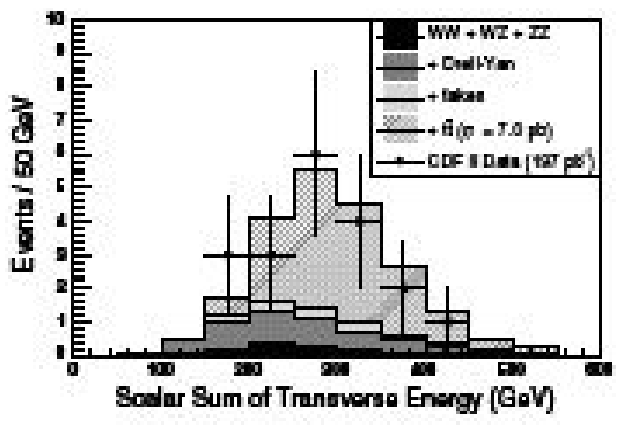,width=0.45\textwidth}
\end{center}
\vspace*{8pt}
\caption{Kinematic characteristics of dilepton events
for CDF analysis in 197 pb$^{-1}$.  The \met\ for explicit dilepton 
events (left) and the event $H_T$ for $\ell+$track events (right) are shown
\protect\cite{cdfr2llcsec}.  Data are indicated by points with error bars,
and estimated signal and background contributions are shown via overlaid
histograms.}
\label{fig:cdfllr2csec} 
\end{figure*}

Remaining backgrounds from $Z/\gamma^*$
were estimated by comparing the number of dilepton events in each jet
multiplicity in the data and using that to normalize a fully simulated 
{\sc Pythia} sample.  Lepton instrumental backgrounds
were estimated by folding fake rates measured in data with a sample selected
to be kinematically signal-like, but without lepton identification selection
on one of the leptons.  
Diboson backgrounds were estimated from events
simulated with {\sc Alpgen} plus {\sc Herwig} or {\sc Pythia}. 
Fig. \ref{fig:cdfllr2csec} shows the \met\ and $H_T$ distributions of
the dilepton events after the application of all the selections.

The Run II explicit dilepton channels were incorporated into a 
\ttbar\ cross section
measurement from all dilepton channels, as shown in Table~\ref{tab:csecs}.  
In this channel, 13 events were observed 
with expected signal of 8.2 events and expected 
background of 2.7 events.  Since the selections overlap, explicit
channels have significant
correlation with the implicit channels described in 
Section~\ref{sec:llImplicit}.  The combined
dilepton cross section for Run II is provided in Table~\ref{tab:csecs}.
The cross section measurement
was repeated for subsamples that have at least one $b-$tag, and that
have two-tight leptons, and were found to be consistent with these results.
Preliminary analyses of 360 pb$^{-1}$ \cite{cdf360llcsec} and 1.2 fb$^{-1}$ 
\cite{cdf1fbllcsec} have been pursued by CDF.  The former combined
both dilepton and $\ell+$track measurements to obtain a cross-section
of $8.5^{+2.6}_{-2.2}(stat)^{+0.7}_{-0.3}(sys)$ pb.  The latter
resulted in a measurement of $6.2\pm 1.1 (stat) \pm 0.7 (sys) \pm 0.4$ pb.

{\subsubsection{Implicit channels}
\label{sec:llImplicit}}

Analysis of a dilepton channel where one lepton is not identified began with
an initial effort by \dzero\ in Run I\cite{d0r1ttcsec} which 
looked for
one isolated, high $p_T$ electron plus exceptionally high \met.  For a concise
review of this analysis, see Ref. \refcite{bhatRev}.  A cross section
for this channel alone was obtained to be $9.1\pm 7.2$ pb, which
is incorporated into the final \dzero\ Run I result given in Table
~\ref{tab:csecs}.



\begin{table}[ht]
\begin{center}
\tbl{Selection cuts for \dzero\ and CDF Run II dilepton and $\ell+$track 
cross section measurements.  Variables are described in the text.}
{\begin{tabular}{lccccc}\toprule
cut & CDF dilepton & CDF $\ell+$track & \dzero\ dilepton & \dzero\ $\ell+$track \\
\colrule
 $N_{\ell}(N_{tracks})$ & 2        & 1 (1)    & 2        & 1 (1) \\
 $p_T^{\ell_2}$ (GeV)   & $>20$    & $>20$    & $>15$    & $>15$ \\
 $|\eta_e|$             & $<2.0$   & $<2.0$   & $<2.5$   & $<2.5$ \\
 $|\eta_{\mu,track}|$   & $<1.0$   & $<1.0$   & $<2.0$   & $<2.0$ \\
 $N_{jets}(N_{tags})$   & $>2 (0)$ & $>2 (0)$ & $>2 (0)$ & $>1 (1)$ \\
 $p_T^{j}$ (GeV)        & $>15$    & $>20$    & $>20$    & $>20$ \\
 $|\eta_j|$             & $<2.5$   & $<2.0$   & $<2.5$   & $<2.5$ \\
 \met\ (GeV)            & $>25$    & $>25$    & $>25$    & $>15$ $(e)$ \\
                        &          &          &          & $>25$ $(\mu)$ \\
 $\Delta\phi(\met,j)$   & $\neq 0$ & $\neq 0$ & & \\
 $\Delta\phi(\met,\ell/track)>$    & $20\deg w/l$ & $5\deg w/track$ & 0.2 $w/\mu (e\mu)$ & \\
                        &          &          & not $180 \deg (\mu\mu)$ & \\
 $H_T$ (GeV)            & $>200$   & ---      & $>140$ ($e\mu$) & --- \\
 Z rejection: & & & & \\
 $M_{\ell\ell}$ (GeV)   & $76-106$ & $76-106$ & $80-100 (ee)$ & $70-100 (e)$ \\
                        &          &          &               & $70-110 (\mu)$ \\
 $\chi^2_{M_Z}$         & ---      & ---      & $>2 (\mu\mu)$ & --- \\
 \met\ near $M_Z$       & $\infty$ & $>40$    & $\infty$      & $>20$ ($e$) \\
                        &          &          &               & $>35$ ($\mu$) \\
\botrule
\end{tabular}
\label{tab:llcscuts}}
\end{center}
\end{table}

In Run II, CDF has performed a cross section measurement in
$197\pm 12$ pb$^{-1}$ using an implicit dilepton selection\cite{cdfr2llcsec}. 
A tight lepton ($e$ or $\mu$) was required with 
strict calorimeter, muon system or tracking
requirements, and an isolation cut.  
Another high $p_T$ track isolated from significant momentum in 
nearby tracks was also required.  This lepton plus track ($\ell+$track) 
approach provides measurement of all final state
particles, aside from $\nu$s.  Thus, top quark properties, such as $m_t$, 
can be measured with techniques appropriate to an explicit
dilepton analysis.  The cost of the looser lepton selection can be
borne by tighter requirements on other quantities, such as jet $E_T$ or
jet $b-$tagging.

As with the explicit channels, $Z/\gamma*$ backgrounds
were estimated by comparing the number of dilepton events in each jet
multiplicity in the data and using that to normalize a {\sc Pythia} plus
full detector simulation sample.  
Instrumental backgrounds were extracted from a kinematically signal-like
sample in data without final identification cuts on one of the lepton 
candidates.  Diboson backgrounds were estimated using
{\sc Alpgen} with {\sc Herwig} or {\sc Pythia}.
The \ttbar\ signal was simulated using {\sc Pythia}.  

\begin{figure*}[!h!tbp]
\begin{center}
\epsfig{figure=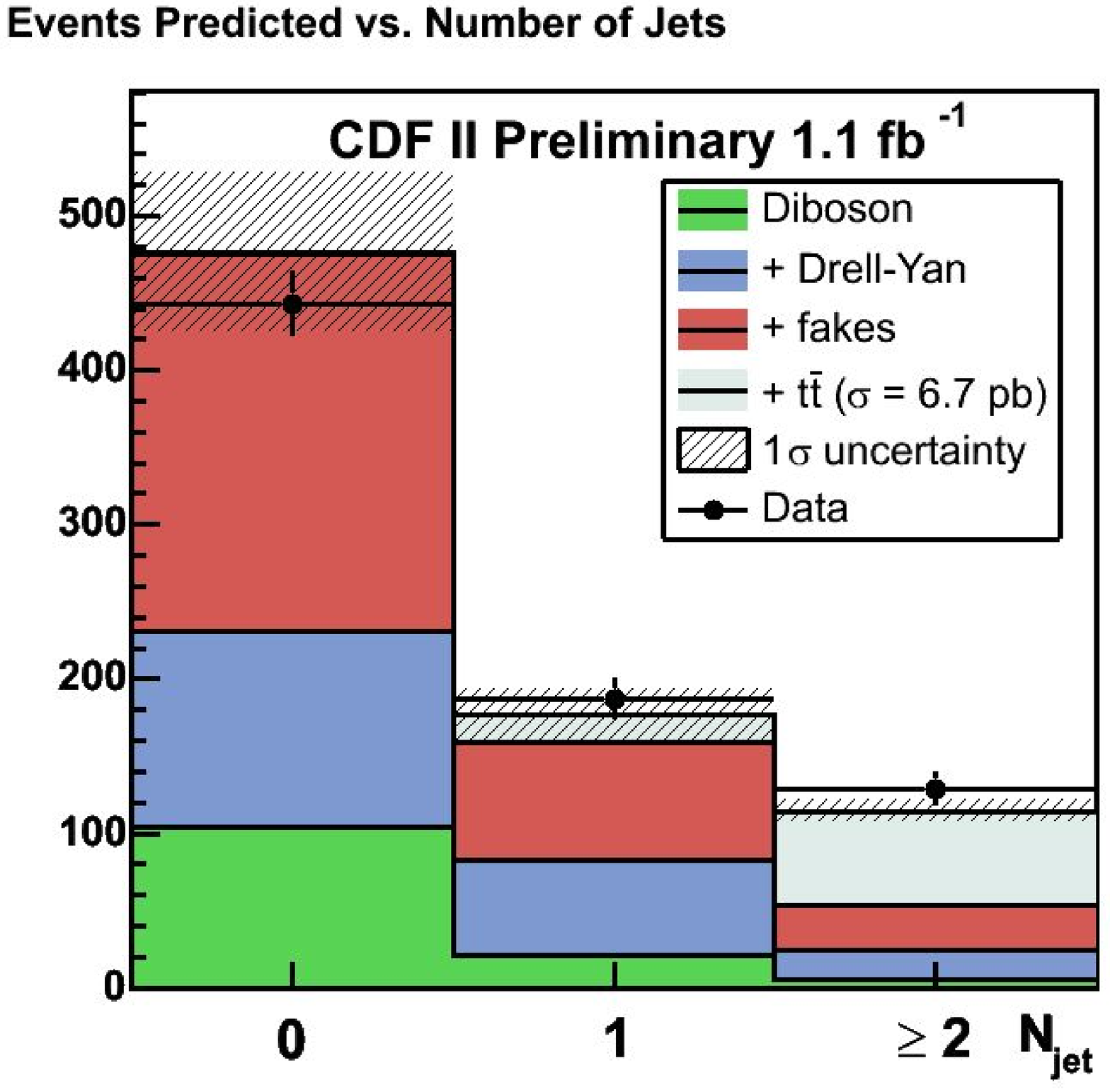,width=0.4\textwidth}
\epsfig{figure=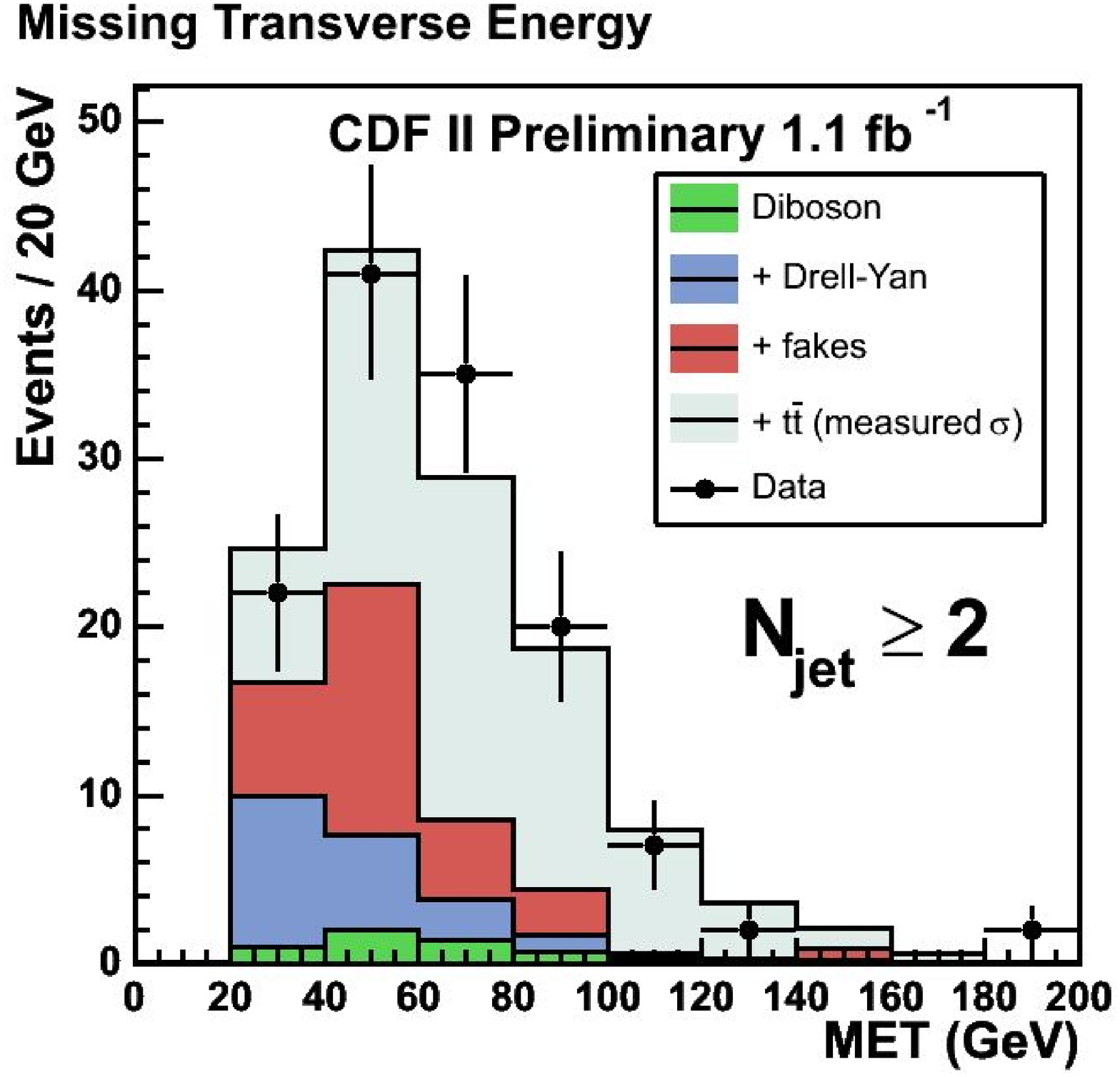,width=0.4\textwidth}
\end{center}
\vspace*{8pt}
\caption{Distributions of jet multiplicity in CDF $\ell+$track events
in the 1 fb$^{-1}$ sample (left).  The \met\ distribution is shown for
two jet events (right).  The data is indicated by points with errors,
and signal and background expecations are overlaid.
\protect\cite{cdf1fbllcsec}.}
\label{fig:cdf1fbltrkcsec} 
\end{figure*}

The data yield 19 events for the $\ell+$track selection
with expected signal of 11.5 events and an expected 
background of 6.9 events.  The implicit
and explicit channels share a significant number of events.  
The combined dilepton cross section measurement 
is given in Table~\ref{tab:csecs}.  The cross section measurement
was repeated for $b-$tagged, and two-tight lepton subsamples
and found to be consistent with these results.  Preliminary
analyses in 360 pb$^{-1}$ and 1 fb$^{-1}$
of data were performed for the $\ell+$track channel. 
The former yielded a combined measurement
with dilepton channels\cite{cdf360llcsec} as shown in Table \ref{tab:csecs}.  
The 1fb$^{-1}$ analysis provided a much more precise measurement of
$8.3\pm 1.3(stat)\pm 0.7(sys)\pm 0.5(lum)$ pb\cite{cdf1fbllcsec}.  Figure
\ref{fig:cdf1fbltrkcsec} shows the agreement between the 
jet multiplicity and \met\ distributions for
the 1fb$^{-1}$ $\ell+$track data and the signal
and background expectations.

The \dzero\ experiment has performed an $\ell+$track 
analysis in 425 pb$^{-1}$ of data\cite{d0425llcsec}.  
At trigger level, single lepton plus jet triggers are employed.
To offset the
higher background resulting from the omission of the lepton identification
cuts, $b-$tagging of jets using a secondary vertex
algorithm is employed to produce a reasonable
$S/B$.  This allows the analysis to have significantly looser kinematic 
selection than \dzero\'s explicit dilepton channels.
Signal and $Z/\gamma^*$ samples are modeled using
{\sc Alpgen} fed into {\sc Pythia}.  Diboson samples are modeled with {\sc Pythia}.
The primary background is instrumental from $Z\rightarrow ee,\mu\mu$ 
with fake \met.  The data and Monte Carlo \met\ distributions 
were observed to be in agreement, and so this background is
extracted from the simulation after normalizing the event yield
to that observed in the data for low \met.  A significant
background from fake leptons or fake tracks exists for this analysis.
This background is estimated in the untagged sample by 
constructing four samples
in data with loose and tight selections on the lepton and track.
From these samples, $b-$tag rates appropriate to $W+$jets events, 
and knowledge of lepton efficiencies and fake
rates, the number of lepton instrumental backgrounds is inferred.
Figure \ref{fig:d0ll425csec} shows the $H_T$ distribution of
$\ell+$track events, with \ttbar\ signal and background estimates
overlaid.  There is good agreement of these expectations with
the data.  The $\ell+$track channel was also analyzed in a preliminary
measurement in 1 fb$^{-1}$\cite{d01fbllcsec}.
The \ttbar\ cross sections determined in combination with the 
explicit dilepton channels are shown in Table~\ref{tab:csecs}.

%% file: ttbar_csec/ljetscsec.tex
Since top quark analyses have been
statistically limited until recently, the extra rate
has caused the $\ell+$jets channel to be key in 
studying the entire top quark sector.  
Generally, backgrounds are easier to control if
the lepton is an $e$ or $\mu$, although $\tau$-based analyses have
been implemented (e.g. Ref.~\refcite{tauCsec}). This section will concentrate
on analyses using leptons of the first two generations.

With four quarks from \ttbar\ decay, top quark events are fairly crowded in
the central region of the detector.  Initial state gluons can land near
these quarks.  Final state gluon radiation can take momentum away from
the quarks.  The resulting jets may overlap and get merged, or may split
into extra jets.  
As a result, a substantial fraction of top quark events will
exhibit only three jets, and a significant number will have more than 
four.  The physics processes which
can produce the $\ell+$jets signature are $W+$jets production in 
association with jets
and, at a much lower level, diboson ($WW, WZ$) production.  Single top quark 
production is also a background to the $\ell+$jets \ttbar\ 
cross section measurement. 
The largest instrumental background comes from multijet production
where the lepton and \met\ are fake.

Strategies for measuring top quark pair production in the $\ell+$jets 
channel have employed three general
techniques.  At the selection level, the balance of
efficiency and background level is crucial.  
A purely kinematic, or `topological',
approach uses the unique kinematic signature of the top quark
to isolate it from background via a multiparameter discriminant.  
Semileptonic decays of $b$-quarks occur at a significant rate.
The backgrounds do not have nearly as high a $b$-jet content as top quark 
events.  So a second strategy involves use of the excellent
capabilities of the \dzero\ and CDF detectors to 
tag jets with leptons from semileptonic decay of $b-$quarks.
The third strategy,
which is the primary approach for the most precise measurements
of the \ttbar\ cross section, involves lifetime tagging $b-$jets to suppress
backgrounds.  

\begin{table}[ht]
\begin{center}
\tbl{Selection cuts for CDF and \dzero\  Run II $\ell+$jets 
cross section measurements.}
{\begin{tabular}{lcc}\toprule
cut                & CDF $\ell+jets$ & \dzero\  $\ell+jets$ \\ 
\colrule
Trigger:           & 1 $e,\mu$    & 1 $e,\mu$ + 1 jet \\
\hline
$b-$tag selection: & & \\
$p_T^{l}$ (GeV)    & $>20$          & $>20$\\
$|\eta_e|$         & $<1.1$         & $<1.1$\\
$|\eta_{\mu}|$     & $<1.0$         & $<2.0$\\
$N_{jets}$         & $>3$           & $>3$\\
$p_T^{j}$ (GeV)    & $>15$          & $>15$\\
$|\eta_j|$         & $<2.0$         & $<2.5$\\
\met\ (GeV)        & $>20$          & $>20$ \\
$H_T$ (GeV)        & $>200$         & $>0$ \\
\hline
topological:       & & \\
$N_{jets}$         & $>3$           & $>4$ \\
$H_T$ (GeV)        & $>0$           & $>0$ \\ 
$\Delta\phi(\met,j_1)\neq$ & $0,\pi$ & \\
\botrule
\end{tabular}
\label{tab:ljcscuts}}
\end{center}
\end{table}

{\subsubsection{$\ell+$jets Channels using kinematic selection}
\label{sec:ljTopo}}

The kinematics of the final state products from top quark decay are quite
striking compared to those produced by other standard model processes.
It is therefore natural that CDF and \dzero\ have devised selections relying 
solely on these
properties.  In the $\ell+$jets channel, the backgrounds are much
higher than the dilepton channel.  As a result, significantly more 
complex approaches have been required to isolate a clean signal.  
A value of such a selection is that it does not rely on the assumption
that $BR(t\rightarrow Wb) = 1.0$.  Sensitivity
is retained for models which produce different final state jet flavor 
content (e.g. Ref.~\refcite{chargedHiggsModels}).

\begin{figure}[thb]
\begin{center}
\epsfig{figure=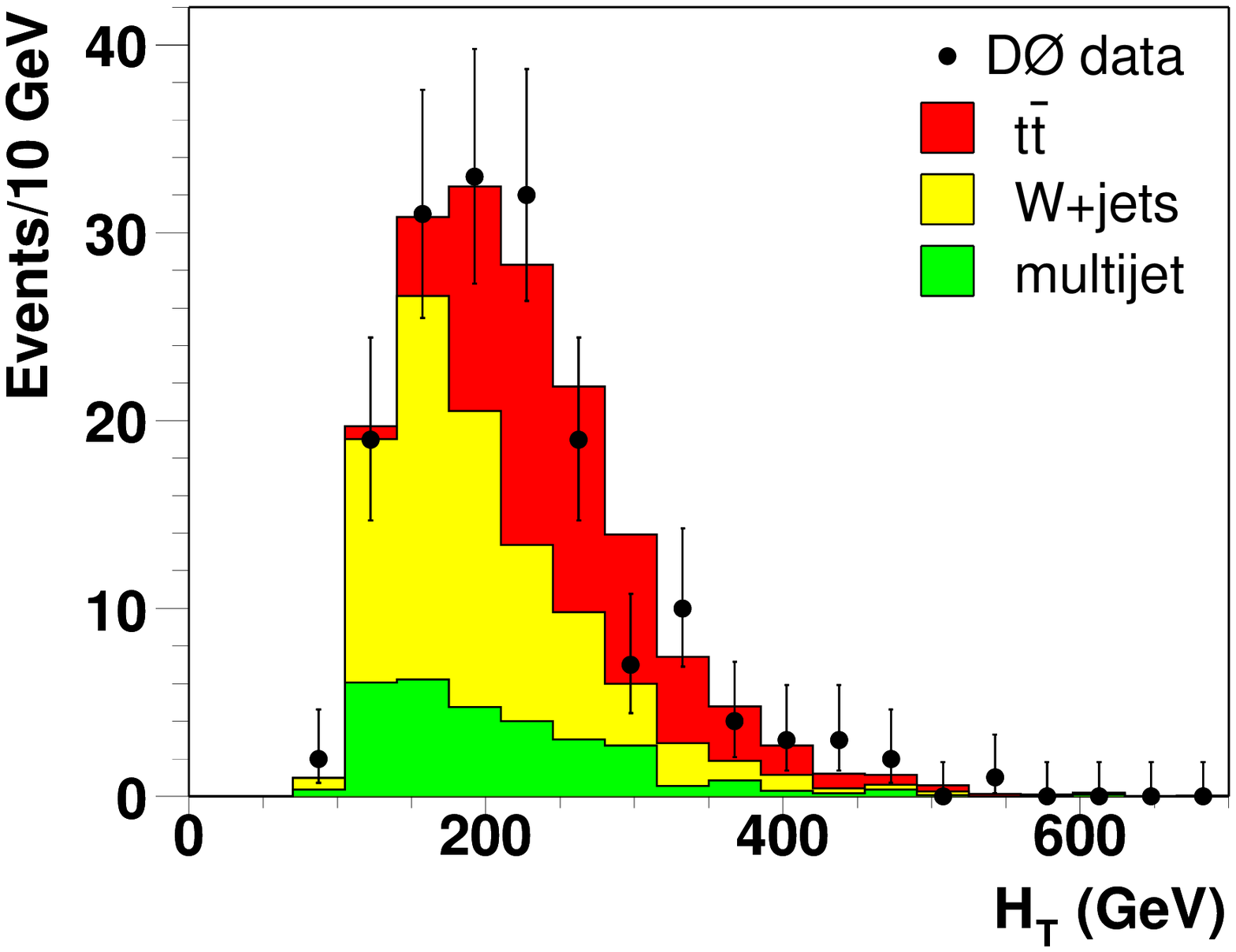,width=5.75cm}
\epsfig{figure=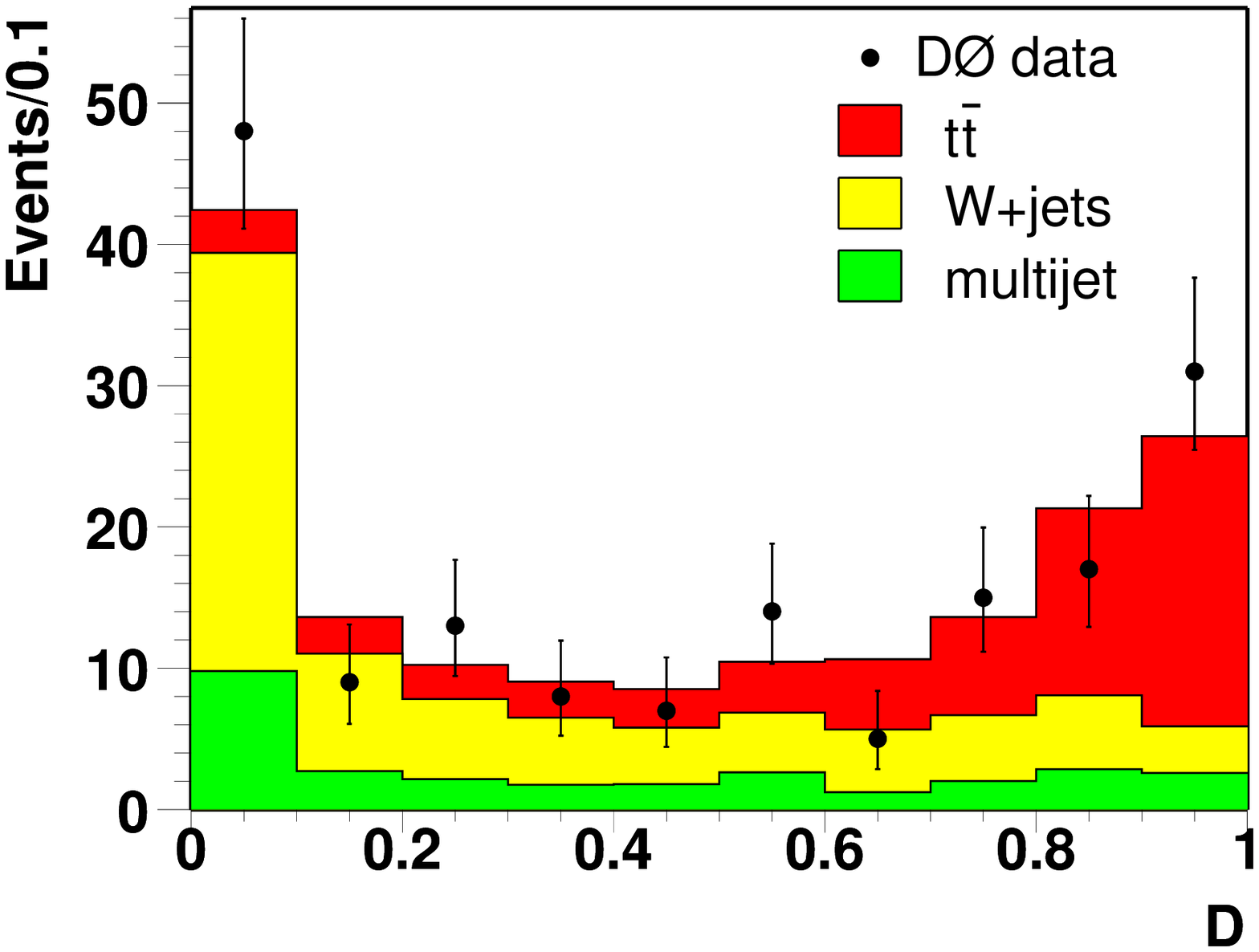,width=5.75cm}
\caption{Distribution of $H_T$ for combined 
$\ell+$jets events in 230 pb$^{-1}$ of \dzero\ Run II data (left).
The sum of \ttbar\ and backgrounds are overlaid.  At right is the
multi-parameter likelihood discriminant for data and also overlaid
with signal and background expectation\protect\cite{ljtopo230}.}
\label{fig:d0ljtopo230csec}
\end{center}
\end{figure}

The initial untagged $\ell+$jets analyses were performed by \dzero\  in 
Run I and contributed the primary significance to their top quark discovery
analysis\cite{d0Obs}.  This laid the basis for a final measurement
at $\sqrt{s}=1.8$ TeV\cite{d0r1ttcsec}, as well as a 
measurement with 226 pb$^{-1}$ of data at $\sqrt{s}=1.96$ TeV\cite{ljtopo230}.
To control backgrounds, four jets were required.
The specific event selection used in the Run II analysis can
be obtained from Table~\ref{tab:ljcscuts}.  The \ttbar\
acceptance and efficiencies were estimated from the Monte Carlo.

Theoretical calculations yield substantial uncertainties
in the production of background processes.  As a result, methods
are needed by which the normalization of the background can be 
calculated from data.  The multijet background kinematic 
shapes were taken from the data by requiring the lepton to fail
the tight identification requirement.  To estimate the normalization of the
background and signal estimates, two steps were employed.  First,
loose and tight cuts for the lepton identification were defined to permit
a variation of the level of QCD multijet background relative to the
$W$ processes ($W+$jets and \ttbar).  Efficiences for real leptons 
and the fake leptons in the QCD sample were measured in data.  Then one
can solve the resulting two event yield equations in two 
unknowns: $N_s$ for top quark  plus $W$ events, and
$N_b$ for the multijet yield.  This is termed the `matrix method'. 
In Run I, the apparent scaling of jet multiplicity 
(sometimes called `Berends scaling' 
\cite{berends}) was used to anchor the $W+$ four jet multiplicity  
normalization with the yields observed for background-rich 
low jet multiplicities.  The
Run II measurement, however, extracts the level of multijet
background from the matrix method and obtains the $W$ normalization
as described in the next paragraph.  The
$W+$jets kinematic shapes are taken from the Monte Carlo.  

A multiparameter discriminant was constructed to identify the top quark 
signal in the data.  For the Run II analysis, this was formed 
from six observables: $H_T$, $\Delta\phi(\met,\ell)$, $\mathcal{C}$, 
$\mathcal{S}$, $\mathcal{A}$
and $K_{Tmin}=\Delta R^{min}_{jj}p_T^{min}/E_T^W$.  The latter is calculated
by determining the jet pair with minimum separation in $\eta-\phi$
($\Delta R^{min}_{jj}$), the $p_T$ of the second leading jet of that
pair, and the scalar sum of the lepton and \met, $E_T^W$.  The discriminant
function is:
\begin{equation}
D = [\Pi_i s_i(x_i)/b_i(x_i)]/[\Pi_i s_i(x_i)/b_i(x_i) + 1]
\end{equation}
where $s_i$ and $b_i$ are the normalized distribution for each
variable, $i$, for signal and background, respectively.  
An assumption was made of uncorrelated variables.  
Figure \ref{fig:d0ljtopo230csec} shows the event $H_T$ and the
likelihood discriminant for data with \ttbar\ signal and background 
superimposed.  The discriminant 
function in data was fit to extract a measurement of $\sigma_{\ttbar}$
and the number of $W$ backgrounds.  
The measurement for $\sqrt{s}=1.96$ TeV was 
$\sigma_{t\bar{t}}=6.7^{+1.4}_{-1.3}(stat)^{+1.6}_{-1.1}(sys)\pm 0.4 (lum)$ pb\cite{ljtopo230}.
This analysis has been updated in 425 pb$^{-1}$
to yield $\sigma_{t\bar{t}}=6.4^{+1.3}_{-1.2}(stat)\pm 0.7(sys)\pm 0.4 (lum)$ 
pb~\cite{d0lj425topo}.  A preliminary result using
900 pb$^{-1}$ has also been obtained \cite{d0lj900topo}.  Statistical uncertainties
of approximately 15\% are being achieved, and the measurement in the latter
sample was $6.3^{+0.9}_{-0.8}(stat)\pm 0.7 (sys)\pm 0.4(lum)$ pb.

\begin{figure}[thb]
\begin{center}
\epsfig{figure=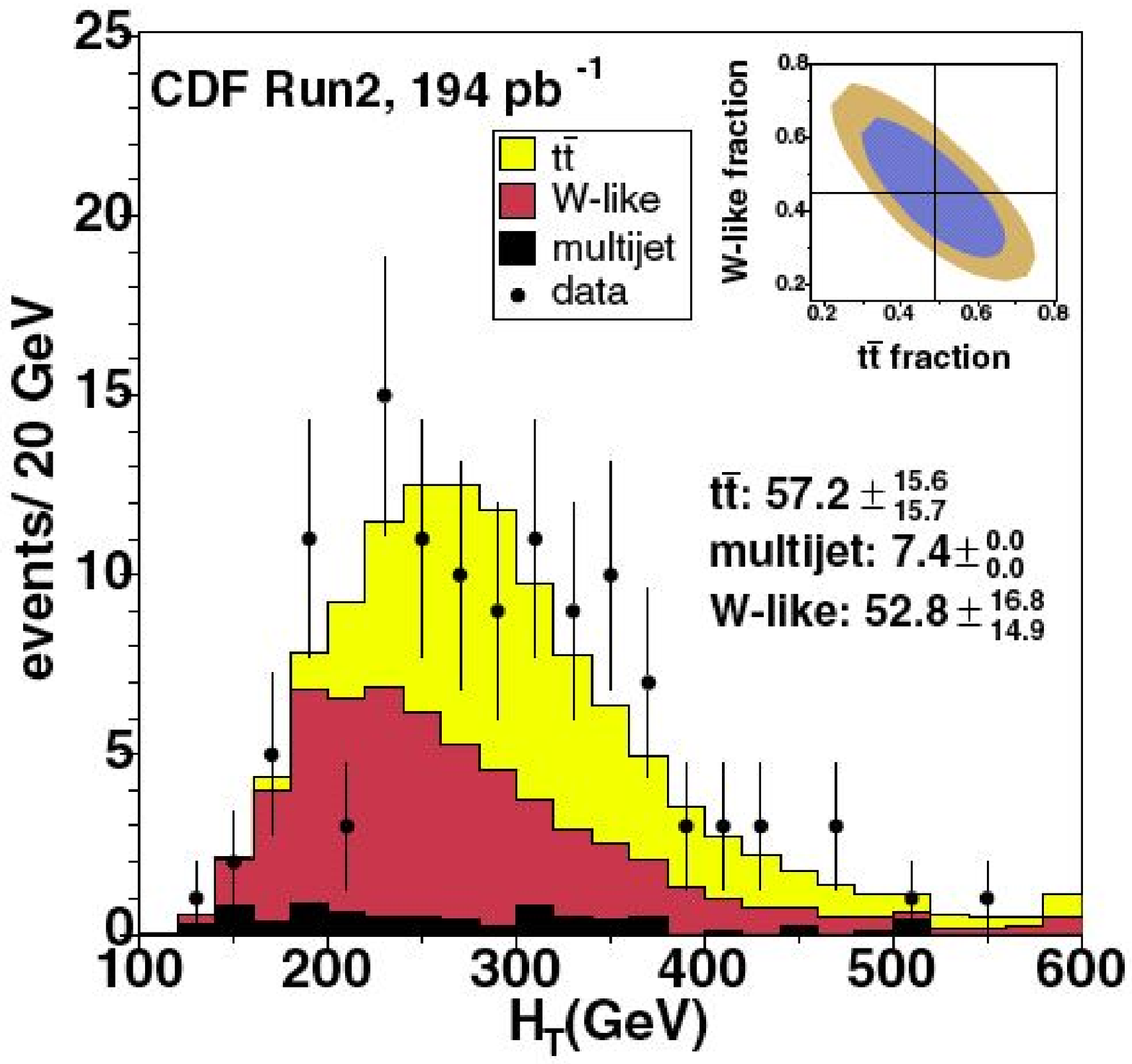,width=5.75cm}
\epsfig{figure=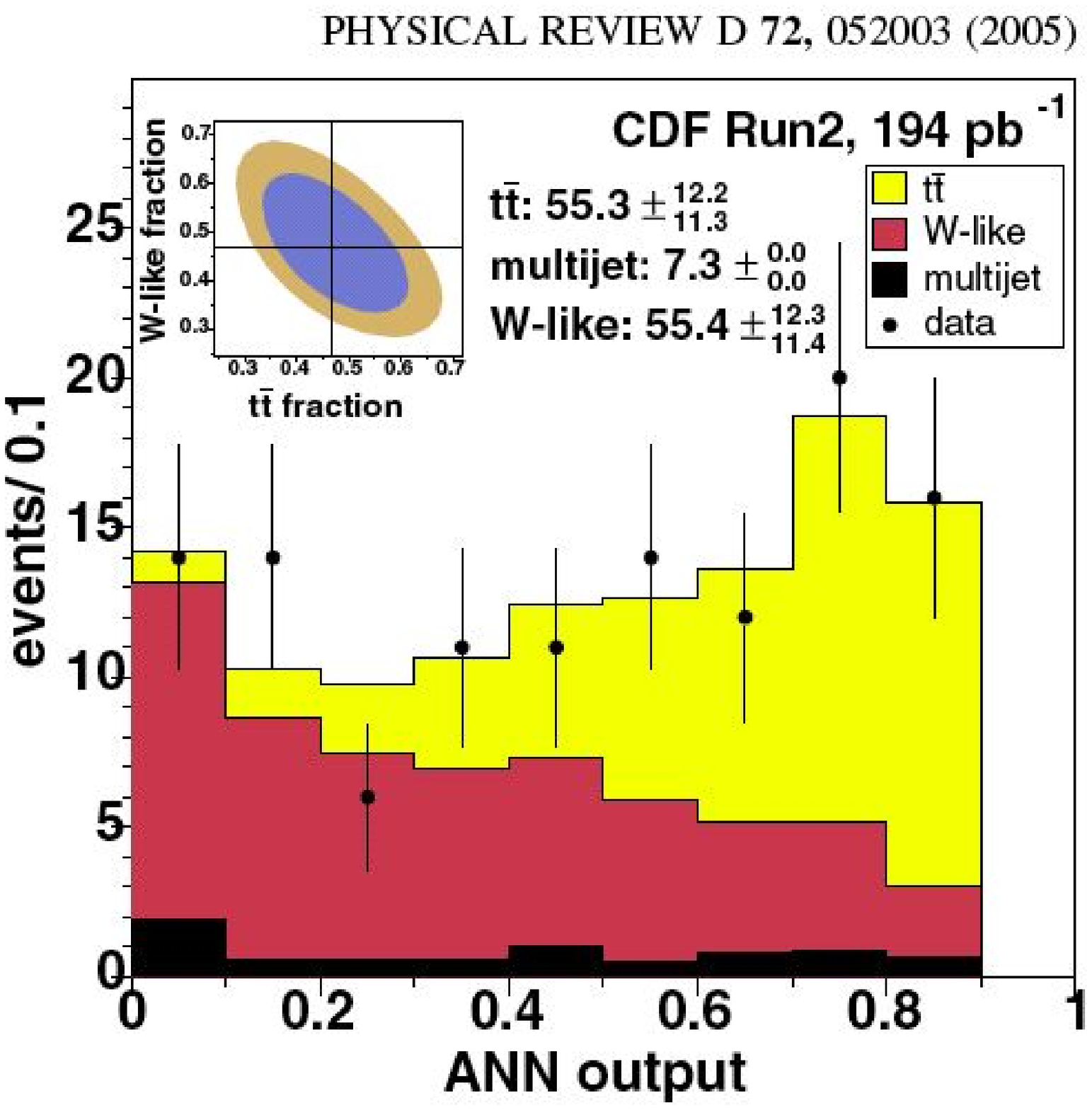,width=5.75cm}
\caption{Distribution of $H_T$ for combined $\ell+$jets events 
in CDF data (left).
The sum of \ttbar\ and backgrounds are overlaid.  At right is the
likelihood for data and also overlaid with signal and background expectation\protect\cite{cdfljtopo194}.}
\label{fig:cdfljtopo194csec}
\end{center}
\end{figure}

\begin{figure}[thb]
\begin{center}
\epsfig{figure=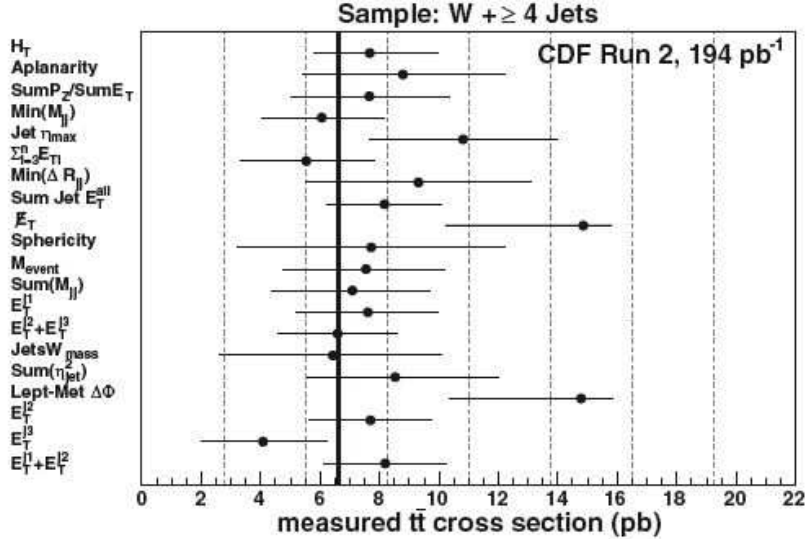,width=11cm}
\caption{Cross section for \ttbar\ in 194 pb$^{-1}$ of $\ell+4$jets events 
considering
topological properties of events\protect\cite{cdfljtopo194}.}
\label{fig:cdfljtopo194checks}
\end{center}
\end{figure}

CDF pursued this `topological' approach with 194 pb$^{-1}$
of collisions at $\sqrt{s} = 1.96$ TeV\cite{cdfljtopo194}.  Only three jets were required.
Events were rejected if $\Delta\phi(\met,j_1)\sim 0$ or $\pi$.
The $W+$jets background shape is extracted from the Monte
Carlo, and the normalization is obtained by maximizing a binned 
likelihood:
\begin{equation}
L(\bar{n}_{t\bar{t}}, \bar{n}_W, \bar{n}_q) = \Pi^{N_{bins}}_{i=1} (e^{-\bar{n}}\bar{n}^{d_i}_i)/d_i!
\end{equation}
where the $\bar{n}$'s are the means for top quark ($\bar{n}_{\ttbar}$), W 
($\bar{n}_W$) and 
QCD ($\bar{n}_q$) yields, and
$\bar{n}_i$ expresses the expected number of events in an $i$th bin given
the probabilities for the signal and background 
contributions to populate that bin.  $d_i$ is the number of events observed
in the $i$th bin.  The value of $\bar{n}_q$ is fixed based on a `sideband' method
where the data are divided 
into quadrants in the lepton isolation versus \met\ plane.  
The number of events with non-isolated
leptons and high \met\ is scaled by the ratio of isolated-to-non-isolated 
events in a $W$-poor, low \met\ region.  
The value of $\bar{n}_{t\bar{t}}$ extracted from the fit
is used to determine $\sigma_{t\bar{t}} = \bar{n}_{t\bar{t}}/\epsilon_{t\bar{t}}\mathcal{L}$.
The cross section is calculated using two different
kinematic discriminants: $H_T$ and a variable from a seven-parameter
artificial neural network technique.  The parameters were:
$H_T$, $\mathcal{A}$, $1/\mathcal{C}$, the minimum dijet mass 
from the three leading jets
$M_{jj}^{min}$, the pseudorapidity of the leading jet $\eta^{j1}$,
the minimum $\eta-\phi$ separation of two jets among the leading three,
and the $H_T$ calculated from the third leading and lower $p_T$ jets. 
The values of both of the
discriminants for the data sample, with signal and background superimposed,
are shown in Fig.~\ref{fig:cdfljtopo194csec}.
The neural network approach included additional information besides
$H_T$ to provide
approximately 30\% better statistical uncertainty than $H_T$ alone.
These variables were each cross checked in the $W+$jets samples. 
The cross section was actually estimated using these and other kinematic
variables individually (see Fig. \ref{fig:cdfljtopo194checks}), 
and a full range of systematic uncertainties
were obtained for the $H_T$ case in the three-jet and four-jet inclusive
bins.  The full analysis provided a measured cross section of
$6.6\pm 1.1 (\rm stat) \pm 1.5 (\rm sys)$ pb.

\subsubsection*{$\ell+$jets Channels using soft lepton tagging}

Hadrons containing $b-$quarks undergo semileptonic decay at a
rate of approximately 17\% for each charged lepton type.  
When decays through $c-$quarks are included, approximately 40\% of 
\ttbar\ events have a soft non-isolated $\mu$, for instance.
In contrast, typical background processes produce
primarily `light flavor' partons ($u/d/s$ quarks and gluons) and the heavy
flavor contribution is small.  So requiring a soft lepton in a jet
can provide a strong background suppression.
While the unbiased flavor selection is important in
its generality, the ability to demonstrate the presence of $b-$quarks
in the top quark candidate sample has also been crucial to validate
whether the observed signal adhered to standard model expectations.

\begin{figure}[thb]
\begin{center}
\epsfig{figure=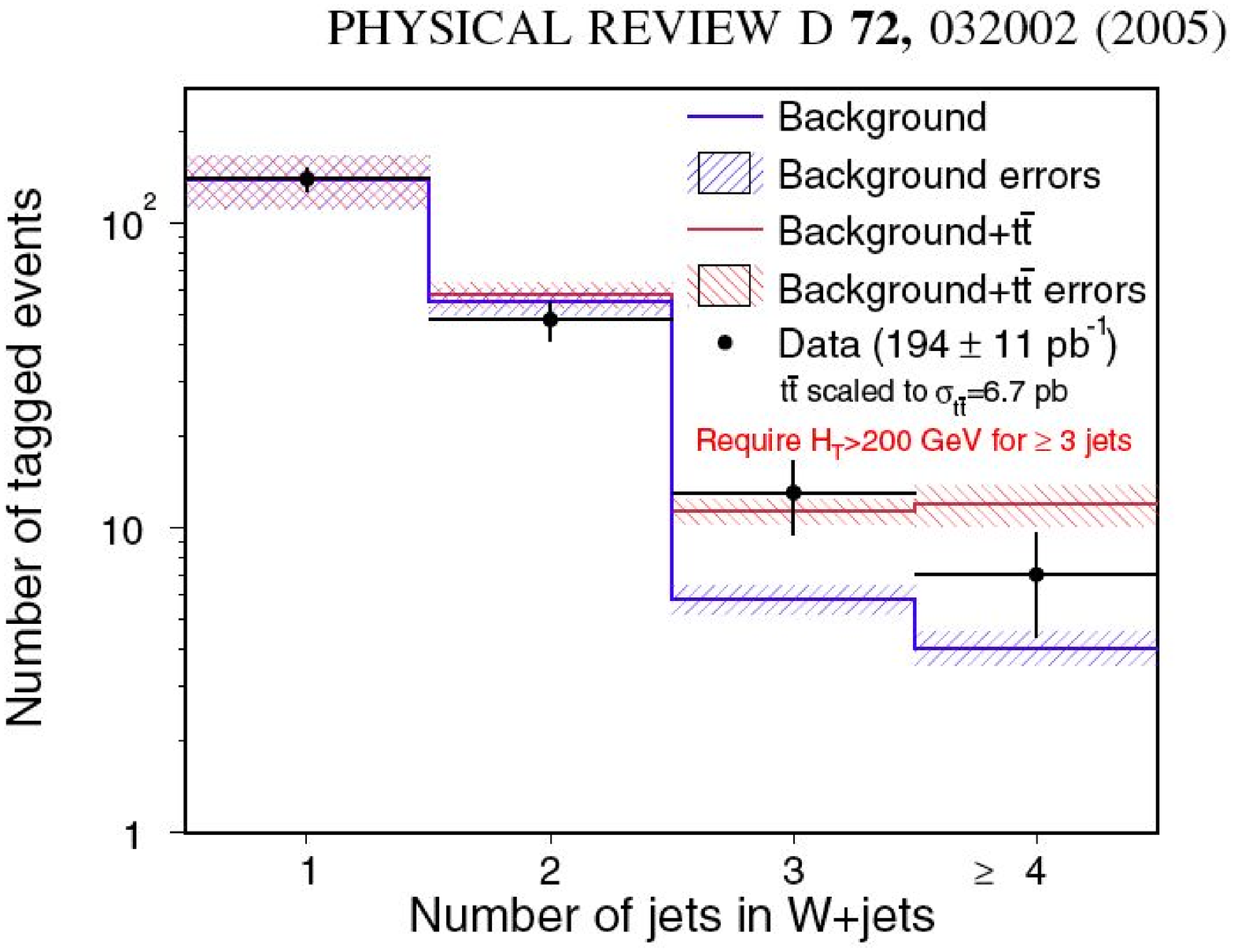,width=6.25cm}
\epsfig{figure=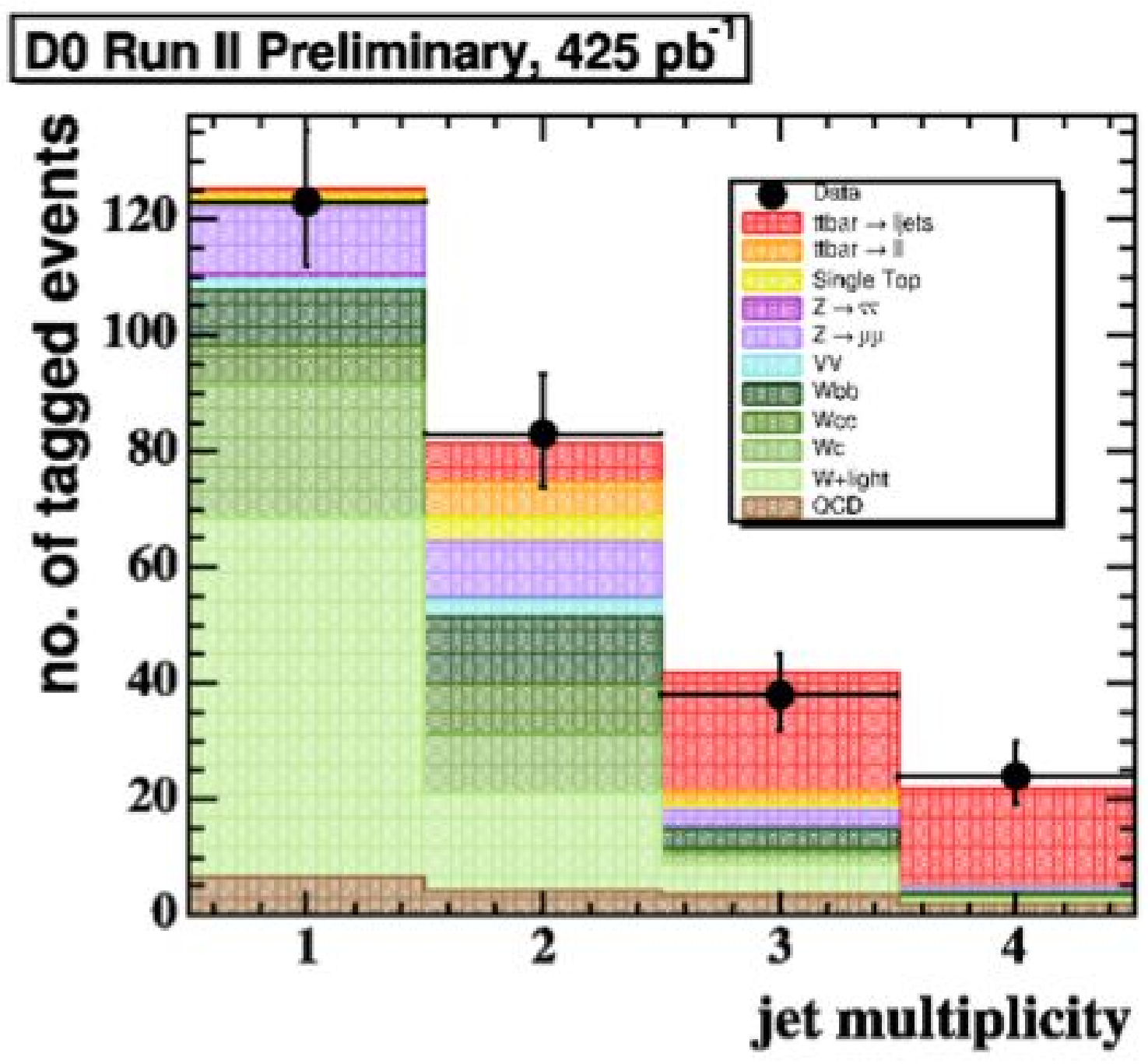,width=5.75cm}
\caption{Jet multiplicity distribution for CDF 194 pb$^{-1}$ 
soft-lepton tagged data sample (left)
\protect\cite{cdfr2ljmutagcsec}.  
Estimated backgrounds are overlaid as a solid histogram.  
At right is the \dzero\ distribution in a preliminary
analysis of 425 pb$^{-1}$ of data\protect\cite{d0r2ljsmt}.}
\label{fig:ljSLTnjets}
\end{center}
\end{figure}

Both Tevatron experiments
have pursued strategies of selecting events where at least one jet is
`tagged' with a soft, non-isolated lepton.  CDF has produced estimates
of $\sigma_{\ttbar}$ using both soft muon and soft electron semileptonic 
decay modes\cite{cdfr1ljsltcsec}.  In Run II, the soft muon variant was 
employed  in 194 pb$^{-1}$ of data\cite{cdfr2ljmutagcsec}.  
In order to optimize their analysis,
CDF chose a cut on $H_T>200$ GeV that maximizes the significance
of signal $=S/\sqrt{S+B}$.  

Signal efficiencies were estimated in Monte Carlo with tag efficiencies scaled
from data.  In the Run I analysis, efficiencies for leptons from heavy flavor, 
as well as backgrounds, were estimated using inclusive jet events from data.  
Cross checks were performed in a sample of $Z\rightarrow \ell\ell + jets$ 
events with soft lepton and secondary vertex tags 
which have a much smaller fraction of top quark signal.
The calculated yield from background in these channels agreed well with 
observed numbers.  In Run II, the  
$W$ and QCD backgrounds are estimated from data in two steps.
The multijet background is estimated 
with the sideband method mentioned above.  The product of this yield 
with the correct tag rate gives the background estimate. 
The applicability of the measured tag rate for the $W$ component of the 
background is verified by using it to estimate the tagged yield in 
several different jet samples.  By comparing this estimate with the observed
tagged event statistics, the rates are validated.  The final
$W+$jets yield is estimated using untagged samples corrected for the expected
QCD contribution and multiplied by the mistag rate from $\gamma+$jet
events.  Other backgrounds are much smaller than the $W$ and QCD
backgrounds.  

Dominant systematic uncertainties arise from the modeling
of top quark acceptance as well as the signal and background tag rate estimates.
The jet multiplicity distribution for the data, as well as the estimated
backgrounds are shown in Fig.~\ref{fig:ljSLTnjets}.
The final cross section estimate for $\sqrt{s}=1.96$ TeV 
is $5.3\pm 3.3 (stat)^{+1.3}_{-1.0}(sys)$ pb.  A preliminary measurement
of the $\sigma_{\ttbar}$ in 760 pb$^{-1}$ of $\mu-$tag events has
also been executed by CDF, yielding a value of
$7.8\pm 1.7(stat)^{+1.1}_{-1.0}(sys)$ pb \cite{cdf760mutagcsec}.

Soft muon tagging has also been utilized by \dzero\  to identify $b-$jets in 
$\ell+jets$ data from both Tevatron runs.  The Run II analysis\cite{d0r2ljsmt} 
constitutes a preliminary result in 425 pb$^{-1}$ of collider data.
These analyses benefited from the 
large coverage of the muon system ($|\eta| < 2.0$ in Run II).  They also
benefited from the depth in interaction
lengths of the material in front of the muon chambers, particularly
the layers outside of the muon iron toroid, which reduced the
instrumental fake muon background to exceptionally low levels.  
Only three jets were required.  The Run II analysis improves over the
earlier one by no longer requiring an $H_T$ cut or a selection for the
$\mu+jets/\mu-tag$ channel of a cut on the $\chi^2$ of a kinematic fit
to a $Z\rightarrow\mu\mu$ hypothesis.  

Signal and physics background ($W+$jets, diboson and single top quark events) 
efficiencies and yields were estimated using {\sc Alpgen} plus
{\sc CTEQ5L} and {\sc Pythia}.  The primary backgrounds are:
$Z\rightarrow\mu\mu$ where one $\mu$ is lost or overlays a
jet and mimics a $b-$tag, QCD multijet background, and $W+$jets.  The
former was obtained from Monte Carlo but the yield was normalized using
the measured $Z$ boson cross section.  The multijet
background in the tagged sample was isolated from the $W-$like events,
which include $W+$jets and \ttbar\ components, using the `matrix method'.
$W+$jets events were estimated by extracting the yield from
the same method in the untagged sample.  The Monte Carlo was used to
ascertain the flavor composition of this sample.  This is an improvement
over the Run I approach in which the event kinematic shapes were also
extracted from the {\sc Vecbos} Monte Carlo.  The appropriate
tagging efficiencies or mistag rates for each flavor component were corrected
to reflect performance in data and applied to the untagged yield.
The jet multiplicity distribution of the selected sample is shown in Fig. 
\ref{fig:ljSLTnjets}.  

The cross section was determined from a maximum likelihood fit to the observed
number of events in three and four jet samples of the $e+$jets and $\mu+$jets
channels.  In each subsample, the backgrounds were constrained by the values
from the matrix method.  A gaussian term was used for each systematic 
uncertainty such that the mean cross section could shift if the expected 
signal and backgrounds were varied over the ranges allowed by the 
uncertainties.  The cross section was measured to be
$7.3^{+2.0}_{-1.8}(stat+sys)\pm 0.4 (lum)$ pb.

\subsubsection*{$\ell+$jets Channels using secondary vertex tagging}

Displaced
vertex $b-$tagging as applied to top quark candidate samples was first
performed by CDF in the analysis of 1.8 TeV collisions\cite{cdfObs}.
An initial cross section measurement from CDF using Run II
data\cite{cdfr2btagkinfitcsec,cdfr2ljbtagtopocsec} employed
a $\ell+$jets selection with at least one jet tagged with
a secondary vertex using $162 \pm 10$ pb$^{-1}$.  
The leading jet $p_T$ was used as a discriminating variable in a likelihood
to extract a top quark fraction in data.
The shape of the $W$ boson $p_T$ spectrum was obtained by selecting a
$W+$jets sample in data where no jets were $b-$tagged to reduce
top quark contamination.  
The instrumental background was obtained in the high \met\ sample by
reversing the isolation cut on the lepton.  
The kinematic distributions were observed to be insensitive to the heavy
flavor composition of the sample. 
To validate the background model, untagged $W+$ one or two jets events were 
scaled by the tag rate.  Agreement with the $b$-tagged sample in data
was observed in the jet $p_T$ distribution.
An unbinned likelihood fit was performed to the observed jet
$p_T$ distribution given the expected signal and background shapes.
A signal fraction of $R_{fit}=0.68^{+0.14}_{-0.18}$ was obtained.  The top quark
cross section was obtained from
$\sigma(\ttbar) = N_{obs}R_{fit}/A_{\ttbar}\epsilon_{\ttbar}L$
where $A$ is the signal acceptance, $\epsilon_{\ttbar}$ is the estimated
efficiency, and $N_{obs}$ is the observed yield of events in data.
The systematic uncertainty for this analysis is dominated
by that from the jet energy scale calibration.  Other significant 
contributors come from the $b-$tagging efficiency and the luminosity estimate.
The estimated cross section and uncertainty are 
$6.0\pm 1.6(\rm stat)\pm 1.2(\rm sys)$ pb and are given in 
Table~\ref{tab:csecs}.

In order to employ $\ell+$jets events in measurements beyond the
cross section, a more general approach is needed to determine backgrounds.
CDF has used a method in 162 pb$^{-1}$ of data which is
more similar with the lifetime tagging approach in Run I
\cite{cdfr2btag162csec,cdfr2ljbtagcsec}.
The background estimation method
was designed to account for the different flavor compositions
of multijet and $W+$jets samples.  The fraction of the $W+$jets background
that contains heavy flavor was estimated from Monte Carlo.
The mistag rate was applied to the QCD background and
to the fraction of $W+$jets background not arising from $\bbbar$ and
$\ccbar$ associated production.  The measured heavy flavor tag
rate was applied to the estimated $W+$ heavy flavor jet fraction.
In Run I, this approach provided good agreement of background estimation
and observed yield in $W+$ one and two jet samples.  In combination
with the soft-lepton tag selection, the $b$-tagged $Z+$jets sample
was also observed to provide agreement between observed and expectation
at all jet multiplicities.  

Signal efficiencies were calculated after correcting for the observed
ratio of the tag rate for jets tagged with soft electrons in
data and a parallel {\sc Herwig} sample.  Systematic uncertainties include
the residual flavor composition of the data sample, as well as
potential differences between soft electron tagged $b-$jets
vs. $b-$jets generally.  
The multijet background was estimated in tagged and untagged
samples using the sideband method in the 
\met\ vs. lepton isolation plane.
The heavy flavor fraction of the $W+$jets background was estimated from 
Monte Carlo.  
This fraction was calibrated using an MC-to-data correction factor
based on data and MC jet samples.  The yield of $W+$ heavy flavor events
was estimated by multiplying the heavy flavor fraction by the
number of events in the untagged sample, and then applying the $b-$tag
efficiencies to that.  Mistagged $W+$jets background
was estimated by weighting each jet
in the untagged sample by the mistag rate.  A correction was applied to 
remove the estimated QCD multijet background in this sample.  
Other small backgrounds were estimated from
the Monte Carlo.  The background estimation method for the $W+$jets sample
was cross-checked by applying similar techniques to a $Z+$jets sample where
there would be little top quark contribution.  Agreement of data with the 
expected background contributions was observed.  The cross section was
estimated in the single tag sample to be 
$5.6^{+1.2}_{-1.1}(stat)^{+0.9}_{-0.6}(sys)$ pb.  A measurement
carried out in the double-tag sample was consistent.

Using a larger data sample of 318 pb$^{-1}$, CDF has produced an extensive 
suite of measurements with different tagging treatments.
The secondary vertexing scheme was updated using a newer tagger
that has higher efficiency, particularly at high $p_T$ 
\cite{cdfr2btag318csec}.
Event selection remained the same.
The signal efficiencies and background estimates were also 
performed similarly to the earlier analysis.  The primary exception 
was that the backgrounds were adjusted with an iterative algorithm to 
correct for the top quark contribution in their control samples until the
cross section measurement varied by less than a percent.  The cross section
result was $8.7\pm 0.9 (stat)^{+1.1}_{-0.9}(sys)$ pb.  A complete measurement
was also performed in the same data sample with a jet probability
tagger\cite{cdfr2jetprobcsec}.  
Two different selections ($<1\%, <5\%$) were used to vary the purity and 
check the consistency of their results.
Event selection was the same as in the secondary vertex analysis with the
addition of a cut on the transverse mass of the $l\nu$ pair $M_T^W>20$ GeV. 
Signal efficiency and background levels were estimated
as in the secondary vertex tag analyses.  For the multijet background,
the definitions of the sidebands were adjusted to provide a 50\% 
systematic uncertainty in both the untagged and tagged samples.
Control samples of events with one  or two
jets validated the background modeling approach.
The background estimate was corrected iteratively to account for top quark 
contamination until the cross section
measurement was stable to within 0.1\%.  For the most sensitive 
jet probability tag selections, the cross section is 
$8.9\pm 1.0(stat)^{+1.1}_{-1.0}(sys)$ pb.  Cross sections
measured in the separated electron and muon channels, and in the 
double-tagged subsample, are all consistent with these results.
The results in 318 pb$^{-1}$ are somewhat higher
than the theoretical value.  
Cross checks in the data using the earlier secondary vertex algorithm,
as well as looking in the double tagged sample provided consistent
values of the cross section.  In 1.1 fb$^{-1}$ of $\ell+$jets events, 
CDF has produced a
preliminary measurement with substantially better statistical
precision: $8.2\pm 0.5(stat)\pm 0.8(sys)\pm 0.5$ pb
~\cite{cdf1fbbtagcsec}.

\dzero\ has published two determinations of the cross section in
collisions at $\sqrt{s} = 1.96$ TeV using
single lepton channels with secondary vertex tagging.  The first
utilized 230 pb$^{-1}$ of collider data\cite{ljbtag230},  
and the second incorporated 425 pb$^{-1}$ \cite{ljbtag425}.  A 
preliminary result based on 900 pb$^{-1}$ has also been pursued 
~\cite{d0900ljbtag}.  Trigger requirements of one lepton and one 
jet are the same as for the $\ell+$jets analysis with topological 
selection.  The data were separated into events with single tags, 
and double tags.  

In the 230 pb$^{-1}$ analysis, signal efficiencies were established by 
applying a tag rate from 
soft muon tagged dijet events corrected from a Monte Carlo study
so that it applied to inclusive $b-$jets. The simulation also provided
the $c-$jet tagging rate, and that was corrected for the semileptonic 
tag efficiency in data vs. Monte Carlo.  For the 425 pb$^{-1}$ 
measurement, separate multijet samples were defined which had different
levels of $b-$jet content.  One required a non-isolated $\mu$ in one
jet of a dijet pair, the other was the subset of this sample which also 
had a secondary vertex tag of the jet opposite this $\mu$.  
Event yields were extracted in subsets of these two samples with and 
without either lifetime or soft-lepton tags.  Eight equations were formed
from these yields which allow the extraction of the $b$-tagging efficiency
for semileptonically decaying $b$-quarks.  The tagging
efficiency for inclusive $b-$jets was
obtained from $\bbbar$ Monte Carlo and scaled to a factor from the 
same eight equation separation scheme.

To estimate the multijet
background, the matrix method was employed in tagged and untagged events.
All other backgrounds, including $Z+$jets and diboson components,
were estimated from the simulation.  NLO calculations of cross sections
were used to normalize the rates of these backgrounds.  The $W+$jets
background was found by subtracting the above contributions from the
untagged sample and then multiplying the remainder by the tag rate
appropriate to the flavor mix in $W+$jets events.  This composition
is evaluated from the {\sc Alpgen} $W+$jets simulation.  In the 425 pb$^{-1}$
analysis, an alternative parton matching scheme was used
to estimate a 20\% systematic uncertainty on the flavor fractions of the
$W+$jets sample.

\begin{figure}[thb]
\begin{center}
\epsfig{figure=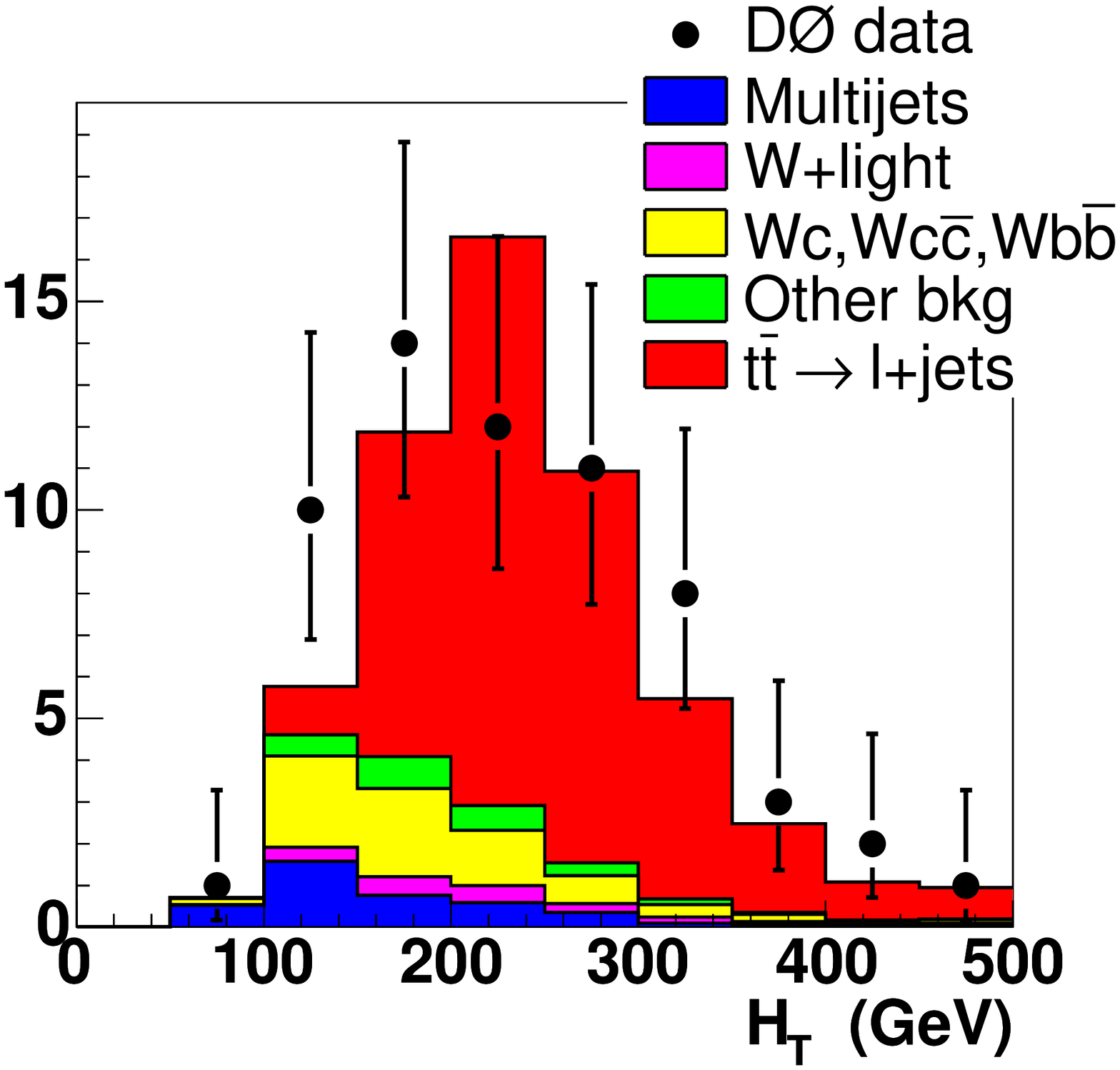,width=5.5cm}
\epsfig{figure=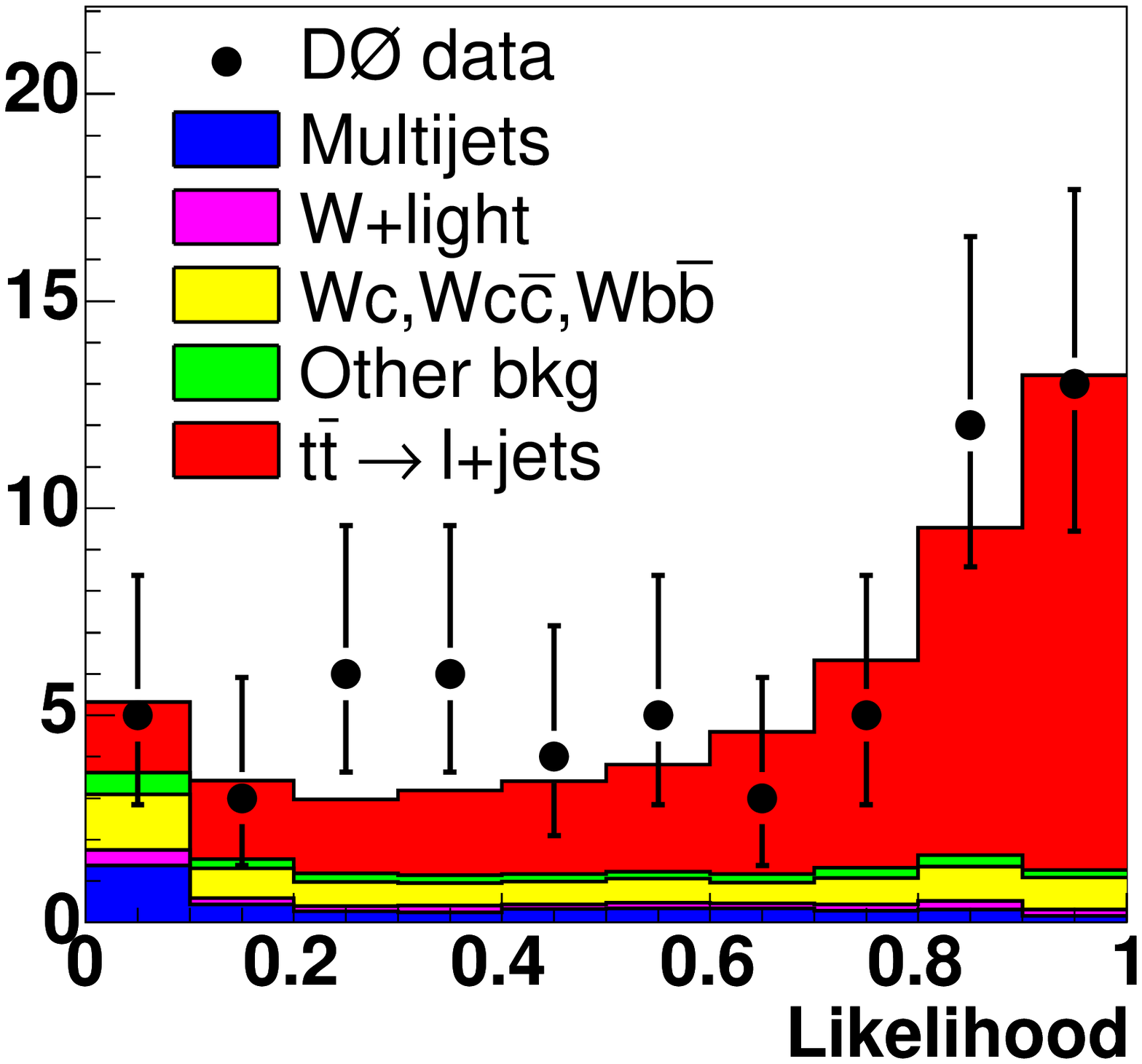,width=5.5cm}
\caption{Distribution of event $H_T$ (left) for the \dzero\
$\ell+$jets 230 pb$^{-1}$ sample with secondary vertex tag,
and likelihood (right)\protect\cite{ljbtag230}.  Data are
indicated with points and error bars, while the expectation
for signal plus background are shown via solid histograms.}
\label{fig:d0ljbt230csec}
\end{center}
\end{figure}

For the 230 pb$^{-1}$ analysis, Fig.~\ref{fig:d0ljbt230csec} illustrates 
the distributions for $H_T$
and a multiparameter kinematic discriminant for data, signal
and background expectations.
\begin{figure}[thb]
\begin{center}
\epsfig{figure=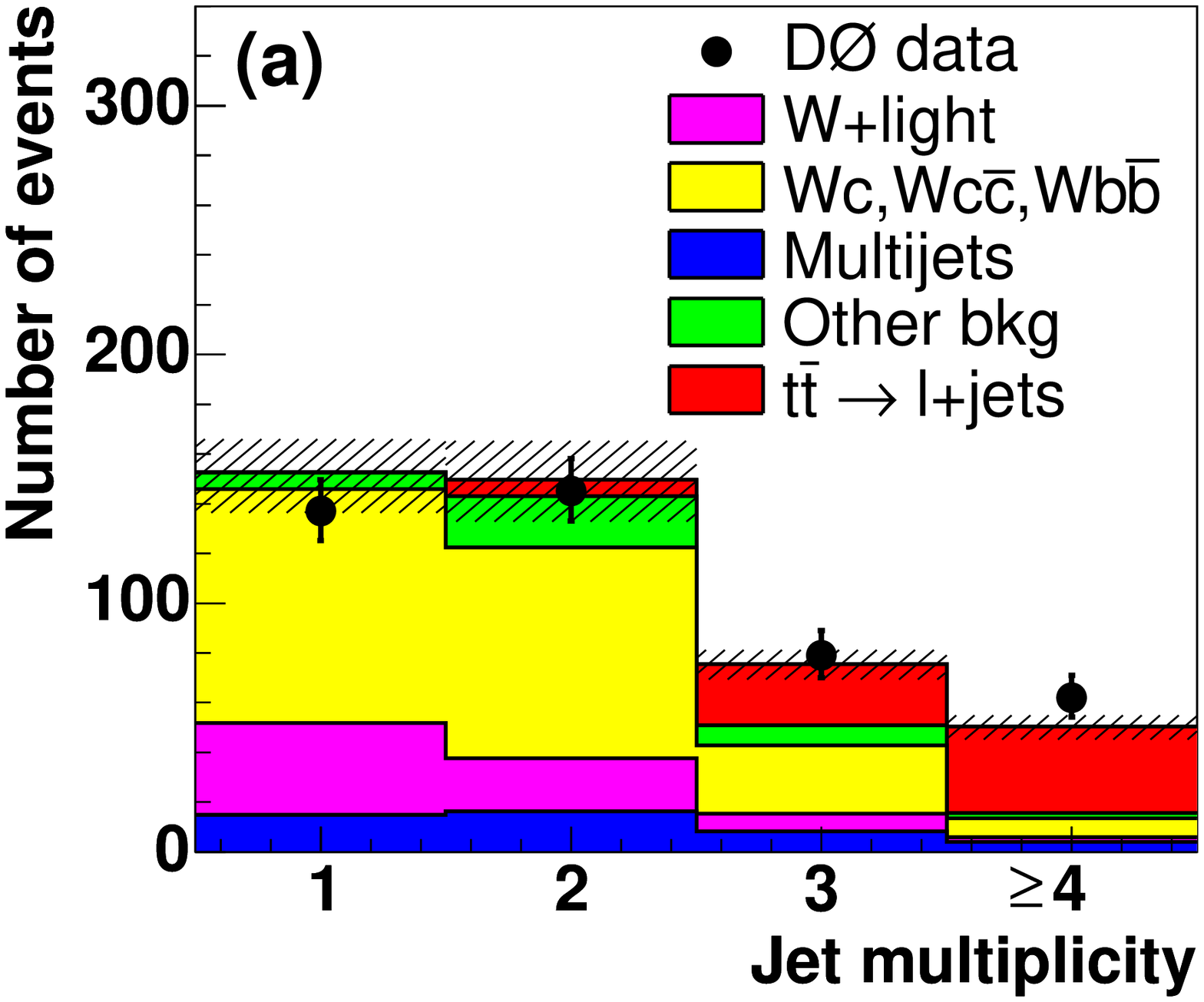,width=5.75cm}
\epsfig{figure=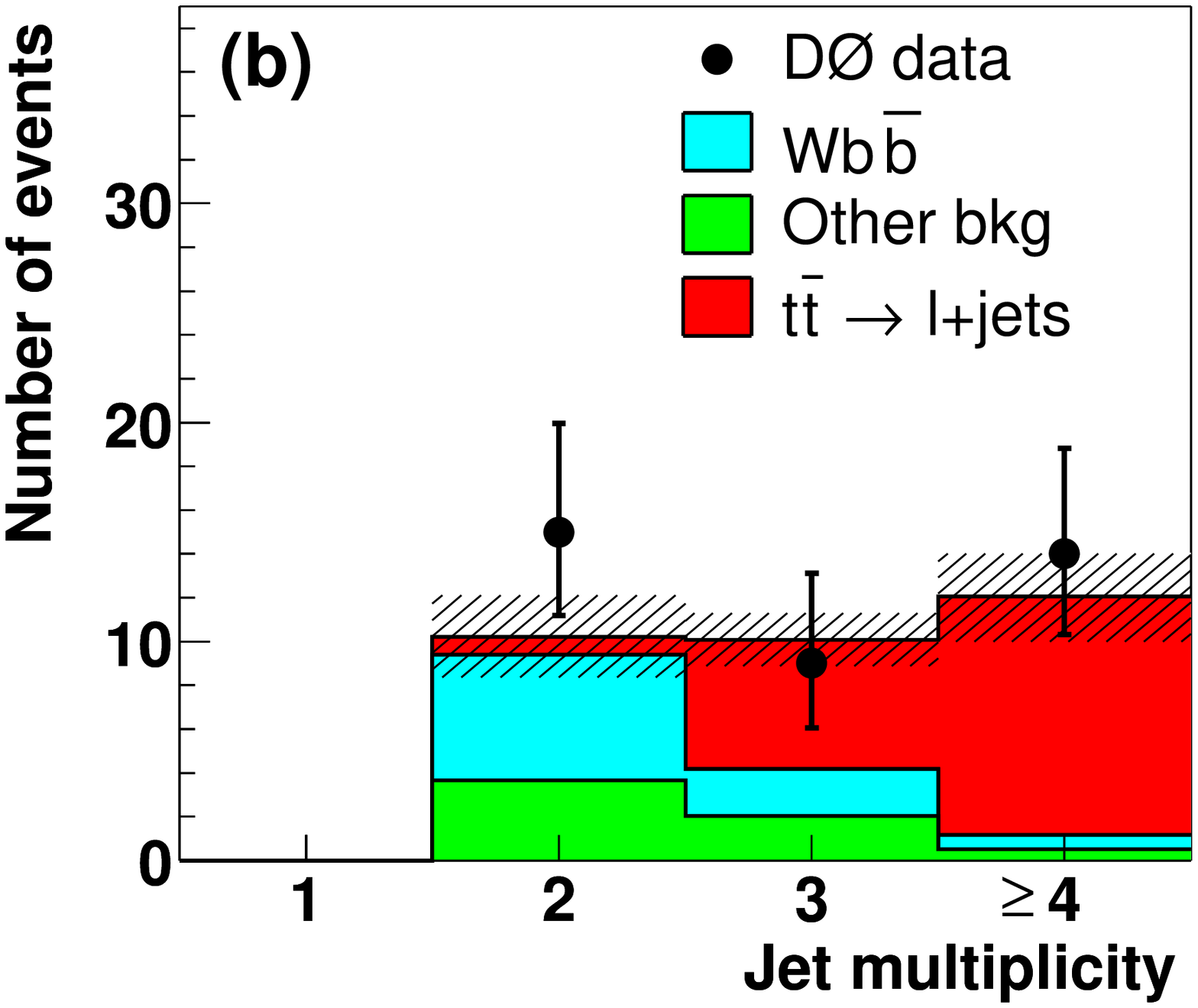,width=5.7cm}
\caption{The jet multiplicity distribution for the \dzero\
425 pb$^{-1}$ $\ell+$jets sample with secondary vertex tag 
for single tagged (left) and double tagged (right) 
events\protect\cite{ljbtag425}.  Data are indicated by points
and error bars, while signal and background expectations are
shown via solid histograms.}
\label{fig:d0ljbt425csec}
\end{center}
\end{figure}
Figure~\ref{fig:d0ljbt425csec} shows the jet multiplicity distribution
for the single tag and double tagged samples in the 425 pb$^{-1}$ 
analysis.  The one-jet and two-jet
bins exhibit good agreement with the expectation, thus validating the
background estimation method.  An excess is observed in the three and four 
jet bins which is attributed to top quark production.  

The measured cross section from the earlier data sample is
$8.6^{+1.6}_{-1.5}(stat+sys)\pm 0.6 (lum)$ pb.  An alternative approach
using an impact parameter tag was also used to calculate the cross 
section.  This method produces very compatible results
with the secondary vertex approach and is actually fairly uncorrelated
with it in terms of event sample.  
As with the soft-lepton tag analysis in 425 pb$^{-1}$, the cross section
was extracted by maximizing a likelihood given the observed number of
events.  Each systematic uncertainty was implemented as a Gaussian which
could alter the fitted cross section.  This provided a cross section 
measurement of $6.6\pm 0.9(stat+sys) \pm 0.4(lum)$ pb.  The measurement
in 900 pb$^{-1}$ uses much the same techniques and gives 
$8.3^{+0.6}_{-0.5}(stat)^{+0.9}_{-1.0}(sys)\pm 0.5 (lum)$ pb.

Substantial uncertainties for all \dzero\ and CDF single lepton mode analyses
arise from tagging efficiency uncertainty, and from the jet energy
scale uncertainty. In addition, the secondary vertex analyses have
significant uncertainty from the background modeling. Both CDF results
have approximately a five percent lepton identification uncertainty. 
The 425 pb$^{-1}$ \dzero\  measurement has significant uncertainties
from $pdf$'s, factorization scale and heavy quark ($b$ and $c$) mass.

%% file: ttbar_csec/all-jets.tex
The all-jets channel is the most copious \ttbar\ final state with a 
branching fraction of $\sim$46\%.
Without any energetic neutrinos in the final state, the all-jets mode
is also the most kinematically constrained, and this allows a full
reconstruction of the $\ttbar$ signal. 
However, the signal suffers from an overwhelming background from QCD 
multijet production, with a cross section many orders of magnitude larger, 
making the extraction of $\ttbar$ events extremely 
challenging. To improve the signal-to-background ratio, 
$S/B$, a set of 
requirements exploiting the kinematic and topological 
characteristics of standard model $\ttbar$ events is 
applied to the data. Unlike the
$\ell+$jets topological selection described in Section~\ref{sec:ljTopo},
kinematic properties alone are not sufficient to isolate the top quark.
They are coupled
with the use of $b-$tagging to separate the heavy-flavor poor
background from the \ttbar\ signal. 
The CDF and D\O\ experiments have previously measured the 
$\ttbar$ production cross section ($\sigma_{\ttbar}$) in the all
hadronic channel in Run~I\cite{d0r1alljetscsec,cdfr1alljetscsec}. 
The strategies employed by their measurements in Run~II 
\cite{cdfr2alljetscsec,d0r2alljetscsec} rely heavily on the
techniques developed during their Run~I
analyses. Table~\ref{tab:all-jets-cscuts} gives an overview of the 
various selection requirements applied by the CDF and \dzero\ 
analysis discussed in this section.\\

\begin{table}[ht]
\begin{center}
\tbl{Selection cuts for Run II all-jets 
cross section measurements by CDF\protect\cite{cdfr2alljetscsec} 
and \dzero\ \protect\cite{d0r2alljetscsec}.}
{\begin{tabular}{lccc}\toprule
cuts & CDF  &  \dzero\  \\
\colrule
preselection & & \\ \hline
primary event vertex ($Z_{vert}<$) & 60 cm & 50 cm\\
lepton veto & yes & yes \\
$p_T^{jets}>$ (GeV) & 15  & 15  \\
$\eta^{jets}<$ & 2 & 2.5 \\
$N_{jets}\geq$ & 6 & 6 \\
$H_T^{jets}>$ (GeV) & 125 & 90 \\\hline
topological selection & & \\ \hline
variables & $H_T^{3j}$, $\sum E_T$ $C$,$A$  & $H_T$, $A$, $E_T^{5,6}$ \\
          &  & $<\eta^2>$, $M_{min}^{3,4}$, $\mathcal{M}$ \\
	  & (kinematic selection) & (neural net selection) \\ \hline
$b-$ tagging & yes & yes \\
\botrule
\end{tabular}
\label{tab:all-jets-cscuts}}
\end{center}
\end{table}

\subsubsection*{Discriminating event characteristics}

\begin{figure*}[!h!tbp]
\begin{center}
\epsfig{figure=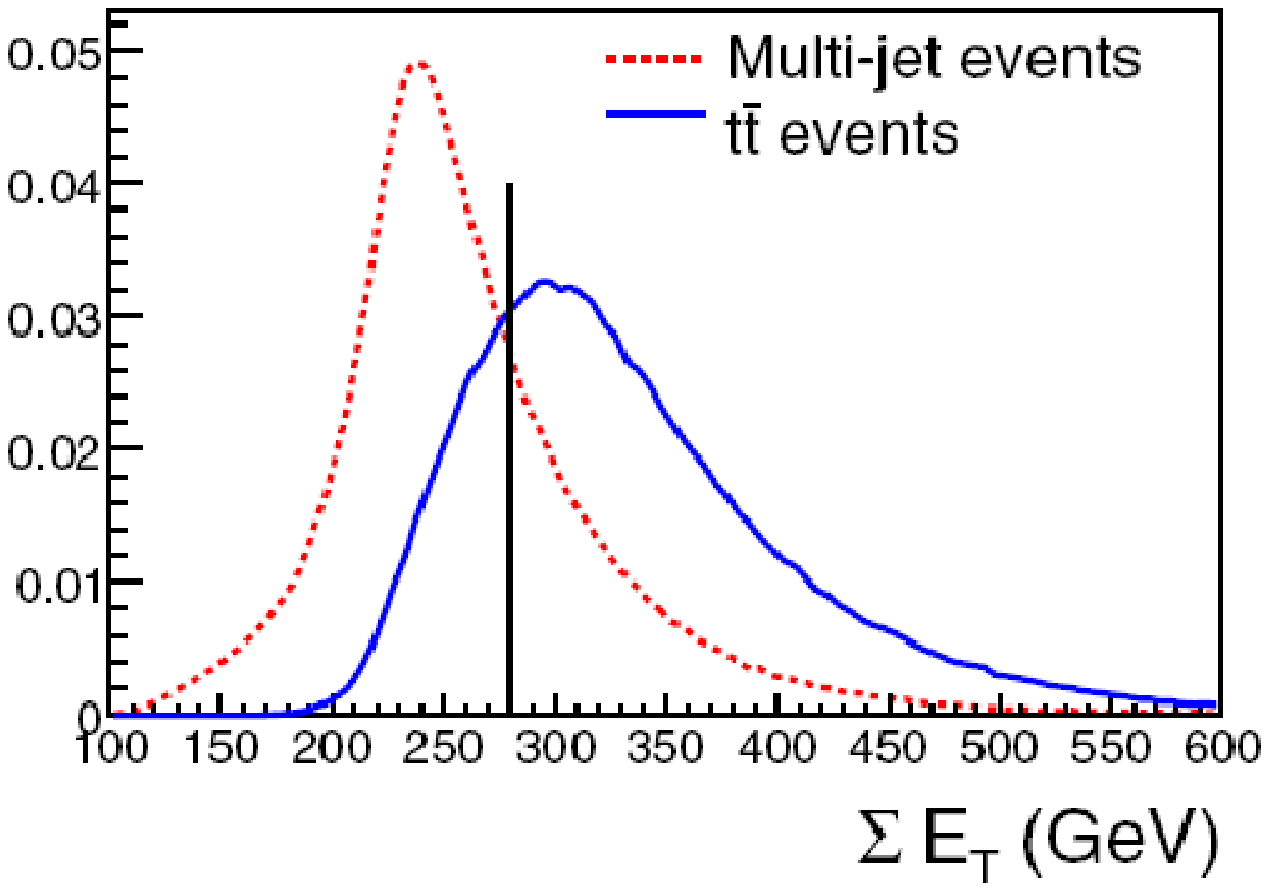,width=0.40\textwidth}
\epsfig{figure=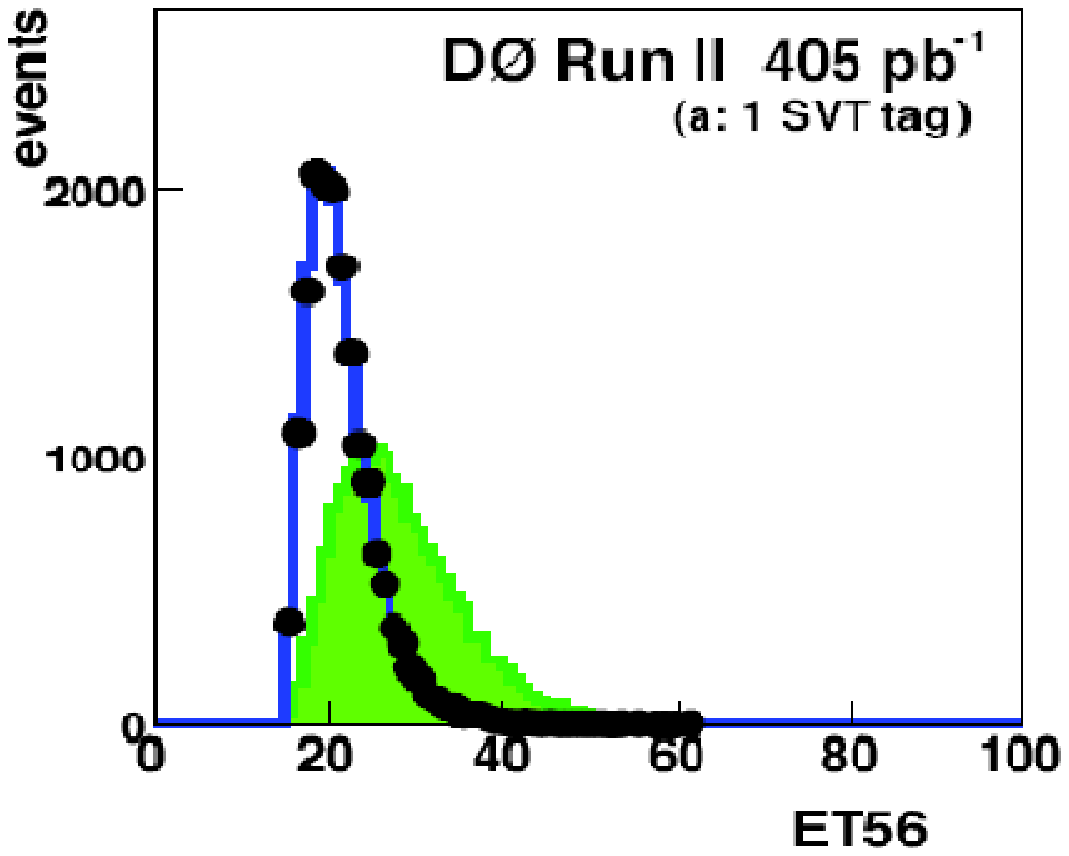,width=0.40\textwidth}
\end{center}
\vspace*{8pt}
\caption{Kinematic variable distributions ($\sum E_T$ (left) used by CDF analysis\protect\cite{cdfr2alljetscsec} and
$\sum E_T^{56}$ (right) used by D\O\ analysis\protect\cite{d0r2alljetscsec}) in multijet data and \ttbar\ Monte Carlo simulation. 
}
\label{fig:CDF-run2-kinem} 
\end{figure*}

In addition to the discriminating variables already discussed 
in section 4.1, the all-jets channels 
also employ some more jet-based variables.  The QCD multijet production 
is dominated by a 
$2\to2$ parton process producing two hard leading jets with extra jets 
produced through gluon radiation. Therefore, the additional jets are expected 
to be softer in QCD background than in \ttbar\ signal. The parameters 
used are $H_T^{3j}$, the scalar
sum of all jet $p_T$'s except the two leading jets, and $E_T^{5,6}$, 
the geometric mean of the transverse energies of the 5th and 6th leading 
jets. The properties which are typical for the 
top quark event structure, owing to the presence of $W$-bosons and $b$-quarks, 
are also used. The variables used are the mass-likelihood
$\mathcal{M}$ and the second-smallest dijet 
mass in the event $M_{min}^{2nd}$.  $\mathcal{M}$ is defined as
$\mathcal{M}=(M_{W_1}-M_W)^2/\sigma^2_{M_W}+(M_{W_2}-M_W)^2\sigma^2_{M_W}+(m_{t_1}-m_{t_2})^2/\sigma^2_{m_t}$
where $M_{W_1}$ ($M_{W_2}$) is the mass of the two jets corresponding to the 
$W$ boson from the first (second) top quark, of mass $m_{t_1}$ ($m_{t_2}$).
The parameters $M_W=79$ GeV and $\sigma_{M_{W}}=11$ GeV 
are the central value and resolution of $W$ boson mass peak, obtained from 
\ttbar\ all-hadronic Monte Carlo along with the resolution of the top quark mass,
$\sigma_{m_t}=21$ GeV. $\mathcal{M}$ is calculated for each possible
assignment of jets to the $W$s and $b$-quarks, while only the permutation
with the smallest $\mathcal{M}$ is used as the discriminator. 
$\mathcal{M}$ provides good 
discrimination between signal and background by requiring two jet pairs that are
consistent with the $W$ boson mass, and two $W+$jet pairs that are consistent
with a single top quark mass of any value. The presence of two $W$ bosons in 
\ttbar\ events provides significant rejection against the QCD background.
A further requirement that the two reconstructed top quarks have equal masses
provides some additional discrimination. The discriminating performance of the 
two chosen kinematic variables, $\sum E_T$ and $\mathcal{C}$ 
can be seen in Fig. \ref{fig:CDF-run2-kinem}.

\subsubsection*{Measurement by CDF:}
In Run II, CDF\cite{cdfr2alljetscsec} has performed the 
measurement of the $\ttbar$ cross section ($\sigma_{\ttbar}$) in this channel 
utilizing approximately 311 pb$^{-1}$ of data sample selected by 
dedicated multijet triggers. Events with at least four jets with 
$p_T \geq$10 GeV and large $H_T\geq$125 GeV are selected 
at the trigger level with $S/B$ $\sim$1/3500. 
The offline preliminary signal selection 
requires events with a large jet multiplicity, 
$6 \leq N_{jets} \leq 8$, with $p_T \geq$15 GeV and $|\eta|<$2. 
Application of a veto on isolated high-$p_T$ leptons 
allow to keep the analysis orthogonal to those involving
leptonic channels. After these requirements 364,006 events are
selected for further analysis.

The $\ttbar$ events are modeled with {\sc Pythia} and {\sc Herwig} 
using $m_{t}=$178 GeV.
In order to improve the $S/B$, the analysis employs a 
kinematic and topological event selection based on  
$H_T$, $H_T^{3j}$, $\mathcal{C}$, and $\mathcal{A}$ optimized 
to achieve the maximum signal significance for $\ttbar$ events,
defined as the ratio between the expected signal and the 
statistical uncertainty on the sum of the signal and background. 
The values for the cuts after optimization are: 
$\mathcal{A}+0.005 H_T^{3j} \geq$0.96, $\mathcal{C} \geq$0.78, and $H_T>$280 GeV.
Application of the kinematic selection yields 3315 events in data with 
an efficiency of 6.7$\pm$1.4\% for the $\ttbar$ signal and a 
$S/B$ $\sim$1/25. In order to further 
enrich the sample with top-like events, 
events are required to have at least one $b-$tagged jet 
which leaves 695 events containing 816 $b-$tags, thus reaching
a $S/B$ of about 1/5.

\begin{figure*}[!h!tbp]
\begin{center}
\epsfig{figure=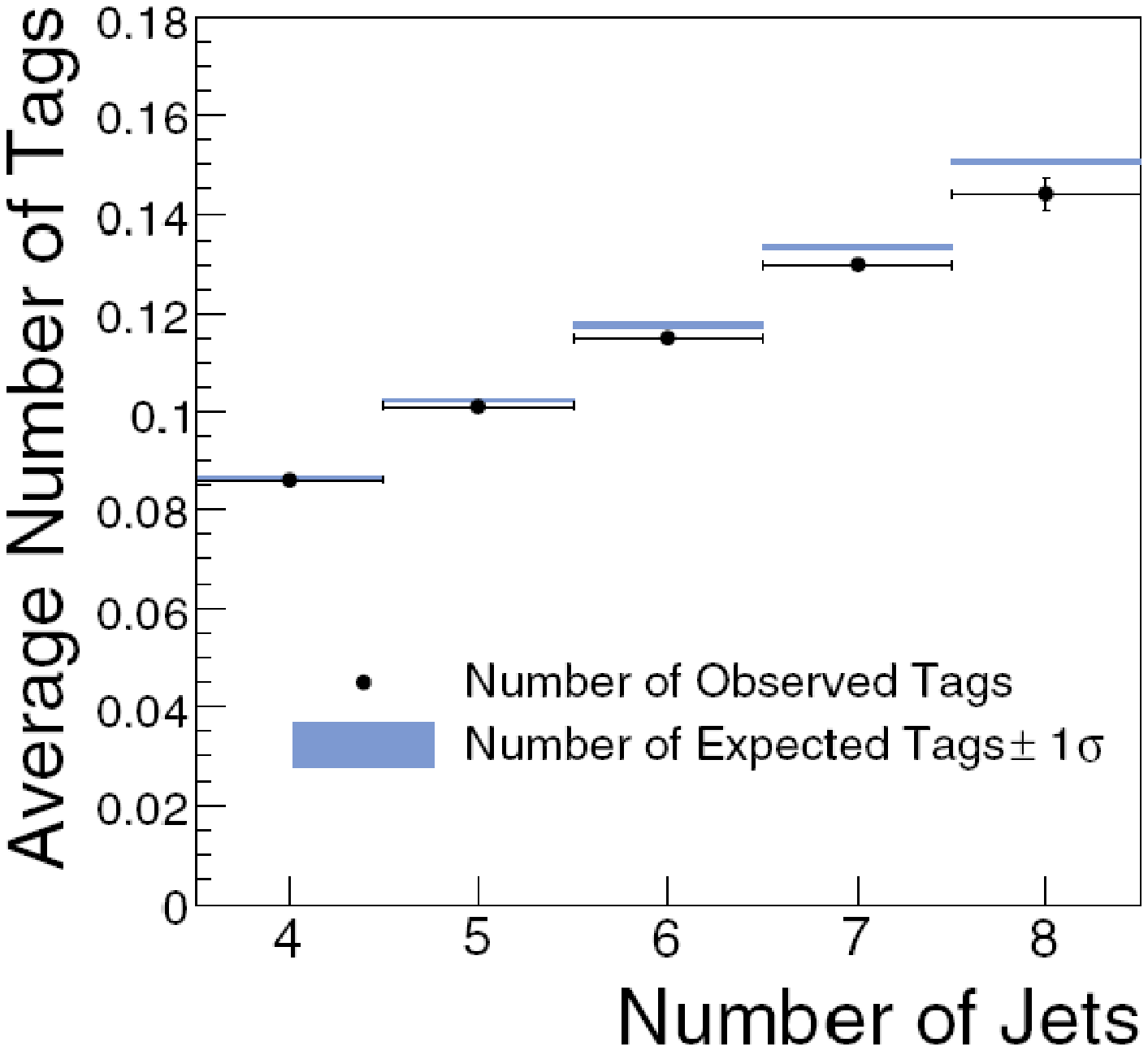,width=0.40\textwidth}
\epsfig{figure=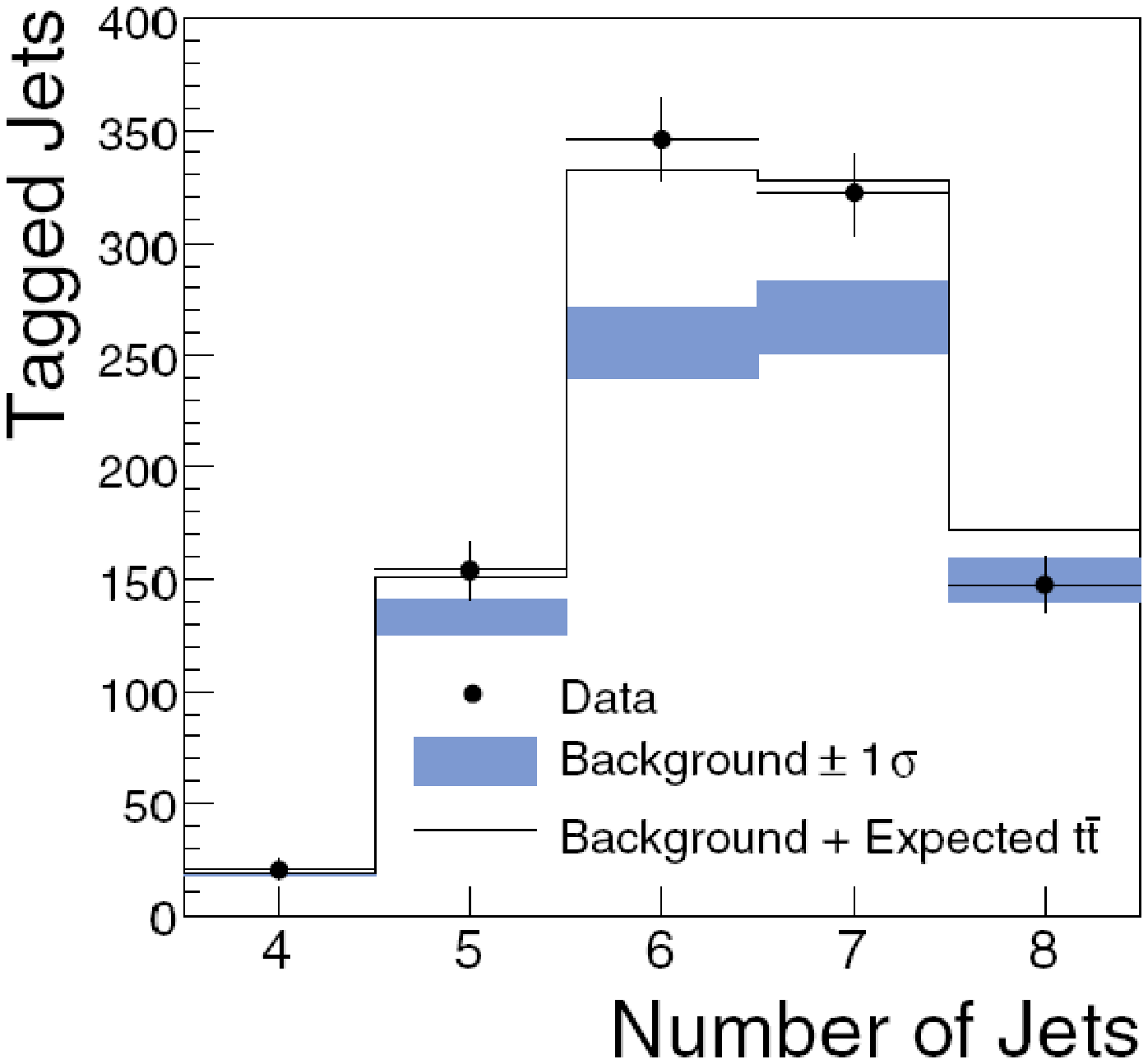,width=0.40\textwidth}
\end{center}
\vspace*{8pt}
\caption{Left: Average number of tags per event observed in the multijet sample before kinematic selection compared with the estimate from the tag rate parameterization in the CDF analysis\protect\cite{cdfr2alljetscsec}. Right: Number of tags observed in multijet data after kinematic selection compared with the expected background in the CDF analysis\protect\cite{cdfr2alljetscsec}. The $\ttbar$ expectation refers to the measured cross-section of 7.5 pb.
}
\label{fig:cdf-run2-311} 
\end{figure*}

The background sources for this final state are due mainly to 
QCD production of heavy quark pairs ($b\bar{b}$ and 
$c\bar{c}$) and false tags from light-quark jets. 
The mistag rate
is evaluated in the exclusive four jet data sample before the kinematic 
selection and is parameterized in terms of jet $E_T$, 
track multiplicity $N_{trk}$, and number of primary vertices 
in the event $N_{vert}$. 
To estimate the background, each multijet event in the signal sample
is weighted by its $b-$tag probability. The sum of these weights 
for the multijet events gives the 
expected number of tags from the non-signal processes.  
Before the kinematic selection, when the multijet sample is still
predominantly composed of background events only, the predicted
and observed number of tagged jets for different jet 
multiplicities agree very well as can be seen in Fig. \ref{fig:cdf-run2-311},
giving confidence in the constructed mistag rate.

The cross section measurement is performed using the total 
number of tagged jets (not events) in order to avoid the 
explicit calculation of the background for double tagged events.
The average  number of tags in a \ttbar\ event passing the 
kinematic selection is $n_{tag}^{ave}=0.846 \pm$ 0.073, measured
in $\ttbar$ simulation by taking into account the 
different tagging efficiencies for $b-$, $c-$ and light-flavored-jets and
correcting for the difference in efficiencies in data and Monte Carlo events.
After application of the kinematic selection, a total of $N_{obs}=$816
candidate tags are observed, whereas the expected number of tags
from background sources in the signal region amounts to 
$N_{bkg}$=684$\pm$38 tags, after correcting for the presence of 
$\ttbar$ events in the pretag sample.
The resulting excess in data of tagged jets in the signal region is
ascribed to \ttbar\ production. The \ttbar\ production cross section is
determined by
$\sigma_{\ttbar}=\frac{N_{obs}-N_{bkg}}{\epsilon_{kin} \times n_{tag}^{ave} \times \mathcal{L}}$
and is found to be $\sigma_{t\bar{t}}=7.5 \pm 2.1(stat.)^{+3.3}_{-2.2}(sys)^{+0.5}_{-0.4} (lum)$ 
pb for $m_{top}$=178 GeV. 
The dominant systematic uncertainty of $\sim$20\% arises from 
the dependence of selection efficiency on the jet energy scale. 
In Fig. \ref{fig:cdf-run2-311}, the distribution of the 
number of observed tags and background is compared to the $\ttbar$ 
signal expectation assuming the measured cross-section 7.5 pb.

\subsubsection*{Measurement by D\O\ :}
In Run II, \dzero\ \cite{d0r2alljetscsec} has
measured the $\ttbar$ cross section ($\sigma_{\ttbar}$) in this channel 
based on $\sim$405 pb$^{-1}$ of data collected using specific
multijet trigger. 
The analysis proceeds on a methodology similar to that used in 
Run~I\cite{d0r1alljetscsec}.
In addition to the single-tagged events, the measurement has
been extended to the double-tagged events. The analysis proceeds
with the preselected data sample composed of events with $\geq$6 reconstructed 
jets with $p_T \geq$15 GeV, $|\eta|<$2.5 and $H_T$ $>$90 GeV. 
The bulk of the background is 
rejected by requiring the presence 
of secondary vertex tagged jets
In order to maximize acceptance and sensitivity, separate samples with
one tagged jet, or two tagged jets were analyzed. 
In the former, a tight requirement ($L_{xy}/\sigma_{L_{xy}}>7$) 
was placed on the tag to control backgrounds. In the latter case,
a looser requirement ($L_{xy}/\sigma_{L_{xy}}>5$) was used to
maximize efficiency.

Further suppression of the multijet background is achieved 
by applying a neural network ($NN$) selection based on six 
discriminating kinematic variables: $H_T$, $\mathcal{A}$, $E_T^{5,6}$, 
$<\eta^2>$, $M_{min}^{3,4}$, $\mathcal{M}$, taken from the equivalent
Run~I analysis\cite{d0r1alljetscsec}.

\begin{figure*}[!h!tbp]
\begin{center}
\epsfig{figure=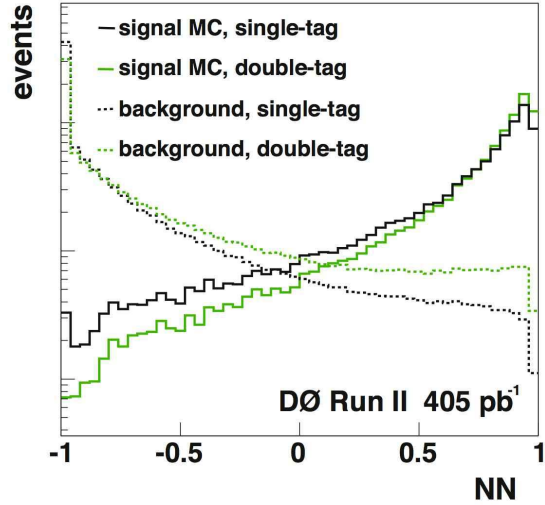,width=0.6\textwidth}
\end{center}
\vspace*{8pt}
\caption{The neural network (NN) discriminant for \ttbar\  signal and multijet background in single- and double-tag analysis by \dzero\ \protect\cite{d0r2alljetscsec}.}
\label{fig:D0-NN} 
\end{figure*}

The $NN$ is trained to force its output near 1 for \ttbar\ events
and near -1 for QCD multijet events, using the multilayer perceptron
network in the {\sc Root} analysis program\cite{ROOT}. 
The very small $S/B$ in the untagged data sample provided the background
input for the training of NN, while 
\ttbar\ Monte Carlo simulation using {\sc Alpgen} and {\sc Pythia} with 
$m_t=$175 GeV was used for the signal.
Fig. \ref{fig:D0-NN} shows the performance of 
the NN in discriminating \ttbar\ signal from the
multijet background for both single- and double-tag samples. 
The NN distributions vary for the two samples due to the variation of 
their flavor content. Overall, the NN displays significant discriminating 
power for the single- and double-tag events, although it exhibits a better
discrimination for single-tag samples.

\begin{figure*}[!h!tbp]
\begin{center}
\epsfig{figure=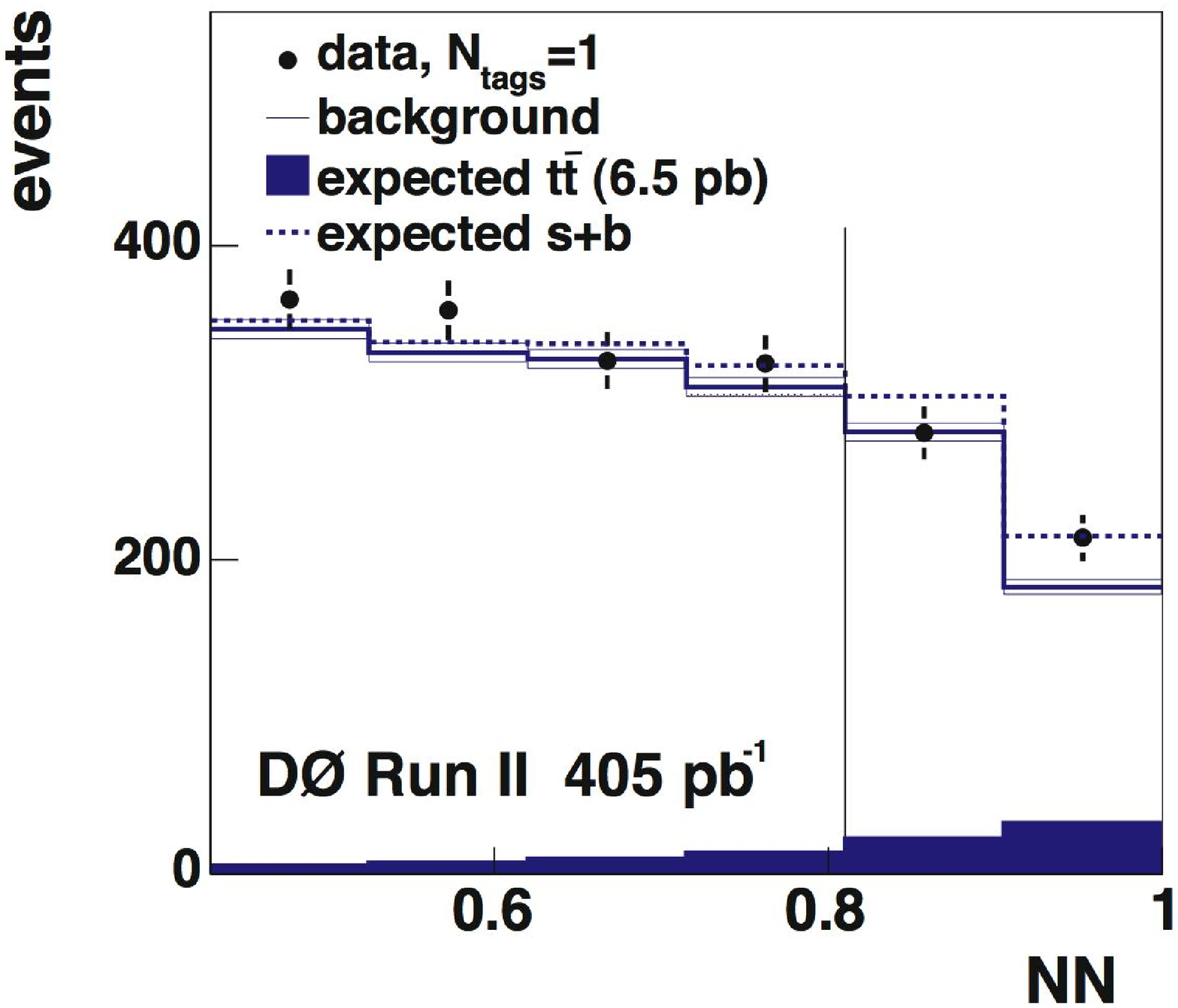,width=0.40\textwidth}
\epsfig{figure=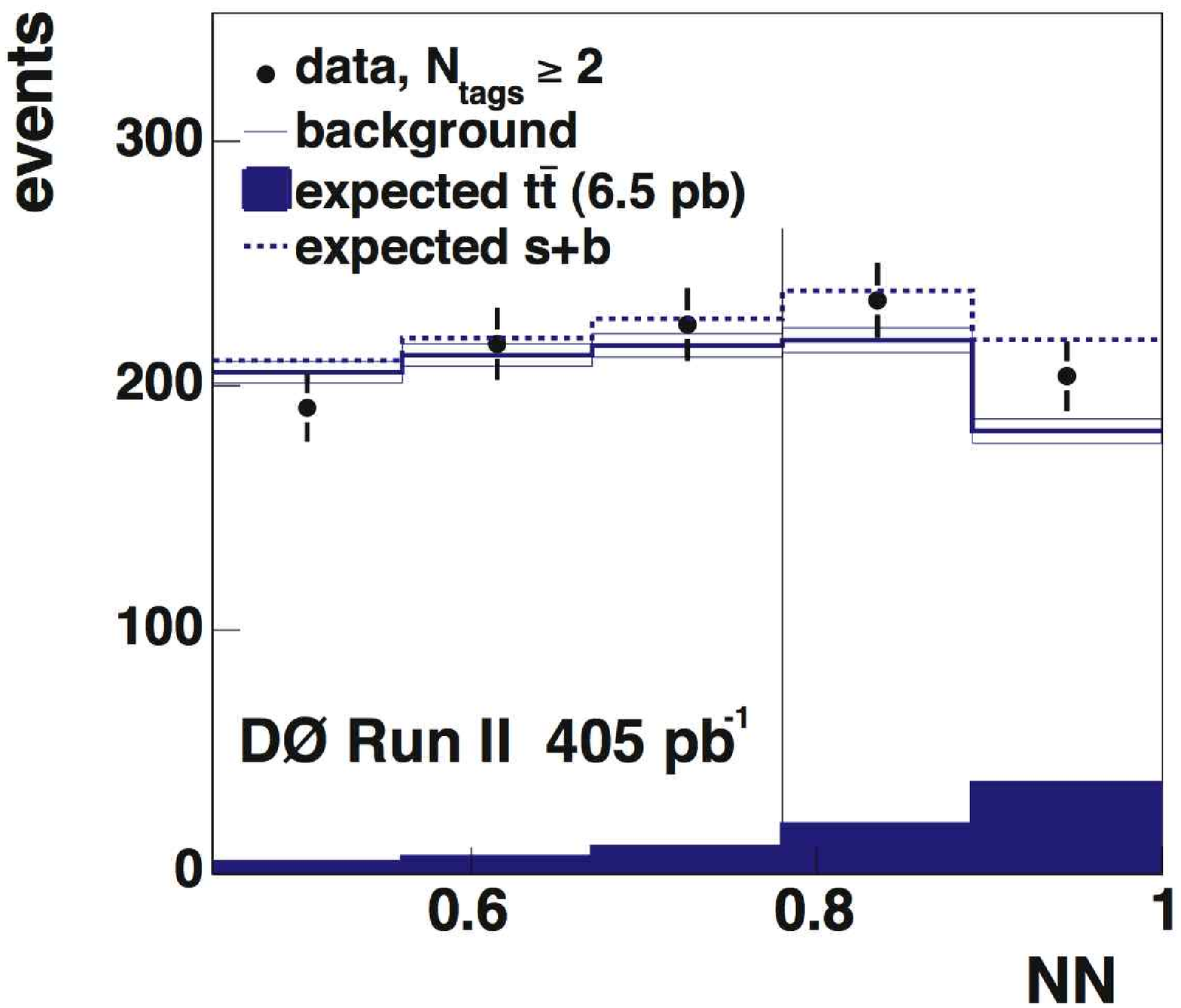,width=0.40\textwidth}
\end{center}
\vspace*{8pt}
\caption{The distribution of NN output for single- (double)-tag events in the D\O\ analysis\protect\cite{d0r2alljetscsec}. Shown are the data (points), background (the line histogram), signal (filled histogram) and signal+background (dashed histogram). The vertical line represents the used cut of NN$>$0.81 (0.78).
}
\label{fig:D0-run2-406} 
\end{figure*}

The top quark cross section is extracted from
the output of the $NN$. 
The background dominated preselected sample is used to estimate 
the background. For the loose and tight tags, the mistag rate  
is derived from the data with $N_{tags} \leq$1. 
It is parameterized in terms of
the $p_T$, rapidity $y$, azimuthal 
angle $\phi$ of the jet and the location of primary vertex along the beam 
direction $z_{PV}$, in four different $H_T$ bins. 
Fig. \ref{fig:D0-run2-406} show the distributions of the 
$NN$ output for the data, the Monte Carlo simulation prediction for 
$\sigma_{t\bar{t}}=$6.5 pb, the predicted background, and the 
predicted signal$+$background. It can be seen that background keeps
dominating over the signal even at large values of $NN$. 
The output of the NN is used to select the sample enriched in 
\ttbar\ signal by applying the cut $NN \geq$0.81 (0.78) for 
the single (double)-tag analysis, optimized to minimize the fractional
statistical error on the cross section measurement.

After the NN requirement, 495 (439) events are observed while the predicted number
of background 
events is 464$\pm$4.6 (400$^{+7.3}_{-6.2}$) in the single (double)-tag 
analyses. The kinematic selection efficiency times branching ratio for 
the $\ttbar$ signal 
in the two analyses are
$\epsilon_{\ttbar} = 0.0242 ^{+0.0049}_{-0.0058}(0.0254 ^{+0.0065}_{-0.0070})$, respectively. 
Based on the clear excess of observed events over the predicted background,
D\O\ has measured a cross-section of 
$\sigma_{t\bar{t}}=4.1^{+3.0}_{-3.0} (stat.)^{+1.3}_{-0.9}(sys) \pm 0.3(lum)$ pb 
for the single-tag analysis and 
$\sigma_{t\bar{t}}=4.7^{+2.6}_{-2.5} (stat.)^{+1.7}_{-1.4}(sys) \pm 0.3(lum)$ pb 
for the double-tag analysis. A combined cross-section measurement of
$\sigma_{t\bar{t}}=4.5^{+2.0}_{-1.9} (stat.)^{+1.4}_{-1.1}(sys) \pm 0.3(lum)$ pb 
is obtained for a $m_t=$175 GeV.
The major systematic uncertainty in the analysis arises from the dependence 
of selection efficiency on the jet calibration and identification.\\

%% file: single_top/SingleTop.tex
Twelve years after the discovery of the top quark via strong pair production,
the first evidence of electroweak production of single top quarks has been
reported by the \dzero\ collaboration\cite{STD0evidencePRL}.  This search is
much more challenging than $t\overline t$ production due to smaller  cross
sections and larger backgrounds.  A general overview of single top quark
production is given in  Sec.\ref{EWKtop}. Here we discuss the signatures and
backgrounds associated with the analyses and review the results by  both the
CDF and \dzero\  experiments.  We review these results with  a historical
perspective and begin with early single  top quark search analyses which
resulted in an upper bound on the production cross section (section
~\ref{secSTsearch}).  Later, in section~\ref{secSTevidence}, we review the
analyses which led to the  evidence for the single top quark production.

\begin{figure*}[!h!tbp]
\begin{center}
\epsfig{figure=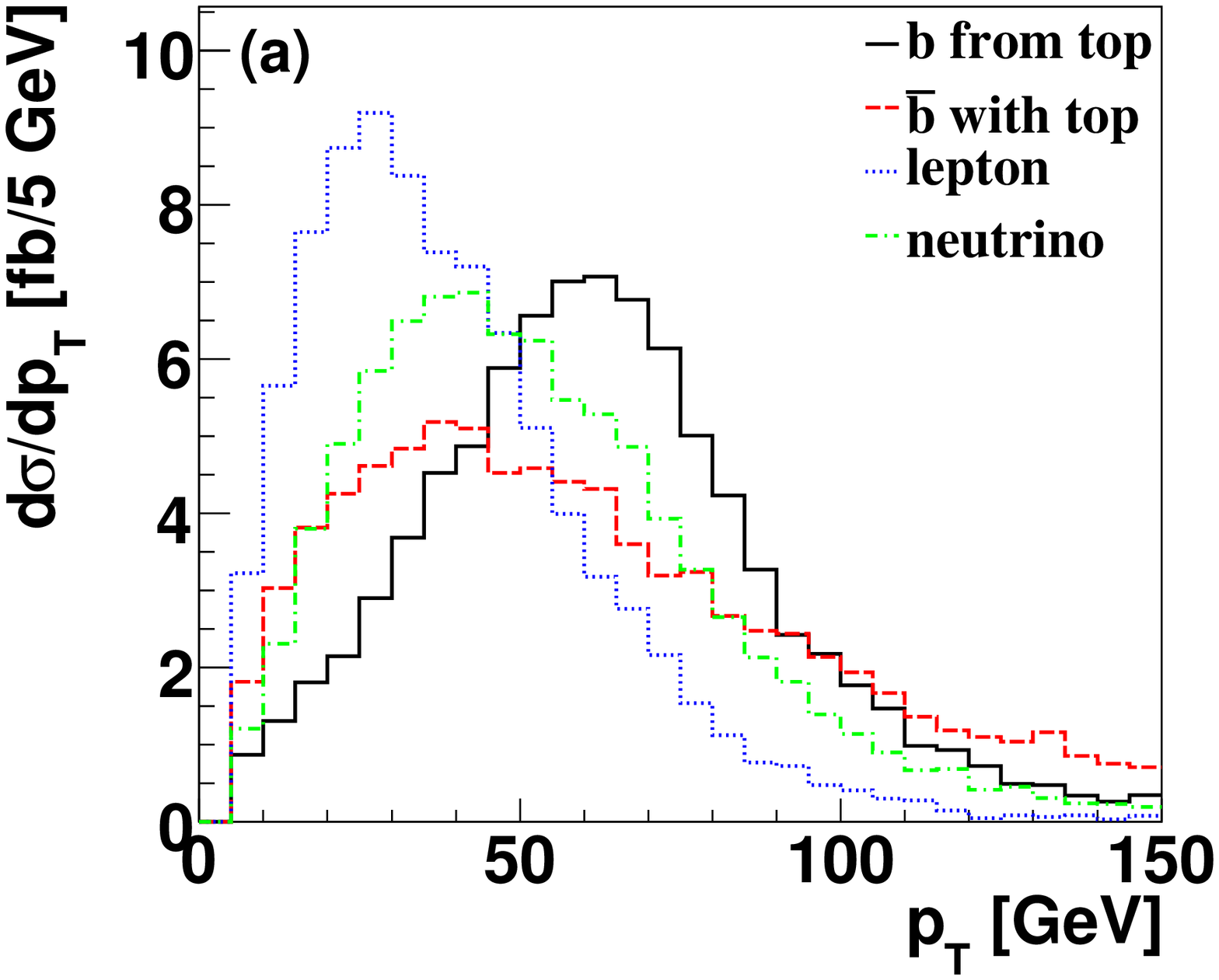,width=0.40\textwidth}
\epsfig{figure=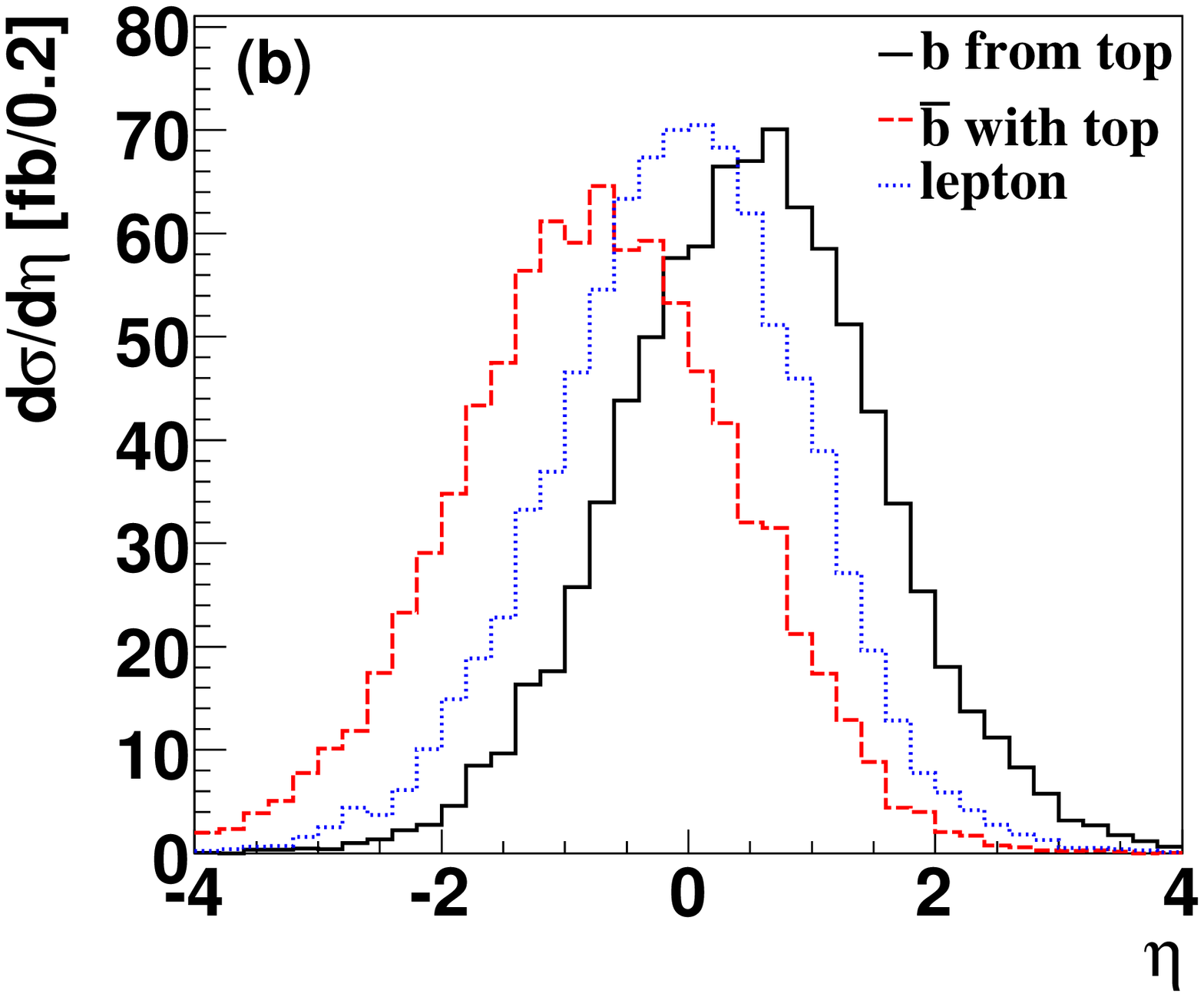,width=0.40\textwidth}
\epsfig{figure=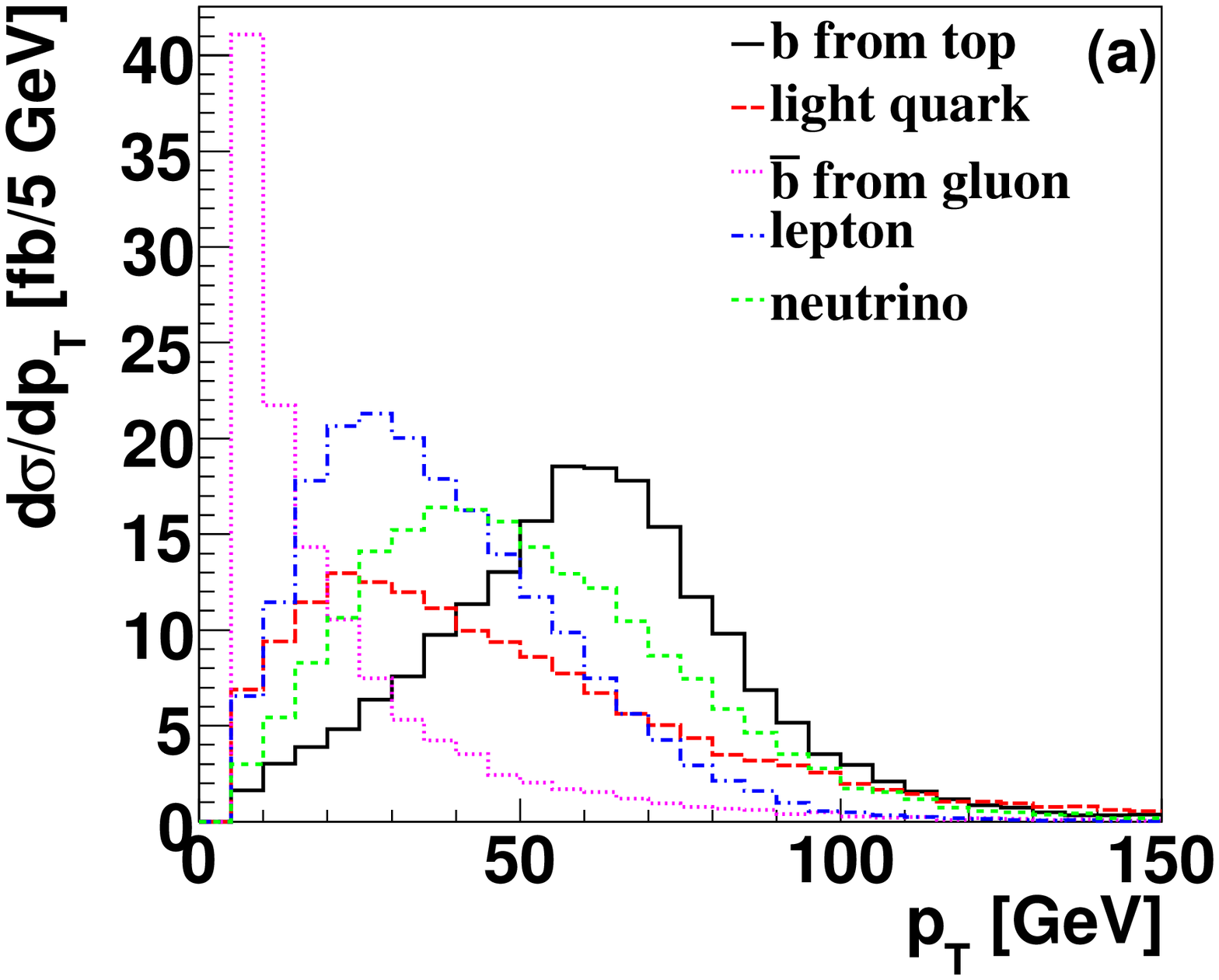,width=0.40\textwidth}
\epsfig{figure=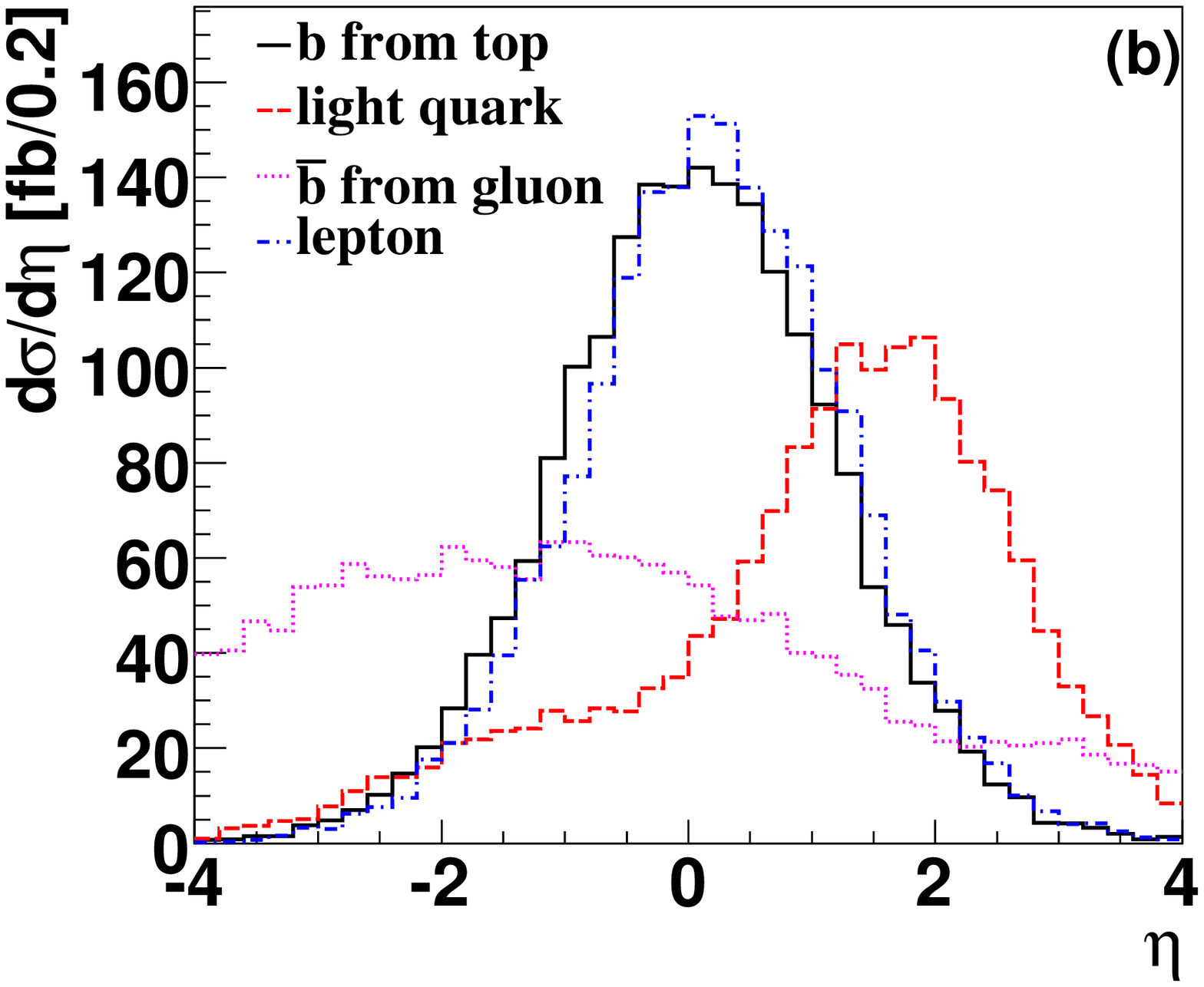,width=0.40\textwidth}
\caption{Distributions of transverse momenta (a) 
and pseudorapidity (b) for the final state partons in 
$s$-channel single top quark events (top row) and for $t$-channel single top
quark events (bottom row).
The histograms only include the final state of $t$, not $\overline{t}$
~\protect \cite{STD0prdone}.}
\label{fig:STkine} 
\end{center}
\end{figure*}

\paragraph{Signal Kinematics}
Figure~\ref{fig:STkine} shows some of the features of the
single top quark events produced in the $s$- and $t$-channels. The plots of
transverse momentum and the pseudorapidities of objects in the single top
quark events shown in this figure reveal that for the $t$-channel process
the $b$-quark has very low transverse momentum and is peaked at large
pseudorapidity (forward region). Thus the $b$-quark in the $t$-channel is
difficult to detect.

\paragraph{Background processes}

Associated production of a $W$ boson with jets ($W+$jets) and  pair
production of top quarks (\ttbar) are the two most dominant  processes which
mimic  the single top quark signatures.
Events with heavy flavor produced in association with a $W$
boson give rise to irreducible backgrounds. These are $Wb \overline b$ and
$Wb\overline b g$ for the $s$- and $t$-channel respectively. In general, the
$t\overline t$ events have  larger jet activity, but they can result in
signatures  similar to the single top quark in case the jets are merged,
mis-reconstructed or are outside the fiducial volume. Leptons can also be
lost in the detector cracks or lie outside the fiducial volume considered.
\ttbar\ decays can look very similar to single top quark events when one $W$
boson decays to $e\nu$ or $\mu\nu$ and the other decays to $\tau\nu$.
Additional multijet background comes from events containing a mis-identified
isolated lepton ($e$ or $\mu$) associated with  hadronic
jets. $Z/\gamma^*+$jets contribute to the backgrounds if  one of the two
leptons from $Z$ decaying to $e^+e^-$ or $\mu^+\mu^-$ is lost.  Diboson
events ($WW,\ WZ$) also contribute to the background, though  at a much
smaller  level.

\paragraph{Modeling single top quark and backgrounds}

The two main sources of background are examined: $t\overline t$ production
and non-top quark backgrounds. The non-top quark backgrounds  include
$W$+jets and mis-identified multijet events.  $W$+jets (including heavy
flavor process $\bar{q}q^\prime\rightarrow Wg$ with $g\rightarrow b\bar{b}$
or  $g\rightarrow c\bar{c}$, and $gq\rightarrow Wc$) are included in their
relative proportions estimated using {\sc Alpgen}\cite{alpgen} and normalized
to the data.

Both the CDF and \dzero\  collaborations have performed searches for  single
top quark production in the data samples gathered in Run~I
(1992-1996)\cite{d0runI,STAcosta}.  Data collected during the first few years
of RunII have led to updated analyses by CDF and \dzero.   All of these
measurements have led to upper limits on the production cross section of the
single top quark as listed in Table~\ref{tabone}.  These are reviewed below.

\begin{table}[ht]
\begin{center}
\tbl{95\% C.L. upper limit on single top quark production cross section (in pb).}
{\begin{tabular}{lccccccccc}\toprule
 & Theory & \multicolumn{2}{c}{Run I } & \multicolumn{6}{c}{Run II } 
\\
&       & CDF & \dzero\  &  \multicolumn{3}{c}{CDF} & \multicolumn{3}{c}{\dzero\ } \\
 \colrule
   &                 & & & $Q\times\eta$ & NN & LD &  Cut  & NN &  LD      \\
Luminosity (pb$^{-1}$) & & & 160 & 695 & 955 & 230 & 230 & 230 \\

 s-channel  &0.88 ($\pm$8\%)   &  $<$ 18 &  $<$ 17 & $<$ 14 & $<$ 3.2 &  0.1$^{+0.7}_{-0.1}$ & $<$ 10.6  & $<$ 6.4 &  $<$ 5.0 \\
  t-channel & 1.98 ($\pm$11\%) & $<$ 13 &  $<$ 22 & $<$ 10 & $<$3.1 &  0.2$^{+0.9}_{-0.2}$ 
& $<$ 11.3 &  $<$ 5.0  & $<$ 4.4 \\
\botrule
\end{tabular}
\label{tabone}}
\end{center}
\end{table}

\subsection{Search strategy ~\label{secSTD0search}}

In this section we review the searches for the single top quark by the CDF and
\dzero\  experiments in Run~II. The CDF  experiment published a search for
the single top quark using a data  sample corresponding to an integrated
luminosity of 160~pb$^{-1}$\cite{RunIIcdf_result} and had provided updated
results with 600-955~pb$^{-1}$ of data in the year 2006. The \dzero\
experiment  performed a search for the single top quark using 230~pb$^{-1}$ of
data\cite{STD0prdone,STPLBD0}.

The strategy of the analyses pursued during Run~II is as follows:  First a
set of events with the signal topology is selected. Next, they are separated
into independent  sets based on the flavor of the lepton ($e$ or $\mu$) and
the number of $b$-quark jets (exactly one $b$-jet or two or more $b$-jets).
The different sets are analyzed independently  by statistical techniques to
enhance the signal.  They are then combined in the final stage. Finally, in
the absence of any significant observed signal, binned likelihood fits are
performed on the outputs of the statistical analysis methods to obtain limits
on the production  cross section for single top quarks.

The single top quark search starts with the selection of events with  at
least one high $p_T$ lepton (electron or muon), significant  \met\ and
between two to four jets.  At least one of the jets is $b-$tagged to separate
signal events with heavy  flavor from the $W+$light jet background event.
Separation of  samples into $s$-channel events and  $t$-channel searches  is
accomplished by requiring at least one untagged jet in the  $t$-channel
analysis. For both the channels, selected samples are further  subdivided
into four orthogonal sets based on the flavor of the lepton ($e$ or $\mu$)
and the number of $b$ quark jets: ``single tagged'' (with exactly one
$b$-tagged jet) or ``double tagged'' (two or more  $b$-tagged jets)
samples. At this initial event selection the  background is still expected to
be about 90\% of the sample.

\subsubsection{Collection of input variables}
To further reduce the backgrounds, both kinematic (object and global event)
and topological information is  used.  Some of the many variables which are
used  for discriminating against the backgrounds are: jet $p_T$  for
different jets; $b-$tag information of the jet; $H$ (total energy); $H_T$
(total transverse energy); $M$ (invariant mass); 
$M_T$  (transverse mass); summing
over various objects in the event; jet-jet separation; jet pseudorapidity
($t$-channel); top quark spin; etc. They are selected based on  extensive
analysis of the  Feynman diagrams of signals and
backgrounds\cite{STvariables}. The list of variables used by  \dzero\
analyses is shown in Table~\ref{tab:STvariables} (CDF
analyses use a subset). The best jet is defined as the jet which together
with the $W$ boson leads to  an  invariant mass  closest to the top quark
mass of 175 GeV.  In the $s-$channel ($t-$channel) analysis, the  top quark
is reconstructed from the $W$ boson and the `best jet' (the leading
$b-$tagged jet).  The $W$ boson is reconstructed from the isolated lepton and
the  \met. The z-component of the neutrino momentum is calculated  by
constraining the lepton and neutrino to the $W$ boson mass, and  the solution
with smaller z-component of the neutrino momentum is chosen  from the two
possible solutions.

\begin{table*}[!h!tbp]
\begin{center}
\tbl{List of discriminating variables.  A check mark in the final four
columns indicates in which  signal-background pair of the neural net analysis
the variable is used~\protect \cite{STD0prdone}.} 
{\begin{tabular}{lp{0.60\textwidth}cccc} \toprule
\multicolumn{6}{r}{Signal-Background Pairs} \\
& & \multicolumn{2}{c}{$s$-channel} &  \multicolumn{2}{c}{$t$-channel} \\
\multicolumn{1}{c}{Variable}&\multicolumn{1}{c}{Description} &      $Wb\overline b$  & ${t\overline t}$ &  $Wb\overline b$ & ${t\overline t}$ \\
\colrule
\multicolumn{6}{c}{\bf{Individual object kinematics}} \\
$p_T({\rm jet1}_{\rm tagged})$     & 
Transverse momentum of the leading tagged jet     & $\surd$ & $\surd$ & $\surd$ & --- \\            
$p_{T}({\rm jet1}_{\rm untagged})$ & 
Transverse momentum of the leading untagged jet   & --- & --- & $\surd$ & $\surd$ \\            
$p_{T}({\rm jet2}_{\rm untagged})$ & 
Transverse momentum of the second untagged jet    & --- & --- & --- & $\surd$ \\            
$p_{T}({\rm jet1}_{\rm non-best})$ & 
Transverse momentum of the leading non-best jet   & $\surd$ & $\surd$ & --- & --- \\            
$p_{T}({\rm jet2}_{\rm non-best})$ & 
Transverse momentum of the second non-best jet    & $\surd$ & $\surd$ & --- & --- \\            
\multicolumn{6}{c}{\bf{Global event kinematics}} \\
$\sqrt{\hat{s}}$ &
Invariant mass of all final state objects 
                                       & $\surd$ & --- & $\surd$ & $\surd$ \\    
$p_T({\rm jet1},{\rm jet2})$        & 
Transverse momentum of the two leading jets& $\surd$ & --- & $\surd$ & --- \\            
$M_T({\rm jet1},{\rm jet2})$        & 
Transverse mass of the two leading jets     & $\surd$ & --- & --- & --- \\
$M({\rm alljets})$           & 
Invariant mass of all jets           & $\surd$ & $\surd$ & $\surd$ & $\surd$ \\ 
$H_T({\rm alljets})$         & 
Sum of the transverse energies of all jets      & --- & --- & $\surd$ & --- \\
$p_T({\rm alljets}-{\rm jet1}_{\rm tagged})$ & 
Transverse momentum of all jets excluding the leading tagged jet    & --- & $\surd$ & --- & $\surd$ \\
$M({\rm alljets}-{\rm jet1}_{\rm tagged})$ & 
Invariant mass of all jets excluding the leading tagged jet   & --- & --- & --- & $\surd$ \\  
$H({\rm alljets}-{\rm jet1}_{\rm tagged})$ & 
Sum of the energies of all jets excluding the leading tagged jet & --- & $\surd$ & --- & $\surd$ \\ 
$H_T({\rm alljets}-{\rm jet1}_{\rm tagged})$ & 
Sum of the transverse energies of all jets excluding the leading tagged jet         & --- & --- & --- & $\surd$ \\ 
$M(W,{\rm jet1}_{\rm tagged})$ & 
Invariant mass of the reconstructed top quark using the leading tagged jet & $\surd$ & $\surd$ & $\surd$ & $\surd$ \\
$M({\rm alljets - jet_{best}})$ & 
Invariant mass of all jets excluding the best jet          & --- & $\surd$ & --- & --- \\            
$H({\rm alljets}-{\rm jet}_{\rm best})$ & 
Sum of the energies of all jets excluding the best jet    & --- & $\surd$ & --- & --- \\      
$H_T({\rm alljets}-{\rm jet}_{\rm best})$ & 
Sum of the transverse energies of all jets excluding the best jet     & --- & $\surd$ & --- & --- \\
$M(W,{\rm jet}_{\rm best})$ & 
Invariant mass of the reconstructed top quark using the best jet  & $\surd$ &
--- & --- & --- \\             \multicolumn{6}{c}{\bf{Angular variables}} \\
$\eta({\rm jet1}_{\rm untagged}) \times Q_{\ell}$ &  Pseudorapidity of the
leading untagged jet  $\times$ lepton charge     & --- & --- & $\surd$ &
$\surd$ \\ $\Delta \cal{R}({\rm jet1},{\rm jet2})$ & Angular separation
between the leading two jets     & $\surd$ & --- & $\surd$ & --- \\
$\cos({\rm \ell},{\rm jet1}_{\rm untagged})_{\rm top_{tagged}}$      & Top
quark spin correlation in the optimal basis for the
$t$-channel\cite{STMahlon:1995zn}, reconstructing the top quark with the
leading tagged jet & --- & --- & $\surd$ & --- \\             $\cos({\rm
\ell},Q_{\ell}$$\times$$z)_{\rm top_{best}}$ & Top quark spin correlation in
the optimal basis for the $s$-channel\cite{STMahlon:1995zn}, reconstructing
the top quark with the best jet  & $\surd$ & --- & --- & --- \\  $\cos({\rm
alljets},{\rm jet1}_{\rm tagged})_{\rm alljets}$          &
Cosine of the angle between the leading tagged jet and the alljets system in
the alljets rest frame     & --- & --- & $\surd$ & $\surd$ \\  $\cos({\rm
alljets},{\rm jet}_{\rm non-best})_{\rm alljets}$  & Cosine of the angle
between the leading non-best jet and the alljets system in the alljets rest
frame  & --- & $\surd$ & --- & --- \\  \botrule
\end{tabular}
\label{tab:STvariables}}
\end{center}
\end{table*}

\subsubsection{Analysis methods}

Several different analysis techniques are employed to separate the signal
from the background: Cut based, Neural Networks (NN), Likelihood and  Matrix
Element based analyses. Generally the analyses are optimized  separately for
$s$-channel and $t$-channel  analyses.  For example, the \dzero\  Neural
Network analysis  focuses on rejecting dominant  backgrounds ($W+$jets and
\ttbar)  by training the neural networks or optimizing the cuts  separately
for each background and for each lepton type. This leads to  eight  separate
sets of cuts or networks (two leptons ($e$, $\mu$) $\times$  number of tagged
$b$~jets ($=1$, $\ge 2$) $times$ signal type ($s$, $t$-channel)).

\begin{itemize}
\item{\bf\underline{Cut Based analysis:}} This is the  traditional method of
applying simple selection requirements on a set of variables. In the \dzero\
analysis,  each of the input variables is assigned a performance metric which
is evaluated by computing the best expected limit for a given  cut on each of
the variables. The variables with the best performance are then combined by a
simple AND in order of an assigned rating based on their relative sensitivity.
 An optimal cut  point is also evaluated
for each intermediate set of combined variables.  Again, the  set with the
best expected limit is chosen for the final analysis. This optimization is
performed for each of the channels separately. For each channel, the optimal
sets of variable and cuts are listed in Ref.~\refcite{STD0prdone}.

\item{\bf\underline{Template Fit Analysis:}} Another technique used by CDF
involves a  maximum likelihood fit to the distribution of $Q\cdot\eta$, where
$Q$ is the  lepton charge and $\eta$ the pseudorapidity eta of the light
quark jet of the events, in order to extract the $s$-and $t$-channel signal
from the data.  The variable $Q\cdot\eta$ is chosen as it shows the largest
difference in kinematics between $s$ and $t-$channel events (see two left
plots of  Fig.~\ref{fig:STCDF}). Templates of  $Q\cdot\eta$ distributions
from expected $s$ and $t-$channel signals, and the two expected background
sources are used for the likelihood fit.  To extract inclusive $s$ and
$t-$channel content  a template fit is carried out using 
the $H_T$ distribution,
as this distribution is similar for both production channels. This is shown
in the right  plot of Fig.~\ref{fig:STCDF}.  The likelihood fit takes into
account the systematic uncertainties from the jet energy scale, the top quark
mass, $b$-tagging efficiencies and the luminosity.  Smaller systematic
uncertainties from  initial state radiation, final state radiation,  parton
distribution functions, the choice of signal Monte Carlo generator,  trigger,
identification are also included.  Some systematic effects can change the
shape of the $Q\cdot\eta$ distribution, which is taken into consideration as
well.

The fit to the $Q\cdot\eta$ distributions using a Bayesian method, leads to a
95\% C.L. upper limit on the production cross sections both in the $s$ and
$t-$channels independently.  The fits using the $H_T$ distribution result
in the combined $s$ and $t-$channel upper limits. These limits are listed in
Table~\ref{tabone}.

\begin{figure}[!h!tbp]
\begin{center}
\psfig{figure=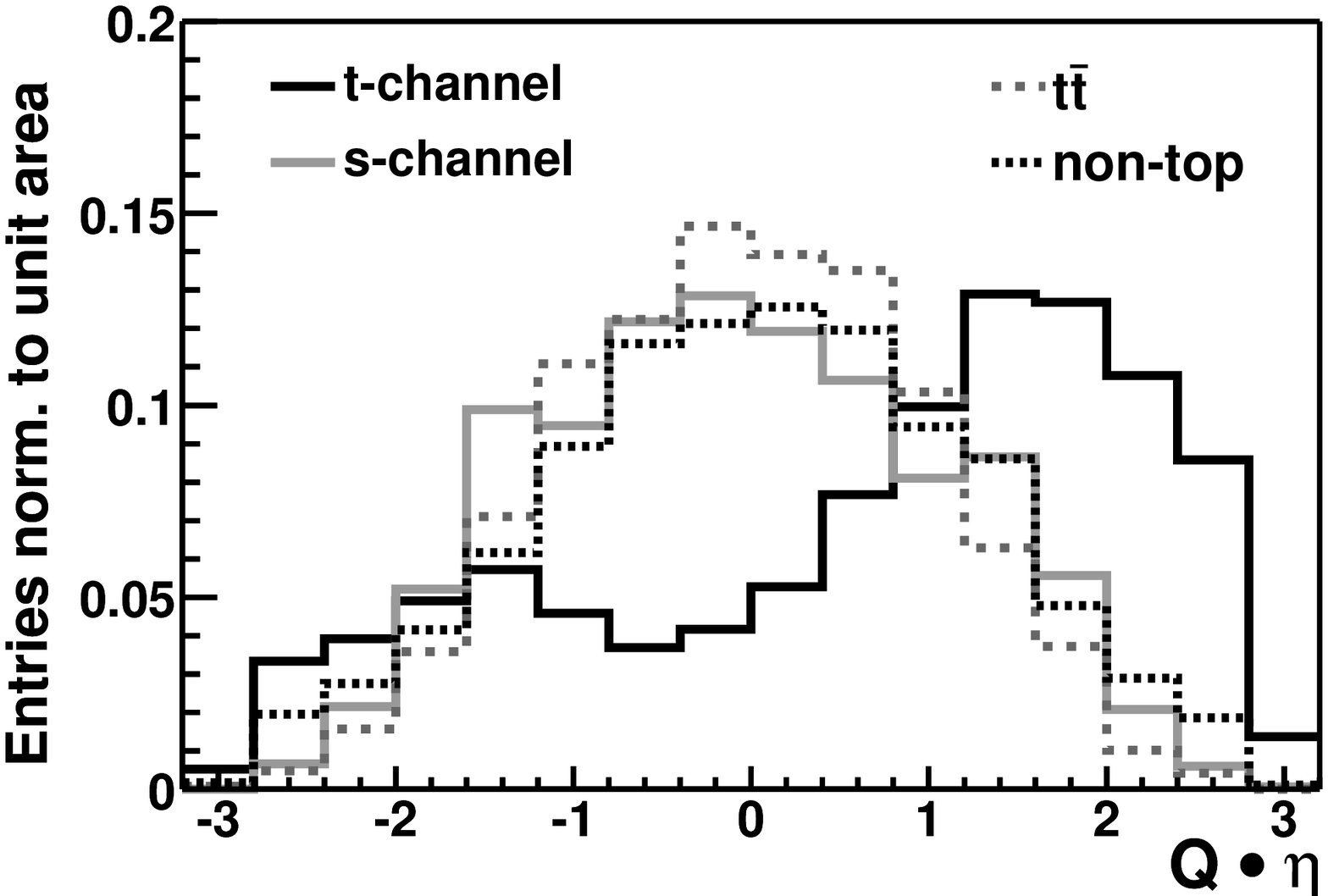,width=0.30\textwidth}
\psfig{figure=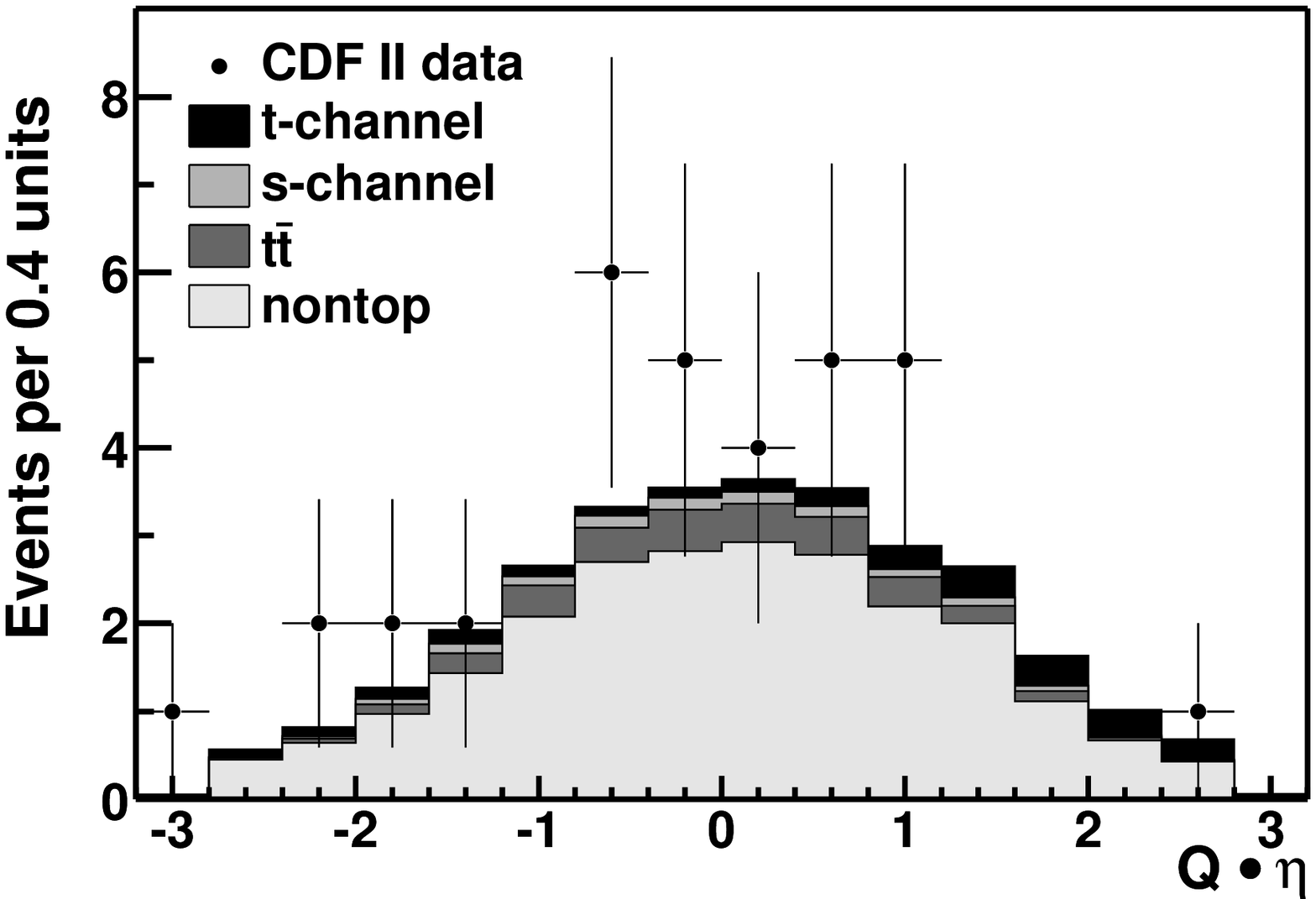,width=0.30\textwidth}
\psfig{figure=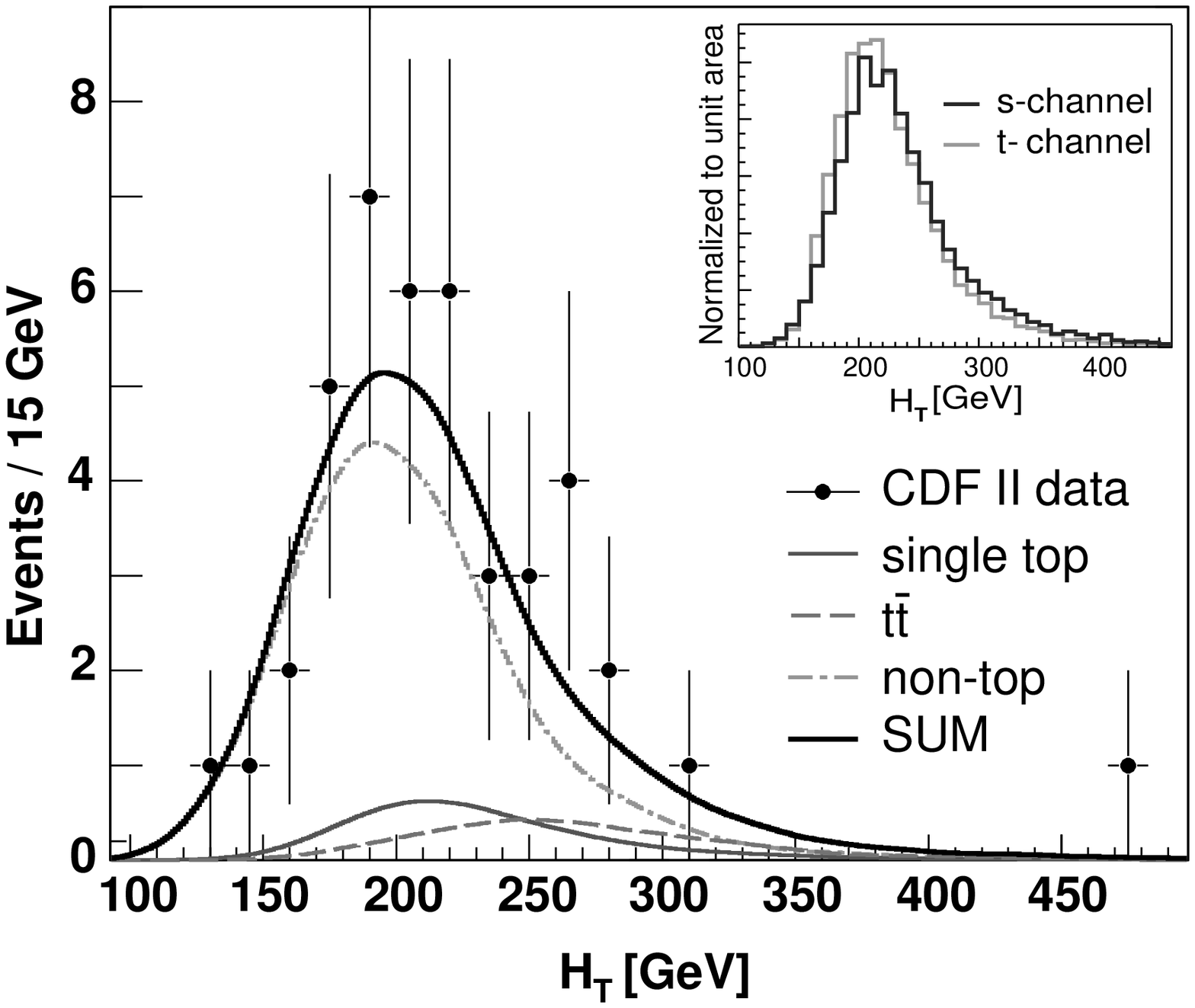,width=0.235\textwidth}
\end{center}
\vspace*{8pt}
\caption{Distributions of $Q\cdot\eta$ (two left plots) and $H_T$ 
(right plot) for the CDF single top search~\protect \cite{RunIIcdf_result}.
}
\label{fig:STCDF}
\end{figure}

\item{\bf\underline{Neural Network Analysis:}} The Neural Network (NN)
``MLPFIT''\cite{STmlpfit} package is used by \dzero\   to analyze  (testing
and training) the events. The ``NeuroBayes''  package\cite{STNNBayes} is used
by CDF for their analysis.  The networks were used with   three layers of
nodes: input, hidden, and output. Monte Carlo simulated events are used for
the training and testing of the networks.  Optimization studies based on the
expected upper limits on the single top quark production cross sections
indicate that each of the channels require  different networks for treatment
of the  dominant backgrounds. \dzero\  separates backgrounds into  two
categories, $W b\overline b$ and $t\overline t$ events. CDF has an additional
category for $W c\overline c$.  This leads to multiple networks (between
three to eight depending on the experiment) for the two  $s$-channel or
$t$-channel analyses.  As an example, the  set of input variables used for the
\dzero\  analysis are listed in  Table~\ref{tab:STvariables}
and denoted by a check mark.  For each network, the set of input
variables could be different (see Table~\ref{tab:STvariables}) 
and the selected combination of variables is
chosen to  give the largest signal to background separation  while keeping
the testing error at its minimum.

The outputs of the NN for the \dzero\  data and the expected signal and
backgrounds are  shown in Fig.~\ref{fig:NNoutputs}.  One notes
that the NN output peaks near one for signal events and at low values (near
zero) for background events.  A good separation between the signal and
$t\overline t$ backgrounds is  seen for the \ttbar\ networks. The $W
b\overline b$ networks  are not as efficient in distinguishing between signal
and backgrounds  since the $W+$jets topology is closer to the signal events
than the $t\overline t$ events.  Similar behavior and separation is  found in
the CDF analysis as well.

\begin{figure}[h!tbp]  
\begin{center}
\psfig{figure=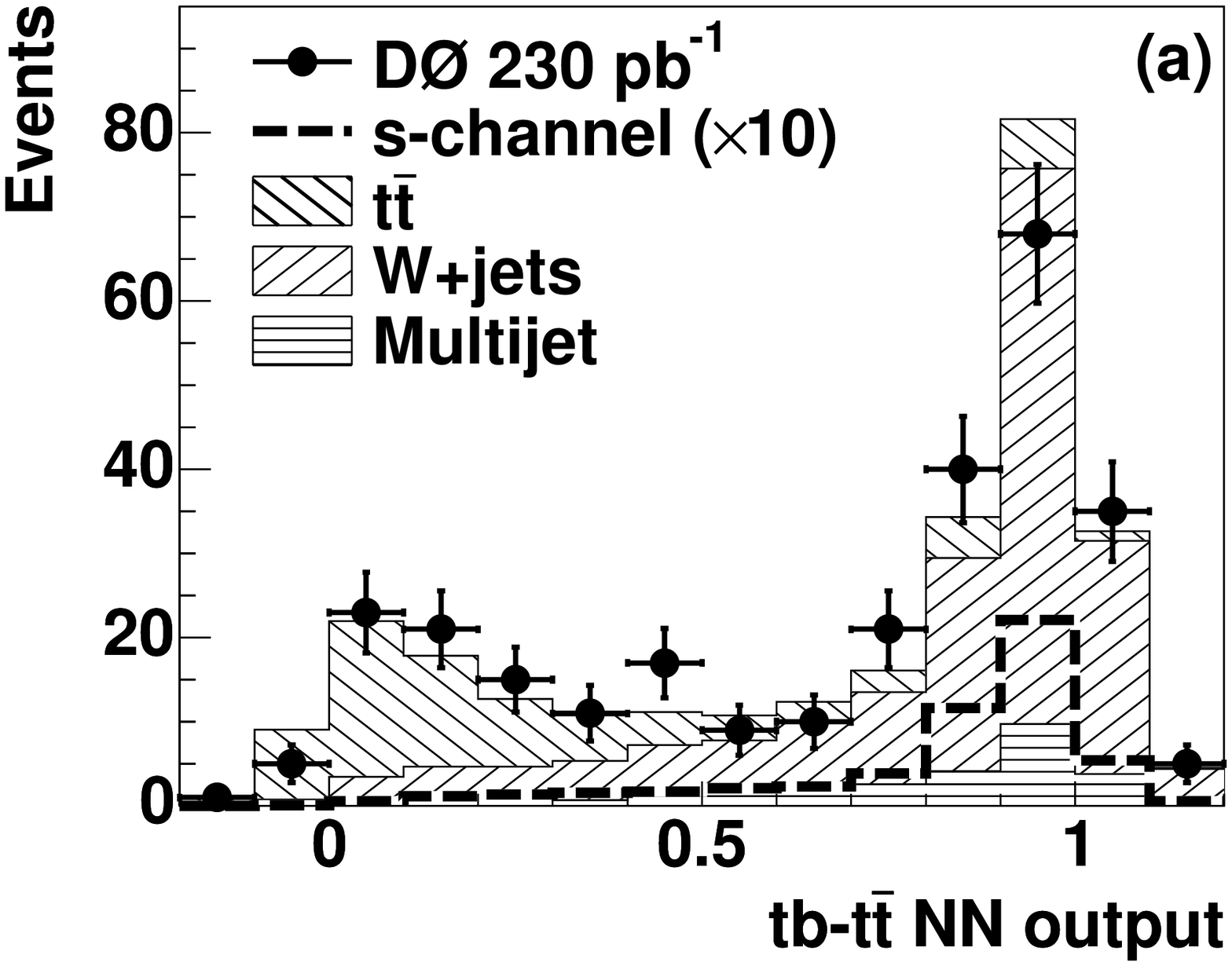,width=0.40\textwidth}
\psfig{figure=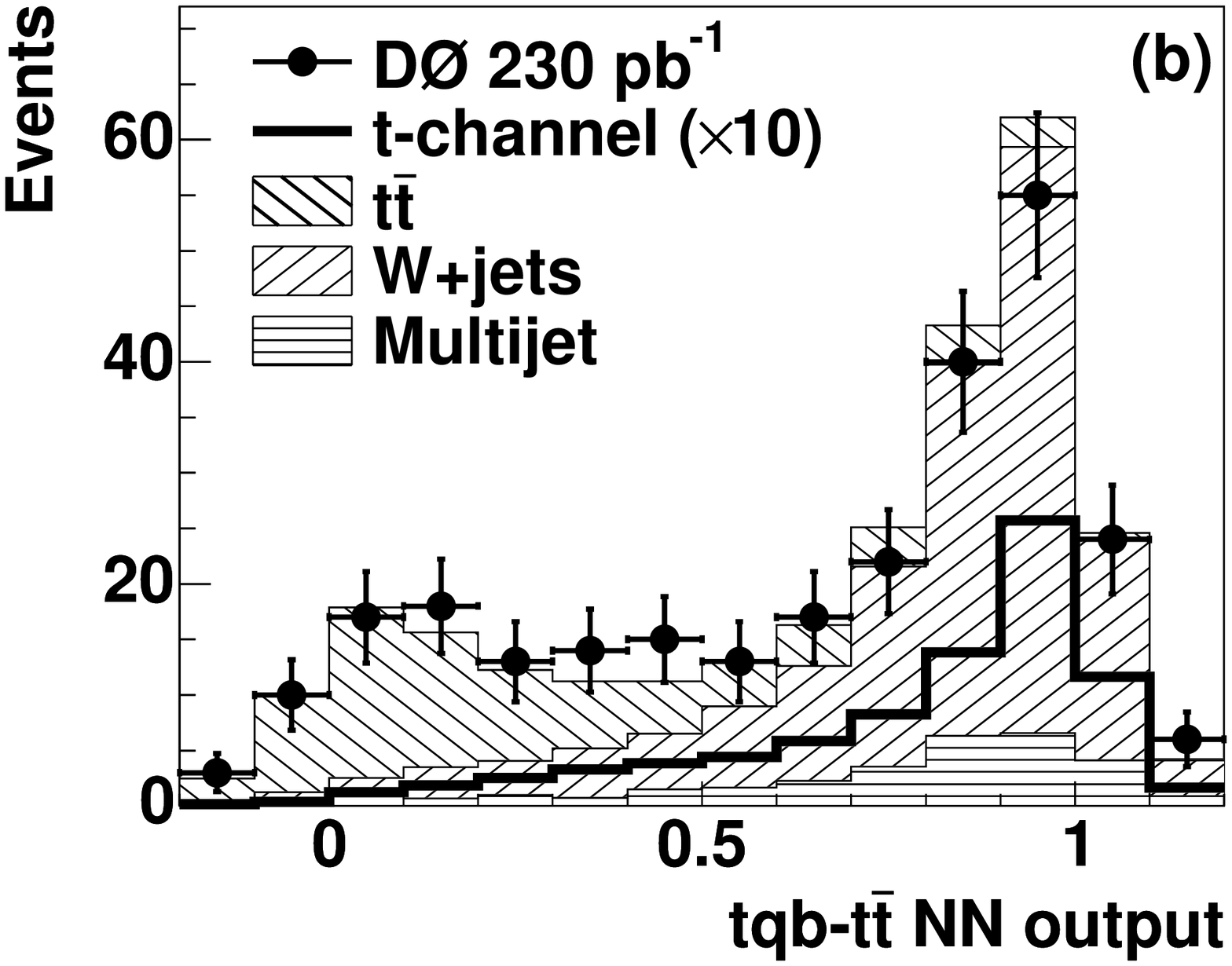,width=0.40\textwidth}
\psfig{figure=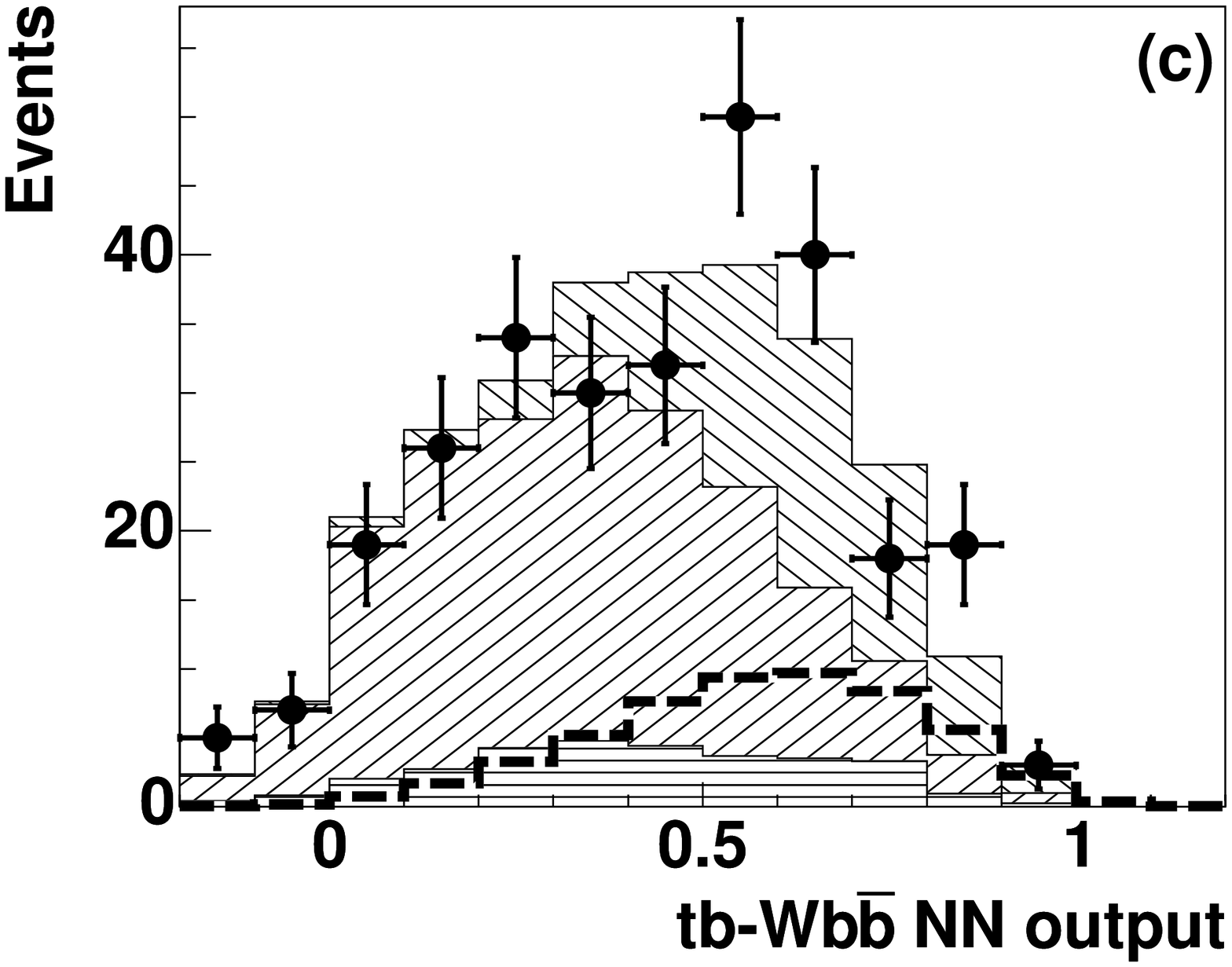,width=0.40\textwidth}
\psfig{figure=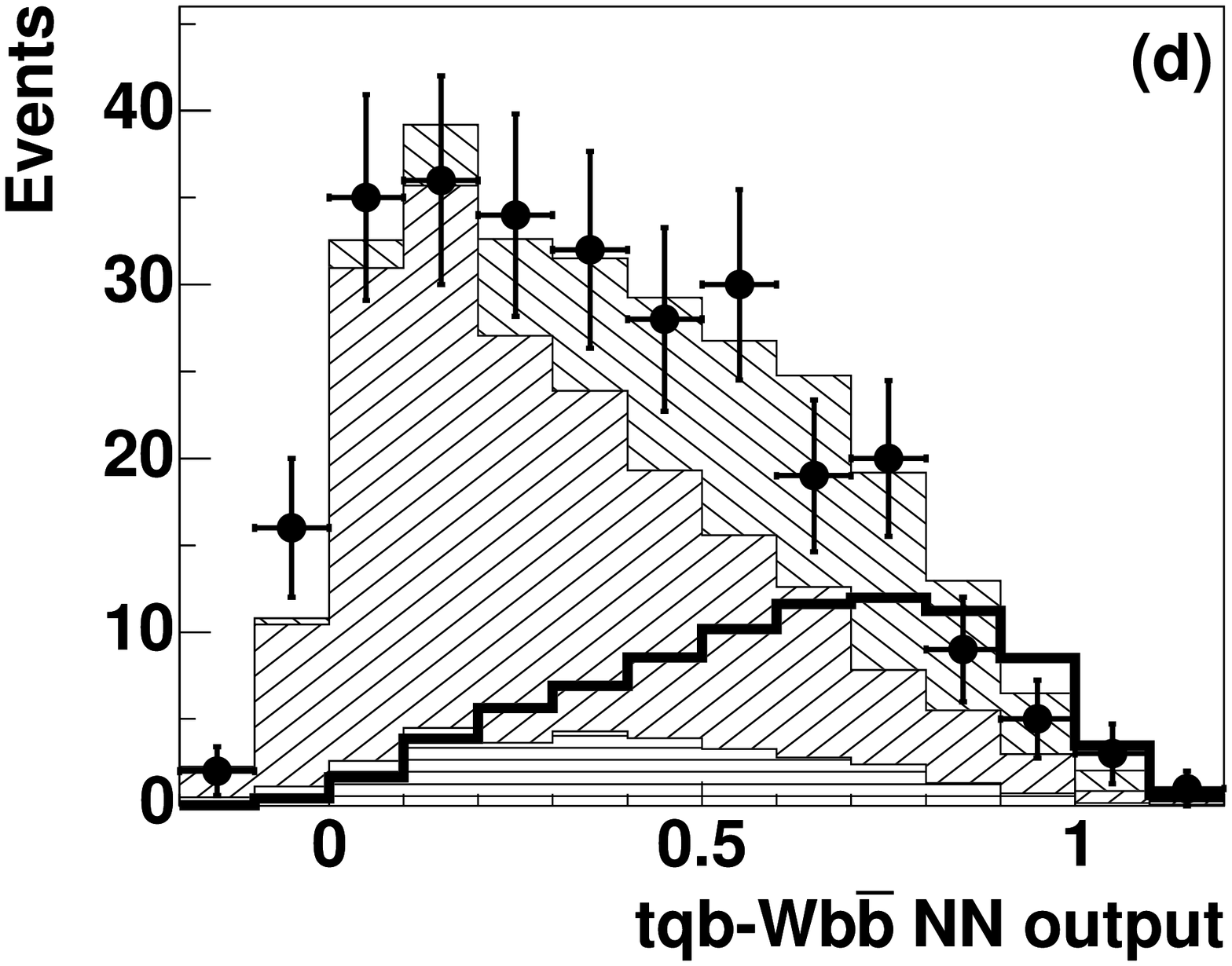,width=0.40\textwidth}
\end{center}
\vspace*{8pt}
\caption{Comparison of \dzero\ background, signal, and 
data for the neural network outputs, for the electron and muon
channels combined, requiring at least one tag, for 
the $tb$-$tt$ filter (a), the $tqb$-$tt$ filter (b), 
the $tb$-$Wbb$ 
filter (c), and the $tqb$-$Wbb$ filter (d). Signals are multiplied by 10 for 
readability~\protect \cite{STPLBD0}.}
\label{fig:NNoutputs}
\end{figure}

\item{\bf\underline{Likelihood Discriminant}}
Another popular technique for extracting small signals from 
large datasets is by  constructing a likelihood discriminant ${\cal L}(\vec
x)$ from a vector of measurements $\vec x$:
\begin{eqnarray}
{\cal L}(\vec x) = {{P_{signal}(\vec x) }\over {P_{signal}(\vec x) +
    P_{background}(\vec x)}}
\end{eqnarray}
where, $P_{signal}(\vec x)$ and $P_{background}(\vec x)$ are the probability
 density for the signal and background events respectively.  The strength of
 this method relies on the use of the difference in the shapes  of the
 distributions for the signal and background events for a given
 variable. Both CDF and \dzero\  analyses separate the signal into  $s$-and
 $t$-channels while constructing the likelihoods.  Monte Carlo events are
 used to construct the one dimensional  probability density functions
 $P_{signal}(\vec x)$ and $P_{background}(\vec x)$ for each of the input
 variables. The final probability density functions of the signal and
 backgrounds are the products of the functions constructed for the individual
 variables. No correlations between variables are used.  One expects the
 value of the likelihood discriminant ${\cal L}(\vec x)$ to be peaked near
 one for signal events and near zero for background events.  Many different
 input variables and combinations  were considered during the optimization
 process to obtain the best expected limit on the cross section
 measurement. The list of input variables used for the \dzero\  and CDF
 analysis are similar to those used by NN.  The Likelihood Discriminant for
 the observed single tagged data  (electron and muon data combined) together
 with the expected signal and backgrounds are shown in
 Fig.~\ref{fig:LDoutput}.

\begin{figure}[h!tbp]  
\begin{center}
\psfig{figure=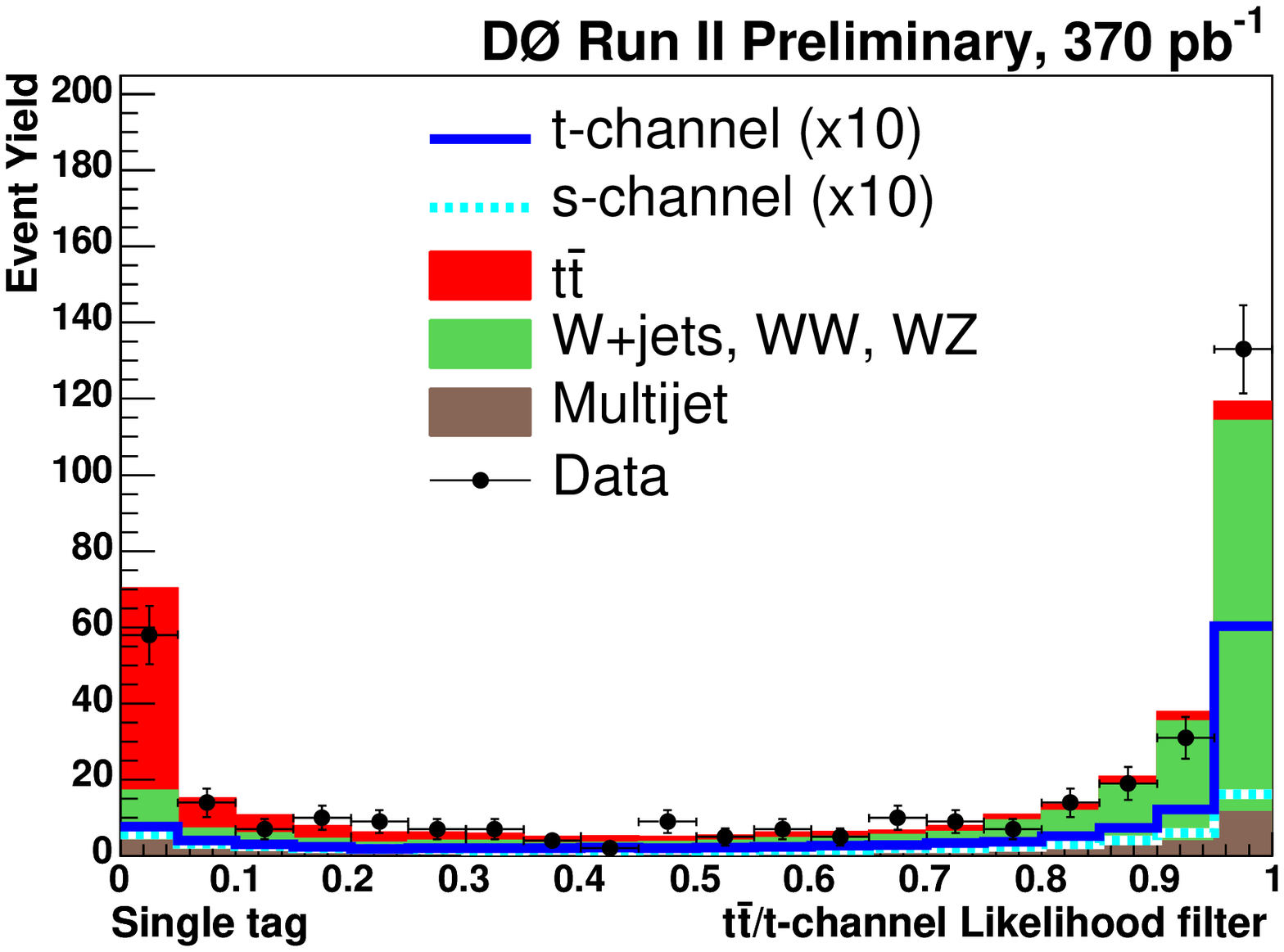,width=0.40\textwidth}
\psfig{figure=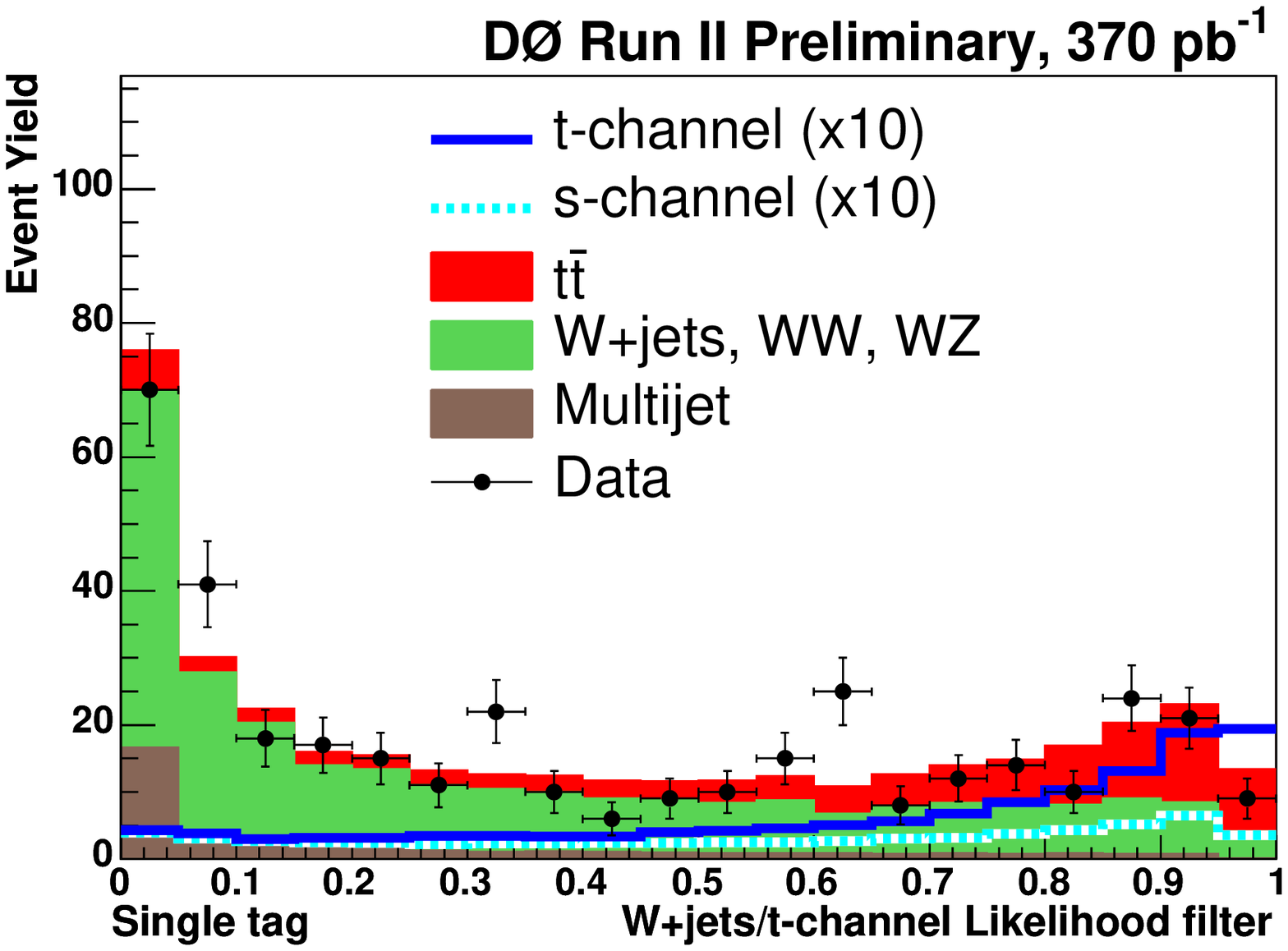,width=0.40\textwidth}
\end{center}
\vspace*{8pt}
\caption{Data to Monte-Carlo comparison for the $tqb/t\overline t$ (left) 
and $tqb/W+jets$ (right) Likelihood Discriminants for single $b$-tagged
 events from the \dzero\  experiment\protect\cite{STD0prdone}.}
\label{fig:LDoutput}
\end{figure}

\subsubsection{95\% C.L. Upper Limit on production cross section}
The cross section for single top quark production is computed  from the
observed data using a Bayesian approach. The probability to observe the
vector of event yields $\bf d$ assuming the single top quark production cross
section $\sigma$ is given by the single top quark acceptance $\alpha$,
integrated luminosity $\cal L$, and the number of events expected from
background $b$, is given by

$$p({\bf d}|\sigma,a,b)=P({\bf d}|a\sigma+b)$$

\noindent where $a=\alpha{\cal L}$ and $P(x|y)$ is the Poisson 
probability to observe $x$ event yield when $y$ are expected.  Using Bayes'
theorem the probability for the single top quark  production cross section to
have the value $\sigma$ is then
$$p(\sigma|{\bf d})=\frac{1}{\cal N}\int\int p({\bf d}|\sigma,a,b) \pi(a,b)
\mbox{d}a\mbox{d}b,$$ where ${\cal N}$ is a normalization constant, a flat
prior for $\sigma$ was assumed and the prior for the other parameters
$\pi(a,b)$ is a product of Gaussians with widths given by the experimental
uncertainties in these parameters. The dependence on all parameters except
$\sigma$ was eliminated by integrating over these so-called nuisance
parameters.

The measured cross section is then given by the value of $\sigma$ for which
$p(\sigma|{\bf d})$ is maximized. If this occurs for $\sigma=0$ an upper
limit $\sigma_{max}$ can be set at confidence level $\beta$ using the
condition
$$\int_0^{\sigma_{max}} p(\sigma|{\bf d})\mbox{d}\sigma=\beta.$$ For a finite
cross section the errors $\delta_+$ and $\delta_-$ are defined by
$$\int_{\sigma_{max}-\delta_-}^{\sigma_{max}+\delta_+} p(\sigma|{\bf
d})\mbox{d}\sigma=0.6827$$ for which $\delta_-+\delta_+$ is minimized.

Since the observed data are consistent with the background predictions for
all analysis techniques used by \dzero\  and CDF,  following the prescription
for cross section computation described above, the 95\% C.L. upper limits on
the single top quark production cross sections are computed. These
measurements are listed in Table~\ref{tabone}.  These upper limits
represented significant improvements over previously published
results\cite{d0runI}, mainly due to the larger data sets analyzed, and the
use of multivariate analysis techniques. It is interesting to note that they
approach the region of sensitivity for models incorporating
fourth quark generation scenario or flavor-changing neutral-currents.

\end{itemize}

{\subsection{Evidence for single Top quark production}~\label{secSTevidence}}

The evidence for single top quark production was reported by the \dzero\
Collaboration in December 2006\cite{STD0evidencePRL}.  This analysis is based
on a large  data sample corresponding to a luminosity of 0.9 fb$^{-1}$
collected between 2002 and 2005. Event sample selection procedures are
similar to  the earlier searches as described in the previous section.
Events are classified into twelve subsamples based on the lepton type ($e$ or
$\mu$), the number of jets (two, three or four jets), and  the number of
tagged $b$-jets (one or two tags) in the event.  The event yields are
tabulated in Table~\ref{tab:STnewevtsel}.

\begin{table}[!h!tbp]
\tbl{Numbers of expected and observed events in 0.9~fb$^{-1}$ for $e$ and
$\mu$, one $b$~tag and two $b$~tag channels combined from the \dzero\
experiment. The total background uncertainties are smaller than the component
uncertainties added in quadrature because of anticorrelation between the
$W$+jets and multijet backgrounds resulting from the background normalization
procedure~\protect \cite{STD0evidencePRL}.}
{\begin{tabular}{lcccccc} \toprule
Source           & \multicolumn{2}{c}{2 jets}
                 & \multicolumn{2}{c}{3 jets}
                 & \multicolumn{2}{c}{4 jets} \\ \colrule
$s$-channel                      &  16  &   3  &   8  &  2  &   2  &  1  \\
$t$-channel                     &  20  &   4  &  12  &  3  &   4  &  1  \\
\hline                                                              
${t\overline t}{\rightarrow}\ell\ell$  &  39  &   9  &  32  &  7  &  11  &  3  \\
${t\overline t}{\rightarrow}\ell$+jets &  20  &   5  & 103  & 25  & 143  & 33  \\
$Wb\bar{b}$               & 261  &  55  & 120  & 24  &  35  &  7  \\
$Wc\bar{c}$               & 151  &  31  &  85  & 17  &  23  &  5  \\
$Wjj$                     & 119  &  25  &  43  &  9  &  12  &  2  \\
Multijets                 &  95  &  19  &  77  & 15  &  29  &  6  \\
\hline                                                              
Total background          & 686  &  41  & 460  & 39  & 253  & 38  \\
Data             & \multicolumn{2}{c}{697}
                 & \multicolumn{2}{c}{455}
                 & \multicolumn{2}{c}{246} \\
\botrule
\end{tabular}
\label{tab:STnewevtsel}}
\end{table}

A multivariate technique known as ``Decision Tree''\cite{STdt-breiman}   is
used for discriminating between signal and background events in the sample.
Decision Trees, DT, are a machine learning technique which are used to
compute the probability of the event  to be a signal event.  Decision trees
provide a way to represent rules underlying data with hierarchical,
sequential structures that recursively partition the data.  It is a binary
tree with a  selection cut implemented at each node such that each event
eventually ends up into a well-separated  class called ``leaves''. A purity
value defined as the ratio of signal  to background events and predetermined
from the training samples is associated with each of the leaves. The output
of the decision tree is the distribution of the final purity values for the
sample and is  discrete. A major improvement in this
technique,  comes from the implementation of the adaptive boosting algorithm
AdaBoost\cite{STDTboost}.  This boosting algorithm modifies the weights of
the  misclassified  events and rebuilds the tree iteratively. The final
decision tree with improved performance is an  average over many trees
produced during  the boosting process.

Forty nine variables (see reference~\cite{STFNALtalk} for details)  are used
as inputs to the Decision Tree. These variables based on individual object
kinematics,  global event kinematics, and angular correlations are
constructed  in order to discriminate between signal and background events.
For each of the three searches $s-$channel, $t-$channel and $s+t$ channel,
decision trees were trained on the twelve event subsamples, leading to  a
total of thirty six decision trees. Thus each one of the 36 decision trees
is trained for a given signal signature against the sum of $W$+jets and
$t\overline t$ backgrounds. The Monte Carlo samples used for training the
trees and testing their performance are independent of each other.

\begin{figure}[!h!tbp]
\begin{center}
\psfig{figure=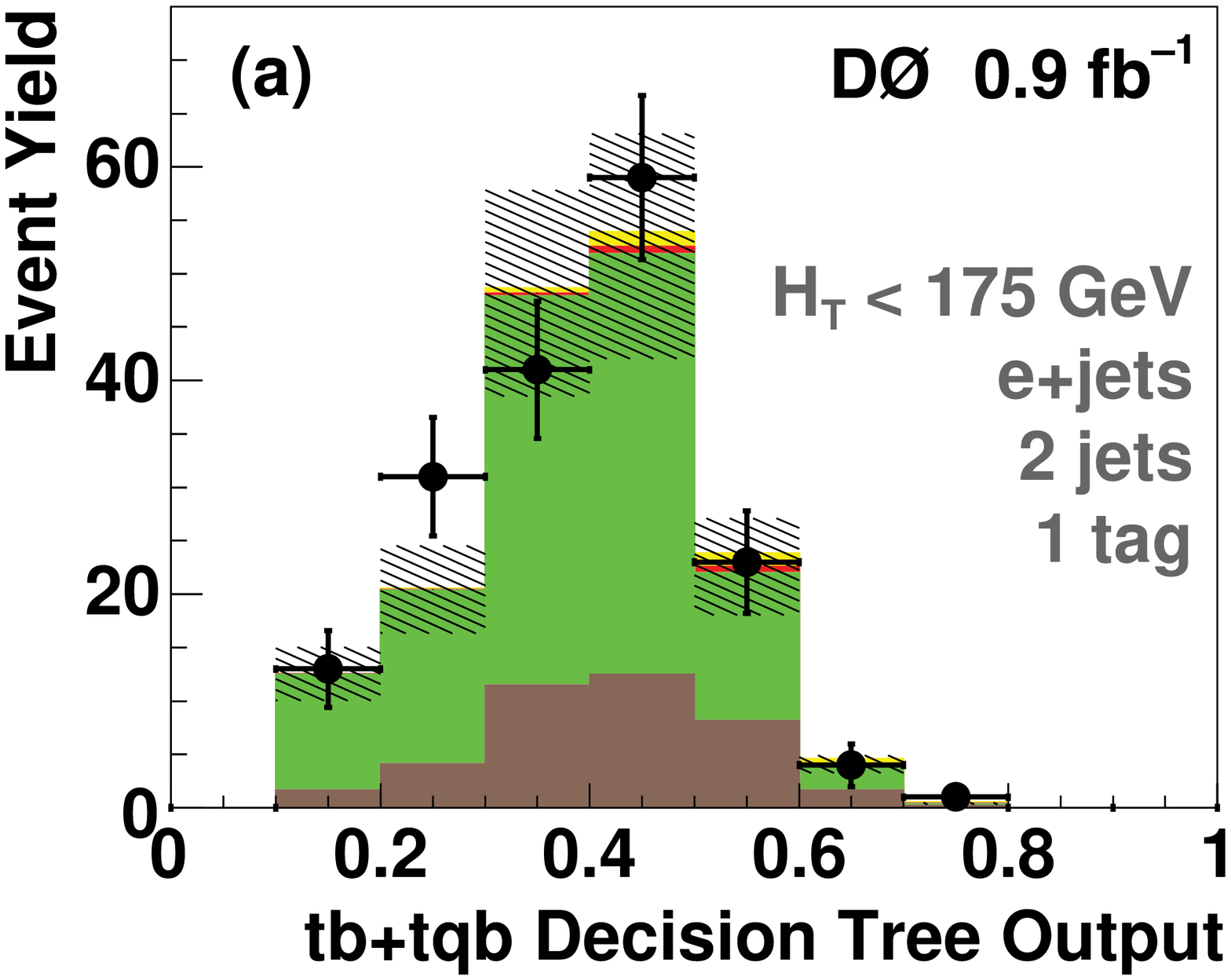,width=0.40\textwidth}
\psfig{figure=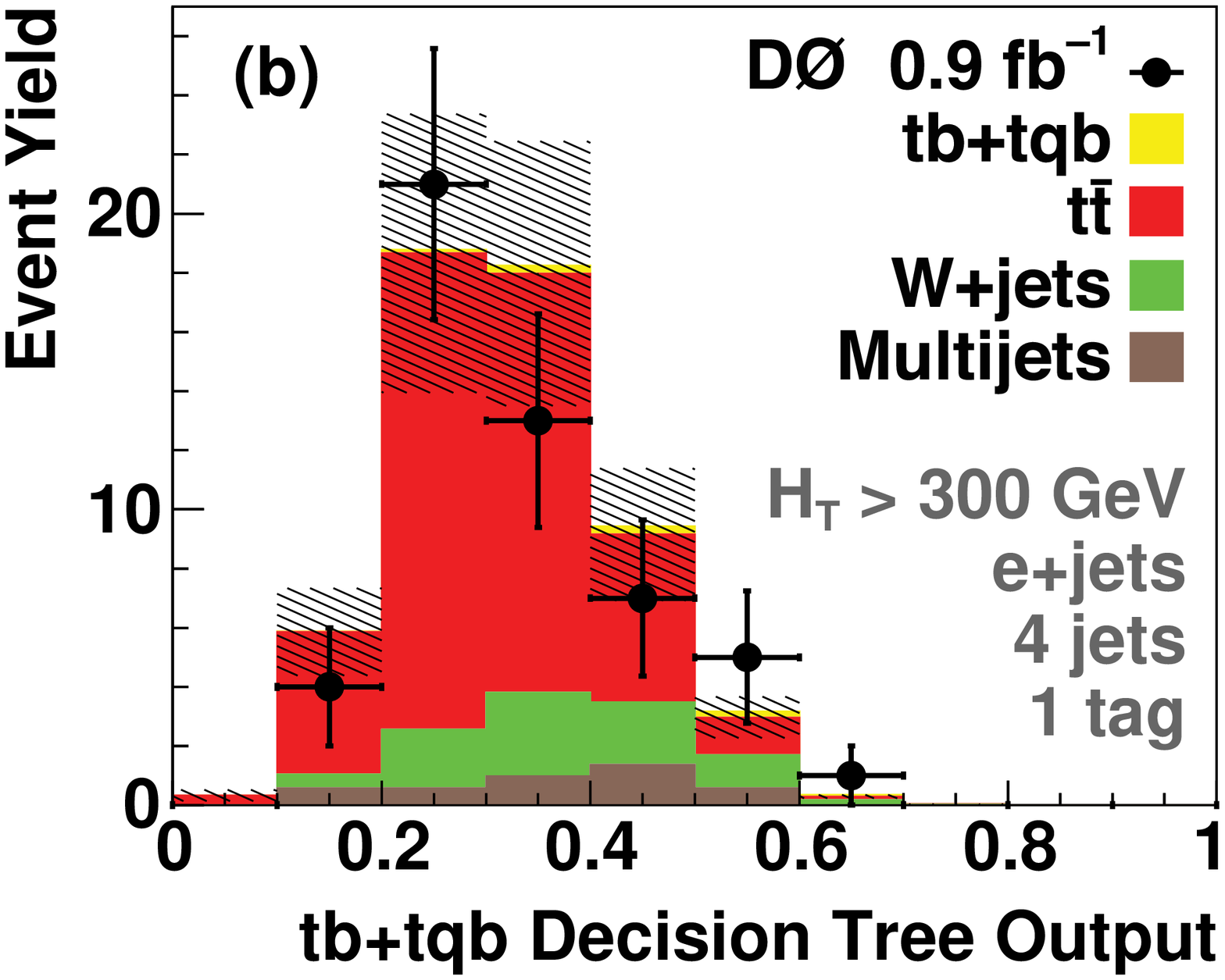,width=0.40\textwidth}\\
\psfig{figure=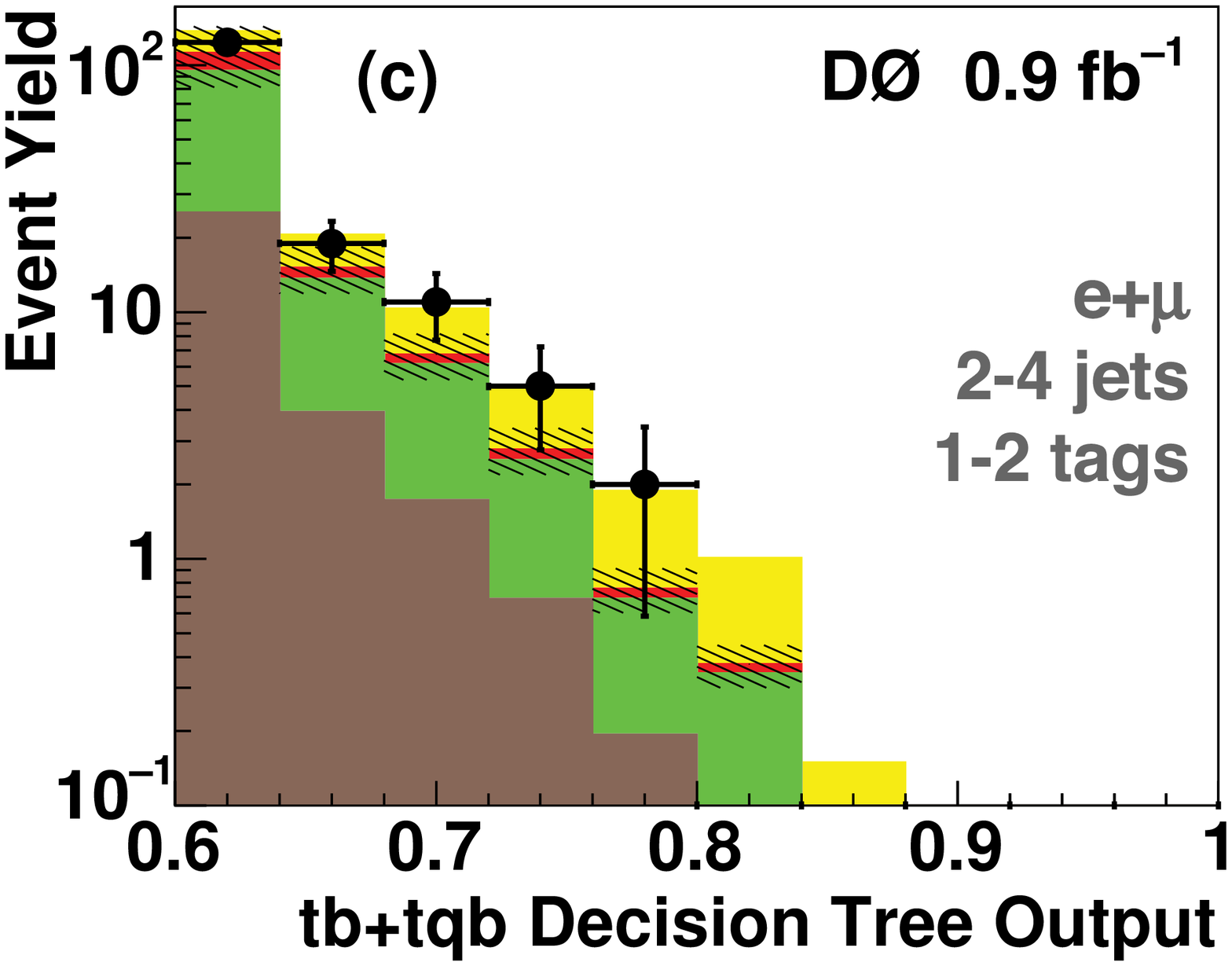,width=0.40\textwidth}
\psfig{figure=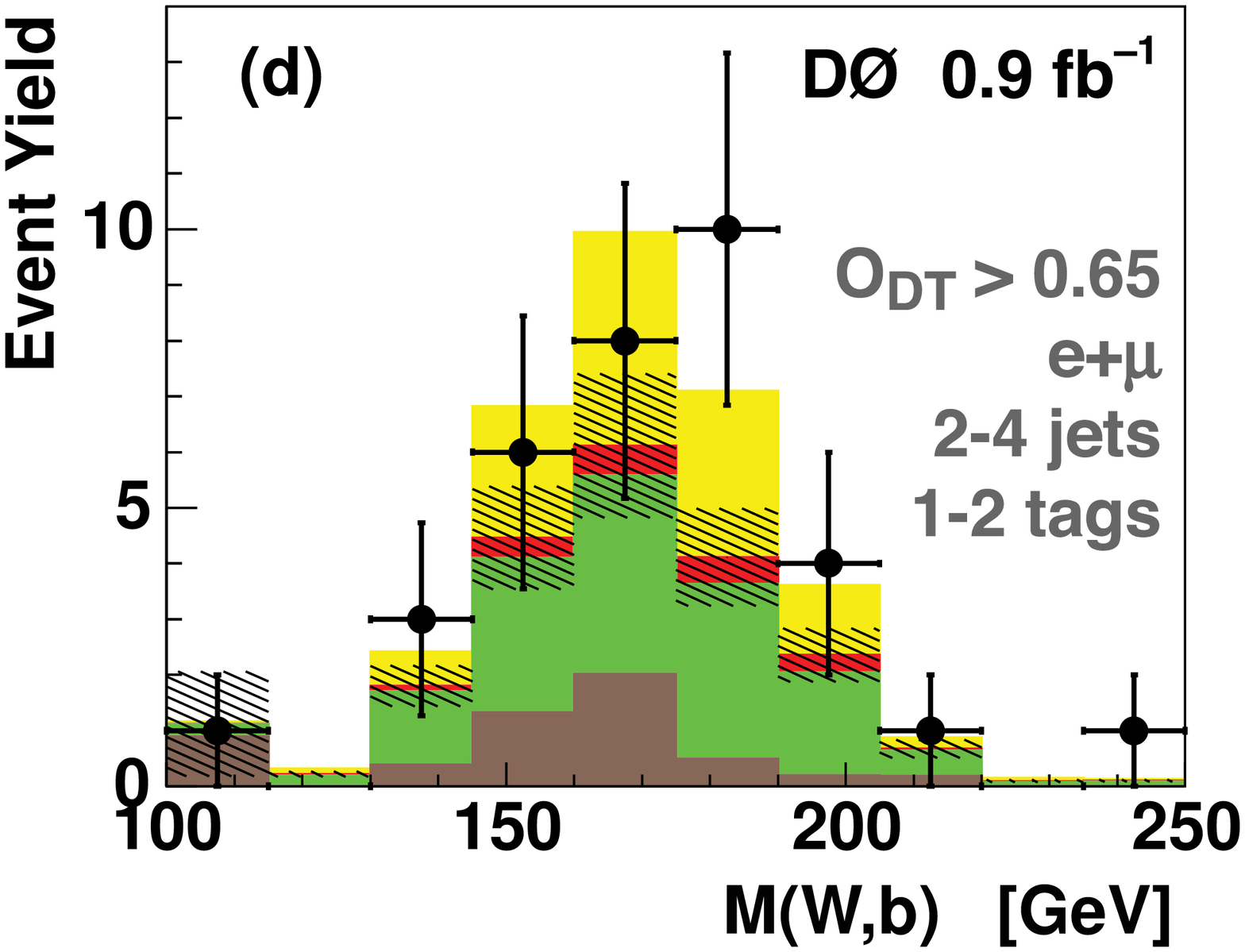,width=0.40\textwidth}
\end{center}
\vspace*{8pt}
\caption{Boosted decision tree output distributions for (a) a
$W$+jets-dominated control sample, (b) a {$t\overline t$}-dominated control
sample, and (c) the high-discriminant region of the sum of all 12
inclusive $tb$+$tqb$ decision trees. For (a) and (b), $H_T = E_T^{\ell} + 
{\not\!\!E_T} + \sum E_T^{\rm alljets}$. Plot (d) shows the invariant mass of the
reconstructed $W$~boson and highest-$p_T$ $b$-tagged jet for events
with $O_{\rm DT} > 0.65$. The hashed bands show the $\pm 1$ standard
deviation uncertainty on the background. The expected signal is shown
using the measured cross section (\dzero\ experiment)
~\protect \cite{STD0evidencePRL}.
}
\label{fig:dt-plots}
\end{figure}

Tests on the performance of the decision trees on the collider data are done
by  studying samples which are rich in background. Two such samples
were created: a $W$+jets sample and  a $t\overline t$ sample. The $W$+jets 
sample was selected by requiring $H_T<175$ GeV for the events with two jets
 and one tagged $b$-jet. The $t\overline t$ sample comprises of events 
with four jets, one tagged
$b$-jet and $H_T> 300$ GeV.  In Fig.~\ref{fig:dt-plots}~(a)
and~(b), the decision tree discriminants, $O_{\rm DT}$,
are shown for both the background enriched data samples.  The prediction of
the background models are overlayed on these  data distributions. Good
agreement is visible between the  prediction and  the observed events in
these background enriched samples.

Output of the decision tree analysis in the high end of the discriminant
spectrum for combined $e$ and $\mu$ data samples is displayed in
Fig.~\ref{fig:dt-plots}~(c).   The expected signal and
backgrounds are overlaid and summed over the twelve decision trees for $s+t$
channels.  A single top quark signal is well accommodated with the data
sample.  The characteristics of the events in the high decision tree
discriminant region  have been examined, by making a selection $O_{\rm DT} >
0.65$. For these selected  events, in
Fig.~\ref{fig:dt-plots}~(d) the distribution
of invariant mass from the best $b-$jet and reconstructed $W$ boson is 
displayed.  Good
agreement is observed between the data and a prediction including background
and a  single top quark signal.

\begin{figure*}[!h!tbp]
\begin{center}
\psfig{figure=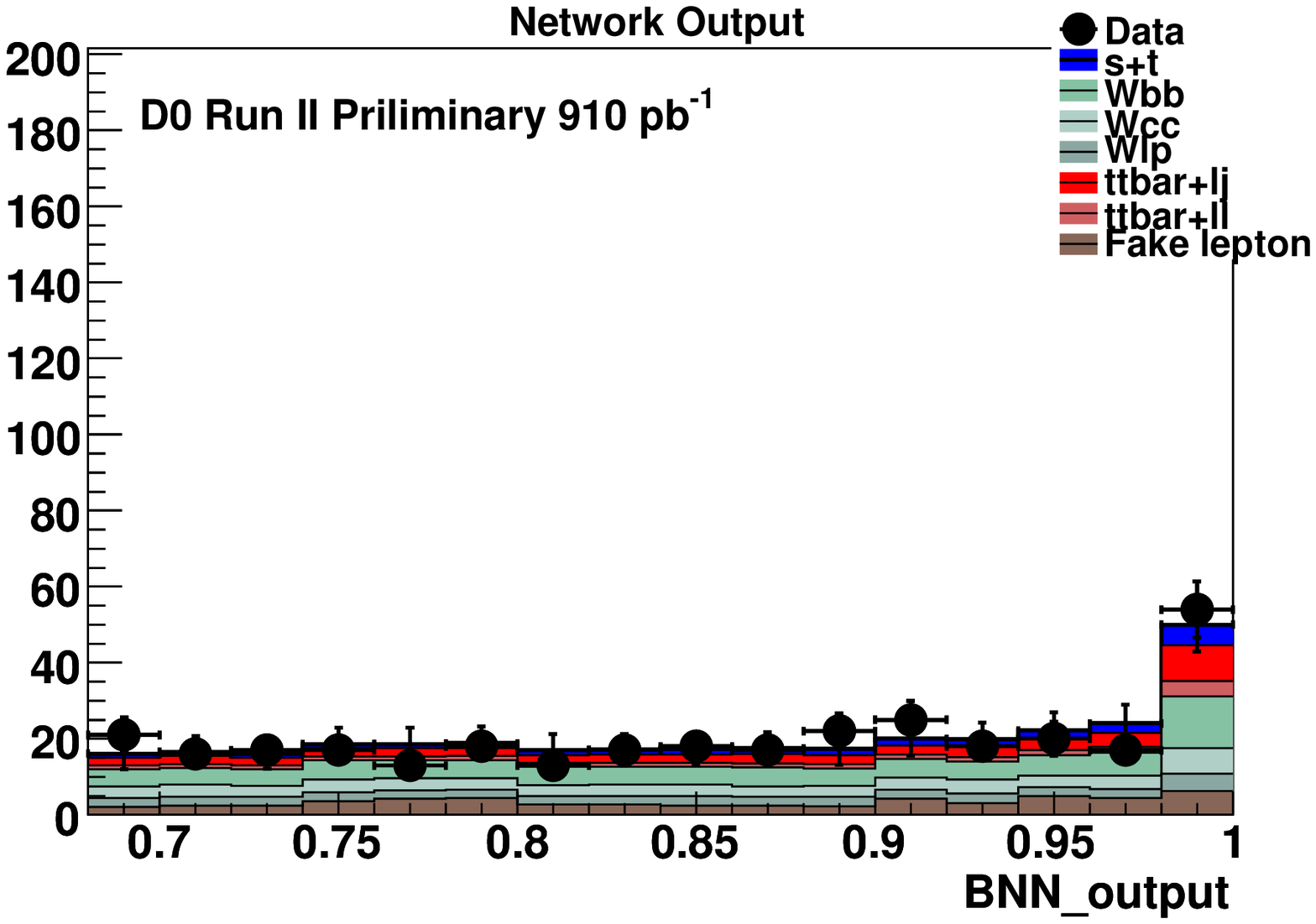,width=0.35\textwidth}
\psfig{figure=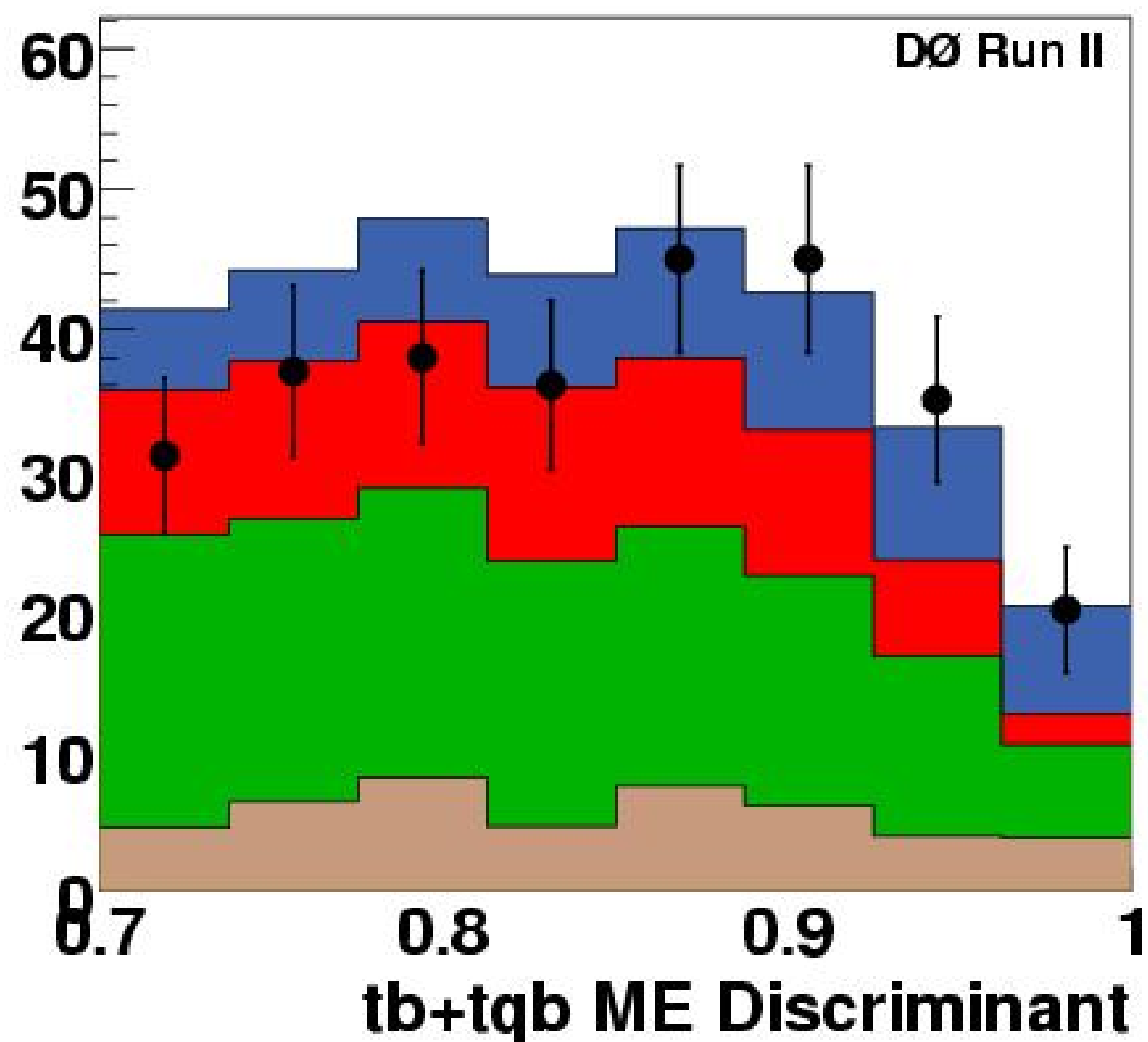,width=0.35\textwidth}
\end{center}
\vspace*{8pt}
\caption{Bayesian Neural Network discriminant(left) and the 
Matrix Element discriminant (right), in the high-discriminant 
region, for the combined event sample (\dzero\ experiment)
~\protect \refcite{STFNALtalk}.}
\label{fig:BNNMEoutput}
\end{figure*}

Two other supporting analyses were carried out. One is based on using
Bayesian Neural Networks and the other is known as the Matrix Element
technique.  The Bayesian Neural network\cite{STbayesianNN}  at the first
stage is similar  to a simple Neural  Network as described in
section~\ref{secSTD0search}. Instead of choosing one set of weights to
characterize the network with Bayesian Neural Networks,  a  posterior
probability density  is computed using  all possible weights.  The final
network is obtained by  computing a  weighted average of  a large number of
networks, obtained after many sets of training cycles, where the  weights are
the probability of each network given the training data.  The BNN uses
a subset of the input variables used in the DT analysis. The training 
samples are the combined  $s$ and $t$ channel Monte Carlo signals and 
the combined background set. The performance of this technique is shown
in Fig.~\ref{fig:BNNMEoutput} (left panel), for events with
electron and two jets,  with one of them tagged as a $b$-jet.

\begin{figure}[!h!tbp]
\begin{center}
\psfig{figure=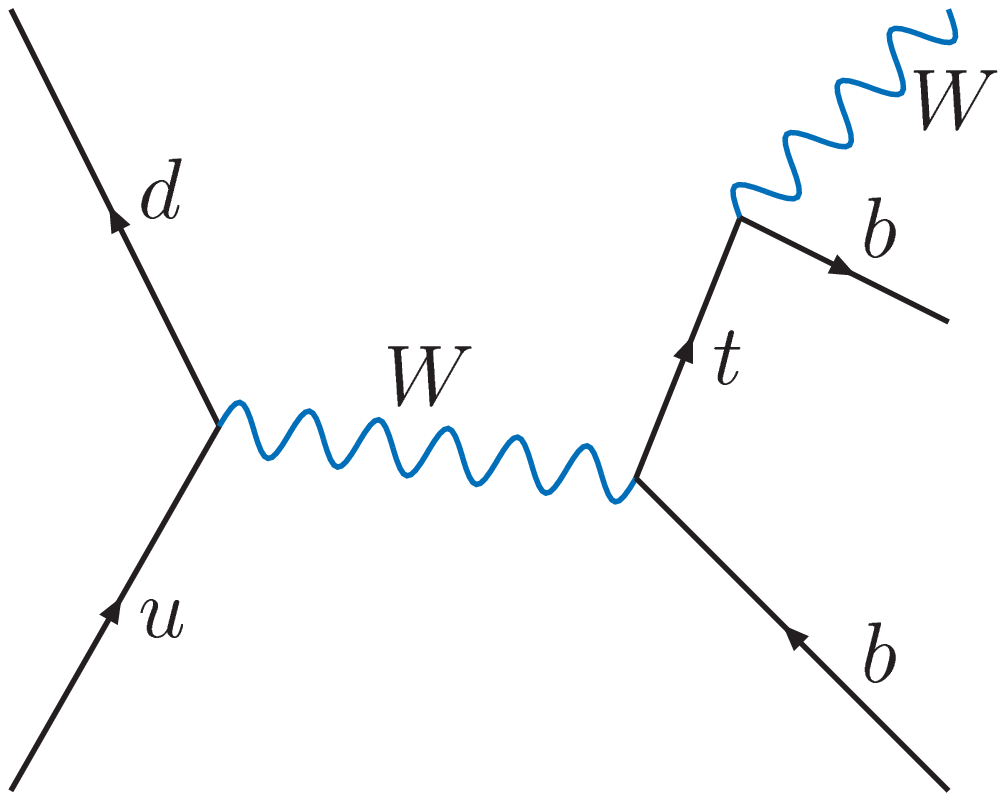,width=0.27\textwidth}
\hspace{0.2in}
\psfig{figure=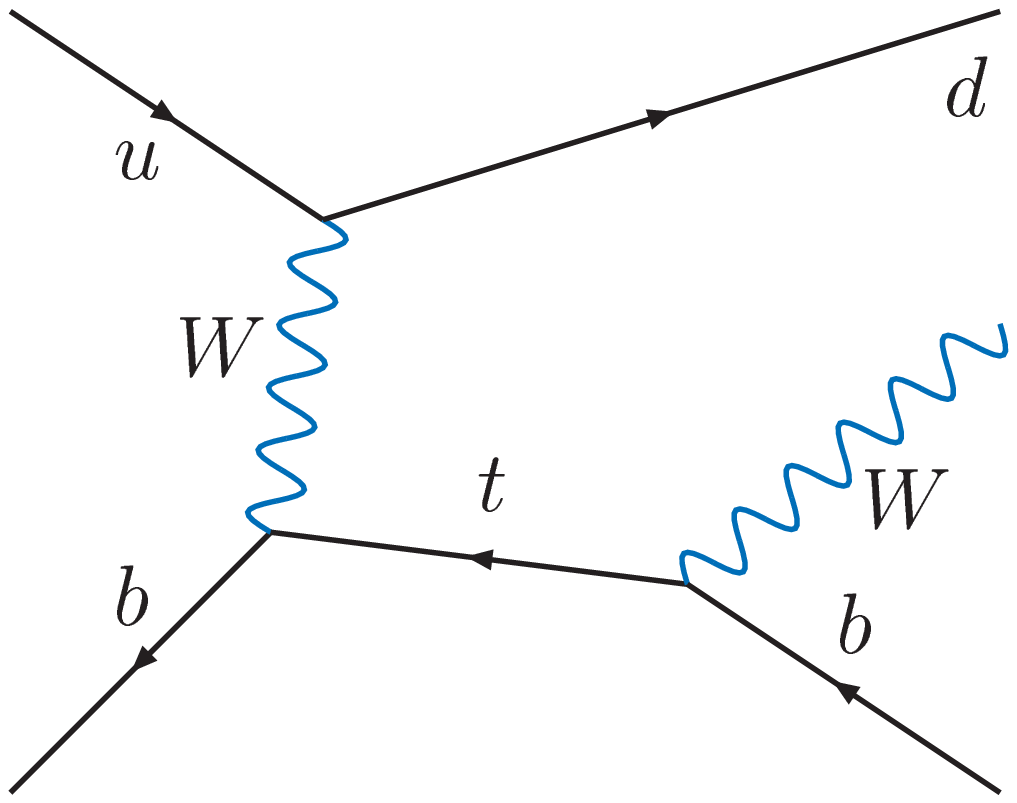,width=0.27\textwidth}\\
\psfig{figure=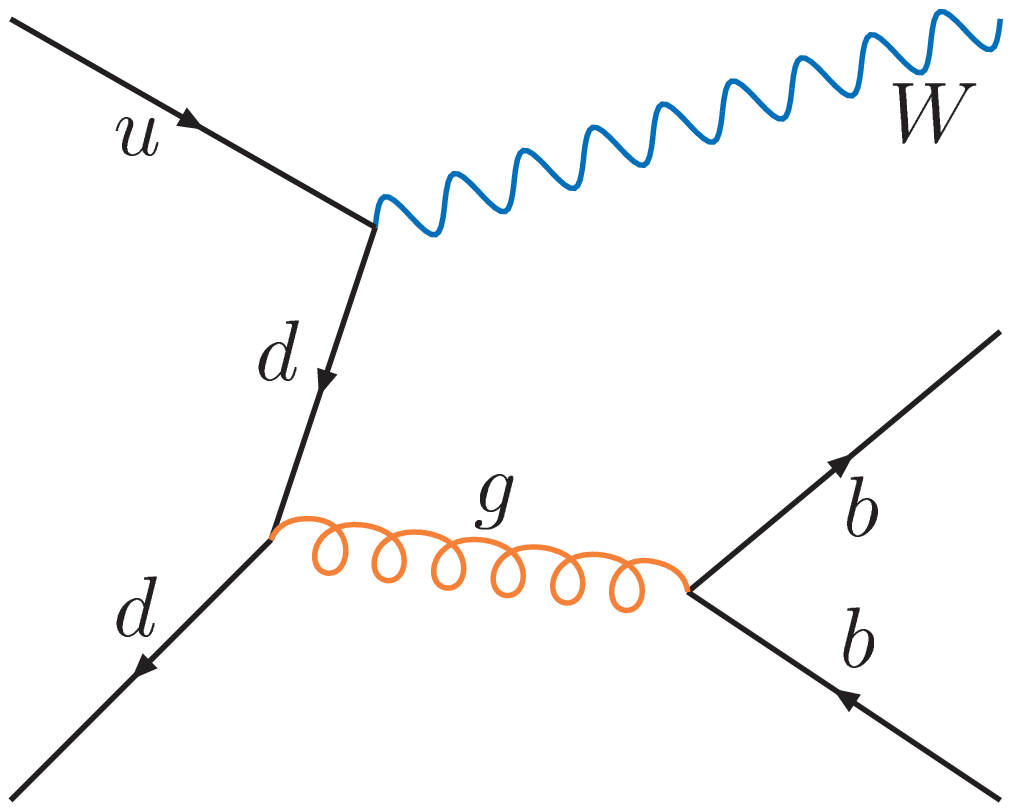,width=0.27\textwidth}
\hspace{0.2in}
\psfig{figure=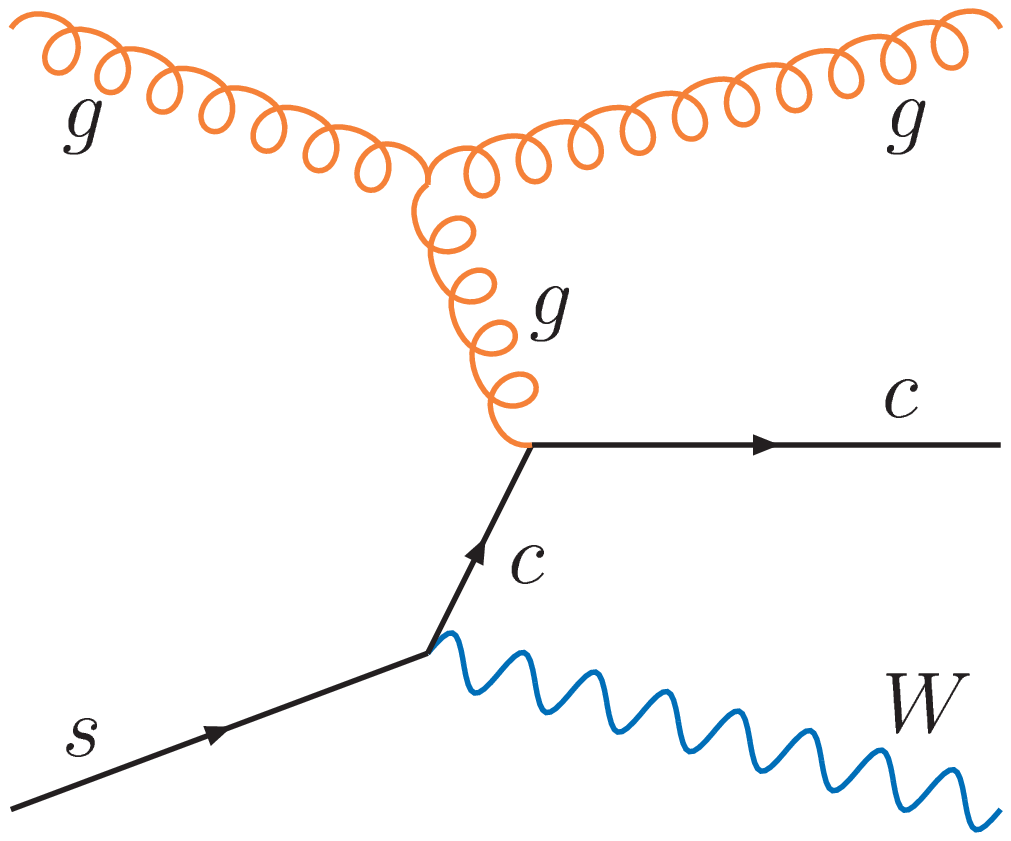,width=0.27\textwidth}
\hspace{0.2in}
\psfig{figure=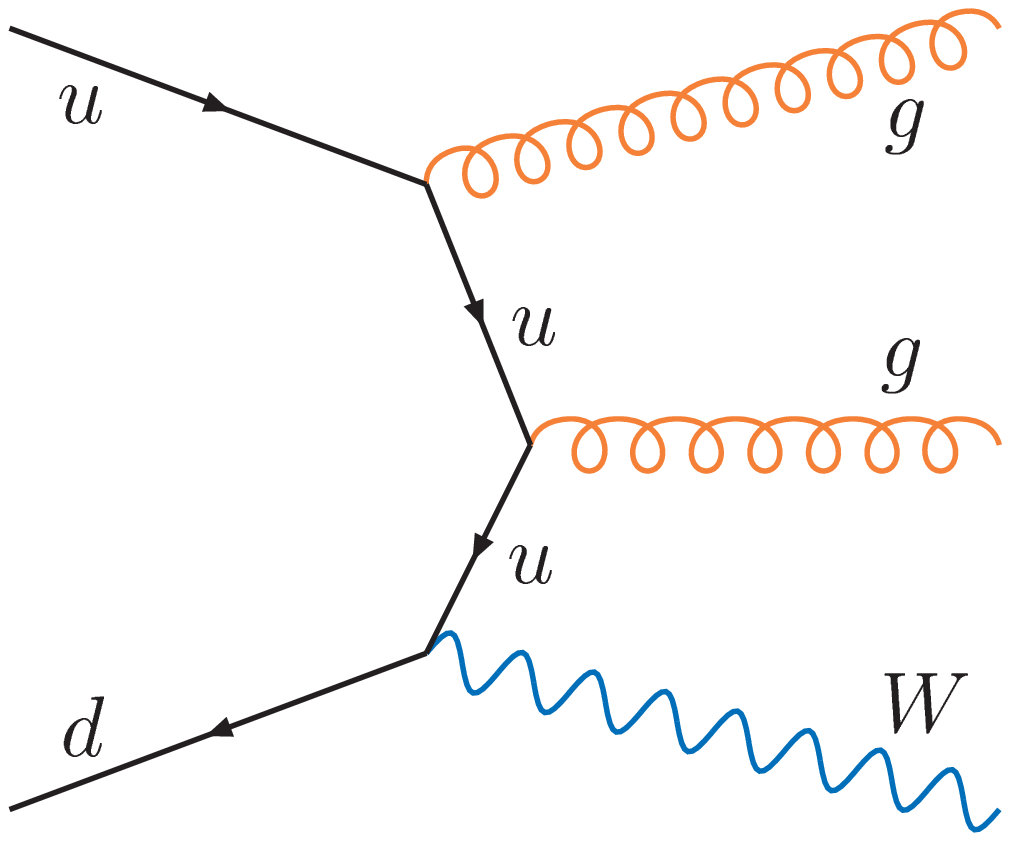,width=0.27\textwidth}\\
\psfig{figure=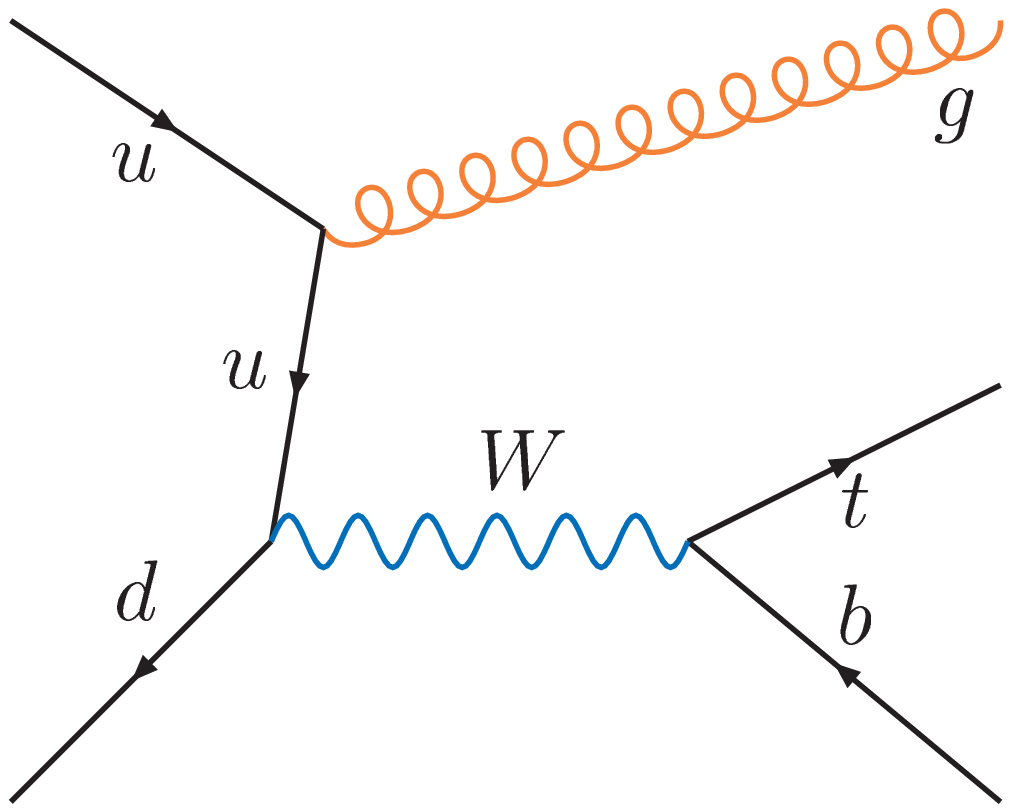,width=0.24\textwidth}
\hspace{0.2in}
\psfig{figure=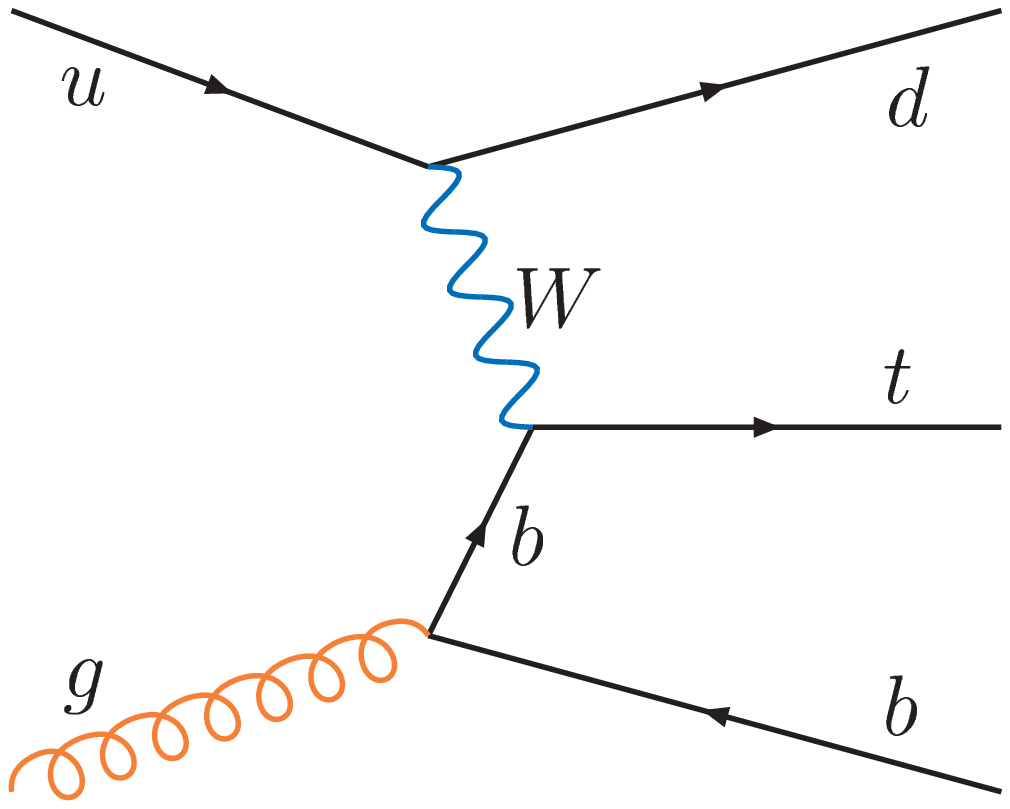,width=0.24\textwidth}
\hspace{0.2in}
\psfig{figure=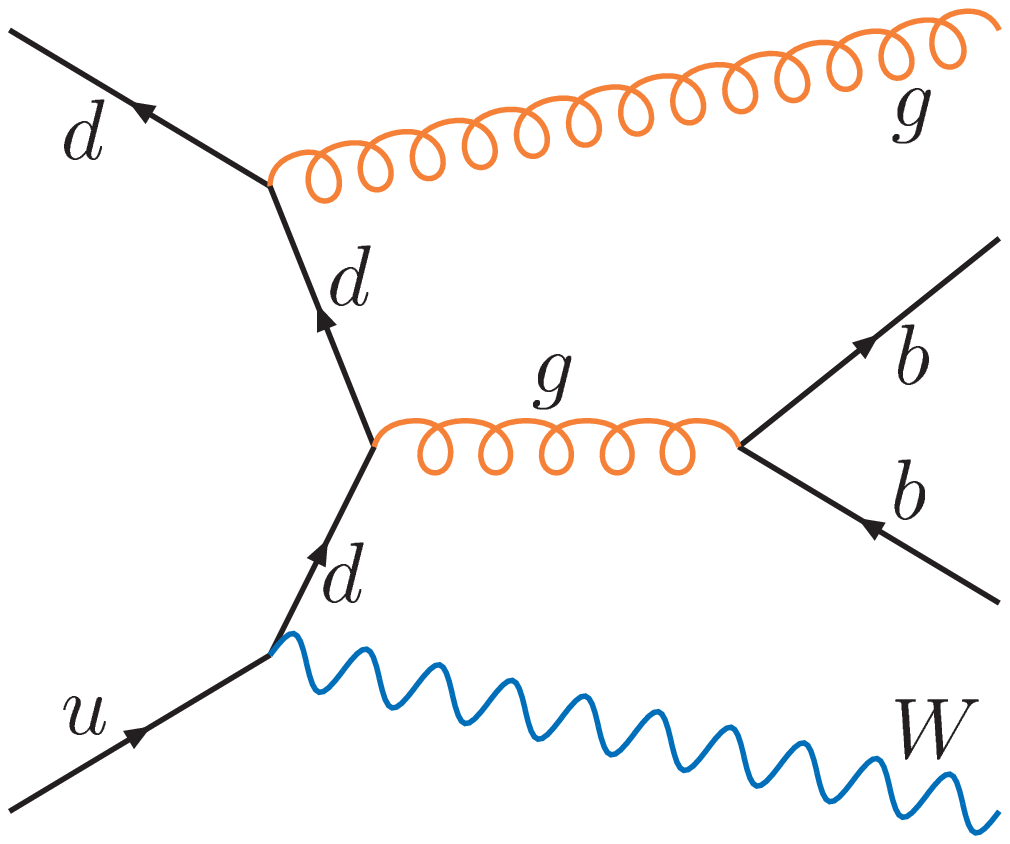,width=0.24\textwidth}
\end{center}
\vspace*{8pt}
\caption{Representative Feynman diagrams corresponding to 
the leading-order matrix elements used for event probability
calculation. The upper and middle row are 
for events with exactly two jets. Upper row, signals:
$ud{\rightarrow }tb$, $ub{\rightarrow }td$; middle row, backgrounds: $ud{\rightarrow }Wbb$,
$sg{\rightarrow }Wcg$, $ud{\rightarrow }Wgg$. The bottom row shows diagrams for
events with exactly three jets. Left two plots:
signals, $ud{\rightarrow }tbg$, $ug{\rightarrow }tbd$; right plot:
background,$ud{\rightarrow }Wbbg$ 
~\protect \refcite{STFNALtalk}.}
\label{fig:feynME}
\end{figure}

The Matrix Element method encodes all kinematic information of the event and
contains all the properties of the interaction. Hence this technique uses
maximal information and is firmly anchored with an understanding of the
underlying physics processes.  Matrix elements of the main signal and
background diagrams are used to compute  an event probability density 
for signal and
background hypotheses.  The signal and  background Feynman diagrams
used for
the subsamples with two jets and three jets are shown in
Fig.~\ref{fig:feynME}. A discriminant $O_{\rm ME}$ is
calculated from the probabilities  for the event to be compatible with the
signal hypothesis, $P_{signal}$  and background hypotheses  $P_{bkg}$.
\begin{equation}
\label{post}
O_{\rm ME} = \frac{P_{signal}}{P_{signal + P_{bkg}}}
\end{equation}
 where $P_{signal}$ is the properly normalized differential   cross section
(${\frac{\partial {\sigma}}{\partial {\vec{x}}}}$)  for an event which
contains objects with  reconstructed four-momenta $\vec{x}$:
\begin{equation}
\label{px}
P_{signal} = \frac{1}{\sigma} \times {\frac{\partial {\sigma}}{\partial
{\vec{x}}}}
\end{equation}
This technique is similar to those employed in the top quark mass measurement
and shares some of the same tools. 
Figure~\ref{fig:BNNMEoutput} (right panel)  shows the
high end region for $O_{\rm ME}$ obtained from the combined event sample  and
it accommodates the single top quark signal reasonably well.

A detailed study of the systematic uncertainty was carried out.  Dominant
sources of systematic uncertainties which are accounted for in the analysis
are: normalization of the three major sources of the backgrounds ($t\overline
t$, $W+jets$ and multijet backgrounds) which includes  a component arising
from the heavy flavor fraction, the $b$-tagging rate functions for the signal
and backgrounds, the jet energy scale uncertainty, and the uncertainty on the
integrated luminosity. Some of these uncertainties are dependent on the shape
of the underlying  spectrum. Uncertainties were assigned for each of the
backgrounds,  as a function of  jet multiplicities, lepton type, and number
of tags. To derive the  affect of each of the uncertainties, the inputs were
shifted  by  $\pm1\sigma$, and the analysis was redone. Systematic
uncertainties on the signal acceptance were also computed in a similar
fashion.

Given that an excess of events compared to the background estimates is observed,
the cross section is computed using the method described in
Sec.~\ref{secSTD0search}.  Before the final cross section results were
computed,  a verification of the cross section computation procedure  was
performed by generating  many sets of pseudo-experiments or `ensembles'.
These ensembles were subjected to the full analysis chain,  including
systematic uncertainties.

Generated ensembles include:
\begin{enumerate}
\item Ensembles at a few different values of total $s$+$t$  channel cross
sections.
\item SM ensemble with inclusive $\sigma$($s$+$t$ channel) =  2.9 pb.
\item Ensembles at the experimentally 
measured inclusive $s$+$t$ channel cross section.
\item zero-signal ensemble
\end{enumerate}

A pool of 1.7 M weighted signal + background events are used to generate
the zero signal ensemble. In the ensemble generation process, the 
relative and total yields of each type of background is 
fluctuated in proportion to
systematic errors. To generate one of the ensembles, a  random sample  is
drawn from a Poisson distribution about the total yield.

All  generated ensembles are used to evaluate the linearity of the cross 
section measurement by comparing the  observed cross
sections with the input cross sections.
The three techniques were tested  
and display a good agreement with a linear response function.

The SM ensemble is used to estimate the compatibility of the  measured value
with the SM expectations.  The ensemble generated at the measured cross
sections were used to verify the estimates of the systematic and statistical
uncertainties. 
The zero-signal ensemble is used to determine the sensitivity of  each
measurement. The expected $p-$value is defined as the  fraction of
zero-signal ensembles in which a SM cross section of  at least 2.9 pb is
measured. The observed $p-$value is computed as a fraction of zero-signal
pseudo-datasets in which at least the  observed cross section is measured. 

Final results of the measured cross sections and the significance are given
in Table~\ref{D0xsec} for the primary DT analysis, together with the two
supporting analyses: BNN and ME.

\begin{table}[ht]
\begin{center}
\tbl{Results from the different analysis techniques for
measurement of the cross sections and significance of the analyses
from the \dzero\ experiment.
SD below denotes the number of standard deviations.
~\protect \cite{STD0evidencePRL}.
}
{\begin{tabular}{lccc}\toprule
  & DT & ME & BNN \\ 
\colrule
$\sigma({p\overline p}{\rightarrow}s+t-{\rm channel})$ & $4.9 \pm 1.4$~pb &  
$4.6^{+1.8}_{-1.5}$~pb & $5.0 \pm 1.9$~pb \\
expected p-value &  1.9\%  & 3.7\%    & 9.7\% \\
observed p-value & 0.035\% & 0.21\%   & 0.81\% \\
observed significance & 3.4 SD & 2.9 SD & 2.4 SD \\
\botrule
\end{tabular}
\label{D0xsec}}
\end{center}
\end{table}

\begin{figure}[!h!tbp]
\begin{center}
\psfig{figure=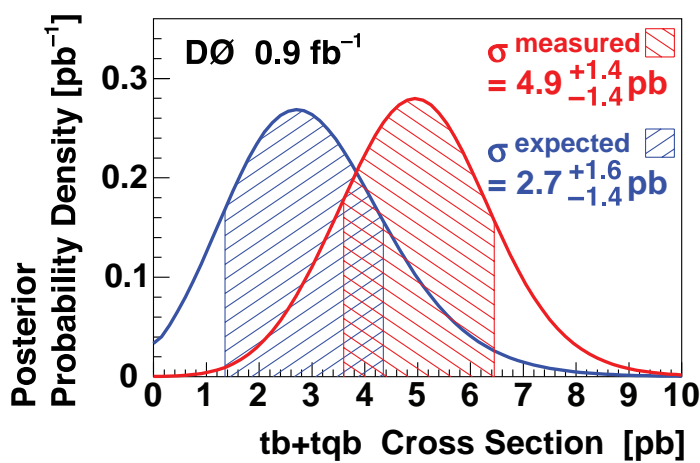,width=0.75\textwidth}
\end{center}
\vspace*{8pt}
\caption{Expected SM and measured Bayesian posterior
probability densities for the $tb$+$tqb$ cross section. The shaded
regions indicate one standard deviation above and below the peak
positions 
(reproduced from Ref.~\protect \refcite{STD0evidencePRL}).
}
\label{fig:STdt-xsec}
\end{figure}

Figure~\ref{fig:STdt-xsec} 
shows the first measurements of the single top quark cross sections
at the Tevatron. The plots are from the DT analyses.

The measured  cross section for single top quark production
$\sigma({p\overline p}{\rightarrow}s+t-{\rm channel}) = 4.9 \pm 1.4$~pb,
which is  consistent with expectations from SM. The observed $p-$value, or
the probability that the background fluctuates to  give the measured cross
section value of 4.9~pb or  greater is 0.035\%, which corresponds to a
3.4$\sigma$ evidence for the  single top quark production at the Tevatron.

The CDF collaboration  confirmed the  \dzero\ evidence for single top 
quark production in  August 2007 by analyzing a data sample corresponding to 
an integrated luminosity of 1.5 fb$^{-1}$. CDF performed analyses using two
different techniques: a multivariate likelihood technique and the matrix
element discriminant technique\cite{CDF_top_SUSY07}. 
These are extensions of the techniques already
described earlier and applied to smaller datasets.  The  single top production 
cross section measured by the matrix element technique is 
$\sigma({p\overline p}{\rightarrow}s+t-{\rm channel}) 
= 3.0 ^{+ 1.2}_{-1.4}$~pb,which is  consistent with expectations from SM
and those measured by the \dzero\ experiment.
This result corresponds to 
a 3.1$\sigma$ excess over SM background. The likelihood discriminant
exhibits a 2.7$\sigma$ excess over SM backgrounds and leads to a compatible
single top quark production cross section measurement.

%% file: mtop/mtop.tex
\subsection{Role in electroweak model}

Mass of the fundamental particles 
is currently understood to be generated from the spontaneous
breaking of electroweak symmetry via the Higgs mechanism. The high
value of the top quark mass\cite{tewg} provides a means to
establish the Higgs mass scale because the two can be related via
radiative corrections to the W mass, $M_W$.  One loop corrections to 
$M_W$ give
$M_{W}^{2}=\frac{\pi\alpha/\sqrt{2}G_{F}}{sin^2\theta_{W}(1-\Delta r)}$
where $\Delta r$ includes a correction depending on $m_t$ 
($\frac{3G_{F}m_t^{2}}{8\sqrt{2}\pi^{2}tan^{2}\theta_{w}}$)
and on the Higgs mass ($\frac{11G_{F}M_{Z}^{2}\cos^{2}\theta_{W}}{24\sqrt{2}\pi^{2}}\ln\frac{M_{H}^{2}}{M_{Z}^{2}}$).

The fact that $m_t$ is 35 times higher than the next heaviest fundamental
fermion gives it a place of importance for another reason.
Given the Higgs vacuum expectation value,
$v\sim 247$ GeV, the current world average $m_t$
points to a top quark Yukawa coupling of $Y_{t}=\frac{m_{t}\sqrt{2}}{v}\sim 1$.
Since the electroweak theory makes no prediction of these couplings
and they range over many orders of magnitude, it seems surprising that
one of them would be unity.
This value raises interesting questions about the precise nature of
the Higgs mechanism,
and some believe that the measured value of $m_t$ may indicate
extra Higgs doublets, as in Minimal Supersymmetric Models (MSSM)\cite{mssmModels}.
For these reasons, the measurement of the top quark mass has rapidly become one of the
most important efforts in the whole top quark sector.  It is in fact
one of the key parameters to come out of the current Tevatron program.

{\subsection{General techniques for mass extraction}
\label{sec:massExtract}}

The top quark mass can be extracted from each of the broad
channels in the \ttbar\ final state: dileptons, $\ell+$jets
and all-jets.  Across these channels, there
are three very general approaches to experimental measurement of $m_t$.
In order of increasingly sophisticated usage of the information in
each event, they are:

\begin{itemize}
\item{\bf\underline{Fitting Kinematic and Decay Parameters:}}  The value of $m_t$ is
   reflected in several individual observables of top quark events.  For instance,
   the direct $t\rightarrow b$ decay means that in the top quark
   rest frame, the $b-$quark has momentum $\sim m_t/2$.  In the lab frame, the
   motion of the top quark and jet specific effects reduce the strength
   of this correlation.  However, the observed $p_T$ spectrum of
   $b$-jets is still anticipated to strongly follow $m_t$.  In addition,
   the decay length distribution associated with $b$-jets
   also correlates with $m_t$.  By fitting 
   one or more of these parameters one can discriminate between
   top quarks of different masses.  Such a fit in a sample of data
   can be performed to pull out a measurement of $m_t$. 

\item{\bf\underline{Kinematic Reconstruction:}}  \ttbar\ final states leave as many 
   as 18 kinematic observables (17 for $\ell+$jets, 14 for dilepton).
   Because of the specific decay chain
   from \ttbar, there are correlations among the kinematics of these
   objects.  For instance, two light-quark jets from a $W$ boson should
   be consistent with a dijet mass $\sim M_W$.  Constraint equations 
   describe these relations and can be used to calculate solutions 
   that provide a mass estimator for each candidate event.  By fitting
   these mass estimators for a set of events, the mass of the top quark
   is extracted.

\item{\bf\underline{Matrix Element Fitting:}}  A special case of kinematic recontruction
   involves the full use of the information about top quark production and
   decay.  By using the leading-order matrix elements, in conjunction with
   a full knowledge of the experimental resolutions of the final state
   object momenta, a fit can be performed to the data.  This
   is used to provide a probability that an observed event configuration is
   consistent with a top quark of a certain mass.

\end{itemize}

Almost all analyses are based on the latter two strategies.
These approaches generally yield substantially better uncertainty than
the first because they use many, if not all, of the measurements made for
each event.  The matrix element approach in particular tends to extract the
maximum information and so has slightly better performance.  Because
each method relies on different information from the complex \ttbar\
event, both
CDF and \dzero\ pursue multiple strategies for each class of final state.
By ascertaining the correlations between these approaches, they 
provide the best overall estimate of the top quark mass.

All analyses that measure the top quark mass generate pseudo-experiments
to test
the performance and calibration of their methods and to determine the effects
of systematic uncertainties. For these tests, simulated event samples
($ensembles$) are generated that correspond to the observed event sample in
number of events and signal and background composition. These ensembles are
then processed in the same way as the collider data. In this way the
experiment at hand can be simulated many times.  The mean $\langle
m\rangle=\sum_{i=1}^Nm_i/N$ of the measured top quark masses $m_i$ from $N$
ensembles, their rms $\sigma(m)=\sum_{i=1}^N \sqrt{(\langle m\rangle -
m_i)^2/N}$, and the mean and rms of their pulls $d=(m_i-m_t)/\delta m_i$
can be determined.  Here $m_t$ is the top quark mass assumed in
the simulation and $\delta m_i$ is the statistical uncertainty in the
measurement from the $i^{th}$ ensemble. A well calibrated analysis method
gives $\langle m\rangle = m_t$, $\langle d\rangle = 0$, and
$\sigma(d)=1$. Ensemble tests can be used to calibrate the methods, i.e.,
determine corrections to be applied to the measurements so that these
relations are satisfied.
Various sources of systematic uncertainties on the mass measurement 
are considered. Some of the most prominent sources which are
carefully evaluated are: 1) jet energy scale, 2) initial state gluon radiation,
3) final state gluon radiation, 4) parton distribution functions,
5) Monte Carlo generators, 6) background model, 7) $b$-tagging,
8) Monte Carlo statistics.

\subsection{The dilepton final state~\label{sec:llmtop}}
\input{mtop/llmtop.tex }

\subsection{Single lepton channels~\label{sec:ljmtop}}
\input{mtop/ljets_mass.tex}

\subsection{All-jets channels}
\input{mtop/alljetsMass.tex}

\subsection{Combined fits and electroweak constraints}

The Tevatron Electroweak Working Group\cite{tewg} has performed
combinations of the top quark mass measurements by \dzero\ and CDF, properly
taking into account correlations between the measurements in different
channels and from both experiments. As of March 2007, 
the most precise value for the top quark mass is 
$170.9\pm1.8$ GeV~\cite{tewg}. 
More details on the combined results  are given in Table~\ref{tab:mtop}.
Their implications are discussed in Section \ref{sec:future}.

The measured value of the top quark mass can be compared to the value of the
top quark mass from the global electroweak fit performed by the LEP
Electroweak Working Group\cite{EWWG}. Based on the $Z$ lineshape measurements
a value of $172.6^{+13.2}_{-10.2}$ GeV is most consistent with the standard
model. If one includes in addition the measured $W$ boson mass and width into
the fit, the best top quark mass goes up to $178.9^{+11.7}_{-8.6}$ GeV. Either value
is completely consistent with the value from the direct
measurement.

\begin{table}[ht]
\begin{center}
\tbl{Summary of Top Quark Mass Measurements.}
{\begin{tabular}{lcc}\toprule
 Channel & Experiment & Measured Mass (GeV$/c^2$) \\  \colrule
$\ell\ell$ & CDF & 164.5$\pm$ 5.6\\
$\ell\ell$ & \dzero\ & 172.5$\pm$ 8.0 \\
$\ell+$jets & CDF & 170.9$\pm$ 2.5 \\
$\ell+$jets & \dzero\ & 170.5$\pm$ 2.7 \\
All-jets & CDF & 171.1$\pm$4.3 \\
Lepton+jets ($L_{xy}$) & CDF & 183.9$\pm$ 15.8 \\
\botrule
\end{tabular}
\label{tab:mtop} }
\end{center}
\end{table}

%% file: mtop/llmtop.tex
It is important to study $m_t$ in all of the final states
in which the top quark can be identified for two main reasons.  Non-standard
decays can impact specific final states differently.  The kinematic
analysis involved in a mass measurement therefore provides an additional
test of the hypothesis that the signal events conform to the
$t\rightarrow Wb$ decay chain.  If this chain is other than
expected, sufficient statistical precision would reveal discrepancies between the measured
$m_t$ as estimated in different channels.  Dilepton channels provide
both an independent statistical sample with which to measure
$m_t$.  As the integrated luminosity at the Tevatron increases or the LHC
turns on, the
statistical limitations of this channel become less relevant.  Systematic
uncertainties are similar in magnitude to those for the single lepton
channels, and so the dilepton-based measurement has an important role
in improving the world-average top quark mass.

\subsubsection{Fitting with kinematic parameters}

The effort to extract a measure of $m_t$ in dilepton events
has involved each of the techniques mentioned in 
Section~\ref{sec:massExtract}.  The most
basic of these involves fitting to kinematic distributions in
candidate events.
Because the top quark is so much more massive than any of the fundamental 
fermions in its decay chain, the magnitude of $m_t$ is directly manifested in the
momenta of these particles. For instance, the transverse momenta of
the $b$-quarks are of order $m_t/2$. 

This approach has been tried by CDF using eight events in their 
explicit dilepton sample.  Two different approaches were combined.
If one ignores $b$-quark
and lepton masses, then the top quark mass can be expressed approximately
as $m_t^{2}=M_{W}^{2}+\frac{2(M_{lb}^{2})}{1-\cos\theta_{lb}}$.
One can rewrite this expression in terms of well-measured quantities,

\begin{eqnarray}
m_t^{2}=\left\langle M_{lb}^{2}\right\rangle +\sqrt{M_{W}^{4}+4M_{W}^{2}\left\langle M_{lb}^{2}\right\rangle +\left\langle M_{lb}^{2}\right\rangle ^{2}}
\end{eqnarray}

\noindent Here the $b$ refers to the jets which are presumed to arise from
the $b$-quarks.  There is some confusion arising from the combinatorics of
mapping the jets and leptons that should come from the same top quark.  
Considering the allowed jet-lepton configurations, each event produces two
$M_{lb}$ measurements. These values are used to provide a value for 
$\left\langle M_{lb}^{2}\right\rangle $
for the sample using a prescription tuned with \ttbar\ events generated
with {\sc Herwig}. A top mass measurement is extracted from this parameter
in the data sample\cite{llmtopMlb}. This is augmented with another
approach which uses the average $p_T$ of the two $b-$jets in the event.
This average is very correlated with $m_t/2$.  A maximum likelihood
fit is applied to templates generated from this parameter for signal
and background samples. A combined mass estimate of 
$161 \pm 17 (stat) \pm 10 (sys)$ GeV was obtained from these 
analyses\cite{llmtopMlb}.

\subsubsection{Fitting using kinematic reconstruction}

While the technique of fitting to kinematic parameters provides useful,
general estimates of $m_t$, it does not use all
of the information in an observed event that might maximize the measurement's
precision.  For instance, momentum correlations between final state particles
which can be used to test whether the event is consistent with two
top quarks of equal mass, or two $W$ bosons with mass equal to $M_W$. 
These capabilities are gained with an explicit kinematic reconstruction
of an event.  Performing such a reconstruction with
dilepton events has the challenge that there are two neutrinos.  There are 18
unknowns for the six initial fermion three-momenta, but only four full
three-momenta measured plus two components for the \met.  Three additional
constraints arise from requiring each $l\nu$ pair to give $M_W$,
and both top quarks must have equal mass.  As a result, a -1C underconstrained
fit results and some information must generally be provided as input to 
solve these events.  Generally, approaches
assume an input value of $m_t$ in steps over the allowed range (e.g.
150 GeV to 200 GeV).  

The specifics of how the events are solved varies with method (see below). 
A two-fold ambiguity results for the solutions for the momentum
of each neutrino.  This ambiguity in kinematic reconstruction is
compounded by the different assignments of lepton
and jet to one of the $t\rightarrow Wb$ decay chains.  A particular 
$configuration$
refers to a particular jet assignment in tandem with one of the
neutrino solutions given that assignment. Generally, a $weight$ is
calculated for each configuration which reflects the relative consistency
of that configuration with some additional observed property in the
event. This weight is summed for all configurations which are solved
for a particular input $m_t$. 

This integration is performed for a range of assumed top quark masses to
provide a probability or `weight' vs. $m_t$ for each event. 
Several
parameters of the weight distribution carry information about the
actual $m_t$. These 'mass estimator' parameters are
used to compare events in data with expectations from top quark signal events
of different masses, and background.  Often, analyses will choose just
the mass for which the weight distribution is maximum (`peak') as the
mass estimator.  The peak does indeed carry most of the sensitivity
to the actual $m_t$ for the two kinematic reconstruction strategies 
described below.  However, more information is available in the rest of
the weight distribution.  Other approaches extract this information by
forming a coarsely binned template from the weight distribution, or
taking instead the first couple moments of this distribution.
In any of these cases, the distribution of mass estimators
is generated for data, and this must be compared via maximum
likelihood fit to templates of distributions for signal and background
samples.

\subsubsection{Neutrino weighting}

One method for solving \ttbar\ events utilizes
the expected neutrino rapidity distributions.
This method is termed `neutrino weighting' ($\nu WT$) and was first
developed by \dzero\ in Run I\cite{d0r1llmtop}. A kinematic fit
is performed by omitting two measured variables, \mex\ and
\mey. To compensate for this
loss of information, one assumes rapidities for the two neutrinos.
An integration over the neutrino rapidity distributions is performed
such that for each choice of neutrino rapidity pair and jet combination, 
the neutrino momentum solutions are calculated.  For each configuration,
a weight is calculated by comparing the measured event \met\
with the kinematically reconstructed \met\ 
\begin{eqnarray}
w={\displaystyle \sum\exp\left(\frac{-(E_{x}^{calc}-E_{x}^{obs})^{2}}{2\sigma_{E_{x}}^{2}}\right)\exp\left(\frac{-(E_{y}^{calc}-E_{y}^{obs})^{2}}{2\sigma_{E_{y}}^{2}}\right)}
\end{eqnarray}

\noindent This weight is summed for all configurations for each value of $m_t$.
Templates using the parameters from the full integrated weight vs. 
$m_t$ distribution can be used to extract a top quark mass.

\begin{figure*}[!h!tbp]
\begin{center}
\epsfig{figure=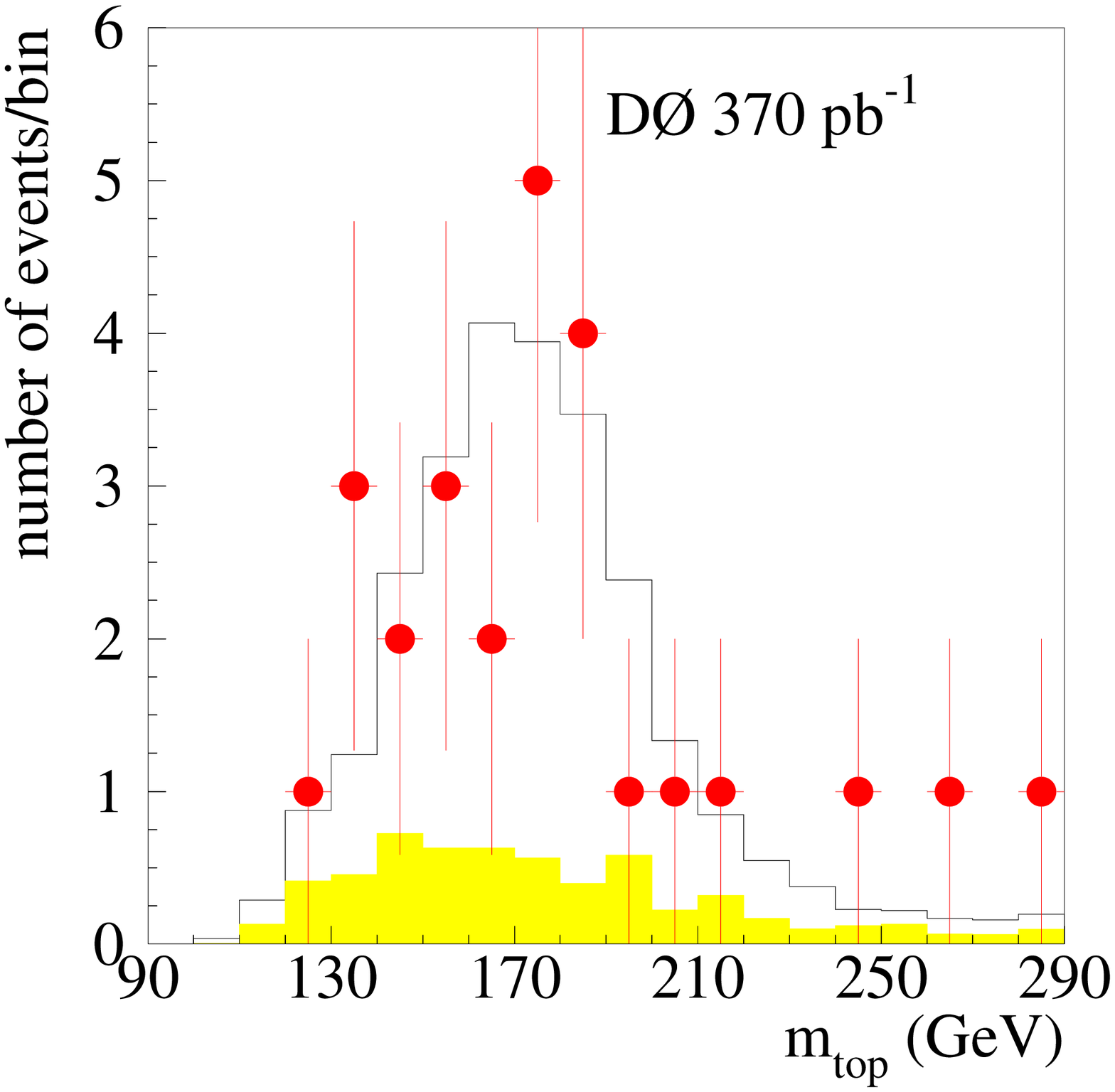,width=0.40\textwidth}
\epsfig{figure=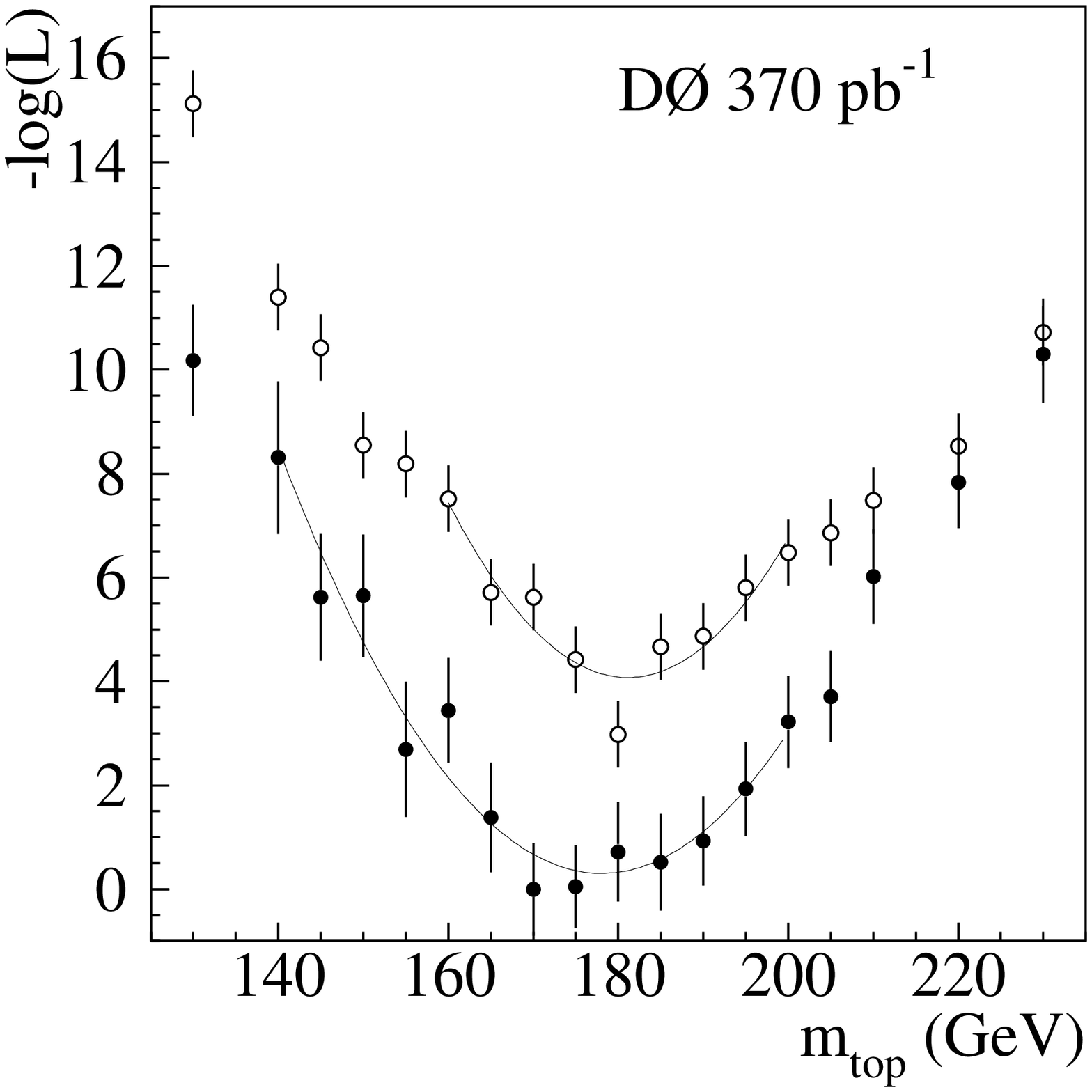,width=0.40\textwidth}
\end{center}
\vspace*{8pt}
\caption{
The $m_{peak}$ distribution from the $MWT$ technique applied to
370 pb$^{-1}$ of dilepton events from \dzero\ (left).  The $\nu WT$ (open circles)
and $MWT$ (closed circles) 
fits are shown (right) for the data.  The former also includes 
the $\ell+$track channel\protect\cite{d0r2llmtPLB}.}
\label{fig:d0r2mt} 
\end{figure*}

\dzero\ has performed this analysis with Run~I\cite{d0r1llmtop} and 
Run~II\cite{d0r2llmtPLB} data by coarsely binning the weight distribution.
The Run II analysis differs primarily in that the $\ell+$track channels
were included.  A maximum likelihood fit of these templates is performed to 
analogous templates for simulated signal and background, and instrumental
backgrounds from data.  A probability density estimator approach is taken 
\cite{PDEref}.  The signal probability for a given $m_t$, 
$f_s(\vec{W}|m_t)$, and background probability, $f_b(\vec{W})$ 
are established by comparison
of data templates ($\vec{W}$) with those for signal and background.  The value
of $m_t$ can be extracted by maximizing the likelihood
\begin{eqnarray}
L(m_t,\bar{n}_b,n) = G(n_b - \bar{n}_b,\sigma)P(n_s+n_b,n)\times
\Pi_{i=1}^n [(n_s f_s + n_b f_b)/(n_s + n_b)]
\end{eqnarray}
where a Gaussian and Poisson constraint are placed on the estimate of
$n_b$ and $n_s+n_b$ given the observed event yield, respectively.
As is typically done with these analyses, the method is tested 
by performing the analysis on independent pseudo-experiments
of simulated signal and background events.  In common with most
of the mass analyses, the relation
of average fitted top quark mass vs. input $m_t$ was verified in each channel 
to have slope near unity and with
small offset.  This helps to keep systematic uncertainties from the method
to a minimum.  Any residual difference of the slope and offset from
unity and zero, respectively, is used to correct the fitted mass.
The statistical uncertainty from a given ensemble is 
extracted from limits in $m_t$ when $-ln(L)$ varies by 0.5 from its minimum
value.  It is cross-checked by verifying that
pull widths are near 1.0.  Statistical uncertainties for the Run II 
analysis were corrected for a small deviation of pull widths from unity.
The plot of the $-ln(L)$ vs. $m_t$ resulting from the maximum likelihood
fit to the Run II data is given in Fig.~\ref{fig:d0r2mt}.
The top mass was measured to be $m_t = 179.5 \pm 7.4 (stat)$ GeV.
A 5.6 GeV systematic uncertainty is dominated by the jet energy scale.
This measurement was combined with the matrix-weighting determination
(see Section \ref{sec:MWT}) to yield
$m_t=178.1 \pm 6.7 (stat) \pm 4.8 (sys)$ GeV\cite{d0r2llmtPLB}.
The Run I measurement was $m_t=170.0\pm14.8(stat)$ GeV~\cite{d0r1llmtop}. 

The dilepton analysis with $\nu WT$ has been pursued in preliminary
analyses in the 1 fb$^{-1}$ data sample \cite{d0r2nuWTmoments,d0r2nuWT1fb}.
One question for these analyses has been what is the most optimal way to use
the output of the weight calculation (i.e. the weight distribution) to
get the smallest expected uncertainty in the top mass.
Ref. \refcite{d0r2nuWTmoments} presents an analysis in $e\mu$ events
of three different approaches
to extract the value of $m_t$ from the weight distribution.  These analyses
attempted to determine an optimal use of the most important parts of a
weight distribution.  One of these three methods
constituted an optimized version of the binned template approach described
above.  Another approach used just the value of the top mass for which the
weight was maximized, and then performed a 2-dimensional fit to this 
peak value and the input $m_t$.  A third approach considered various 
properties of the weight distribution (e.g. moments, integral weight in
high and low mass bins, etc.) and settled on the first two moments as
the minimal number of parameters that provided the 20\% improvement in sensitivity
usually associated with the use of the full binned template.  The 2-dimensional
fit and moments approaches each were found to match the $a priori$ expected sensitivity
of the binned-template method, 
and were combined into a measurement using dilepton events ($ee, e\mu, \mu\mu$)
in 1 fb$^{-1}$ \cite{d0r2nuWT1fb}.  A value of $m_t = 172.5 \pm 5.8(stat) \pm 3.5 (sys)$ 
GeV was obtained.

\begin{figure*}[!h!tbp]
\begin{center}
\epsfig{figure=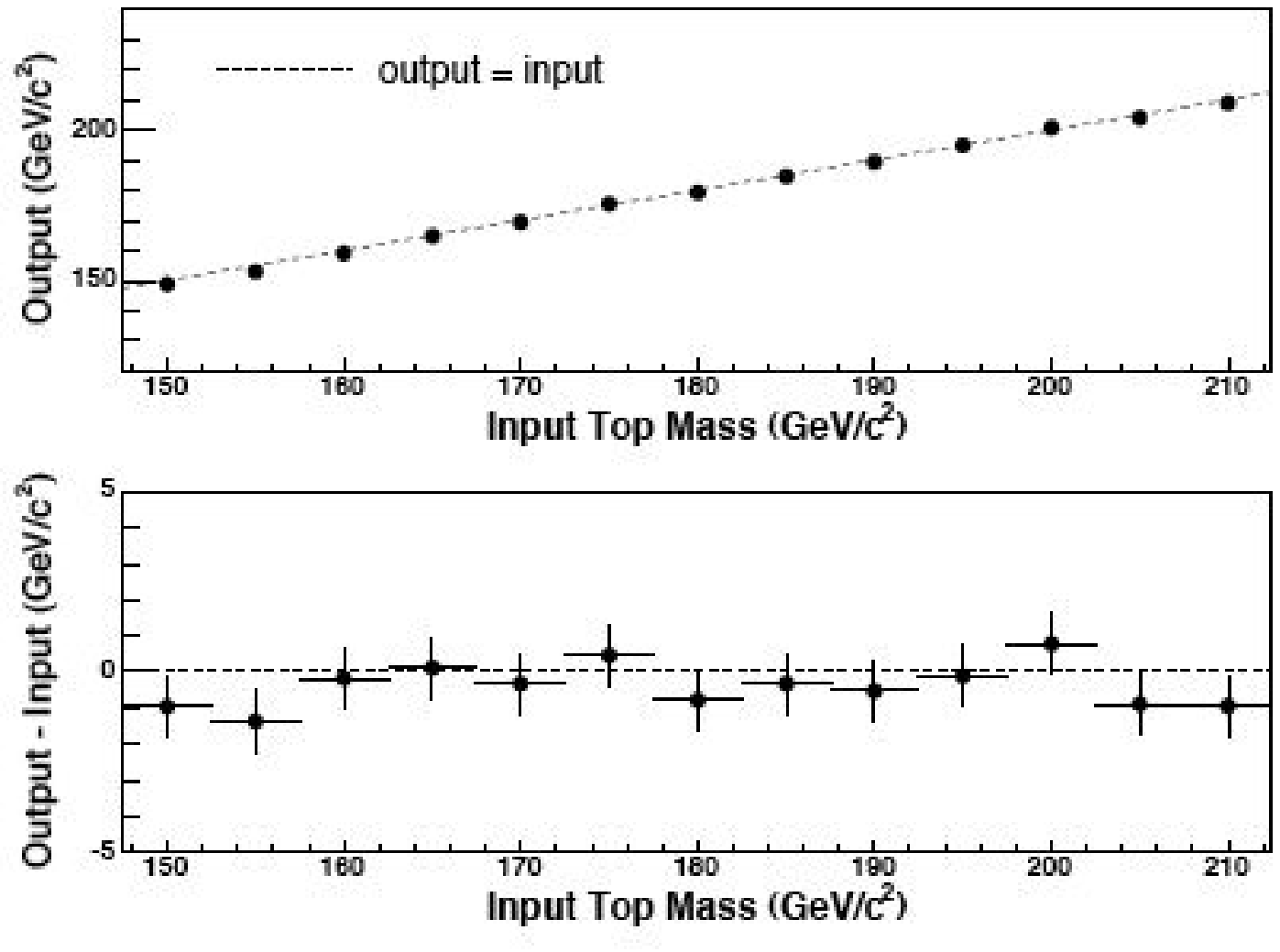,width=0.80\textwidth}
\epsfig{figure=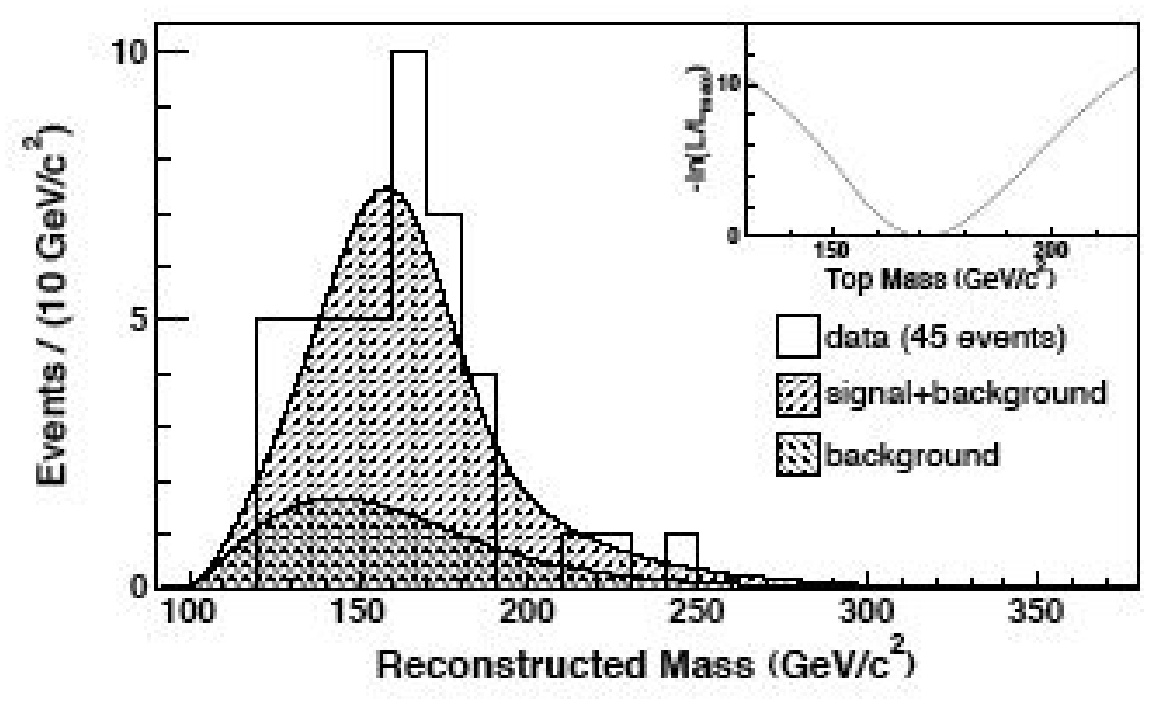,width=0.80\textwidth}
\end{center}
\vspace*{8pt}
\caption{$\nu WT$ results from the 359 pb$^{-1}$ CDF Run II $\ell+$track 
sample. The linearity
of the fit (left) and data overlaid with signal and background templates
(right) are shown\protect\cite{nuWTcdf2}.
}
\label{fig:d0r2ltrmt} 
\end{figure*}

CDF has pursued the $\nu WT$ approach in 359 pb$^{-1}$ of data in 
Run II. The basic kinematic
reconstruction is performed similarly to that used by \dzero\ .  
In Run~II, they applied the technique to their $\ell+$track
sample of 46 events. CDF chose the mass estimator 
as the peak of the weight distribution.  Thus the data $m_{peak}$
distribution is compared to the expectation 
from signal and background (see Fig.~\ref{fig:d0r2ltrmt}).  The maximum 
likelihood fit on data gives $m_t=170.7^{+6.9}_{-6.5} (stat) \pm 4.6 (sys)$ 
GeV\cite{nuWTcdf2}. 
Systematic uncertainties were estimated for jet energy scale, gluon radiation
(ISR and FSR), background template shape, $pdf$s, and showering Monte Carlo 
generator.  A further improvement for CDF is obtained by combining this 
analysis with two others of lesser statistical power.  These use an assumption
of the longitudinal momentum of the top quarks, or the azimuthal angle of
the neutrinos, to solve each candidate event.  The final result from this
combination is $m_t=170.1\pm 6.0 (stat) \pm 4.1 (sys)$ GeV\cite{cdfR2lltemp}.
In Run I, which used an explicit dilepton selection, the
mass estimator for each event was taken as the weighted mean
in a window around the mass with highest weight. This approach generated
a value of $m_t=167.4 \pm 10.3 (stat) \pm 4.8 (sys)$ GeV\cite{nuWTcdf1}. 
CDF performed
a consistency check with the fitting procedure garnered in single
lepton double-tagged events by treating the two light-quark jets
from the W has a lepton and a neutrino. By mimicing the dilepton signature
in this way they obtain a mass estimate in these events which was
within the expected resolution.

{\subsubsection{Matrix weighting}
\label{sec:MWT}}

Another approach to kinematic reconstruction of top quark events has been
proposed~\cite{dalitz,kondo}.  In this method, the measured
\met\ is not omitted from the kinematic reconstruction, thereby
permitting a \ttbar\ event to be solved to an eightfold ambiguity if 
$m_t$ is assumed.   Unlike the $\nu WT$ approach, both 
the $t$ and $\bar{t}$ are solved simultaneously.  A weight
is calculated for each solved configuration with an assumed mass, $m_t$:
\begin{eqnarray}
w=f(x)f(\overline{x})P(E_{l1}|m_{t})P(E_{l2}|m_{t})
\end{eqnarray}
\noindent where $f(x)$ is the $pdf$ taken at the Feynman $x$ obtained
for the solution.  The
$P(E^*_l | m_t)$ are the probabilities for a lepton to have energy $E_l$
in the solved top quark rest frame.  This approach
is termed `matrix weighting' ($MWT$). 

The \dzero\ Collaboration has pursued this approach in both Run I and Run
II.  Using the 370 pb$^{-1}$ data sample from Run II, they considered the
explicit dilepton channel~\cite{d0r2llmtPLB}.  
The candidate sample was broken down into
the low background $b$-tagged subsample, and the higher background sample
with strictly topological selection.  This topological selection
loosens kinematic and electron identification requirements on the 
$e\mu$ channel relative to what was done for the $\nu WT$ analysis 
described above.  This separation by $b-$tagging was done to enhance the
sensitivity to the signal events in data. The maximum likelihood is performed
by taking the peak of the weight distribution:
\begin{eqnarray}
L(m_t) = \Pi_{i=1}^{n_{bin}} ((n_s s_i(m_t) + n_b b_i)/(n_s+n_b))^{n_i}
\end{eqnarray}
\noindent where the product is taken over the $i$ bins of the $m_{peak}$
distribution from data ($n_i$) as compared to a signal template 
($s_i$) and a background template ($b_i$).  This constitutes a different
approach than used for Run I since there are single templates from the
$m_{peak}$ distributions of the data, signal and background rather than 
templates from full weight distributions for each event in data, signal 
and background samples. 
From 26 dilepton events, a value of $m_t = 176.2 \pm 9.2 (stat) \pm 3.9 (sys)$
GeV was determined.  
The plot of the $-ln(L)$ vs. $m_t$ is given in Fig.~\ref{fig:d0r2mt}.
For comparison, the Run I measurement was
$m_t=168.2 \pm 12.4 (stat)$ GeV\cite{d0r1llmtop}.  
For the Run II measurement, the 
dominant systematic uncertainty was from
the uncertainty in jet energy calibration.  The matrix-weighting
approach has been applied to the 1 fb$^{-1}$ sample in dilepton
events and yielded $m_t = 175.2 \pm 6.1 (stat) \pm 3.4 (sys)$ GeV
\cite{d0r2MWT1fb}.

\subsubsection{Matrix element fit}

Use of the full matrix-element to encompass the expected correlations
of particle kinematics in a mass analysis was first applied by \dzero\ 
in complete
form with single lepton events\cite{MEmtop}. The CDF collaboration
has performed such an analysis with dilepton events\cite{llMEcdf2}.
The analysis uses the explicit dilepton event selection from the
cross section measurement described above. This gives 33 events in
340 pb$^{-1}$.

This analysis differs from those described above which integrate over
just one class of parameters to perform the fit, such as neutrino
rapidity or Feynman $x$. Instead the leading order matrix element for
$q\bar{q} \rightarrow \ttbar$ production is used.
Some initial assumptions are implemented which permit a more
tractable calculation. Lepton momenta are assumed to be perfectly
measured. Quark directions are assumed to be perfectly indicated by
observed jet angles. An integration is then performed over the neutrino
and quark energies as calculated from the LO matrix element. This
integration is constrained by a $transfer$ $function$ which expresses
the mapping of quark energies to observed jet energies obtained from
a sample created with {\sc Herwig} fed into {\sc GEANT}. The probability
density can then be expressed as

\begin{eqnarray}
P_{s}(x|m_{t})=\frac{1}{\sigma(m_{t})}{\displaystyle \int d\Phi\left|\mathcal{M}_{tt}(q_{i},p_{i};m_{t})\right|^{2}\times\prod_{jets}W(p_{i},}j_{i})f_{PDF}(q_{1})f_{PDF}(q_{2})
\end{eqnarray}

\noindent where the matrix element is performed as in Ref.~\cite{MEref}. When
performing this integration, the contribution from the leading order
gluon-gluon production is omitted. A similar probability density calculation
is performed for those background processes arising from Z and $WW$
production, as well as W + jets in which one jet fakes a lepton. One
can then obtain a joint likelihood for an event sample by multiplying
the likelihoods for each event from the background and signal probability
densities. 

\begin{figure*}[!h!tbp]
\begin{center}
\epsfig{figure=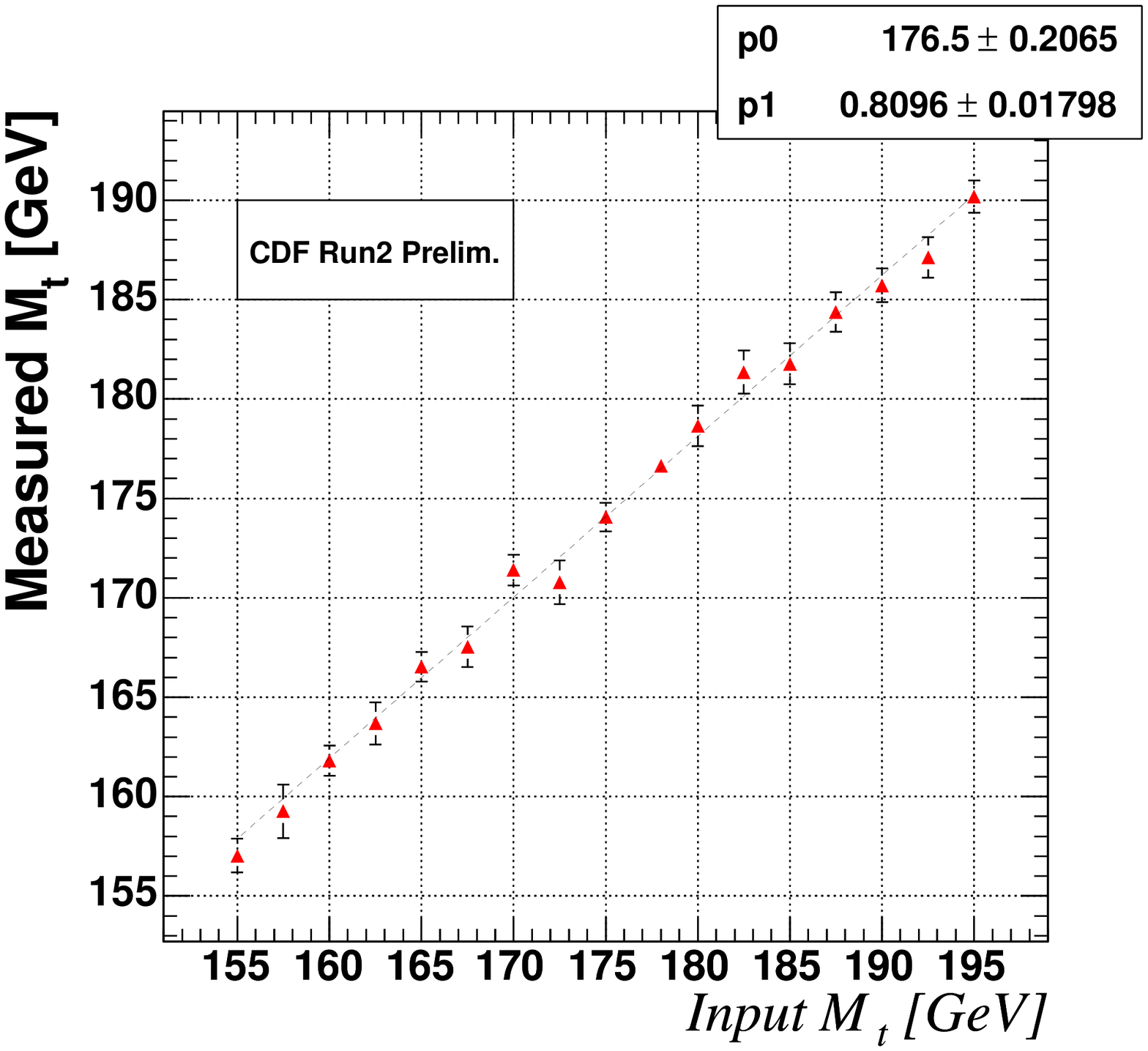,width=0.40\textwidth}
\epsfig{figure=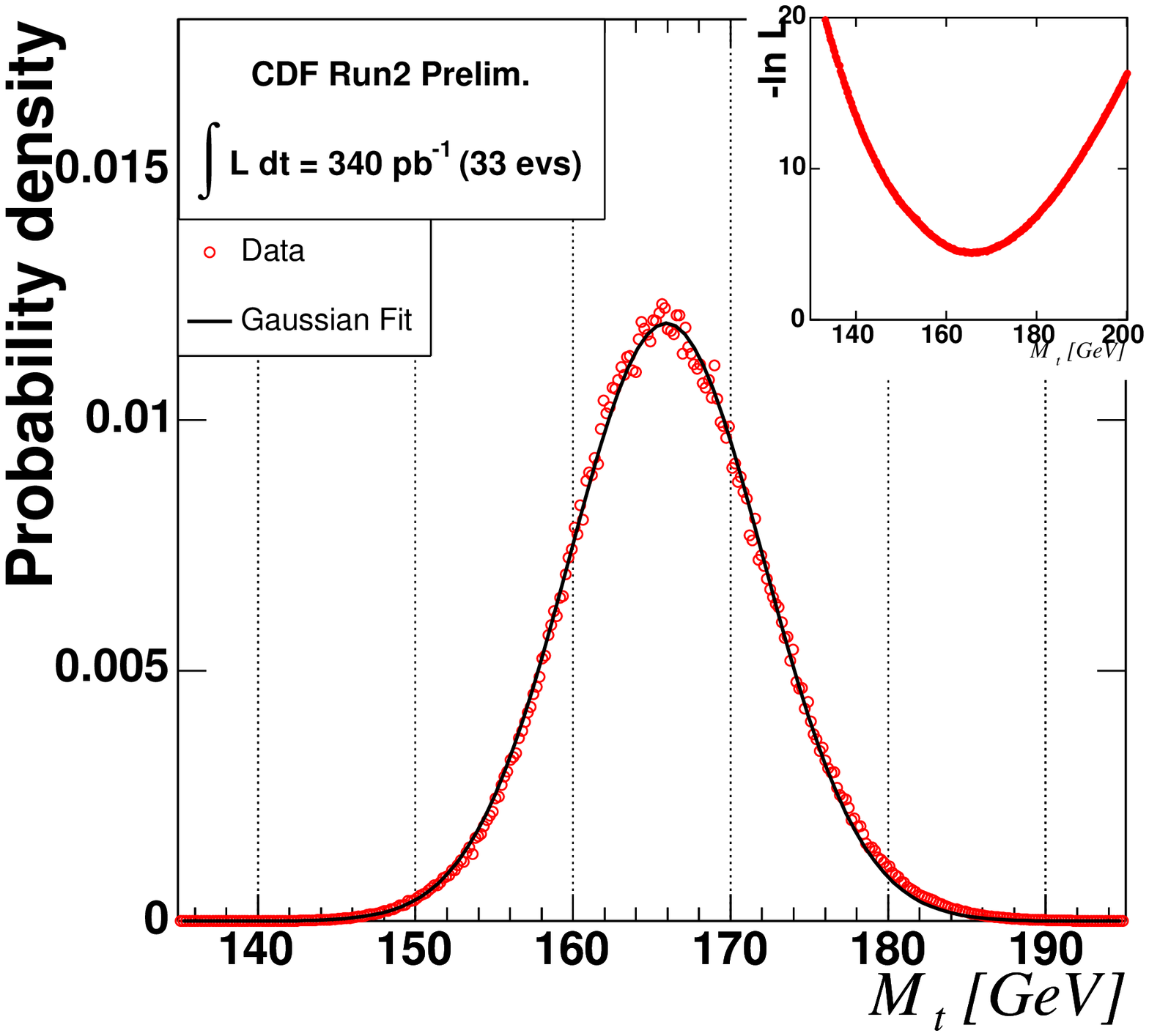,width=0.40\textwidth}
\end{center}
\vspace*{8pt}
\caption{CDF matrix element results in 340 pb$^{-1}$ dilepton events. 
The scale between fitted and input top
mass is shown (left) and the probability density as a function of $m_t$
is shown at right\protect\cite{llMEcdf2}.}
\label{fig:cdfllME} 
\end{figure*}

In order to test the validity of the result, ensembles are constructed
from simulated signal and background samples which were allowed to
fluctuate with Poisson fluctuations. The instrumental background was
obtained from data. The resulting slope vs. input (pole) $m_t$,
shown in Fig.~\ref{fig:cdfllME}, is not unity primarily because 
of the simplification
of the background components in the probability density calculation.
The measured top quark mass and estimated statistical uncertainty are corrected
for this slope. Additionally, the resulting pulls are 
1.51, so the statistical uncertainty is corrected for this. The resulting
top mass measurement in data is $m_t = 165.2 \pm 6.1 (stat)$ GeV\cite{llMEcdf2}.
Figure.~\ref{fig:cdfllME} shows the probability density vs. $m_t$, as
well as the $-ln(L)$ vs. $m_t$ (inset).
Systematic uncertainties are estimated for several sources. The chief
one is the jet energy scale (2.6 GeV) which is obtained by varying
this calibration within its allowed range. The effect of uncertainties in
$pdf$s
is quantified in several ways, for instance comparing CTEQ5L
and MRST72, and by exploring the range given in the CTEQ6M set of
functions. Combining all sources gives a 1.0 GeV $pdf$ uncertainty.
Initial and final state radiation are varied giving a 0.5 GeV and
0.7 GeV uncertainty, respectively. {\sc Pythia} and {\sc Herwig} generated events
are used to provide a Monte Carlo generator uncertainty (0.8 GeV).
Other uncertainties are calculated for background modeling and the fitted
mass slope correction. The total uncertainty tallies to $3.4$ GeV. 
When combined with the template-based methods, the value of $m_t$
was determined to be $167.9 \pm 5.2 (stat)\pm 3.7 (sys)$ GeV.
The matrix element method was also applied in 1.03 fb$^{-1}$ of data 
and yielded $164.5\pm 3.9 (stat)\pm 3.9 (sys)$ GeV \cite{cdfllME1fb}.

%% file: mtop/ljets_mass.tex
$\ell+$jets events the top quark mass is reconstructed from
the momenta of the charged lepton ($e$ or $\mu$),  the
transverse momentum of the neutrino which is inferred from the \met\ in the
event and the momenta of the four jets in the detector, 
two from the fragmentation of the $b$ quarks and two from the 
hadronic $W$ decay. 
There are two unknown parameters in the kinematic description of lepton+jets
events:  $\mbox{$\slash\kern-5.5pt p_z$}$ and $m_t$. 
On the other hand the event kinematics have to
satisfy four constraints. The charged lepton and neutrino momenta must add up
to a system whose mass equals the mass of the $W$ boson. Adding the momentum
of the $b$-jet from the semi-leptonic top quark decay must give a system with the
mass of the top quark. The invariant mass of the di-jet system from the
hadronic $W$ decay must equal the $W$ boson mass and the three-jet system
from the hadronic top quark decay must have the mass of the top quark.
We can thus perform a 2-C kinematic fit of the $\ell+$jets decay hypothesis
to these events to determine the most likely value of the top quark
mass. This simple picture is complicated by the presence of jets from
gluons radiated from 
the initial state and by the splitting of jets from the top
quark decay. Moreover it is not possible to uniquely assign the jets in the
event to the partons from \ttbar\ decay. If only the four jets with
the largest transverse momentum values are considered, there are 12 different
ways to assign jets to partons. If one jet is identified as originating from
the fragmentation of a $b$-quark and is only assigned one of the $b$-quarks
this reduces to eight permutations. If two jets are identified as $b$-jets,
this further reduces to two permutations.

The experiments have applied several techniques to extract the top quark mass
from $\ell$+jets events. 
Almost all are based on the kinematic reconstruction of
the top quark mass. An exception is a measurement by the CDF collaboration that
used the decay length distribution of $b-$jets to determine the
top quark mass. The simplest kinematic reconstruction technique
is the template method which has been used by CDF and \dzero. 
The matrix element
method and the ideogram method were both developed at \dzero\ and are now used by
both experiments.  We will
describe these methods in the following sections.

The kinematic reconstruction of the top quark mass requires that all energy
and momentum measurements are well calibrated. The electron energy scale and
the muon momentum scale can be calibrated precisely enough with $Z\rightarrow
\ell^+\ell^-$
 decays that the residual uncertainty is negligible. The calibration
of the jet energy scale, however, is more difficult and gives rise to the
dominant systematic uncertainty in the top quark mass measurement.
In addition to the $a priori$ calibration of the jet energy scale, most recent
measurements of the top quark mass in the $\ell$+jets channel also make use
of the hadronic $W$ boson decays in these events to gain an additional
constraint on the jet energy scale. These analyses use the known $W$ mass as
an input parameter and then determine simultaneously the top quark mass
$m_{t}$ and an overall jet energy scale parameter $\alpha_{jes}$ that
multiplies all jet energies.

\subsubsection{Template method~\label{sec:template}}

For the template method one chooses an estimator for the top quark mass,
typically the best fit mass $m_{fit}$ from a kinematic fit of the event. A
simulation is used to compute the expected distribution of this estimator
based on the expected number of signal and background events as a function of
the top quark mass. These distributions are referred to as templates. One
then performs a fit of the templates to the distribution of the estimator
observed in the data to determine the value of the top quark mass that best
predicts the data distribution.

The early measurements of the top quark mass by CDF and \dzero\ 
in 1995\cite{d0Obs,cdfObs} used this method. \dzero\ has used this 
method for its first
top quark mass measurement from Run~II using of approximately 230 pb$^{-1}$
of data\cite{KB_thesis}. CDF also performed a 
measurement of the top quark mass using this method on 318 pb$^{-1}$
of data from Run~II\cite{CDF_template}.

The CDF analysis divides the sample of events with a high-$p_T$ electron or
muon, \met\, and at least four high-$p_T$ jets into four subsamples
of different combinatorics and background contamination levels defined
by the number of $b$-tagged jets: at least two $b$-tagged jets, at least 
one `tight' $b$-tagged jet, at least one `loose' $b$-tagged jet, and
no $b$-tagged jets.  For every event with
$\chi^2<9$ for the kinematic fit the hypothesized top quark mass $m_{fit}$
that minimizes $\chi^2$ is entered into a histogram. The masses of all pairs
of jets that are not tagged as $b$-jets are entered in another
histogram. Both quantities are histogrammed separately for the four
subsamples. 

\begin{figure}[htp]
\begin{center}
\includegraphics[width=3in]{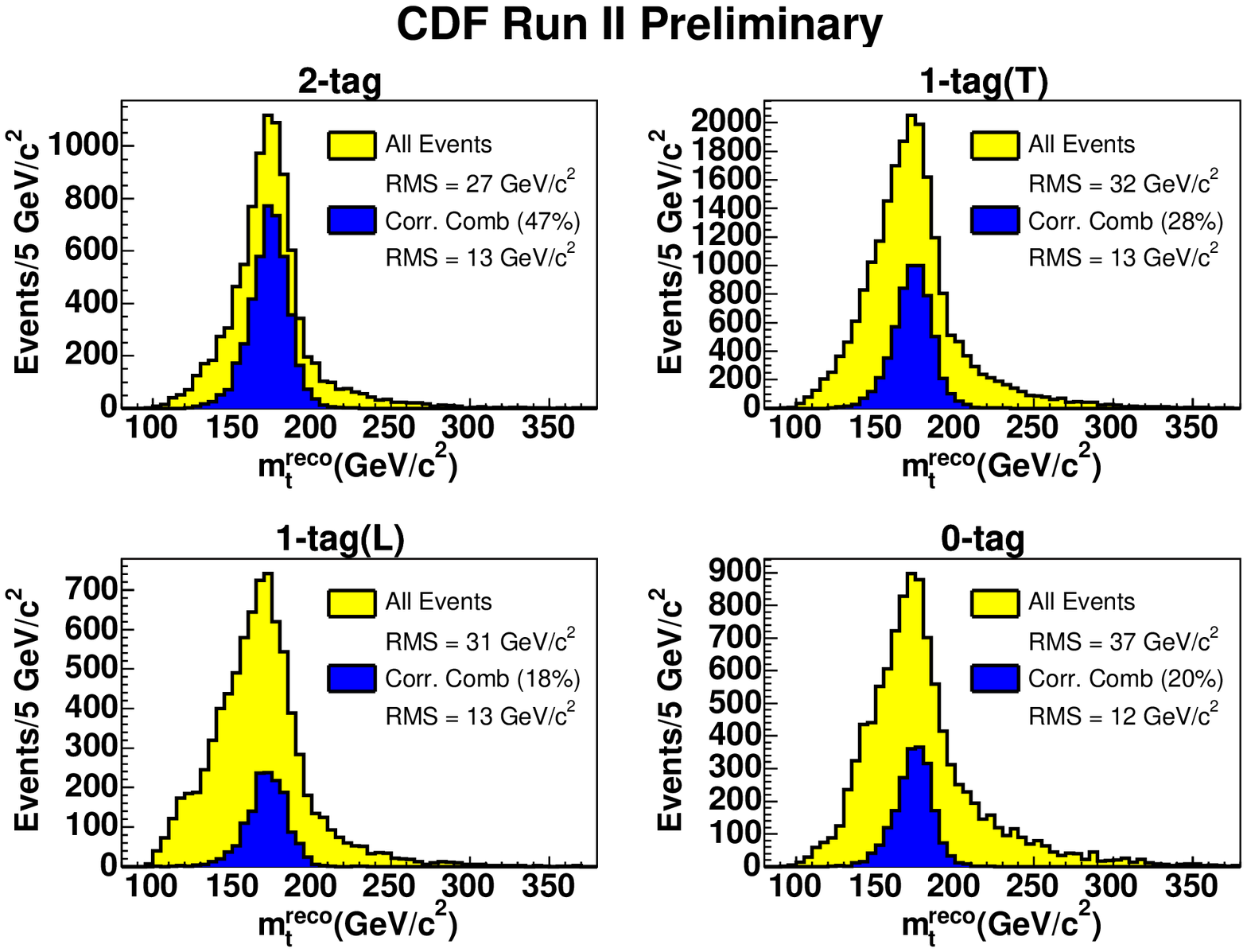}
\includegraphics[width=3in]{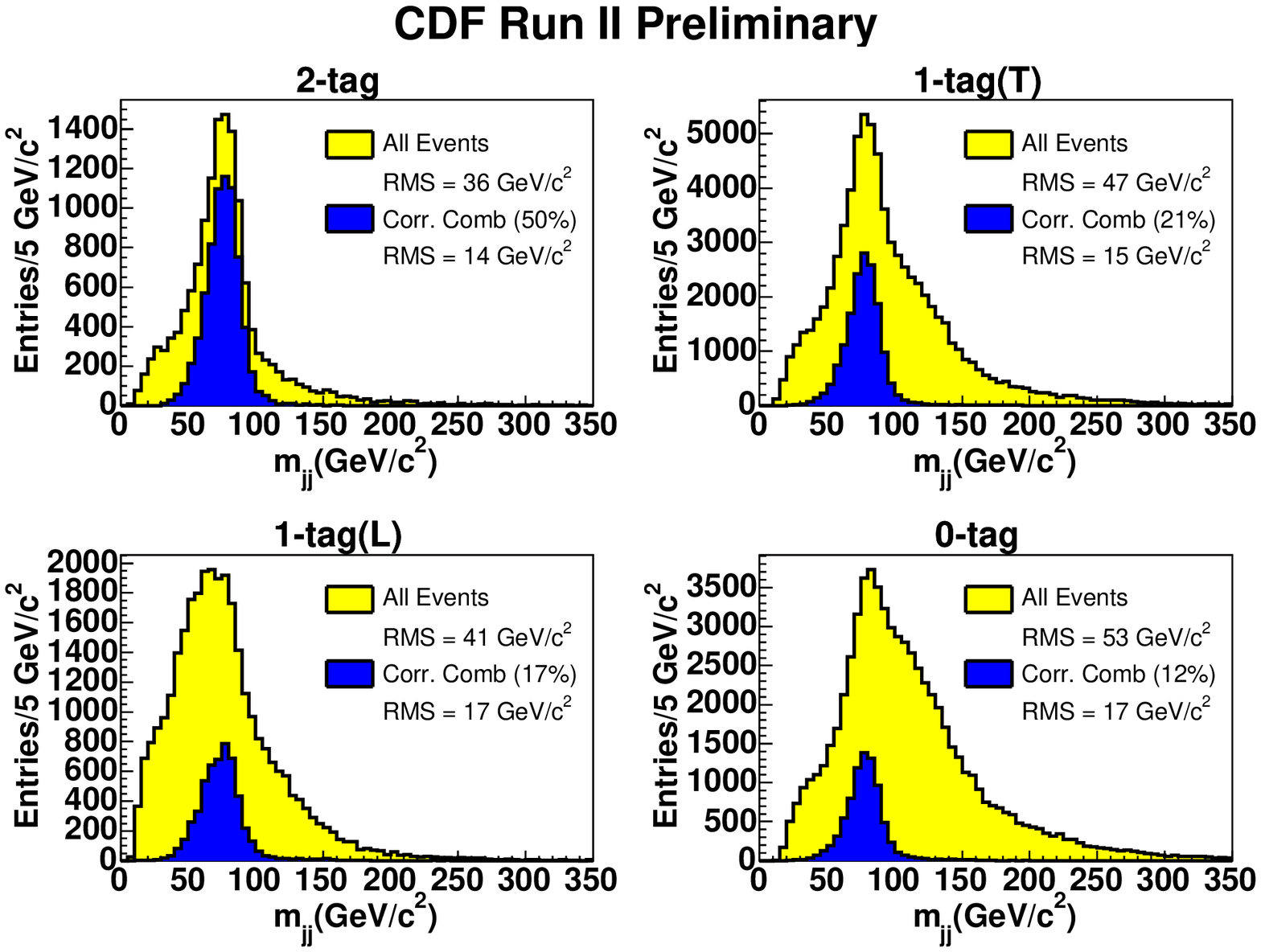}
\caption{\label{fig:CDF_templates} Sample templates for $m_{fit}$ and $m_{jj}$ for the CDF $\ell$+jets data\protect\cite{CDF_template}.}
\end{center}
\end{figure}

The templates are histograms of $m_{fit}$ and $m_{jj}$ from Monte Carlo
simulations, parameterized in terms of the hypothesized top quark mass
$m_{t}$ and the jet energy scale parameters $\alpha_{jes}$ and separately
for each of the subsamples. Sample templates are shown in Fig.~\ref{fig:CDF_templates}.

\begin{figure}[htp]
\begin{center}
\includegraphics[width=3in]{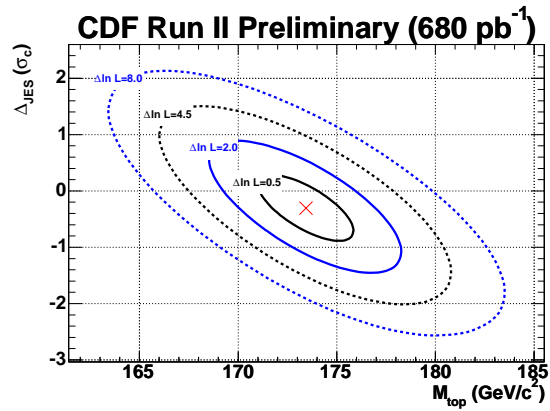}
\caption{\label{fig:CDF_template_fit} Contour plot of the likelihood for the
  template fit to the CDF $\ell+$jets data\protect\cite{CDF_template}.  This
shows the measured shift in jet energy calibration as it is correlated
with the measured $m_t$.}
\end{center}  
\end{figure}

A simultaneous fit of the data to the parameterized templates gives best
agreement for
$m_{t}=173.5^{+3.7}_{-3.6}(\mbox{stat}\oplus\mbox{jes})\pm1.3(\mbox{sys})$~GeV. The $a priori$
calibration of the jet energy scale $\alpha_0\pm\delta\alpha$ 
enters
into this fit as a Gaussian constraint on $\Delta\alpha=0\pm1$, where
$\Delta\alpha = (\alpha_0-\alpha_{jes})/\delta\alpha$. Best agreement is
achieved for $\Delta\alpha = -0.10^{+0.78}_{-0.8}$. 
Figure \ref{fig:CDF_template_fit} shows
a contour plot of the likelihood in the $m_{t}-\Delta\alpha$ plane.

An identical analysis is carried out on a  factor of two larger data sample 
corresponding to a luminosity of 680  pb$^{-1}$~\cite{CDF_template_2}, 
leading to a measurement of 
$m_{t}=173.4\pm 2.5(\mbox{stat}\oplus\mbox{jes})\pm1.3(\mbox{sys})$~GeV. 

A variation of this analysis which uses a 2-dimensional templates of
top quark mass and hadronic $W$ boson mass constructed using a Kernel Density
Estimate (KDE)  has been carried out by the CDF collaboration.
With this technique and a data sample with integrated luminosity of 
1.7 fb$^{-1}$ the top quark mass is measured to be 
$m_{t}=171.6\pm 2.1(\mbox{stat}\oplus\mbox{jes})\pm1.1(\mbox{sys})$~GeV~\cite{CDF_top_SUSY07}.

\subsubsection{Matrix element method~\label{sec:ME}}

The matrix element method was developed by \dzero\ between Run~I and Run~II.
It was
used first on the Run~I data set to obtain a top quark mass measurement of
$m_{t} = 180.1 \pm 3.6 (\mbox{stat}) \pm 3.9 (\mbox{sys})$~GeV\cite{MEmtop}.
A significantly improved
statistical precision was achieved compared to the previous 
measurement  $m_{t}=
173.3 \pm 5.6 (\mbox{stat}) \pm 5.5 (\mbox{sys})$
GeV\cite{D0_ljmass_RunI} on the same data set which used the 
template technique.

The idea is to compute for each event the probability density to observe the
event as a function of the top quark mass, based on the full kinematic
information from the event. This probability density is given by
\begin{equation}
p_{evt}(x|m_{t},\alpha,f) = f p_{top}(x|m_{t},\alpha)+(1-f) p_{bkg}(x|\alpha).
\label{eq:p}\end{equation}
Here $x$ stands for all measured quantities in the event, i.e. the momenta of
the charged lepton, the jets and the \met.  $\alpha$ is a scale
parameter for the jet energies, and $f$ represents the fraction of top quark
decay events in the data sample. The event-by-event probability densities are
combined into a joint likelihood for all events in the sample,
\begin{equation}
-\ln L(x_1...x_n|m_{t},\alpha,f) = -\sum_{i=1}^n\ln  p_{evt}(x|m_{t},\alpha,f). 
\label{eq:L}\end{equation}
The measurement of the top quark mass is then the value of $m_{t}$ that
maximizes $L$ for any value of the parameters $\alpha$ and $f$.

The signal probability density is given by the differential cross section
normalized to the total cross section $\sigma_{t\overline t}(m_{t})$ for all events
accepted by the analysis cuts. It is given by
\begin{equation}
 p_{sig}(x|m_{t},\alpha) = \frac{1}{\sigma_{t\overline t}(m_{t})}\int dz
 d\overline z f(z) f(\overline z)d\sigma_{t\overline t}(y,m_{t})W(x|y,\alpha),
\end{equation}
where $z$ and $\overline z$ are the fractions of the proton and antiproton
momenta carried by the initial partons, $f(z)$ is the parton distribution
function for the proton, $y$ represents the momenta of all partons taking
part in the hard scatter event. The transfer function $W(x|y,\alpha)$ gives
the probability to measure the observables $x$ for the parton momenta $y$ and
the jet scale parameter $\alpha$. The parton cross section $d\sigma_{t\overline t}$
is calculated based on the leading-order matrix element $\cal M$ for the
process $q\overline q\rightarrow t\overline t\rightarrow \ell\nu b q\overline
q'\overline b$:
\begin{equation}
d\sigma_{t\overline t}(y,m_{t}) = \frac{|{\cal M}|^2}{x\overline x s}  d\Phi_6,
\end{equation}
where  $d\Phi_6$ is the Lorentz-invariant six-particle phase space element and
$s$ is the parton center of mass energy. In order to make $p_{sig}$
calculable with reasonable computing power, a number of simplifying
assumptions must be made. Typically these are assuming that $p_T({t\overline t})=0$,
that all angles are well measured and that the transfer function factorizes
into contributions from each final state object. After these assumptions the
remaining six integrations can be carried out using Monte Carlo integration
techniques. The calculation is carried out for all possible jet-parton
assignments and for both solutions of $p_z$ of the neutrino. These
contributions are then added together for the final value of $p_{sig}$.  A
similar calculation is carried out for $p_{bkg}$ except that this quantity
does not depend on the top quark mass.

The power of this method originates from the use of all kinematic information
from the events and the use of signal and background probabilities for each
event. This effectively weights events in the event likelihood $L$ that are
more likely to be from top quark decay stronger than events that are more likely to
be background. The downside is that the phase space integration is very
computationally intensive.

\begin{figure}[htp]
\begin{center}
\includegraphics[width=2in]{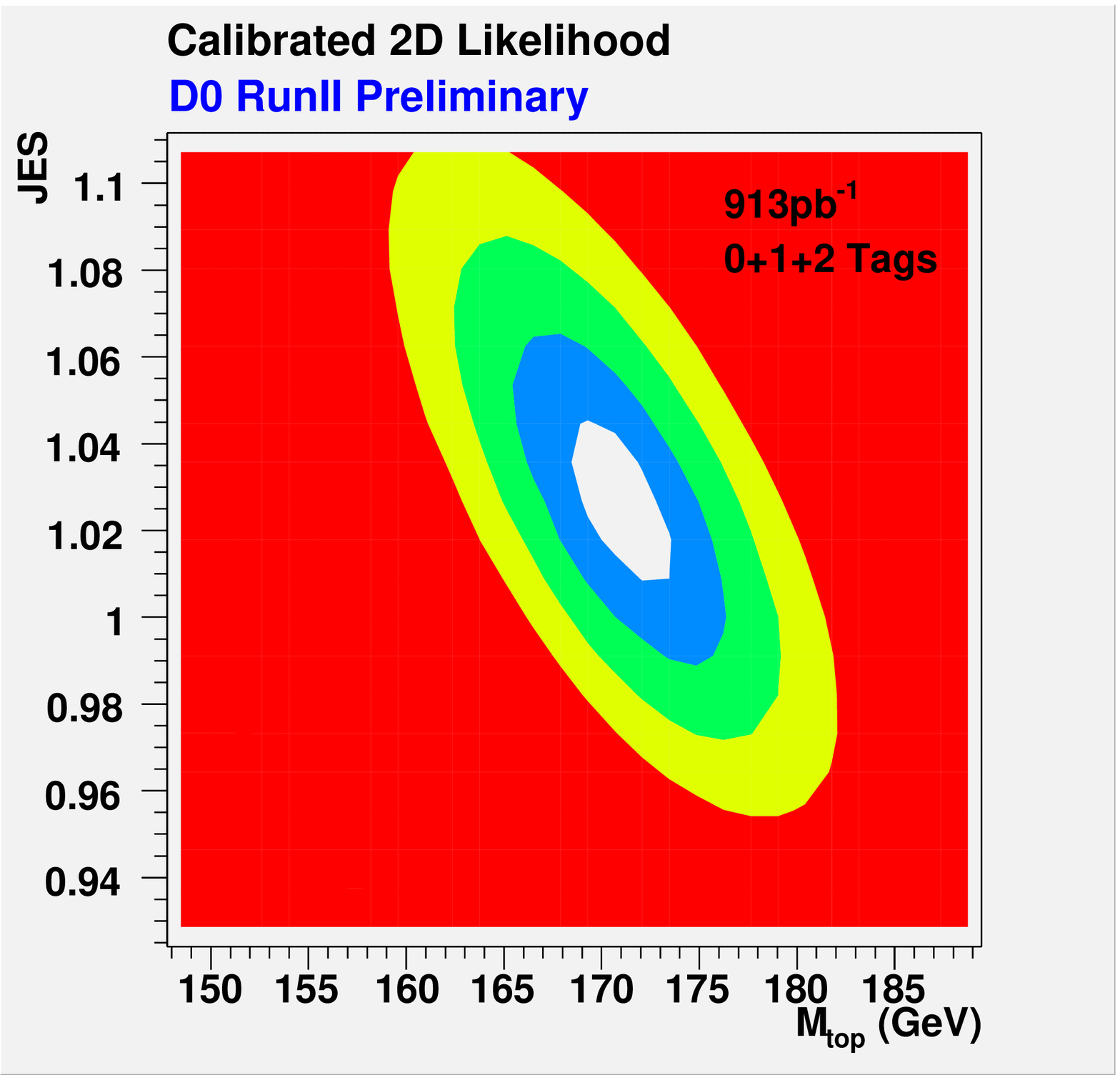}
\includegraphics[width=2in]{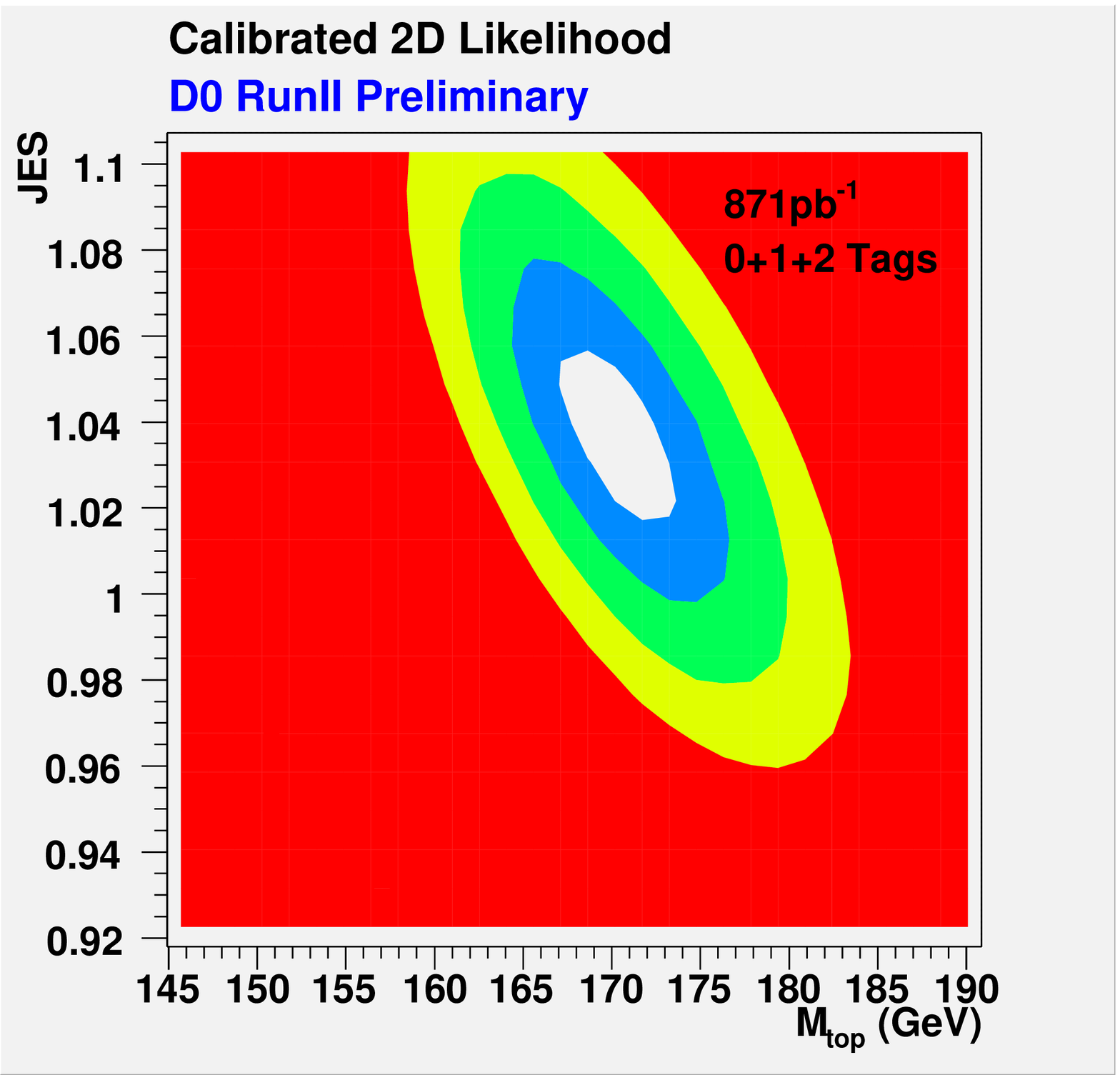}
\caption{\label{fig:D0_ME_fit} Contour plot of the likelihood for the matrix 
element analysis of the D0 $e$+jets data (left panel) and $\mu$+jets data
(right panel) \protect\cite{D0_ME_W07}.}
\end{center}
\end{figure}

\begin{figure}[htp]
\begin{center}
\includegraphics[width=2in]{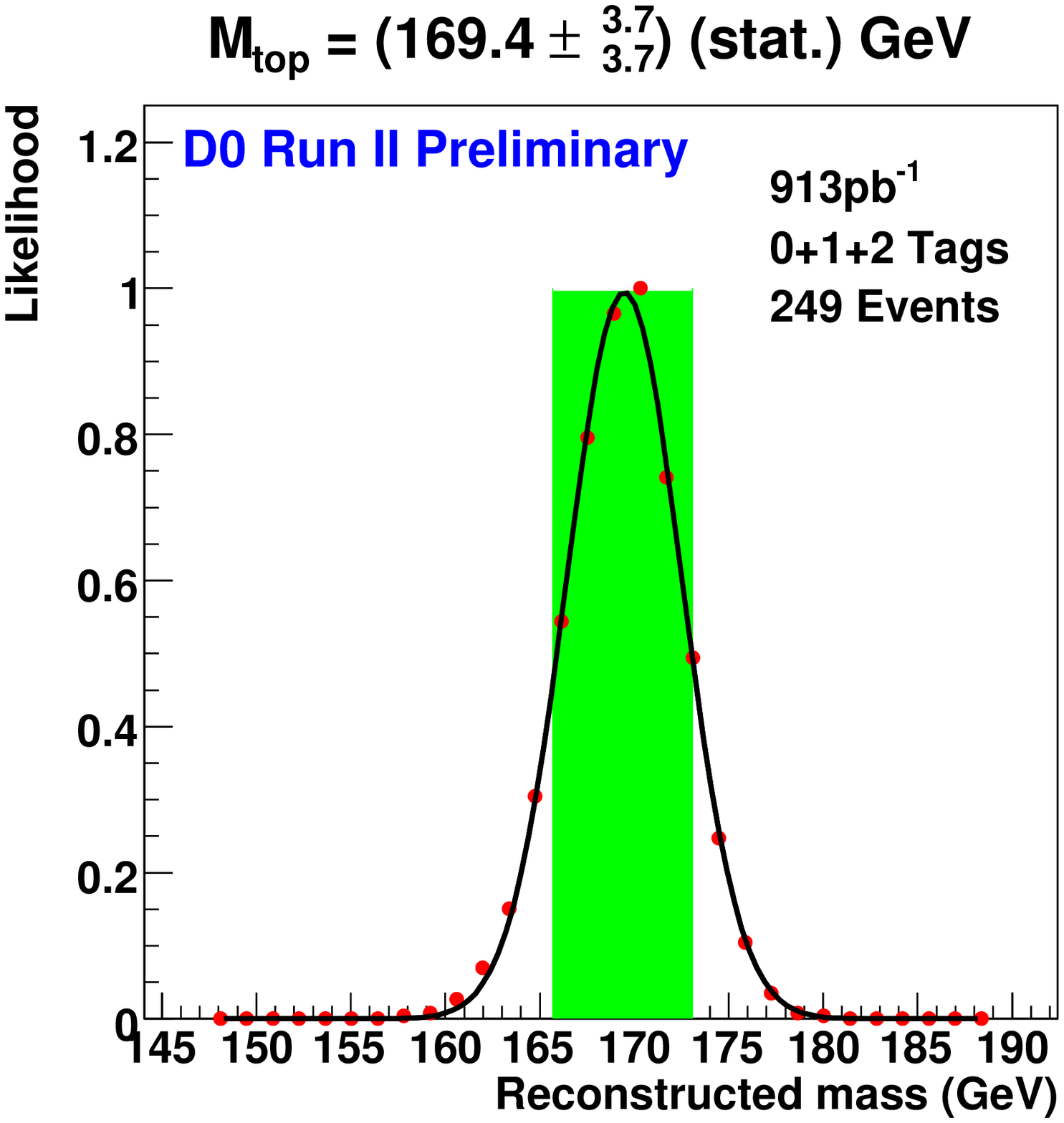}
\includegraphics[width=2in]{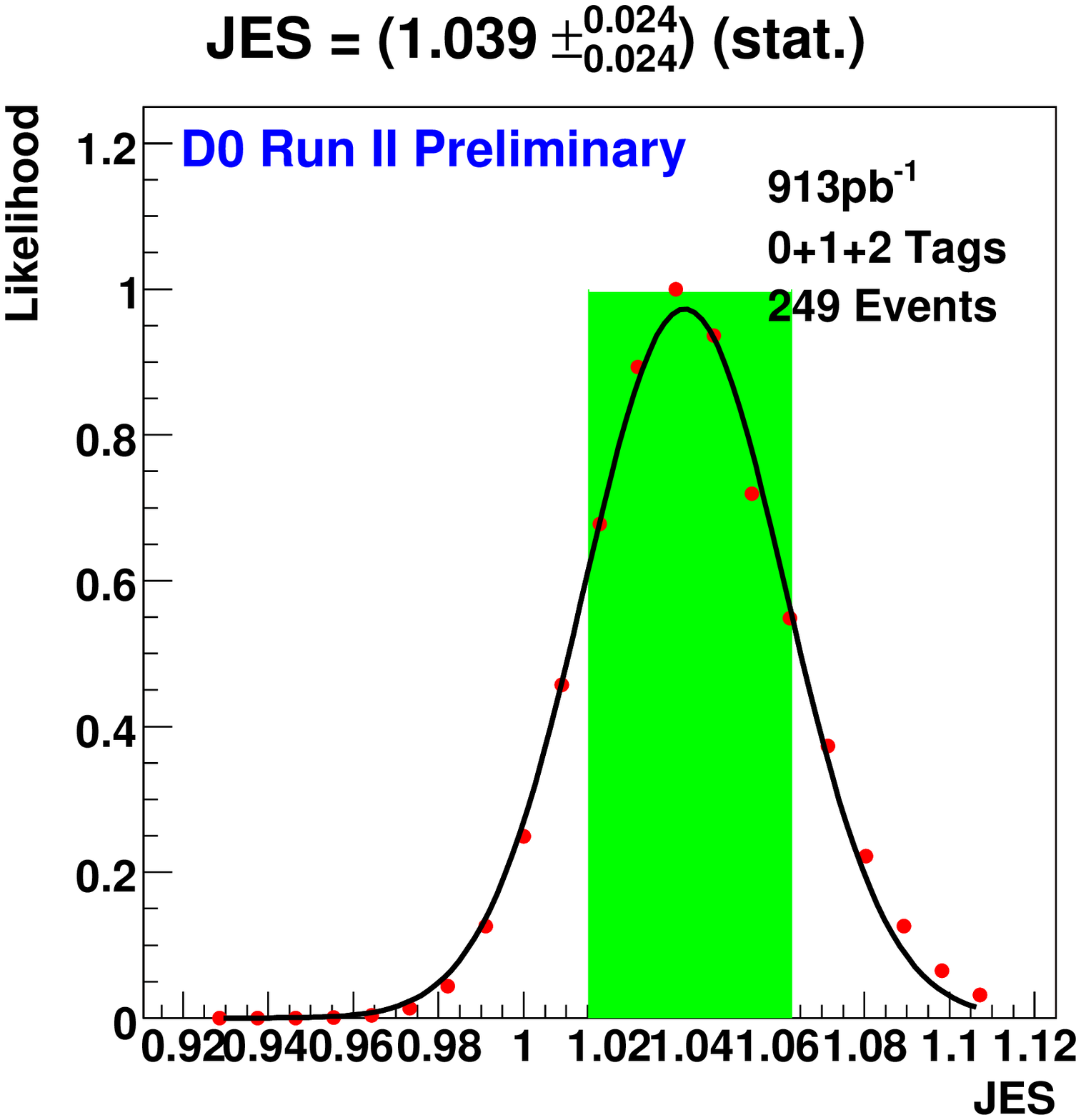}
\includegraphics[width=2in]{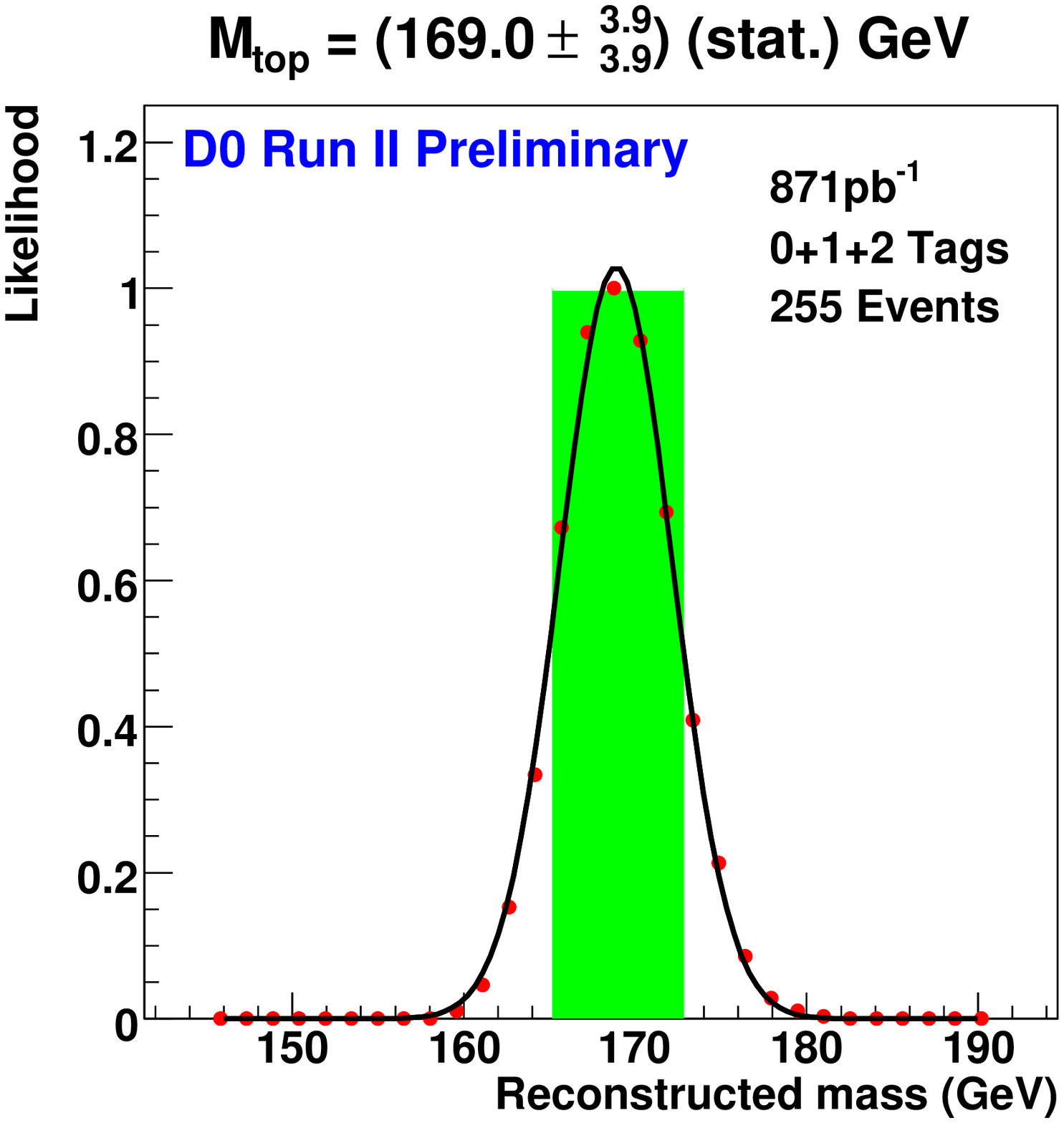}
\includegraphics[width=2in]{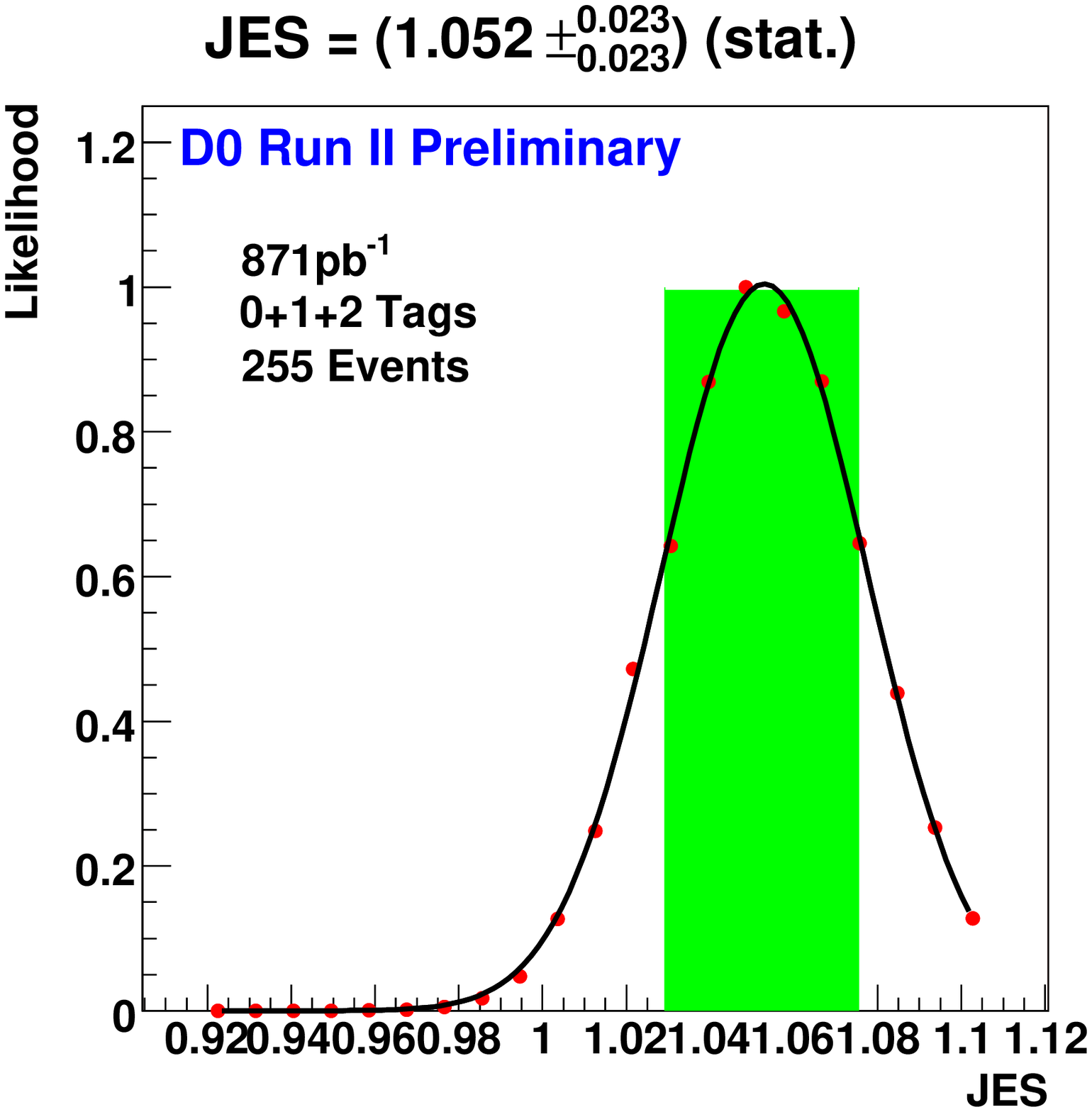}
\caption{\label{fig:D0_ME_mfit} Likelihood versus top quark mass (left), 
and jet energy scale (right) for the
  matrix element analysis of the \dzero\ $e$+jets data (top plots) and
  $\mu$+jets data (bottom plots) \protect
\cite{D0_ME_W07}.}
\end{center}
\end{figure}

\begin{figure}[htp]
\begin{center}
\includegraphics[width=2in]{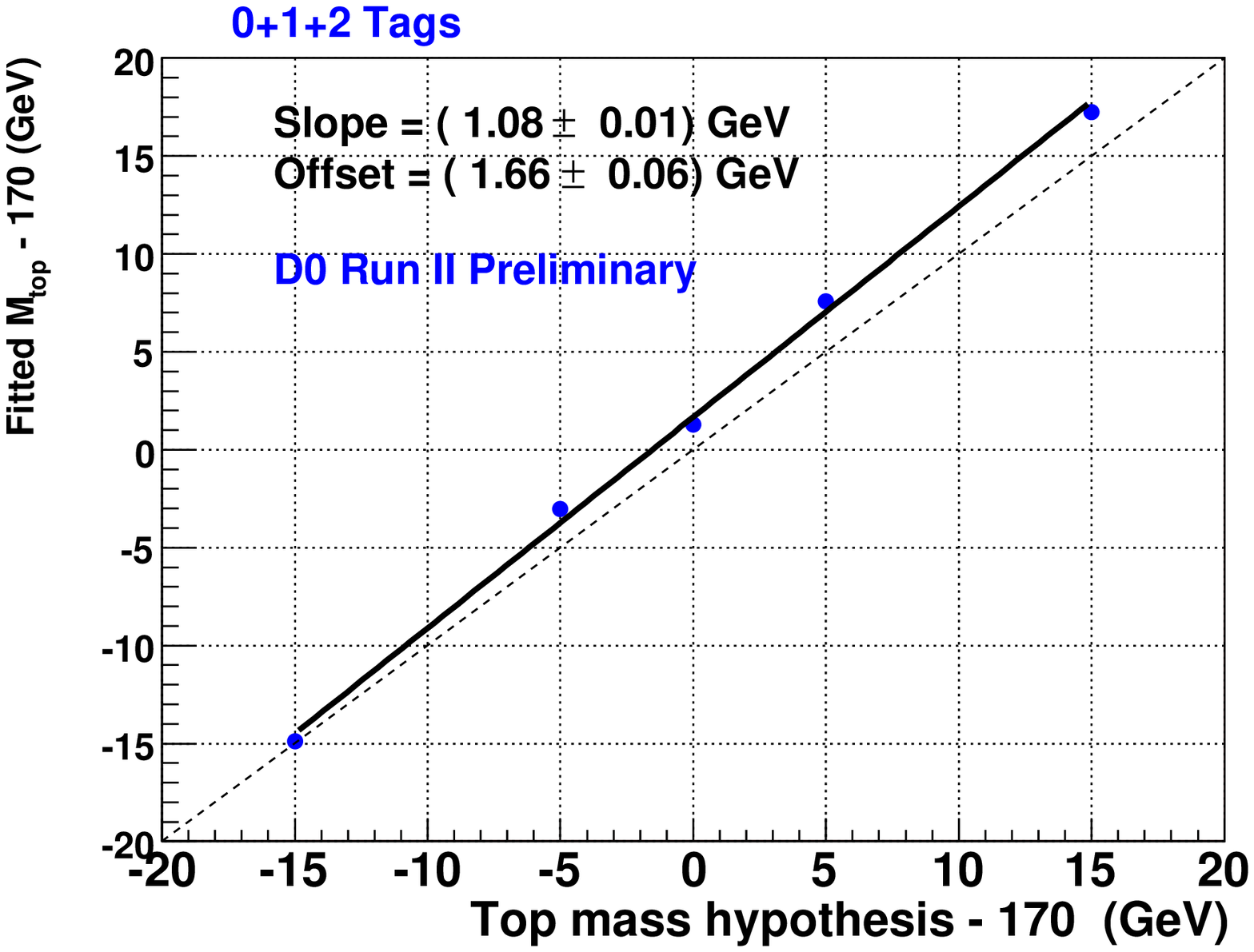}
\includegraphics[width=2in]{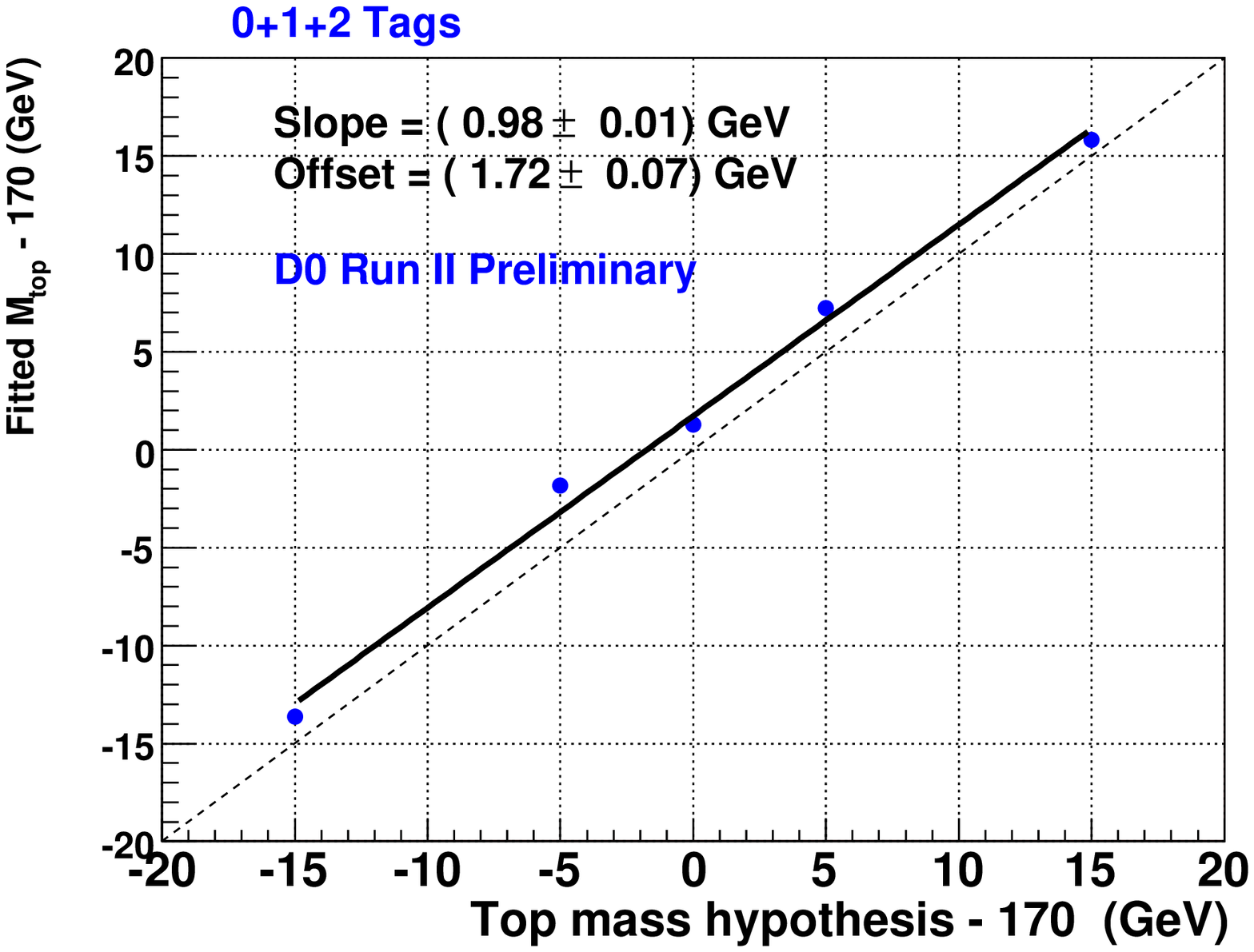}
\caption{\label{fig:D0_ME_ensembles} Measured top quark mass from ensemble
  tests versus input top quark mass for matrix element analysis of 
 the \dzero\ $e$+jets data (left) and   $\mu$+jets data 
(right)\protect\cite{D0_ME_W07}.}
\end{center}
\end{figure}

The \dzero\ Collaboration published a measurement of the top quark
mass using the matrix element method based on 370 pb$^{-1}$ of data from
Run~II\cite{D0_ljmass_RunII}.  In this measurement \dzero\ makes use of
$b$-tagging to give higher weight to jet permutations that assign $b$-tagged
jets to $b$-quarks. This leads to a reduction in the combinatoric background
and improves the statistical precision of the
measurement. The result of this measurement is $m_{t} =
170.3^{+4.1}_{-4.5} (\mbox{stat}\oplus\mbox{jes})^{+1.2}_{-1.8}
(\mbox{sys)}$ GeV.  
The  measurement was updated using a data sample
corresponding to an integrated luminosity of 1 fb$^{-1}$\cite{D0_ME_W07}. 
Figure~\ref{fig:D0_ME_fit} shows a contour plot of the event
likelihood $L$ defined in equation~\ref{eq:L} as a function of the assumed
top quark mass $m_{t}$ and the jet energy scale factor
$\alpha$. Plots in Fig.~\ref{fig:D0_ME_mfit} show the projections onto
the two axes. 
The updated measurement  applied to $b$-tagged events
yield a top quark mass of  $m_{t} =
170.5 \pm 2.4 (\mbox{stat}\oplus\mbox{jes}) \pm 1.2 (\mbox{sys)}$ GeV.
The first uncertainty is derived from the width of the
likelihood and accounts for both the statistical uncertainty and the
uncertainty from the jet energy scale calibration. Ensemble tests show
that the matrix element algorithm results in a measured value of the top
quark mass that tracks the input top quark mass with a small offset.
Figure~\ref{fig:D0_ME_ensembles} shows a plot of measured top quark mass
versus input top quark mass for this analysis. The result is corrected for
the observed offset.

An analysis using the same technique is carried out by CDF
using a sample of  940  pb$^{-1}$ of data~\cite{CDF_ME}. The events
are required to have a lepton, \met\ and only four jets, 
out of which one is expected to be tagged as a $b$-jet. This sample
yields a measurement of
$m_{t}=170.9\pm 2.2(\mbox{stat}\oplus\mbox{jes})\pm1.3(\mbox{sys})$~GeV.

By combining the signal probability computed using the matrix element technique
with a Neural network discriminant to reject backgrounds, CDF experiment
has measured 
$m_{t}=170.9\pm 1.3(\mbox{stat})\pm 1.2(\mbox{jes})\pm1.2(\mbox{sys})$~GeV
This measurement is based on a data sample with integrated luminosity 
of 1.7 fb$^{-1}$~\cite{CDF_top_SUSY07}.

\subsubsection{Quantized dynamical likelihood method}

CDF has developed another method, based on Ref.~\cite{Kondo}, called
the quantized dynamical likelihood method. In this analysis only events with
exactly four jets are used. The four jets are assigned to partons in every
possible way with the restriction that $b$-tagged jets are only assigned to
one of the $b$ quarks.

For each permutation a likelihood is computed as 
\begin{equation}
L_1(m_{t}) = \frac{d\sigma}{d\Phi_6},
\end{equation}
\noindent where $d\Phi_6$ is the Lorentz-invariant six particle phase space element and
$d\sigma$ is the differential $t\bar{t}$ production cross section integrated
over the momentum fractions $z$ and $\bar{z}$ of the initial state partons
and the transverse momentum $p_T(t\bar{t})$ of the $t\bar{t}$ system:

\begin{equation}
d\sigma = \int \int \int dz d\overline z f(z) f(\overline z) d^2p_T(t\bar{t}) f(p_T(t\bar{t}))d\sigma_{t\overline t}(x,m_{t}),
\end{equation}

\noindent where $f(p_T(t\bar{t}))$ is the expected $p_T$ distribution of the $t\bar{t}$
system, $d\sigma_{t\overline t}(x,m_{t})$ is the parton cross section defined in
section~\ref{sec:ME} and all other parameters are also as defined in
the same section.

The likelihood $L_1$ is typically averaged over 50{,}000 random generations
of the parton momenta $x$ from the observed jet momenta using Monte Carlo
derived transfer functions. These functions are taken over 
all possible jet-parton assignment and over
the two possible values of the $z$-component of the neutrino momentum to
obtain the event likelihood $L(m_{t})$. This joint likelihood for all
events is computed by multiplying together all the event likelihoods. The top
quark mass is measured as the hypothesized top quark mass for which the joint
likelihood is maximized. Finally, this result is corrected for the effects of
backgrounds in the sample. This correction is determined from ensemble tests
using Monte Carlo data.

Using a data sample of 63 ${t\overline t}$ candidates in the lepton+jets channel with
at least one $b$-tagged jet from 318 pb$^{-1}$ of data, CDF measures
$m_{t}=173.2^{+2.6}_{-2.4}\mbox{(stat)}\pm3.2\mbox{(sys)}$ GeV.

\subsubsection{Ideogram method}

The ideogram method\cite{D0_ideogram_RunII} 
is based on ideas that were used by the Delphi
Collaboration to measure the $W$ boson mass\cite{W_Delphi}. It was first
adapted for the measurement of the top quark mass by the \dzero\ Collaboration. It
is based on the same general idea that was developed for the matrix element
method in equations~\ref{eq:p} and~\ref{eq:L}. The difference lies in the
definitions of $p_{sig}$ and $p_{bkg}$. In order to reduce computing
requirements, this analysis method makes use of the same kinematic fit as the
template method.

\noindent
The signal probability is defined as
\begin{equation}
p_{sig}(o|m_{t},\alpha) = \tilde p_{sig}(D) \sum_{i=1}^{24}
\exp\left(-\frac{\overline\chi_i}{2}\right) \left[f\int G(m_i,m',\sigma_i)
B(m',m_{t})dm' + (1-f) S(m_i,m_{t})\right].
\end{equation}

\begin{figure*}[!h!tbp]
\begin{center}
\epsfig{figure=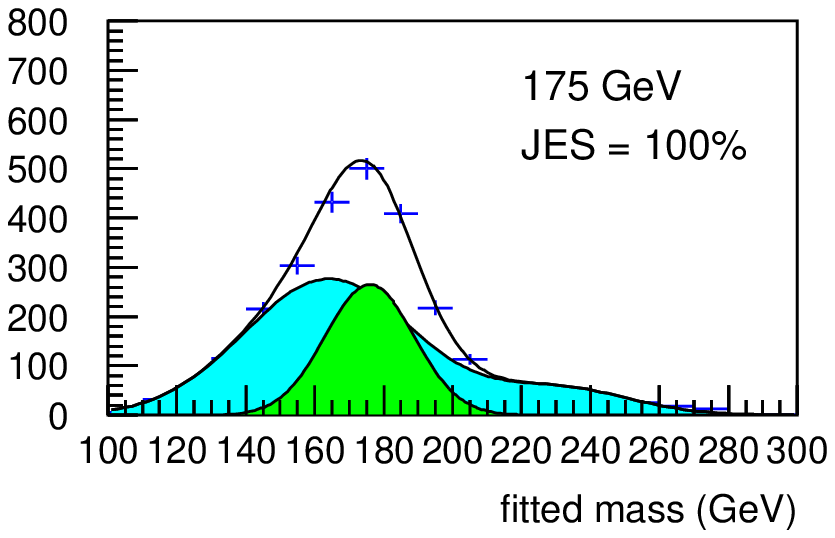,width=0.30\textwidth}
\epsfig{figure=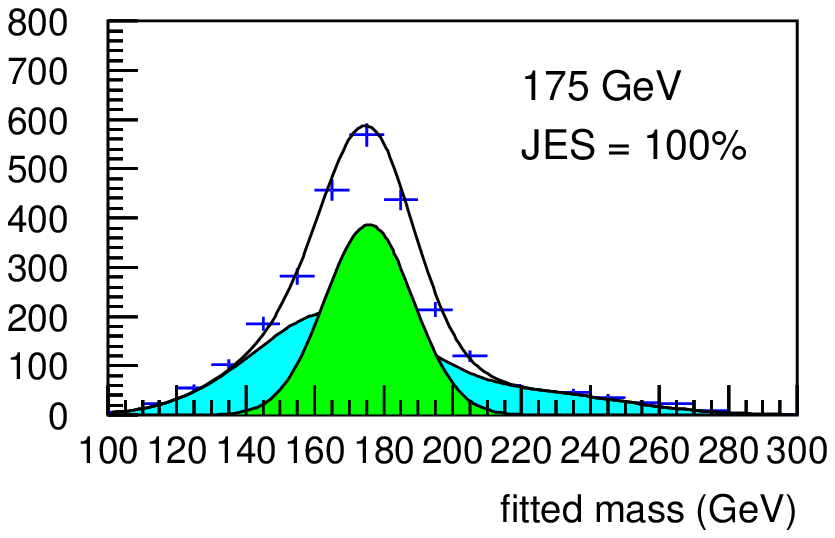,width=0.30\textwidth}
\epsfig{figure=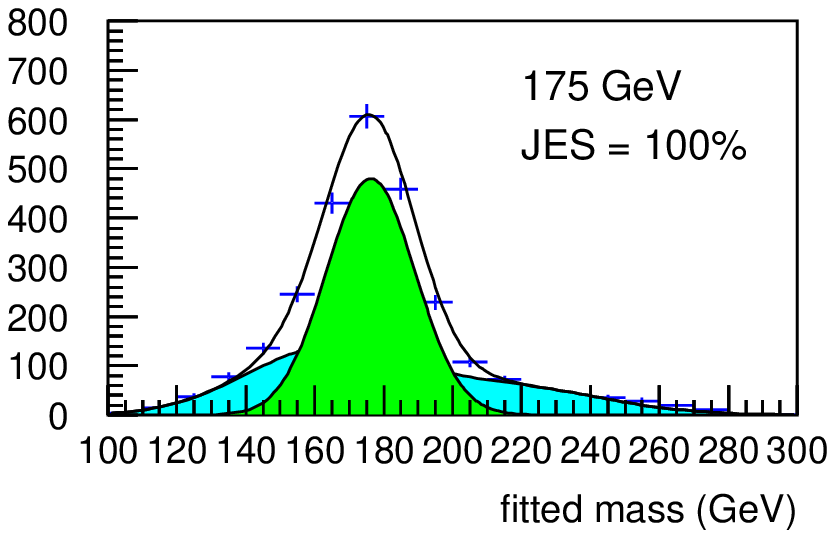,width=0.30\textwidth}
\end{center}
\vspace*{8pt}
\caption{Event likelihoods for simulated \dzero\ $\ell+$jets events with 0, 1, or 2 tags\protect\cite{D0_ideogram_RunII}.
}
\label{fig:D0_ideograms} 
\end{figure*}

\noindent
Here $D$ is a discriminant based on the kinematic observables in the event
that is constructed such that most top quark like events have $D\approx1$ and
most background like events have $D\approx0$ and $\tilde p_{sig}(D)$ is the
probability density for signal events with discriminant value $D$. The sum
over $i$ runs over all 12 jet permutations and the two solutions for $p_z$ of
the neutrino. $\overline\chi_i^2$ is the minimum value of the $\chi^2$
goodness of fit variable for the kinematic fit described in 
section~\ref{sec:template} and $m_i$ is the corresponding value of 
the hypothesized
top quark mass. $G(x,x_0,\sigma)$ is a Gaussian bell curve with mean $x_0$
and width $\sigma$ and $B(x,x_0)$ is a Breit-Wigner function of mean $x_0$. 
The integral  is a convolution of the Breit-Wigner line
shape of the top quark with a Gaussian resolution function and represents the
contribution from jet permutations with the correct jet-parton assignments
and $f$ indicates the probability that a jet permutation corresponds to the
correct jet-parton assignments. $S(m_i,m_{t})$ represents the contribution
of jet permutations with the wrong jet-parton assignments. 
Figure~\ref{fig:D0_ideograms} shows the event likelihood curves 
for simulated events
with zero, one, or two tags. The background probability density $p_{bkg}$ is
determined using a Monte Carlo simulation.


\begin{figure*}[!h!tbp]
\begin{center}
\epsfig{figure=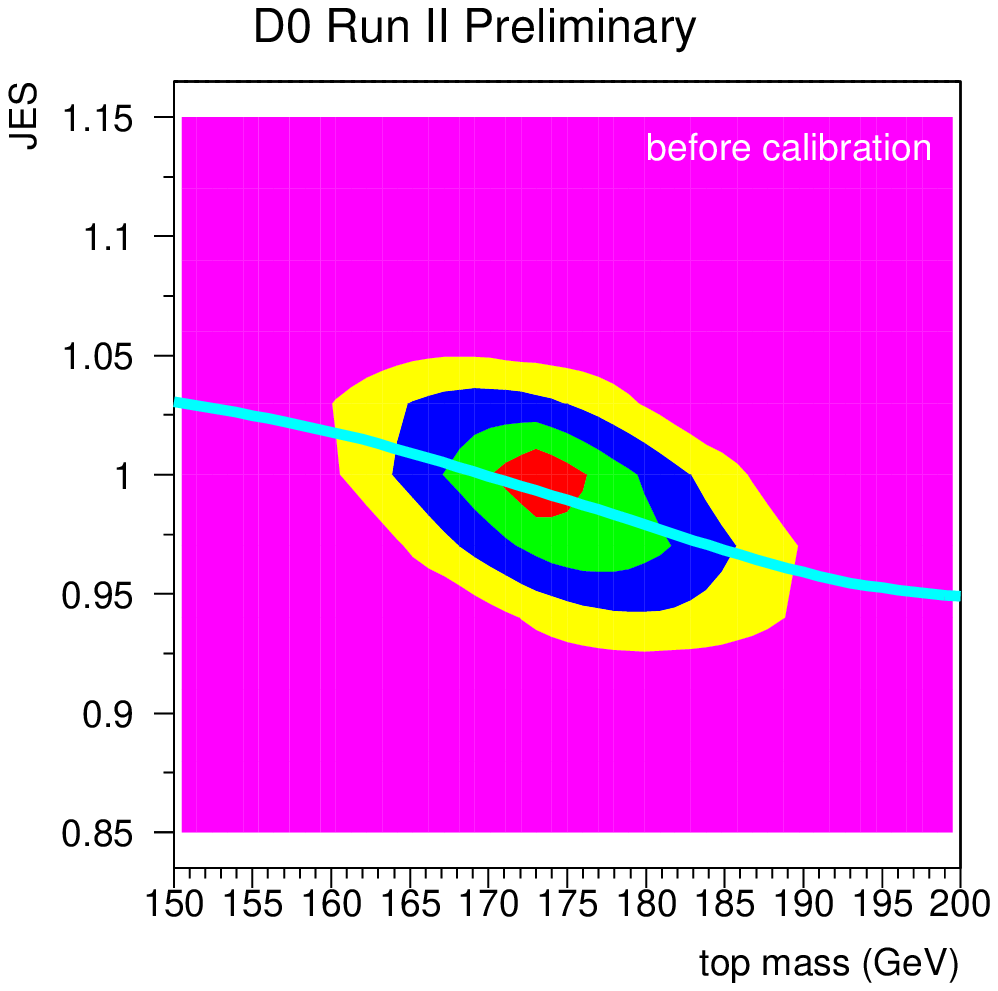,width=0.45\textwidth}
\epsfig{figure=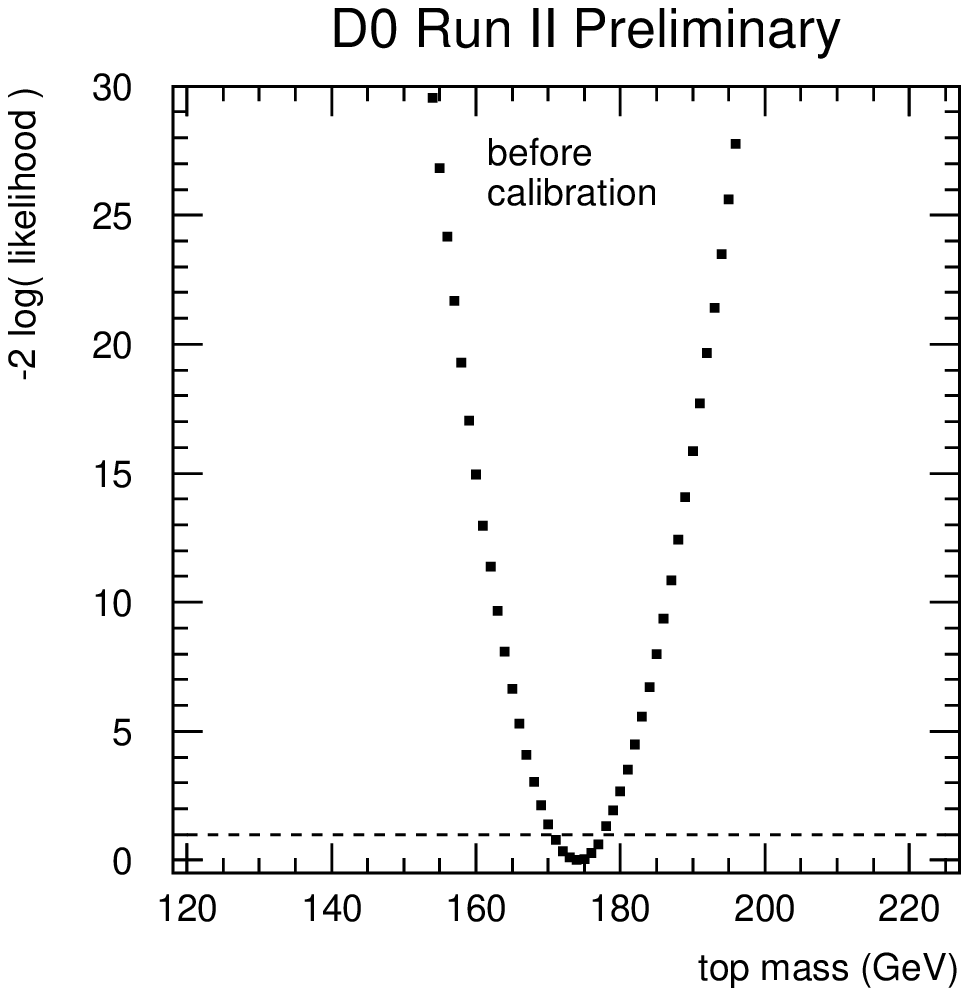,width=0.45\textwidth}
\end{center}
\vspace*{8pt}
\caption{Contour plot of the likelihood (left) 
and Likelihood versus top quark mass for the
  ideogram analysis of  the \dzero\ $\ell$+jets data\protect\cite{D0_ideogram_RunII}.
}
\label{fig:D0_ideogram_fit} 
\end{figure*}

The \dzero\ collaboration has applied this technique to the same data set as was
used by the matrix element analysis described in section~\ref{sec:ME} except
that the ideogram analysis uses events with four or more jets. 
Figure~\ref{fig:D0_ideogram_fit} (left plot) shows a contour plot of the 
event likelihood $L$
defined in equation~\ref{eq:L} as a function of the assumed top quark mass
$m_{t}$ and the jet energy scale factor
$\alpha$. Figure~\ref{fig:D0_ideogram_fit} (right plot) 
shows the likelihood versus top
quark mass. The result of this measurement is $m_{t} = 173.7\pm4.4
(\mbox{stat}\oplus\mbox{jes})^{+2.1}_{-2.0}(\mbox{sys)}$ GeV. The first uncertainty
is derived from the width of the likelihood and accounts for both the
statistical uncertainty and the uncertainty from the jet energy scale
calibration. 
Ensemble tests show that the ideogram
algorithm results in a measured value of the top quark mass that tracks the
input top quark mass with a small offset. The final quoted result  
includes this calibration.

\subsubsection{Decay length distribution}

In addition to the measurement techniques described in the past few sections,
the CDF collaboration has measured the top quark mass from the decay length
distributions of $b$-hadrons from top quarks decays. The mean momentum of the
$b$-quarks from the decays of top
quarks 
increases with the top quark mass. A
harder momentum spectrum leads to a higher Lorentz-factor and a longer decay
length in the lab frame. Thus the decay length spectrum of $b$-hadrons from
top quark decays can yield a measurement of the top quark mass, albeit with
significantly poorer statistical sensitivity than the methods that are based
on the kinematic reconstruction of the top quark decay. The advantage is that
this method has different systematic uncertainties. Most importantly, it does not depend
strongly on the jet energy scale calibration. The result of this measurement
is $m_{t} = 180.7^{+15.5}_{-13.4}(\mbox{stat})\pm8.6(\mbox{sys)}$ GeV~\cite{cdf_lxy}.

%% file: mtop/alljetsMass.tex
Events in which both $W$ bosons 
decay to quark pairs provide another avenue with which
to measure the top quark mass.  The lack of a neutrino simplifies the kinematic 
reconstruction by removing quadratic ambiguities and providing for 
an overconstrained fit from the momenta of six jets.  The large 
all-jets branching fraction provides a potentially substantial
source of data that is essentially uncorrelated with the leptonic modes.  On
the other hand, a very high background persists even though $b-$tagging
is used.  This substantially impacts the statistical power of 
this channel.  The Tevatron experiments have pursued several analyses
to measure $m_t$ in the all-jets channel: three template-based
measurements, an ideogram analysis and a matrix element analysis.

\subsubsection{Template methods}

Templated approaches for the all-jets mass extraction have been
tried by CDF in Run II\cite{cdfr2mt6jtemp} and I\cite{cdfr1alljetscsec}, and by 
\dzero\ using Run I data published in 2005\cite{d0r1alljetsmtop}.  
The most sensitive of these is the 
Run II CDF analysis which is a preliminary result that used 1.02 fb$^{-1}$.  
All three use event kinematic selections
very similar to those described in Section~\ref{sec:csec6j}.  
At least six jets are required in all cases, and CDF relaxed its initial
$\Sigma E_T$ cut to minimize the bias to the mass measurement.
The \dzero\ analysis used soft $\mu$-tagged all-jets events, while
CDF used secondary vertex $b$-tagged events.

To model signal, top quark events were modeled by both experiments
using {\sc Herwig}.  CDF considered 
masses from 150 GeV to 200 GeV in their Run II analysis.
Backgrounds were estimated similarly by both experiments. An untagged
multijet sample was selected just as the signal sample.  Known
mistag rates were then applied to properly weight events when constructing
templates.  Estimatation of signal and background levels gave
334 events and 573 events, respectively, for CDF's Run II sample.  
The Run I analyses had 65 and 136 events total for \dzero\ and CDF,
with 48 and 108 expected background events, respectively.

\begin{figure*}[!h!tbp]
\begin{center}
\epsfig{figure=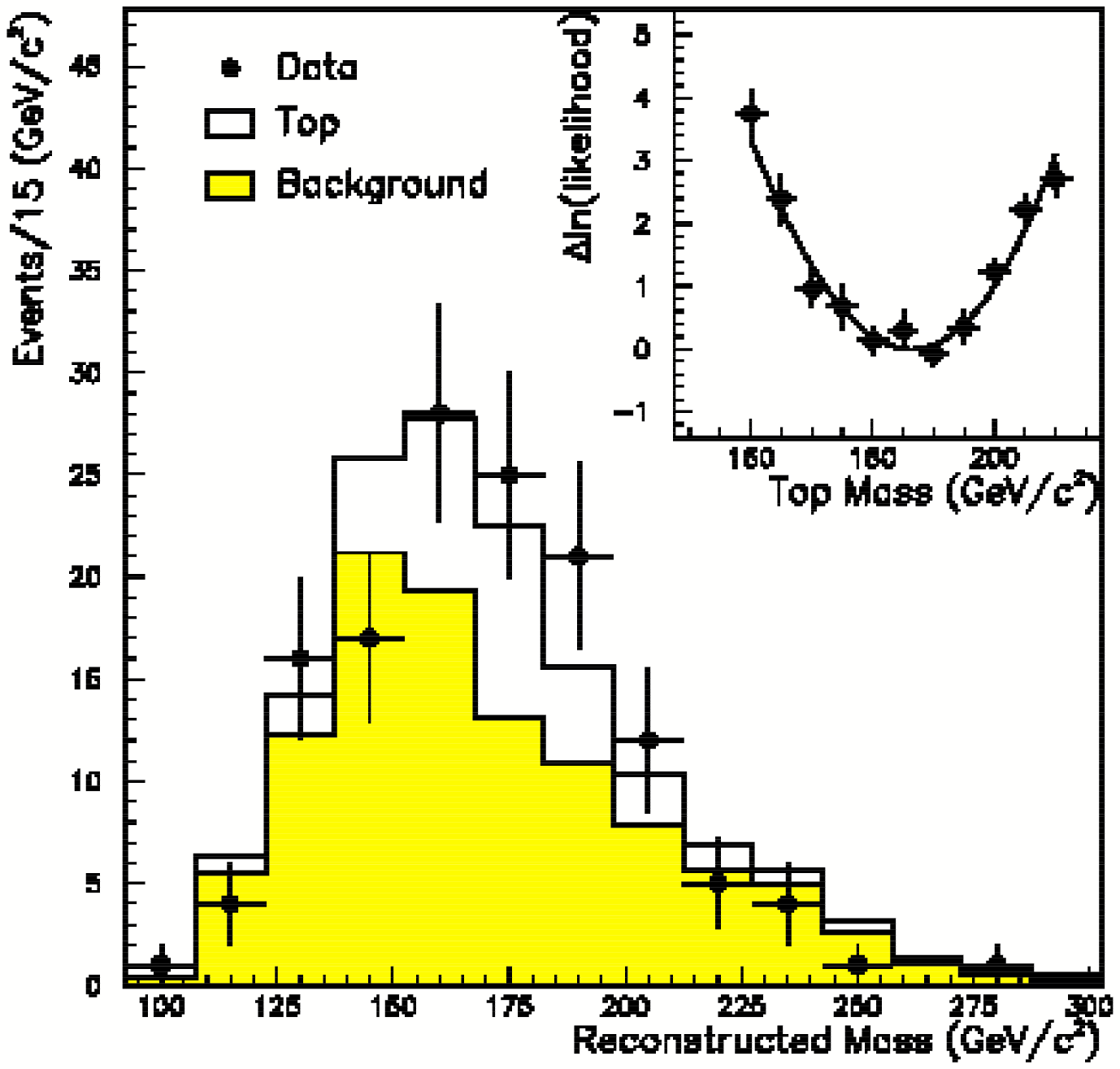,width=5cm}
\epsfig{figure=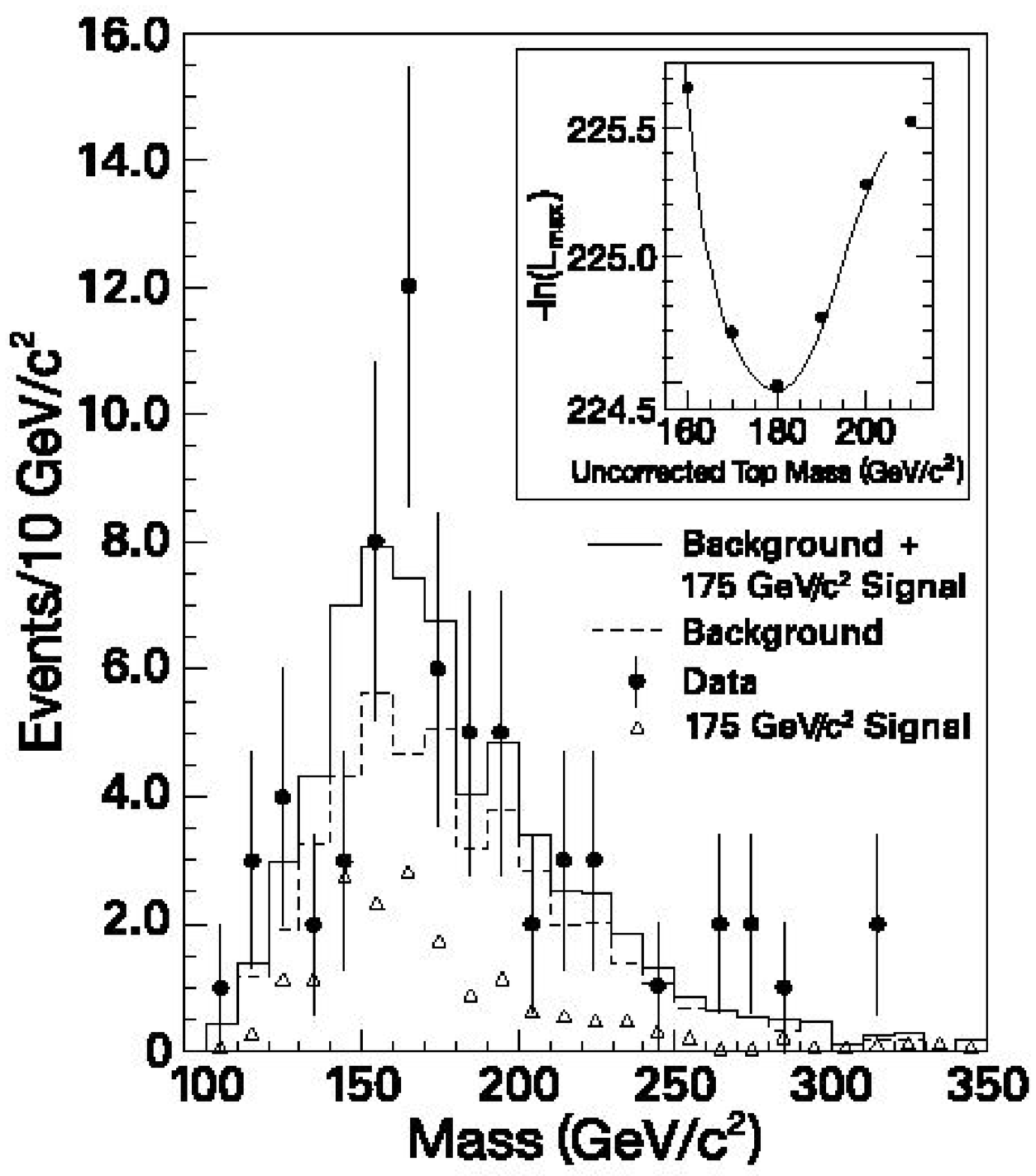,width=4.25cm}
\caption{
The CDF measurement of the top mass in the all-jets final state using
Run I data (left)\protect\cite{cdfr1alljetscsec}.  Expected background and 
top quark signal are shown in solid,
and data points provide data.  At right is the \dzero\ distribution
\protect\cite{d0r1alljetsmtop}.  Insets at
upper right of both plots provide the fit used to determine the mass.}
\label{fig:6jMass}
\end{center}
\end{figure*}

\begin{figure*}[!h!tbp]
\begin{center}
\epsfig{figure=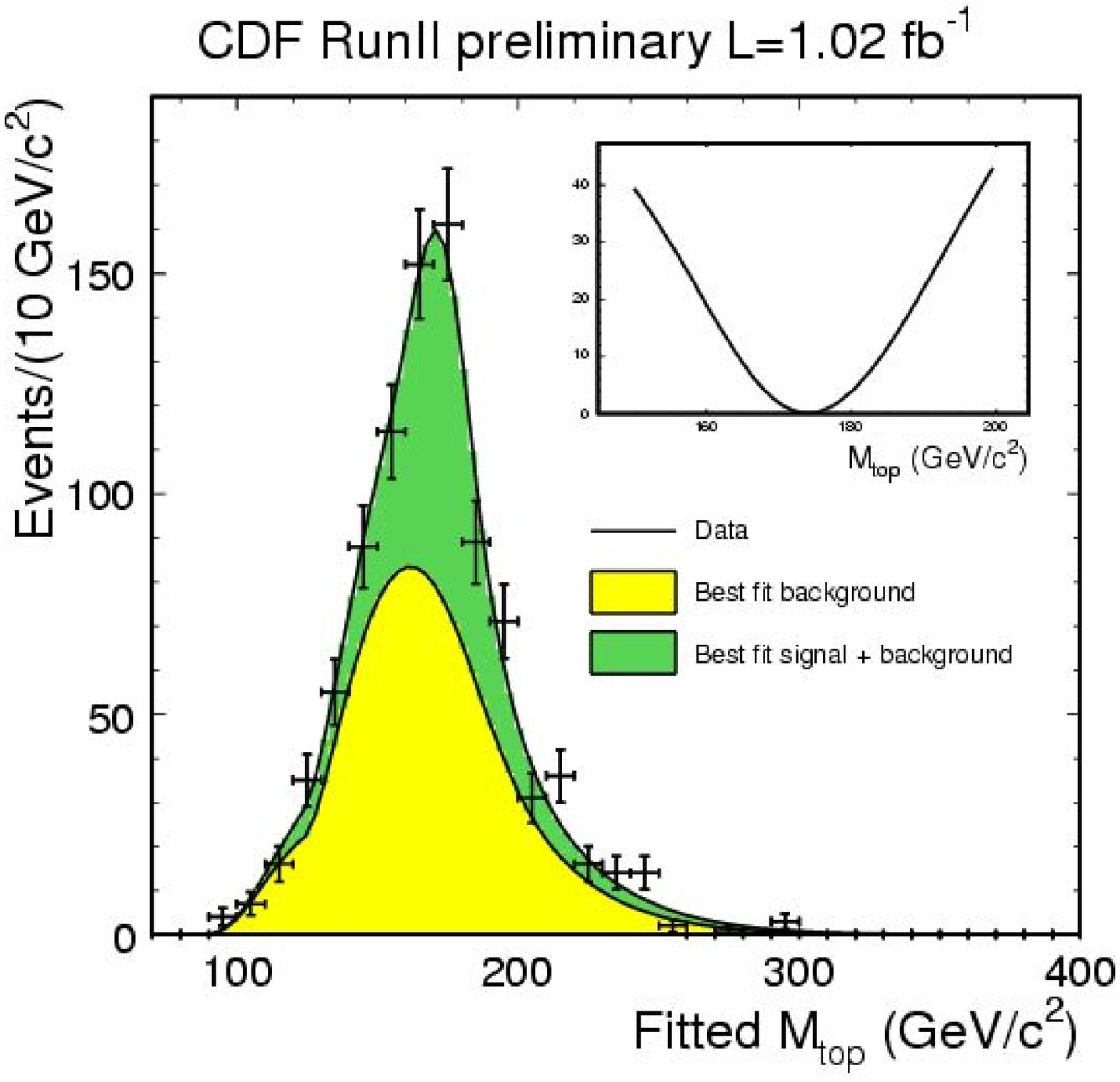,width=4cm}
\epsfig{figure=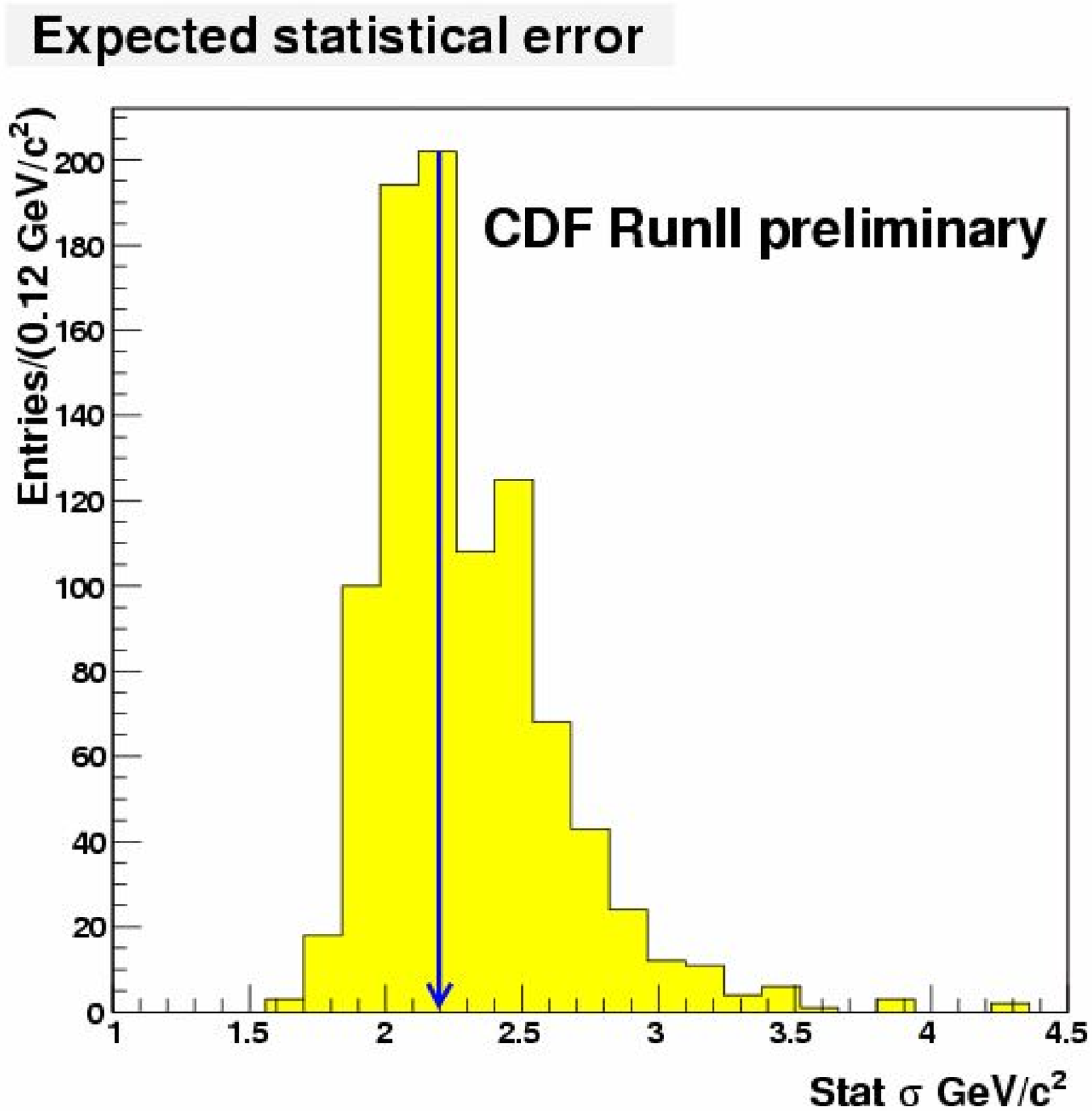,width=4cm}
\caption{The result of the template analysis of 
1 fb$^{-1}$ of CDF Run II data
is shown with signal and expected background components (left).
The inset at upper right provides the resulting $-ln(L)$ vs.
$m_t$.  The distribution of expected statistical uncertainties 
(right) \protect\cite{cdfr2mt6jtemp} is also provided showing
the result from data (see arrow).}
\label{fig:r2mt6jtempl}
\end{center}
\end{figure*}

Kinematic reconstruction of the candidate events was performed.
For this reconstruction, each of the six highest $E_T$ jets was 
associated with one of the quarks from a top quark or $W$ boson 
decay.  Tagged jets were specifically associated with $b-$quarks.  There is
an ambiguity in how the jets are assigned in pairs to $W$ bosons and
how these are associated with $b-$jets to form each top quark.  
Resolving this situation involved considering all possible configurations 
and calculating a $\chi^2$ with respect to a correct \ttbar\ configuration. 
Each configuration presents two jet triplets purported to originate from
each top quark.  For example, \dzero\ defined a $\chi^2$ for
each configuration,

\begin{eqnarray}
\chi^2={\displaystyle \left(\frac{m_{t_1}-m_{t_2}}{2\times\sigma_{m_{t}}}\right)^2
+\left(\frac{M_{W_1}-M_{W_0}}{\sigma_{M_W}}\right)^2
+\left(\frac{M_{W_2}-M_{W_0}}{\sigma_{M_W}}\right)^2}
\end{eqnarray}

\noindent where $M_{W_0}(=77.5)$ GeV is the mass reconstructed in \ttbar\ simulation for
the two jets from a $W$ boson.  The values of $M_{W_i}$ and $m_{t_i}$
correspond to the masses calculated from the jet pair or triplet
for the $i$th $W$ boson or top quark, respectively, given the configuration
being considered.  All analyses resolved the combinatoric ambiguity by
selecting the one with the lowest value of the $\chi^2$.
Using this combination, the invariant masses of three jet groupings
were calculated.  They are shown in Fig.~\ref{fig:6jMass} for the Run I
measurements, and Fig.~\ref{fig:r2mt6jtempl} for the analysis in 
1 fb$^{-1}$.

This distribution is fit using expectations for backgrounds and for a 
top quark of varying mass and background. CDF in Run II
has used a convolution of one gaussian and two gamma functions to 
represent the signal shape, and two gaussians plus a gamma function
for background.  The background shape was adjusted for a small estimated
top quark contamination.  CDF obtained measurements of 
$m_t = 174.0\pm2.2(\rm stat)\pm4.8(sys)$ GeV and $186 \pm 10$ GeV in
Run II and Run I, respectively.  The primary systematic uncertainties
in the former arose from the jet energy scalibration and the Monte Carlo
modeling of the top quark signal.  When 
applying the maximum likelihood fit, \dzero\ allowed both signal and background 
levels to float.  The fitted mass is $178.5\pm 13.7$ (stat) GeV.  The 
systematic uncertainties for jet energy calibration, $b$-tag rate, and 
background statistics totalled 7.7 GeV.

\subsubsection{Matrix element analyses}

As with other channels, matrix element techniques have now been
exploited in the all-jets channel, to date by CDF
\cite{cdfr2mt6jIdeo,cdfR2alljetsMass}.  The former is actually
a variant of the ideogram method used in $\ell+$jets samples but
also incorporating the matrix element approach to perform the
$m_t$ determination.  This preliminary result was obtained from
310 pb$^{-1}$ of $b$-tagged all-jets data.  
Each event presents 90 potential configurations of grouping
jets to form $W$'s and top quarks.  All combinations are considered
when performing a kinematic fit.  Goodness of fit and the probability
that two jets are $b$-jets were used to pick the right configuration.
A two dimensional likelihood was performed in terms of the masses
of the two top quarks, $m_{t1}$ and $m_{t2}$.
The shape of the signal was obtained from {\sc Pythia} and {\sc Herwig} using
Gaussian resolution functions for jets obtained from data.  The QCD
background was modeled with {\sc Alpgen}.  The likelihood yielded
a measurement of $m_t = 177.1 \pm 4.9 (\rm stat) \pm 4.7 (sys)$ GeV.
The leading systematic uncertainties were the jet energy calibration
and the background shape.

\begin{figure*}[!h!tbp]
\begin{center}
\epsfig{figure=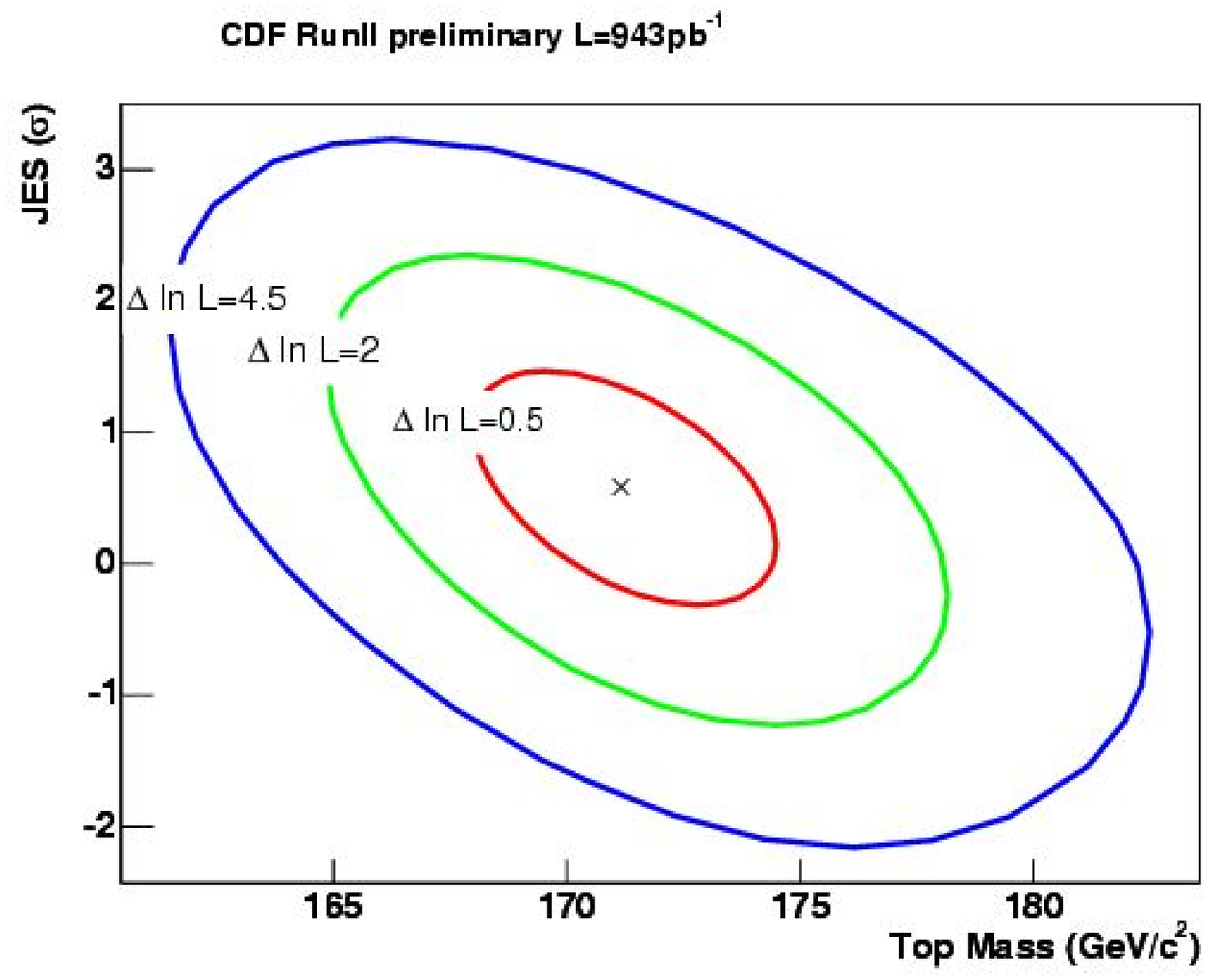,width=5cm}
\epsfig{figure=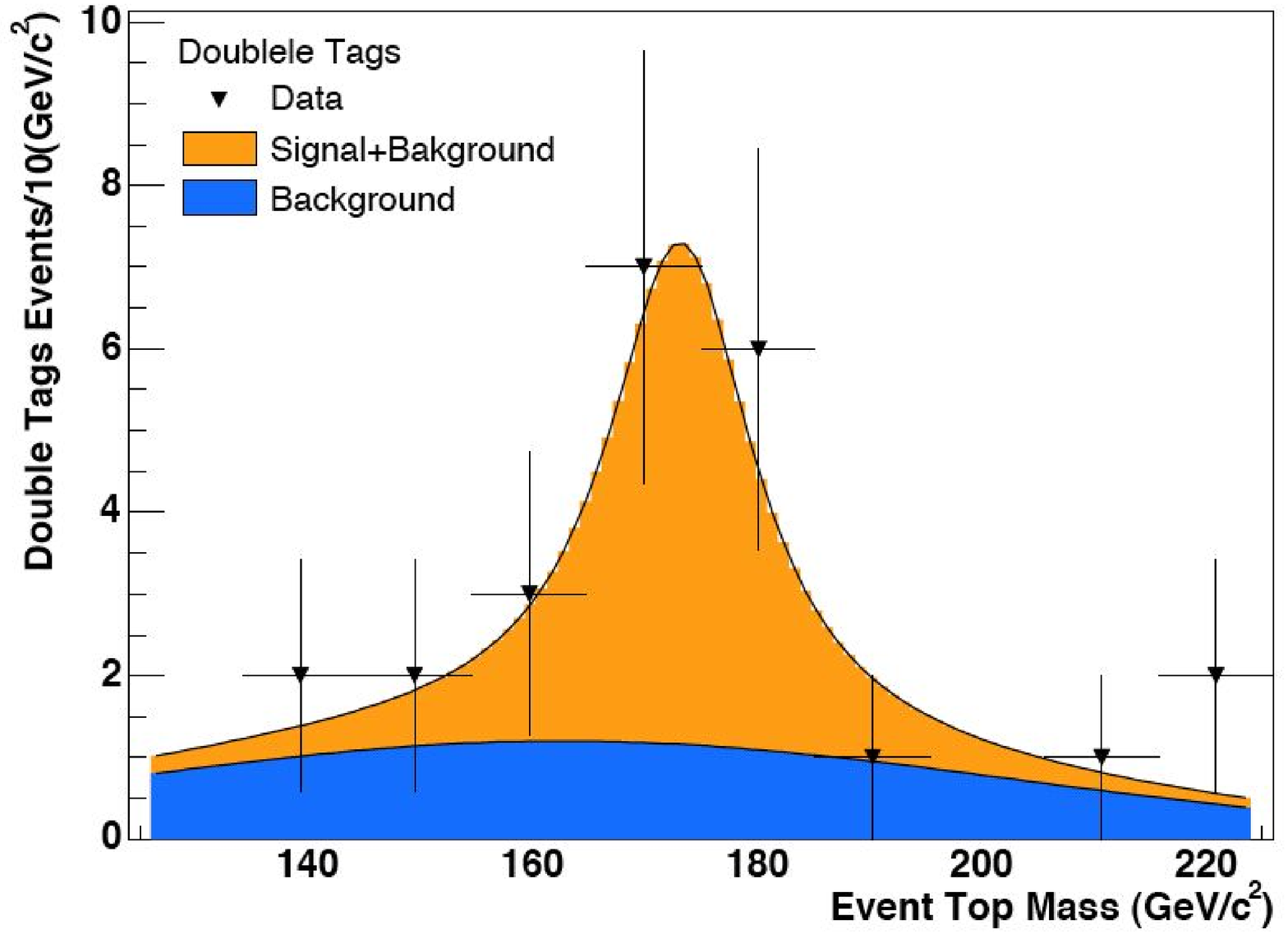,width=5cm}
\caption{Contour plot of the favored region in the jet energy scale
vs. $m_t$ for the CDF matrix-element analysis of 943 pb$^{-1}$ of data
\protect\cite{cdfR2alljetsMass}(left).  
The distribution of fitted top mass for the double tagged sample 
is shown at right.  Expected signal and background are shown.}
\label{fig:RII6jMass}
\end{center}
\end{figure*}

Using a pure matrix-element approach, CDF has also generated a
preliminary result\cite{cdfR2alljetsMass} in 943 pb$^{-1}$ which is an 
indicator of things to come.  The event selection requires six jets with
a scalar sum of their $E_T$ of more than 280 GeV.  Spherically distributed
events which tend to have central jets are kept if they have at least one
$b$-tagged jet.  Extraction of the top quark
mass proceeds by first estimating a probability vs. top quark mass using
a matrix-element calculation.  Transfer functions were used to relate 
reconstructed jet spectra to parton level momenta.  This weight distribution
is then used as a template and the data is compared to templates generated
from signal and background Monte Carlo.  By relating the fitted dijet
masses from $W$ bosons according to each jet-quark combination chosen to the
expectation from $W\rightarrow jj$, an estimate can be made of the residual
jet energy scale as well as the measured correction to the
top quark mass.  This relation is shown in Fig.~\ref{fig:RII6jMass}.
This serves to validate
the method and provide a systematic uncertainty for the residual jet 
energy scale.  The fitted mass distribution for double-tagged
candidate events is also shown in Fig.~\ref{fig:RII6jMass}.
The CDF analysis yielded a result of $m_t = 171.1\pm 4.3$ GeV where 2.1 GeV
of the uncertainty is due to systematic effects.

%% file: proper/properties.tex
As with the other fundamental fermions, the ability to study the
top quark presents new opportunities to test our current understanding.
Measuring the charge of the top quark helps to confirm if the 
particle discovered in 1995 is in fact the weak isospin partner
to the $b$-quark.  Determining $V_{tb}$ will either confirm or refute
whether we live in a world confined to three quark doublets.
Several properties of the top quark have been measured so far and the
increasing sizes of data samples are placing real constraints on
fundamental physics.

\subsection{$W$ boson helicity measurements}

\input{proper/Whel.tex}

\subsection{Measurement of $B(t \to Wb)/B(t \to Wq)$ and $|V_{tb}|$}
\input{proper/BR.tex}

\subsection{Top Quark Charge}
\input{proper/charge.tex}

%% file: proper/Whel.tex
The high mass of the top quark raises the question whether there are new 
interactions near the electroweak symmetry breaking energy scale.
Measuring the helicity of the $W$ boson from top quark decay probes 
the $tWb$ vertex and provides a stringent test of 
the standard model. A general form of the Lagrangian for the $tWb$ 
interaction is given in Ref.\cite{tWb}. It contains four form factors: 
$f_1^L$, $f_1^R$ that parameterize the $V-A$ and $V+A$ interactions, and 
$f_2^L$, $f_2^R$ which reflect the strength of an anomalous weak magnetic 
moment.  In the SM, the coupling of $W$ bosons to fermions is purely $V-A$, 
and therefore the only non zero form factor is  $f_1^L$.
If the $tWb$ couplings are standard, then top quarks decay to  
left-handed $W$ bosons ($W_-$) or to longitudinal $W$ bosons ($W_0$). 
In the presence of non-standard couplings such as $V+A$, some admixture of 
right-handed $W$ bosons ($W_+$) is expected.

\subsubsection{Sensitive Variables}

Measurement of the $W$ boson helicity can be performed using any leptonic
top quark decay. 
CDF and \dzero\ have analyzed both their dilepton and $\ell+$jets samples
to extract limits on the $W$ boson helicity using three sensitive variables.
\begin{itemize}
        \item the helicity angle $\theta^*$, defined as the angle between the 
        charged lepton and the top quark directions in the $W$ boson rest frame,         \item the transverse momentum of the lepton in the laboratory frame,
        \item the invariant mass, $M_{\ell b}$, of the charged 
        lepton and $b-$jet thought to come from the same top quark.
\end{itemize}

If we define $\theta^*$ as the angle of the decay positron in the $W$ boson 
rest frame, with the polarization axis defined by the direction of the $W$ 
boson in the top rest frame, then the angular distribution of the lepton 
with respect to the polarization of the $W$ boson is given by 
$$w(cos(\theta^*)) = f_- \frac{3}{8} (1-cos(\theta^*))^2+ f_0 \frac{3}{8}
(1-cos^2(\theta^*))+ f_+ \frac{3}{8} (1+cos(\theta^*))^2$$ 
where $ f_-,  f_0$ and
$f_+$ are the fractions of left-handed, longitudinal, and right-handed  $W$
bosons, respectively. In the SM  
$f_-, f_0$ and $f_+$ are expected to be 0.7, 0.3
and 0, respectively.

Computation of the helicity angle requires the reconstruction of the top quark
and the pairing of the lepton with a jet in the event. Reconstruction of the 
top quark using semileptonic decays is performed using a constraint kinematic 
fit to the $t\overline t$ hypothesis (see Section \ref{sec:ljmtop}). 
In the fit, $M_W$ and $m_t$
are fixed to their respective measured values. The fit is 
performed using all possible jet assignments, 12 in general, six if there 
is a single $b$-tagged jet, or two if there are two $b-$tagged jets. 
The permutation 
which gives the lowest $\chi^2$ is chosen. Once the event is reconstructed, 
the helicity angle is computed after a boost to the rest frame of the 
reconstructed $W$ boson.

When dilepton events are used, each event has two leptons and hence 
contributes twice to the measurement. Kinematic reconstruction of dilepton events is described in Section \ref{sec:llmtop}. Here, the 
presence of the two neutrinos kinematically underconstrains the system. 
For an assumed top quark mass, the kinematics are solved algebraically with a 
four-fold ambiguity in addition to the two-fold ambiguity which arises from 
the pairing of the lepton with the jet. Once the detector resolution effects 
are folded into the measurement, an average value of the helicity angle for 
each lepton is available.

Because the $M_{\ell b}$ and lepton $p_T$ are directly determinable
from the lab frame, using them greatly simplifies the analysis.  They can
also be readily applied in dilepton events as well as $\ell+$jets events 
where one of the four final state jets might not have been reconstructed.
The $M_{\ell b}$ approach relies on the fact that the helicity angle cosine,
$cos(\theta^*)$, can be approximated by the expression\cite{tWb}
\begin{equation}
        cos(\theta^*) \simeq \frac{M^2_{\ell b}}{m_t^2 - M_W^2} - 1.
\end{equation}
Each lepton-jet pairing proves a measurement of the angle.  
The lepton $p_T$ method relies on the correlations in $W$ boson momentum and
that of its decay lepton.
Left-handed $W$ bosons tend to emit the charged lepton in the 
direction opposite to their direction of flight, longitudinal $W$ bosons 
perpendicularly to their direction of flight, and right-handed $W$ bosons 
emit the leptons preferentially along their direction of flight, leading to 
increasingly harder lepton $p_T$ spectra. However, this method is less 
powerful than measuring the helicity angle.

\subsubsection{CDF Results}

The CDF experiment has pursued measurements of the polarization fractions 
of the $W$ boson via both the $M_{\ell b}$
and lepton $p_T$ measurements.  These measurements have used three 
channels: dilepton, $\ell +$ jets with one $b-$tagged jet, and $\ell+$jets 
with two tagged jets.  In each case, templates are generated for \ttbar\ 
decays with $V+A$ and $V-A$ couplings as well as background events. A 
likelihood calculation is used to extract constraints on the non-standard 
couplings.

In an analysis published in 2005, the $\ell+$jets sample was used in 
$\sqrt{s}=1.8$ TeV collisions to provide a measurement of the $V+A$ decay 
rate at the $tWb$ vertex \cite{cdfWhelR1}. They utilized the $M_{\ell b}$ 
method to  place a limit on $f_+$. To strengthen the 
limits, they combined these results with earlier measurements of the $W$ 
boson polarization using the lepton $p_T$, $f_0 = 0.91\pm 0.37(\rm stat) 
\pm 0.13 (syst)$\cite{cdfWpolR1}. A limit of  $f_+ < 0.18$
was obtained at 95\% C.L.

This was  followed with a measurement in $\sqrt{s}=1.96$ TeV
collisions \cite{cdfWhelR2a}.  A 162 pb$^{-1}$ sample of $\ell+$jets events 
was used in
both single and double tagged channels.  Both the $M_{lb}$ and
lepton $p_T$ measurements were employed for these channels.  Dilepton
events from 193 pb$^{-1}$ were also used to obtain $f_0$ and
$f_+$ via the lepton $p_T$ technique. 
Each dilepton event provides two measurements for each event.
These samples are described in Section \ref{sec:llcsec} and \ref{sec:ljcsec}. 
The primary difference in $\ell+$ jets event selection involved an additional 
requirement
of a fourth jet with $p_T > 8$ GeV and $|\eta| < 2$.  This allowed
the events to be kinematically reconstructed with the \ttbar\
hypothesis.  $m_t$ taken to be 175 GeV and the lepton-jet
matching was performed based on the result of this fit.

In order to extract a measurement, templates were generated for the
different samples.  In all cases, one $W$ boson was simulated
with SM expectations and the other was forced to a chosen helicity.
Leptons from the decay of this second $W$ boson were used to
generate $M_{\ell b}$ and $p_T^{\ell}$ templates for \ttbar\ decays with 
$V-A$ and $V+A$ couplings and for background events.

These templates were used in a fit to data. To extract $f_0$, $f_+$ was fixed to zero, and to extract $f_+$, $f_0$ was fixed to the standard model expectation of 0.7. Interestingly, the value obtained in the
dilepton sample ($f_0 = -0.54^{+0.35}_{-0.25}\pm 0.16$) was unphysical
because of the soft $p_T$ spectrum of the observed events.
This result is consistent with $f_0 = 0.7$ only at the 0.5\% level.  
It differs by 2 $\sigma$ from the $\ell+$ jets measurement of 
$f_0 = 0.95^{+0.35}_{-0.42}\pm0.17$.  As indicated in
section \ref{sec:mTop}, the dilepton sample was analyzed and
determined to be marginally consistent with the standard model.
All results were combined to obtain $f_0 = 0.74^{+0.22}_{-0.34}
(\rm stat + sys)$ and $f_+ = 0.00^{+0.20}_{-0.19}(\rm stat+sys)$.  

This gives a limit of $f_+ < 0.27$ at 95\% c.l.
To account for correlations among
uncertainties, Monte Carlo experiments were conducted to obtain
the necessary correlation coefficient.  Significant sources
of systematic uncertainty came from background modeling, the
uncertainty in the value of $m_t$, and the calibration of jet 
energies.

\begin{figure*}[!h!tbp]

\begin{center}

\epsfig{figure=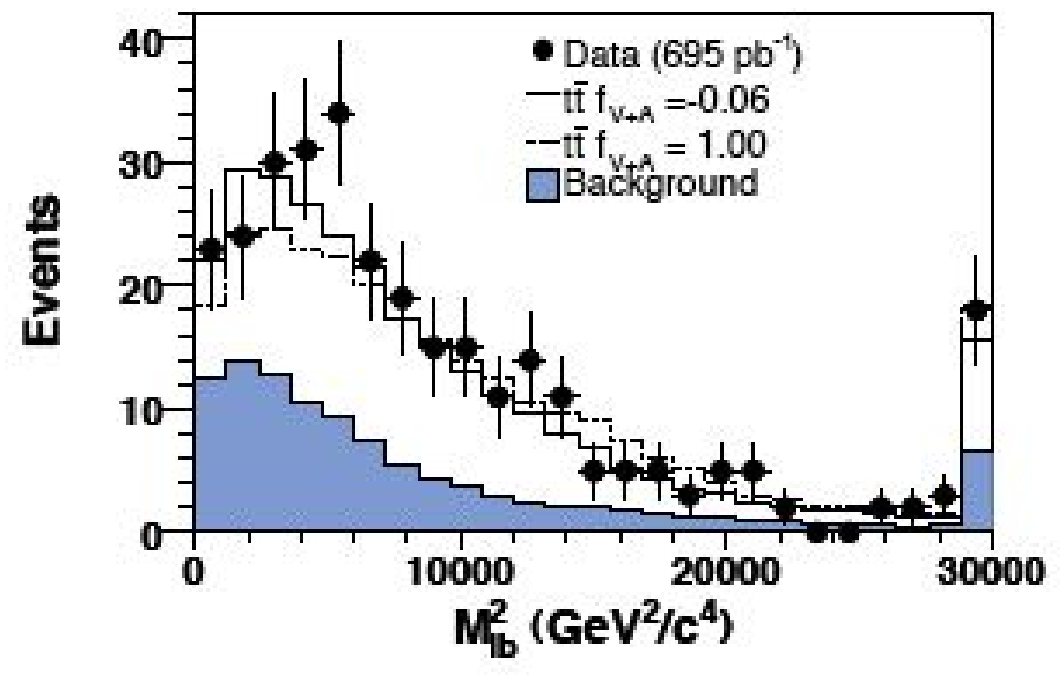,width=0.45\textwidth}
\epsfig{figure=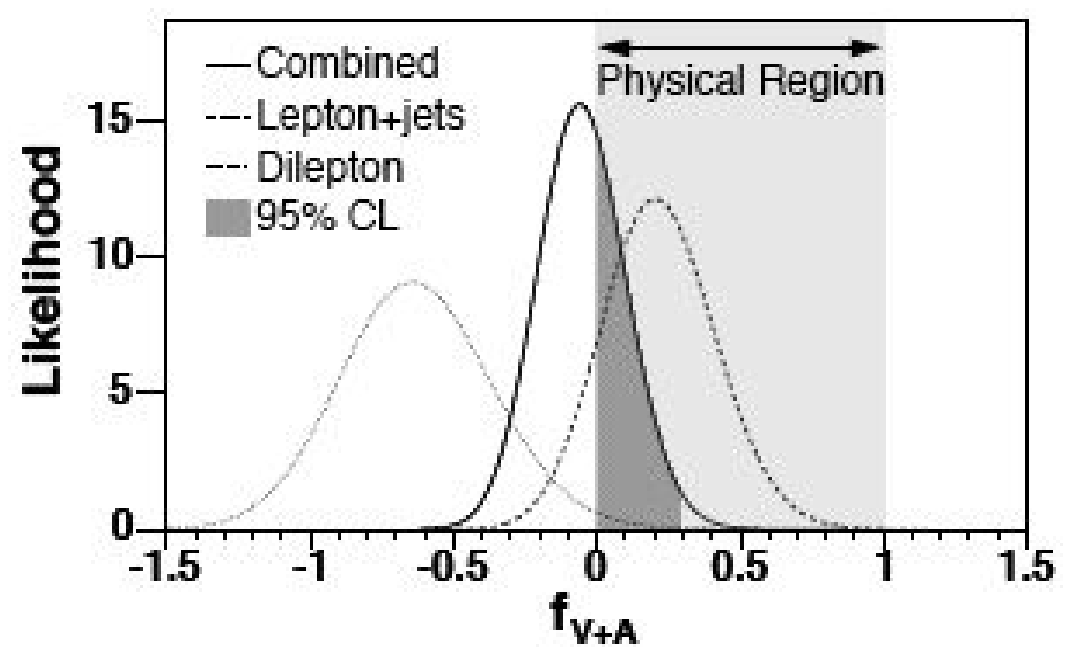,width=0.45\textwidth}

\end{center}

\vspace*{8pt}

\caption{Distribution of $M_{\ell b}$ in 700 pb$^{-1}$ of CDF 
$\ell +$ jets events, 
with expected \ttbar\ and background overlaid (left).  Likelihood
for dilepton, $\ell+$jets and combined analyses indicating
estimated $f_{V+A}$ (right) \protect\cite{cdfWhel700}.}
\label{fig:cdfWhel}
\end{figure*}

In 700 pb$^{-1}$ of Run II data, CDF has used the $M_{\ell b}$ technique
in dilepton, $\ell +$ jets with one $b-$tag and $\ell+$ jets with
two tags \cite{cdfWhel700}.  The event selection follows Section
\ref{sec:llcsec} and \ref{sec:ljcsec}.  A fourth jet was not explicitly
required in the $\ell+$jets sample.  
In constructing templates for the $\ell+$jets samples, the following
observations can be made.  In the single tag case, the tagged
jet comes from the same top quark as the identified lepton in
about 50\% of the cases.  For the double tag sample, there are two possible
combinations of $b-$tagged jet with charged lepton.  This led
to the use of a one-dimensional $M_{\ell b}$ template for the former,
and a two-dimensional template for the latter.  Templates for
background were obtained from {\sc Alpgen} $W\bbbar$ events and, for the
single tag case, a 15\% admixture of multijet instrumental background from
data.  The dilepton analysis considers each charged lepton as an 
independent measurement.  In each case, there are two possible matches
to the leading two jets in an event, which are assumed to be the
$b-$jets.  Like the double-tagged $\ell+$jets analysis, a two-dimensional
template of $M_{\ell b}$ is generated for signal.  The background
was modeled by a 50\%:30\%:20\% mix of $Z\rightarrow \ell\ell$,
$W+$jets, and diboson processes.  The $W+$jets instrumental
background was obtained from data according to the prescriptions 
described in Section \ref{sec:llcsec}.  The expectations
from backgrounds for $\ell+$jets (dilepton) channels were validated
in background-dominated one and two (one) jet samples in data.  Figure
\ref{fig:cdfWhel} indicates the distribution of $M_{\ell b}$ for
the $\ell+$jets data sample.  Both $V+A$ and $V-A$ expectations
are shown for comparison.

To obtain an estimate of $f_{V+A}$, a binned log likelihood fit was
employed to the data using templates from \ttbar\ decays with $V+A$ and
$V-A$ couplings and from background events. The \ttbar\ and background
cross sections were treated as nuisance parameters in the fit.
The fitted values of these parameters were near the input values
(e.g. $\sigma_{\ttbar} = 7.3 \pm 0.9$ pb).  The validity of the
fit method was explored by using ensemble testing where the
\ttbar\ and background composition was allowed to vary by
the estimated uncertainty in these estimates.  Ensembles were
also constructed with a varying $V+A$ component.  The
measurement from the $\ell+$jets analysis yields
the fraction of $V+A$ current 
$f_{V+A} = 0.21\pm0.28$.  Dilepton events yield $f_{V+A} = -0.64\pm0.37$.
These results are approximately $1.8 \sigma$ apart.  The
combined result gives $f_{V+A} = -0.06\pm 0.25 (\rm stat+sys)$.
The likelihood vs. $f_{V+A}$ for each channel, and the combined
result, are shown in Figure \ref{fig:cdfWhel}.
The primary systematic uncertainty came from the uncertainty
in the jet energy scale.  By using a Bayesian approach,
an upper limit of $f_{V+A} < 0.29$ was obtained at 95\% c.l.
These results translate into a value for the fraction of right-handed
$W$ bosons of $f_+ = -0.02\pm 0.07(\rm stat)\pm 0.04(sys)$ and
a limit of $f_+ < 0.09$ at 95\% C.L.

The most recent analysis of the $W$-boson helicity fractions
from CDF\cite{CDF_top_SUSY07} was performed using a dataset corresponding 
to an integrated
luminosity of 1.7fb$^{-1}$, The distribution of 
the  helicity angle cosine, $cos(\theta^*)$, was used to obtain the
fractions of longitudinal $f_0$ or right-handed $f_+$ $W$-bosons. 
Fully reconstructed  $\ell+$jets $t\overline t$ events were used for
this analysis. While all combinations of jet-parton assignments were 
processed via the kinematic fitter, the one with the smallest fit 
$\chi^2$ was chosen as the final event configuration and was used
for reconstructing the variable $cos(\theta^*)$. An unbinned likelihood
fit to the  observed $cos(\theta^*)$ distribution for the data was performed
using templates for the longitudinal, right- and left-handed signal events 
and the  background events. From this fit  a measurement of 
the uncorrected longitudinal and right-handed fractions of $W$-boson events 
in the data was obtained. To convert this measurement to the 
true longitudinal and right-handed 
fractions of events, a correction for the selection and reconstruction
efficiencies was applied. Three different types of fits were employed.
In all three fits the constraint $f_- + f_0 + f_+ = 1$  was imposed.
Two 1-D fits correspond to the measurements of the fraction
$f_+$ ($f_0$) with  $f_0$ ($f_+$) fixed  to the value
expected from SM. If one constrains the right-handed fraction $f_+ = 0$, then 
the measurement of $f_0 = 0.57\pm 0.11 {\rm (stat)}\pm 0.04 {\rm (syst)}$.
Constraining $f_0=0.7$ lead to a measurement $f_+ = -0.04\pm 0.04 {\rm
  (stat)}\pm 0.03 {\rm (syst)}$ or $f_+ <  0.07 $ at 95\% C.L..
The third fit was a simultaneous $2D$ fit for both 
parameters $f_+$ and $f_0$ and lead to a measurement of
$f_0 = 0.61\pm 0.20 {\rm (stat)}\pm 0.03 {\rm (syst)}$ and
$f_+ = -0.02\pm 0.08 {\rm   (stat)}\pm 0.03 {\rm (syst)}$.

\subsubsection{\dzero\ Results}

During Run II, \dzero\ has performed measurements of the $W$ boson helicity 
using the $cos(\theta^*)$ distribution. In the most recent 
publication\cite{Whel06Dzero}, both dilepton and $\ell +$ jets events are 
used for the measurement. The event selection in both channels is similar 
to that used for the $m_t$ measurements. The backgrounds for the  
$\ell$+jets events are mainly $W$+jets production and the instrumental 
background from misidentified multijet events. Discrimination between 
$t\overline t$ events and background events is obtained by constructing a 
discriminant which is close to one for signal $t\overline t$ events and near 
zero for background events. This discriminant is computed from variables 
that exploit the differences in the event kinematics and jet flavor of the 
signal and background events. 

The kinematic variables used for building the discriminant include $H_T$,
$\Delta\phi(\ell,\met)$, aplanarity ($A$), sphericity ($S$) and
the minimum dijet mass $m_{jjmin}$.  These are described
in Section \ref{sec:ttSelVars}.  The $\chi^2$ from the kinematic fit was 
also used.In addition a variable which distinguishes the flavors of the jets
is obtained by using the impact parameters of the tracks in the jet
with respect to the primary vertex. The impact parameters of tracks are
used to form a probability for the jet to
originate from the primary vertex. The  average of the two 
smallest probabilities are used to characterize the event. 
This variable has smaller 
values for top quark events than for background events.
For the dilepton channel, backgrounds arise mainly from $WW$+jets or $Z+$jets 
processes. After the selection of the events using kinematic quantities, 
a fairly good signal to background ratio is obtained and no further cut 
on an event discriminant is required. 

\begin{figure*}[!h!tbp]

\begin{center}

\epsfig{figure=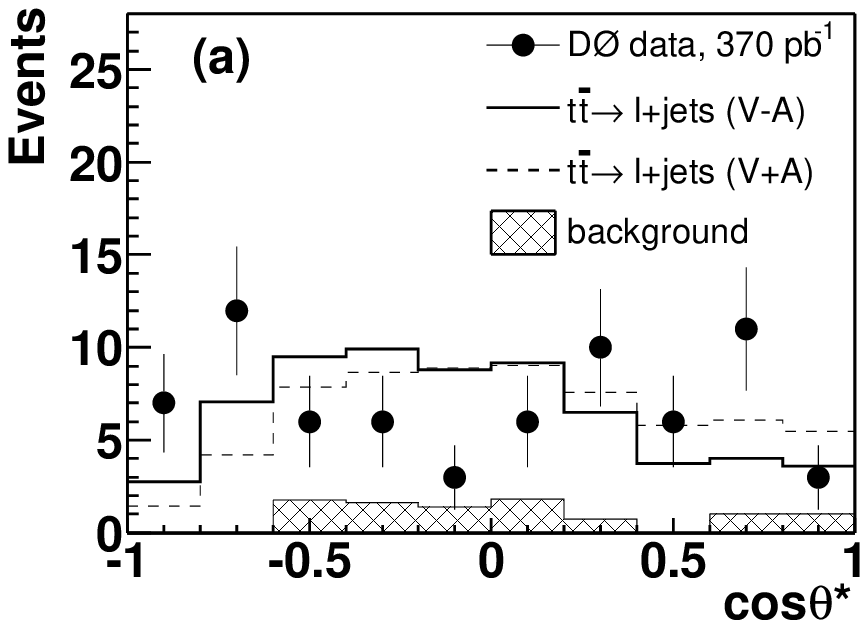,width=0.45\textwidth}
\epsfig{figure=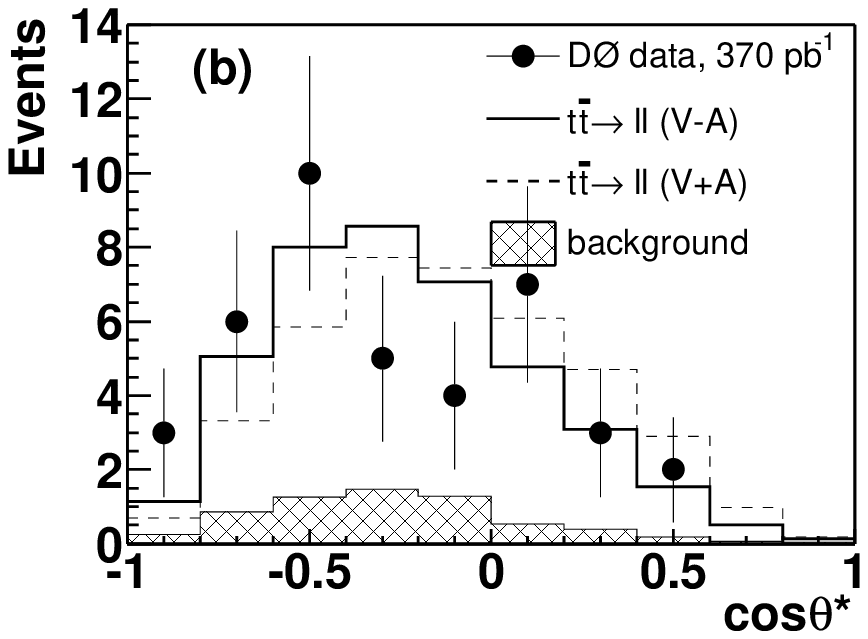,width=0.45\textwidth}

\end{center}

\vspace*{8pt}

\caption{Distribution observed in \dzero\ $\ell+$jets events (left) and 
dilepton events (right)\protect\cite{Whel06Dzero}. 
The standard model predictions are shown as the solid lines. A model with a 
pure $V + A$ interaction would result in the distribution given by the 
dashed lines.}

\label{fig:D0Whel}
\end{figure*}

Measurement of the $W$ boson helicity is performed by fitting
the $cos(\theta^*)$
distributions observed in both the dilepton and $\ell +$ jets data samples.
The observed distribution is fit to two components: the expected
backgrounds and  a $t\overline t$ signal sample which is 
generated at specific values of $f_+$. 
During the fit, $f_0$ is fixed at 0.70 and $f_+$ is varied. 
The $cos(\theta^*)$ templates for signal samples
are constructed by generating  $t\overline t$ events with two different 
values of $f_+$=0 and $f_+$=0.30 and using a linear interpolation to 
generate a value of $f_+$. The backgrounds with real leptons are generated
using {\sc alpgen} and {\sc pythia}, and the multijet background templates
are extracted from the background control samples 
(see Section \ref{sec:ljmtop} for details).
Figure~\ref{fig:D0Whel} shows the $cos(\theta^*)$ distribution observed in 
$\ell+$ jets and  dilepton events for a luminosity corresponding to 
370 pb$^{-1}$. Overlaid on the figure are the expected distributions from
the standard model prediction and a model with a pure $V+A$ 
interaction. A binned Poisson likelihood $L(f_+)$ is computed for the 
data to be consistent with the sum of signal and background templates 
at each of the chosen $f_+$ values in the range $0<f_+<0.3$. 

A parabola is fit to the $-ln[L(f_+)]$ points to extract the likelihood 
as a function of $f_+$. Systematic uncertainties are folded in prior to the
determination of the final result. The dominant contributions to the
systematic uncertainties arise from jet energy scale measurements and 
the input top quark mass. They are assumed to be fully correlated between 
the dilepton and $\ell +$ jets events. With the assumption that $f_0$ is 
fixed at 0.7, \dzero\ measures 
$f_+ = 0.109 \pm 0.094(\rm stat) \pm 0.063(\rm syst) $
using $\ell+$ jets events, and $f_+ = 0.089 \pm 0.154(\rm stat) \pm 0.059(syst)$using dilepton events. Combining the two yields $f_+ = 0.056 \pm 0.080(\rm stat) \pm 0.057(syst)$. A Bayesian confidence interval is computed using a flat
prior distribution which is non-zero only in the physically allowed
region of $0<f_+<0.3$ leading to $f_+ < 0.23$ at 95\% C.L.

An earlier analysis with 230 pb$^{-1}$ data was also carried out
by \dzero\ \cite{Whel05Dzero}. The analysis strategies and techniques were
similar to the ones described above.  The main difference involved the
further subdivision of the $\ell+$ sample into two categories 
depending on whether a tagged $b-$jet was found in the event. 
A likelihood analysis of the angular distribution of the leptons, 
leads to the result $f_+=0.0\pm0.13(stat)\pm0.017(syst)$ and $f_+<0.25$ at 
95\%\ C.L. All measurements are consistent with the SM prediction 
$f_0$=0.7, $f_+$=0.

\dzero\ also has carried out a first model independent  measurement of 
the $W$ boson helicity fractions on a larger dataset corresponding to 
an integrated luminosity of 1 fb$^{-1}$\cite{Whel07Dzero}. 
As described above, all 
earlier measurements of helicity fractions $f_+$, $f_0$ have been 
performed by keeping one 
of them fixed to the value predicted by the SM and measuring the other 
by a fit to one of the kinematic variables sensitive to the $W$-boson
helicity.  The model independent study is based on a simultaneous
measurement of $f_+$, $f_0$,  with the condition that $f_+ + f_0 + f_- = 1$.
This analysis also benefits from better event selection efficiency. 
The event selection efficiency and the background rejections compared
to analyses on the smaller datasets  are improved by using a 
likelihood based event discriminant. The likelihood discriminant 
is based on the following variables: $H_T$,  centrality $C$, 
$k_T(min)$,  the sum of all jet and charged lepton energies $h$, 
the minimum dijet mass of the jet pairs $m(jj)_{min}$, aplanarity $A$, 
sphericity S, \met\, and the dilepton invariant mass (for definitions
of these variables see Section\ref{sec:topPairs}). The likelihood
discriminants $\cal D$, are computed independently for each of the different
channels considered in the analysis ($ee$, $e\mu$, $ee$, $e+$jet,
and $\mu$+jet) and
an appropriate selection on $\cal D$ is applied to maximize 
the signal to background for each channel. In addition, the neural 
network $b$-tagging
algorithm was used to identify the $b$-jets.

\begin{figure*}[!h!tbp]
\begin{center}

\epsfig{figure=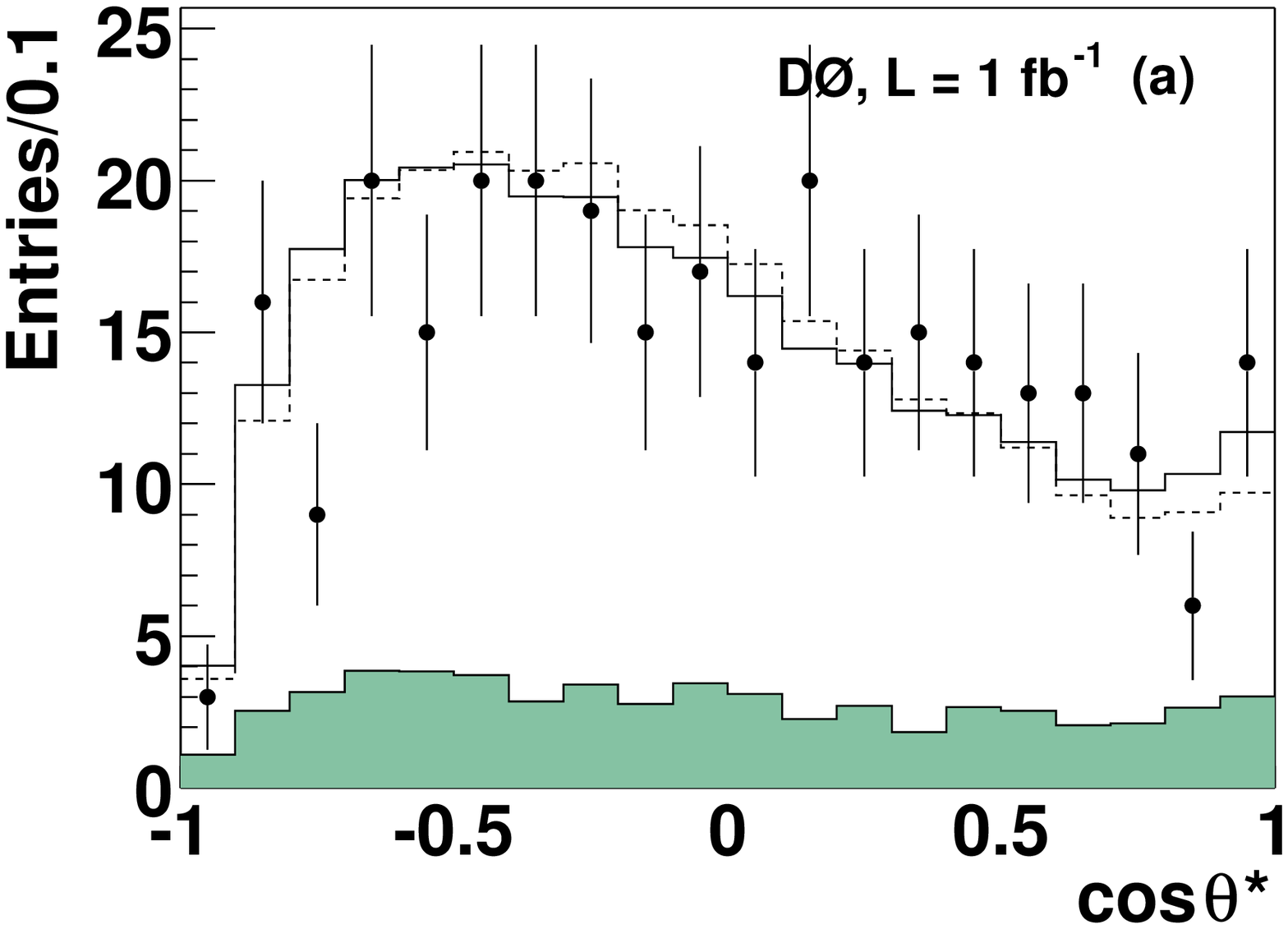,width=0.45\textwidth}
\epsfig{figure=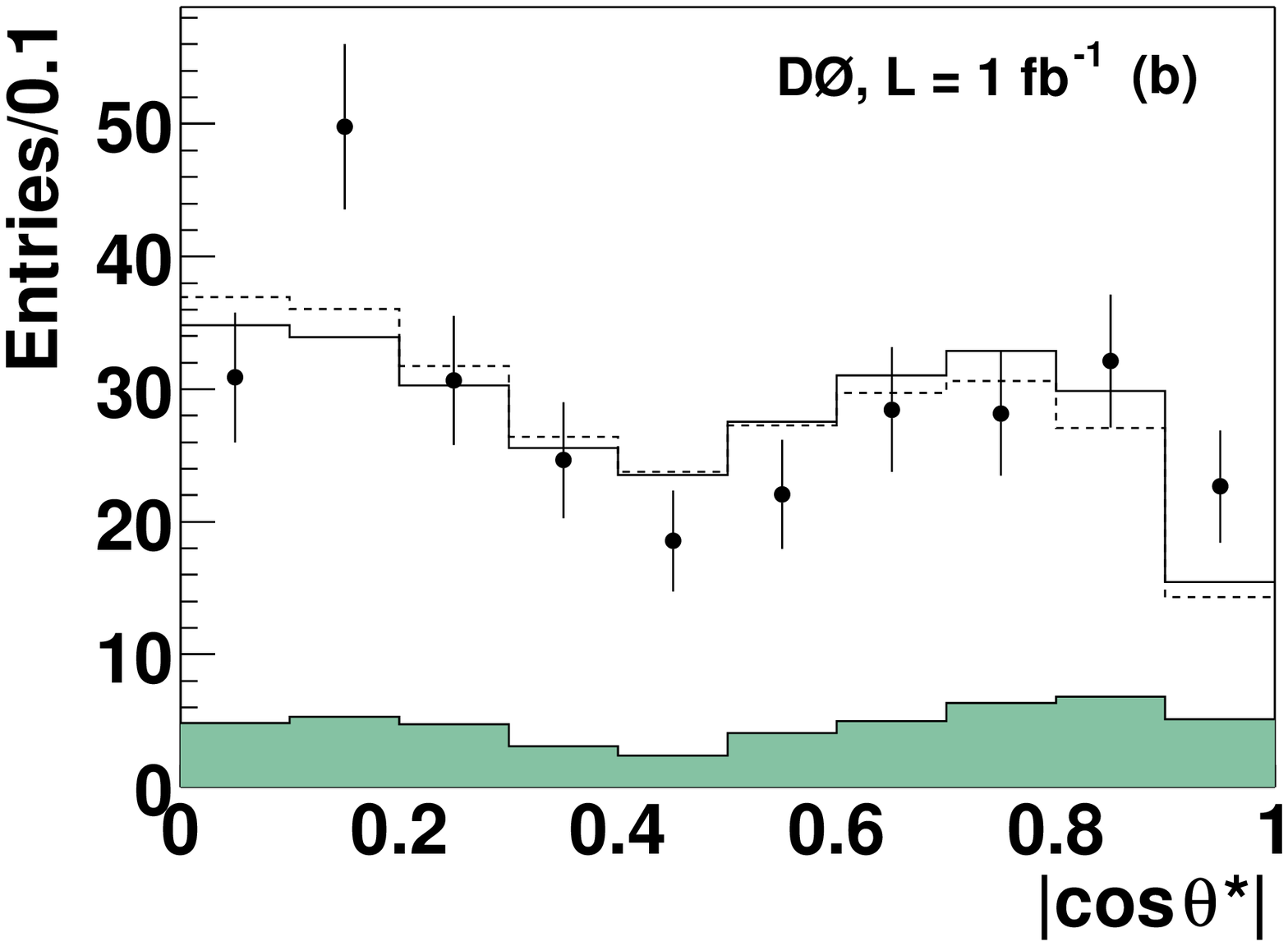,width=0.45\textwidth}

\end{center}
\vspace*{8pt}

\caption{Distribution of  $cos(\theta^*)$ 
for  the leptonic $W$ bosons (left) and  hadronic $W$ 
bosons (right) obtained in the  $\ell+jets$
sample.\protect\cite{Whel07Dzero}. 
The data are represented by dots, the expected signal is shown as
the dashed histogram and the shaded histogram corresponds to the
expected background. }

\label{fig:D0Whel07costh}
\end{figure*}

The simultaneous measurement of $f_0$ and $f_+$ is performed using 
the distribution of $cos(\theta^*)$. A further enhancement in this analysis
compared to previous ones is the use of the hadronicaly decaying $W$ 
boson in the  $cos(\theta^*)$ distribution in the $\ell+jets $ events. 
The distributions of  $cos(\theta^*)$ for the leptonic and hadronic $W$
bosons obtained in the  $\ell+jets $ events is shown 
in Fig.\ref{fig:D0Whel07costh}.

\begin{figure*}[!h!tbp]
\begin{center}

\epsfig{figure=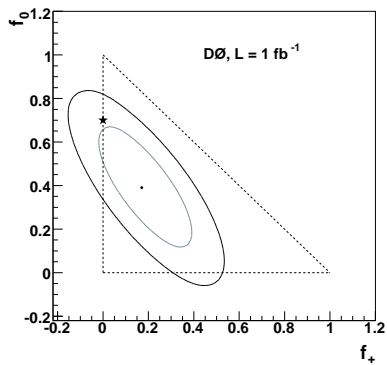,width=0.45\textwidth}

\end{center}
\vspace*{8pt}

\caption{Result of the model-independent simultaneous fit to the
$W$ boson helicity fractions $f_0$ and $f_+$. The star denotes
the SM prediction, the ellipses represent the  measured 
68\% and 95\% C.L.  contours. The  triangle indicates the
physically allowed region.\protect\cite{Whel07Dzero}.}

\label{fig:D0Whel07res}
\end{figure*}

The simultaneous fit for $f_0$ and $f_+$  is performed using a 
binned Poisson likelihood ${\cal L}(f_0; f_+)$. The consistency 
of the data  with the sum of signal and background templates 
as a function of the helicity fractions is computed, keeping the
background normalization constrained to the expected value 
within uncertainties. With this procedure \dzero\ obtains
$$f_0 = 0.425\pm  0.166 ({\rm stat.}) \pm 0.102 ({\rm syst.}) {\rm and}
f_+ = 0.119  \pm 0.090 ({\rm stat.}) \pm 0.053 ({\rm syst.}).$$
In Fig.\ref{fig:D0Whel07res}, the 68\%, and 95\% C.L. contours from the fit are
shown compared to the prediction from SM. While the measurement
may suggest a smaller fraction of longitudinal $W$ bosons and
a larger fraction of right-handed $W$-bosons as compared to 
SM predictions, they are statistically consistent with the SM. 
Given the uncertainties in the measurement, \dzero\ computes  
a 27\% chance for observing the discrepancy. 

\vspace*{8pt}

Measurements of $f_0$ and $f_-$ can also be performed using the
earlier prescription of fixing one of the them to the SM value. 
Fixing $f_+$ to  the SM value gives $f_0 = 0.619\pm 0.090 
({\rm stat.}) \pm   0.052 ({\rm syst.})$ and fixing $f_0$ to the
SM value leads to a value of  $f_+ = -0.002 \pm 0.047 (\rm {stat}.) 
\pm  0.047 ({\rm syst.})$.

%% file: proper/BR.tex
The top quark decays predominantly via $t \to Wq$, where $q$
can be a $d$, $s$, or a $b$-quark.
Flavor changing neutral current (FCNC) decays of the type $t \to Vq$, 
where $V=g,Z,\gamma$ and $q=u,c$ are of the order of $10^{-10}$ or 
smaller within the SM\cite{fcnc}. In the SM, the ratio of top quark 
branching fractions, $R=BR(t \to Wb)/BR(t \to Wq)$, can be expressed in 
terms of the CKM matrix elements,
$R=\frac{|V_{tb}|^2}{|V_{tb}|^2+|V_{ts}|^2+|V_{td}|^2}$.
Based on the assumption of three quark generations and unitarity of the
CKM matrix, and using the very small experimentally measured values 
of $|V_{ub}|$ and $|V_{cb}|$, one obtains 0.9990$<|V_{tb}|<$0.9992 
at 90\% C.L.\cite{ckm_measurement}. This also constrains $R$ to be 
in the interval 0.9980--0.9984 at the 90\% C.L.\cite{ckm_measurement}.
Thus, in the SM,  the top quark 
is expected to decay nearly 100\% of the time into a $W$ boson and a $b$-quark.
A significant deviation from the SM prediction of $BR(t \to Wb)$ $\simeq$ unity 
could arise in the presence of additional quark generations or non-SM 
processes in the production or decay of the top quark. 
The identification of \ttbar\ events at the Tevatron allows us to 
check for such a possible deviation which would imply the appearance 
of new physics.  In Run I, CDF measured $R=0.94^{+0.31}_{-0.24}$(stat+sys) 
\cite{cdf-run1-br} and set a lower limit of $R>$0.61 (0.56)  
at 90\% ( 95\%) C.L., in agreement with SM predictions. 
In Run II, both CDF\cite{cdf-run2-br} and D\O\ \cite{d0-run2-br}
have reported direct measurements of $R$ by examining the candidate \ttbar\ events
in data samples of $\sim$160 pb$^{-1}$ and 
$\sim$230 pb$^{-1}$, respectively. These analyses assume that $t \to Xq$, 
where $X \neq W$ is negligible.

\paragraph{Analysis strategy:}
In the SM case with $R \approx 1$, the \ttbar\ event signature contains  
two $b-$jets from the two top quark decays. 
However, with $R$ of an unknown value, a \ttbar\ event might have zero, one or two $b-$jets 
since the top quark can also decay into a light quark ($d$ or $s$) and a $W$ boson.
The ratio $R=BR(t \to Wb)/BR(t \to Wq)$ determines the fraction of \ttbar\ events
with zero, one and two $b-$jets and therefore how \ttbar\ events are distributed 
among samples with zero, one or two-tags. Thus $R$ can be extracted from the 
relative rates of identified $b-$jets in 
\ttbar\ events. If each of the two top quarks in the event has a probability 
$R$ to decay to a $b$-quark, and there is an efficiency $\epsilon_b$ to tag 
the $b-$jet, the efficiencies to detect zero, one or two $b-$jets 
in the event are

\begin{eqnarray}
\epsilon_0 = (1-R \epsilon_b)^2, \epsilon_1 =  2R \epsilon_BR(1-R \epsilon_b), \epsilon_2 =  (R \epsilon_b)^2 \\
\Longrightarrow 
R \epsilon_b = \frac{2}{\epsilon_1/\epsilon_2+2} = \frac{1}{2\epsilon_0/\epsilon_1+1} = \frac{1}{\sqrt{\epsilon_0/\epsilon_2}+1} \\
R \epsilon_b = \frac{2}{N_1/N_2+2} = \frac{1}{2N_0/N_1+1} = \frac{1}{\sqrt{N_0/N_2}+1}
\end{eqnarray}

\noindent where $N_i$ are the number of \ttbar\ events observed with $i$ tags, 
where $i$ can be zero, one or two.
Any two tagging rates determine $R\epsilon_b$ uniquely, while all three 
tagging rates jointly overdetermine $R\epsilon_b$. 
The ratios of tag rates can only determine the product $R\epsilon_b$, 
as it is not possible to distinguish missing tagged jets due to 
tagging inefficiency ($\epsilon_b$ less than unity) from missing tagged jets
due to the absence of $b$-quarks ($R$ less than unity). $R$ can be extracted 
by measuring $\epsilon_b$ independently, from \ttbar\ 
simulation calibrated with complimentary data samples. 
As $R$ depends only on relative tag rates, this measurement is 
independent of any assumptions of the overall \ttbar\ cross section
($\sigma_{\ttbar\ })$. 

The measurement thus follows three basic steps. After identifying 
samples enriched in \ttbar\ events, the background level in those 
events is estimated for different $b-$jet multiplicity. 
Then the expected tag rates (and, implicitly, their ratios)
in \ttbar\ events are predicted as a function of $R\epsilon_b$. 
Finally, the observed tag rates are compared to the expectations, 
yielding the most likely value of $R\epsilon_b$, and allowing to 
set a lower limit on $R$. This strategy
was pioneered by CDF's Run I analysis\cite{cdf-run1-br}. 

\paragraph{Measurement by CDF:}
Based on $\sim$160 pb$^{-1}$ of Run II data, CDF followed the same 
technique and determined $R$ from the relative numbers of \ttbar\ events 
with different multiplicity of $b-$tagged jets in the $\ell+$jets 
and dilepton channels\cite{cdf-run2-br}. 
The event selection, $b-$jet identification and the associated 
background estimation are 
essentially equivalent to the ones in the cross section analyses 
discussed in earlier sections. In particular, in the one-tag and two-tag 
subsamples of the $\ell+$jets sample, $a priori$ estimates of the background are
made with a collection of data-driven and simulation techniques. In the zero-tag
$\ell+$jets sample, where the $W+$jets production rate 
dominates over the \ttbar\ pairs, the event rate is determined by
exploiting the unique kinematics of \ttbar\ events with an artificial 
neural network (ANN). A binned maximum likelihood fit of the ANN 
output distribution is performed for the \ttbar\ fraction in the 
subsample.  In the dilepton channel, 
most of the jets in the background events arise from generic QCD radiation. 
To determine the background distribution across the subsamples 
with zero, one and two b-tags, a parameterization of the probability to tag a 
generic QCD jet, 
derived from jet-triggered data samples, is applied to the jets in 
the dilepton sample, correcting for the enriched \ttbar\  content of 
the sample. Table~\ref{had-xsec} shows the observed number of events ($N^{obs}$)
in the $\ell+$jets and dilepton samples, and
the corresponding estimates of background levels ($N^{bkg}$) and 
expected event yields ($N^{exp}$).
The \ttbar\ event tagging efficiency $\epsilon_i$ depends on the fiducial 
acceptances for jets that can potentially be $b-$tagged, and the efficiency
to tag those jets. These efficiencies in turn depend on the species of 
the jet's progenitor quark. The efficiency $\epsilon_i$ thus depends
strongly on $R$, as $R$ not equal to unity implies fewer $b-$jets available for $b-$tagging, 
and more light-quark jets instead. The jet acceptances and tagging 
efficiencies are used to parameterize $\epsilon_i(R)$ using 
\ttbar\ simulated events. The nominal values of $\epsilon_i$ for $R$ equal to unity are
also given in Table~\ref{had-xsec}.
\\

\begin{table}[ht]
\begin{center}
\tbl{Summary of observed and expected event yields as a function of $b-$jet multiplicity in the $l+$jets and dilepton samples in the CDF\protect\cite{cdf-run2-br} and D\O\ \protect\cite{d0-run2-br} analyses.}
{\begin{tabular}{lccc|ccc}\toprule
          & 0-tag & 1-tag & 2-tag & 0-tag & 1-tag & 2-tag \\ \hline  
CDF ($l+$jets) & & & & CDF (dilepton) & & \\     
 \colrule
$\epsilon_i$(R=1) & 0.45$\pm$0.03 & 0.43$\pm$0.02 & 0.12$\pm$0.02 & 
0.47$\pm$0.03 & 0.43$\pm$0.02 & 0.10$\pm$0.02\\
$N_i^{bkg}$ & 62.4$\pm$9.0 & 4.2$\pm$0.7 & 0.2$\pm$0.1 & 2.0$\pm$0.6 & 0.2$\pm$0.1 & $<$0.01\\ 
$N_i^{exp}$ & 80.4$\pm$5.2 & 21.5$\pm$4.1 & 5.0$\pm$1.4 & 6.1$\pm$0.4 & 4.0$\pm$0.2 & 0.9$\pm$0.2\\ 
 $N_i^{obs}$ & 79 & 23 & 5 & 5 & 4 & 2\\ 
 \hline
 D\O\ (l+3 jets) & & & & D\O\ (l+$\geq$4 Jets) & & \\ \hline
 $N_i^{\ttbar\ }$ & 32.4$\pm$1.6 & 32.3$\pm$1.6 & 8.2$\pm$0.5 & 35.6$\pm$2.8 & 41.5$\pm$3.3 & 13.5$\pm$1.4\\ 
$N_i^{exp}$ & 1275$\pm$44 & 79$\pm$5 & 11.4$\pm$0.8 & 297$\pm$19 & 56$\pm$4 & 14.4$\pm$1.4\\ 
 $N_i^{obs}$ & 1277 & 79 & 9 & 291 & 62 & 14\\ 
 \botrule
\end{tabular}
\label{had-xsec}}
\end{center}
\end{table}

\begin{figure*}[!h!tbp]
\begin{center}
\epsfig{figure=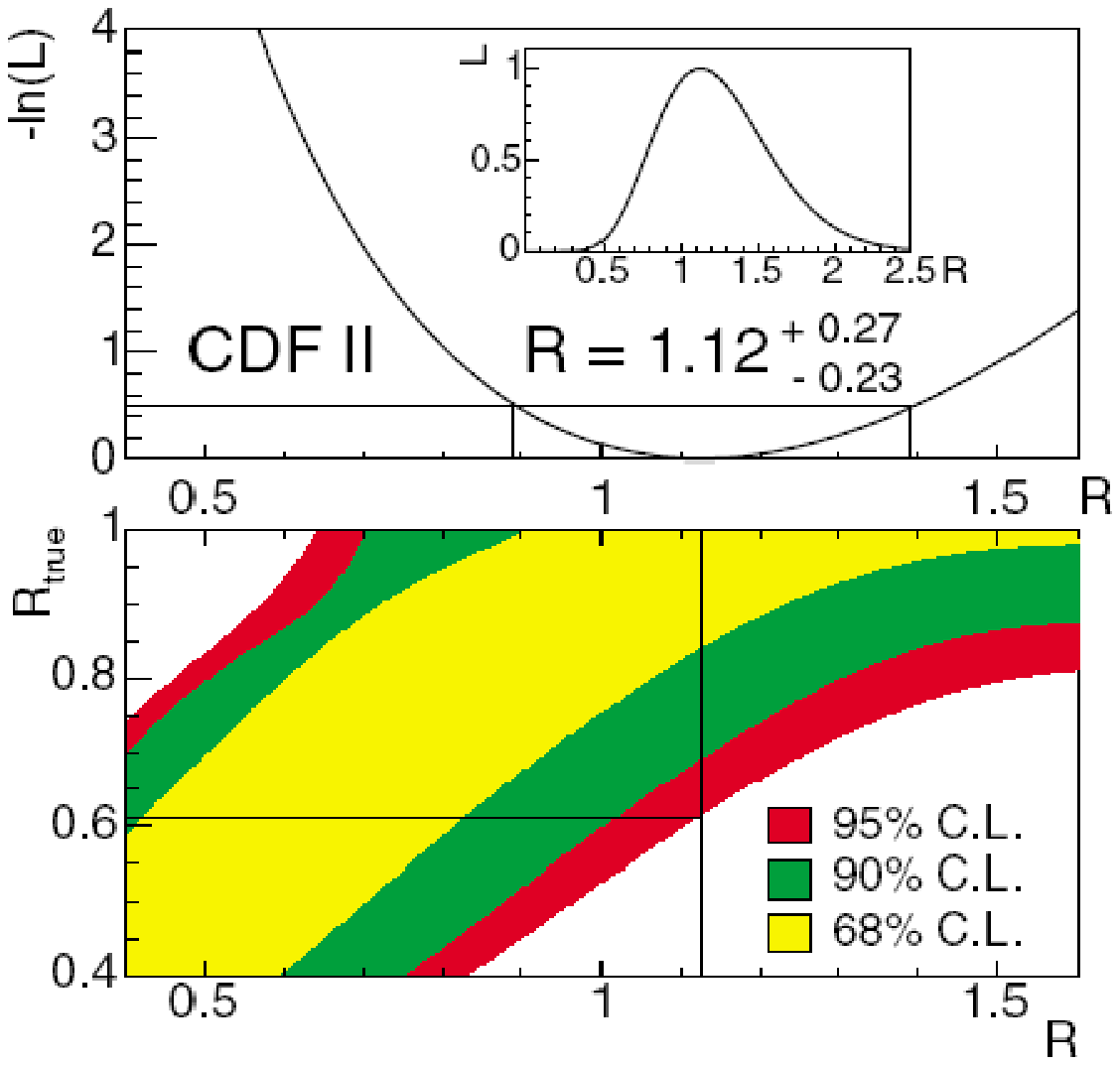,width=0.40\textwidth}
\epsfig{figure=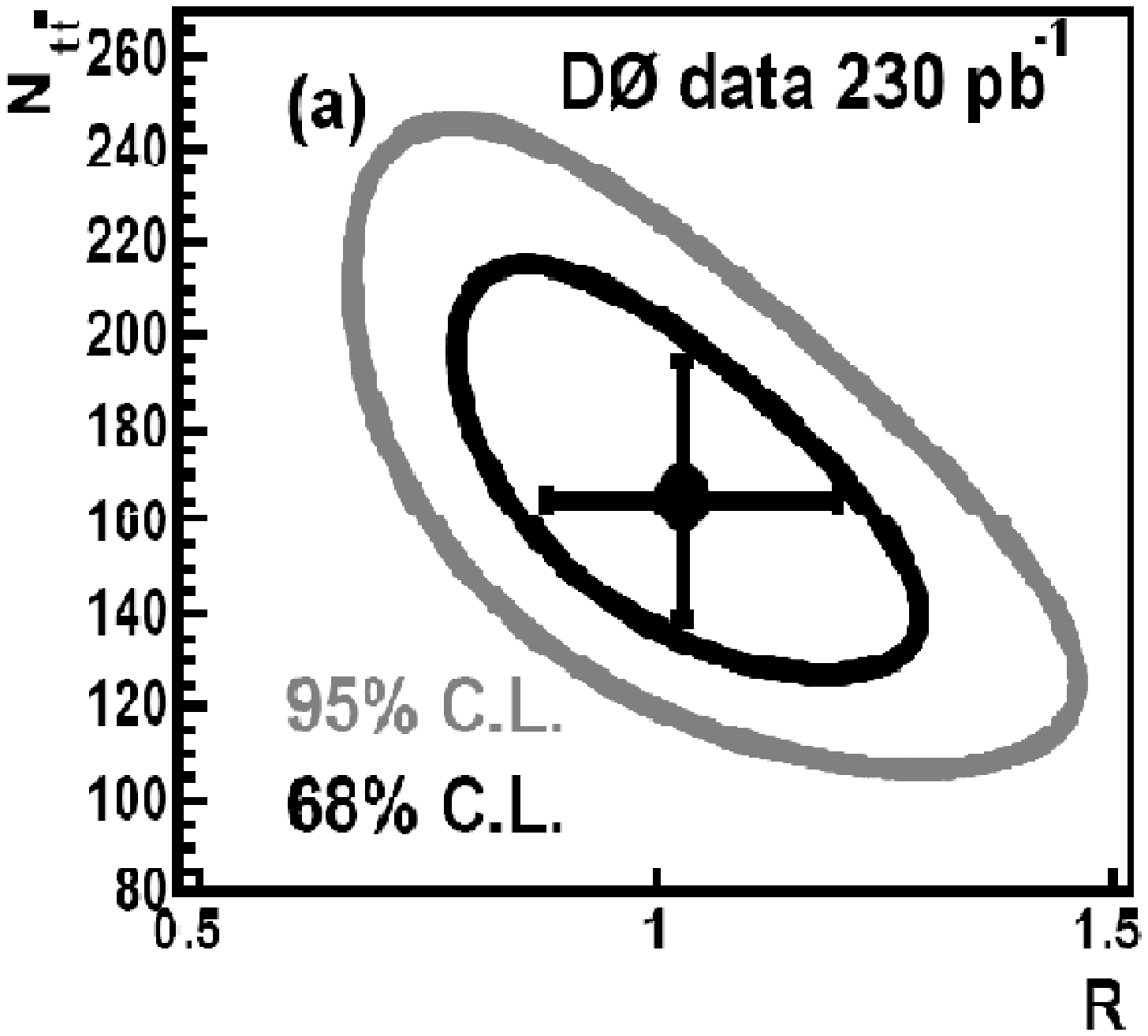,width=0.40\textwidth}
\end{center}
\vspace*{8pt}
\caption{Left: The upper plot shows the likelihood as a function of $R$ (inset) and its negative logarithm. The intersections of the horizontal line ln(L)=-0.5 with the likelihood define the statistical 1$\sigma$ errors on $R$. The lower plot shows 95\% (outer), 90\% (central), and 68\% (inner) C.L. bands for $R_{true}$ as a function of $R$ by CDF\protect\cite{cdf-run2-br}. Right: The 68\% and 
95\% statistical confidence contours in the ($R$,$N_{\ttbar\ }$) plane by 
\dzero\ \protect\cite{d0-run2-br}. The point indicates the best fit to the data.
}
\label{fig:topbranching} 
\end{figure*}

The expected event yield in each of the three tagged subsets of 
$\ell+$jets and dilepton samples is 
\begin{eqnarray}
N_i^{exp} = N_{inc}^{\ttbar\ } \epsilon_i(R) + N_i^{bkg}
\end{eqnarray}
where $N_{inc}^{\ttbar\ } = \sum_i(N_i^{obs} -N_i^{bkg})$ is an 
estimate of the inclusive number of \ttbar\ events in the sample. 
The analysis constructs a likelihood function using this information to 
get the best estimate of $R \epsilon_b$. 
The full likelihood is a product of independent likelihoods for the 
$\ell+$jets and dilepton samples. Each likelihood function is built as 
the product of the Poisson functions comparing $N_i^{obs}$ to $N_i^{exp}$
for each value of $i$, multiplied by Gaussian functions which 
incorporate systematic uncertainties 
in the event-tagging efficiencies and backgrounds, taking into account the 
correlations across different subsamples.  The resulting likelihood as a 
function of $R$ is shown in Fig. \ref{fig:topbranching}, along with 
the negative logarithm of the likelihood, which gives a central value of
$R=1.12^{+0.21}_{-0.19}(stat)^{+0.17}_{-0.13}(sys)$ most consistent 
with the observation. 
\paragraph{Measurement by \dzero\ :}
The \dzero\ experiment 
measures simultaneously the ratio of branching fractions $R$ together
with the \ttbar\ production cross section, using 230 pb$^{-1}$ of data 
and selecting a sample enriched in \ttbar\ events which are consistent with 
$\ell+$jets final states\cite{d0-run2-br}. The \ttbar\ enriched sample is divided into
$\ell+3$jets and $\ell+\geq 4$jets, which are further categorized on the 
basis of zero, one and two or more $b-$jets. 
The analysis fits simultaneously $R$ and the total number of \ttbar\ events 
($N_{\ttbar}$) in the zero, one and two-tag samples to the number of observed one-tag 
and two-tag samples, and, in the zero-tag events, to the shape of a discriminant 
variable ($D$) that exploits kinematic differences between the background and
the \ttbar\ signal. The event selection, background determination and 
secondary vertex $b-$tagging algorithm are essentially equivalent to that
of the corresponding cross section analyses described in the section 4.4.

In an analysis based on the SM, with $R \approx$ unity, the \ttbar\ event tagging 
probabilities are computed assuming that each of the signal events contain 
two $b-$jets. In the present analysis with $R$ not equal to unity, the
\ttbar\ event tagging 
probability becomes a function of $R$ since a \ttbar\ event  might have 
zero, one or two $b-$jets. To derive the event tagging probability as 
a function of $R$, the event tagging probability is determined for the three 
following scenarios (i) $t \bar{t} \to W^+bW^-\bar{b}$, (ii)  
$t \bar{t} \to W^+bW^-\bar{q_l}$, and (iii) $t \bar{t} \to W^+q_lW^-\bar{q_l}$,
where $q_l$ denotes either a $d-$ or $s-$ quark. The probabilities $P_{i-tag}$ 
to observe $i-tag=$ zero, one or $\geq$ two $b-$jets are computed 
separately for the three
types of \ttbar\ events, using the probabilities for each type of jet ($b$, $c$ or light
flavor jet) to be $b-$tagged. The probabilities $P_{i-tags}$ in the three scenarios are
then combined in the following way to obtain the \ttbar\ tagging probability as a function of $R$:

\begin{eqnarray}
 P_{i-tag}(\ttbar\ ) = R^2 P_{i-tag}(i)
 + 2R(1-R)P_{i-tag}(ii)
 + (1-R)^2P_{i-tag}(iii)
\end{eqnarray}
where the subscript $i-tag$ runs over zero, one and at least two tags.
Table~\ref{had-xsec} compares the observed number of events with the sum of the predicted 
backgrounds and the fitted number of \ttbar\ events as a function of the number 
of $b-$tags. To measure $R$ and $\sigma_{\ttbar\ }$, a binned maximum likelihood
fit is performed to the data. The values of $R$ and $\sigma_{\ttbar\ }$ that
maximize the total likelihood function are $R=1.03^{+0.19}_{-0.17}$(stat+sys) and
$\sigma_{\ttbar\ }=7.9^{+1.7}_{-1.5}$(stat+sys)$\pm$0.5 (lum) pb, respectively,
and in good agreement with the SM expectation. 
The result of the 2-dimensional fit is shown in Fig.~\ref{fig:topbranching}  in the plane 
($R$, $\sigma_{\ttbar\ }$), along with the 68\% and 95\%  
confidence level contours.
A preliminary simultaneous measurement of $\sigma_{\ttbar}$ and $R$ was performed
in 900 pb$^{-1}$ of data~\cite{900pbRmeas}.  This gives a measured value of 
$R=0.97^{+0.09}_{-0.08}(\rm stat+sys)$.


With their measured values of $R$, CDF and D\O\ have set a 
lower limit on $R$. Assuming  that non-$W$ decays of the 
top quark can be neglected, 
that only three-generations of fermions exist, and that the CKM matrix is unitary,
implying $R=|V_{tb}|^2$, they have extracted the CKM matrix element $|V_{tb}|$.
The results of this measurement are summarized in the Table~\ref{Btb-Vtb}. All the measurements of $R$ and $|V_{tb}|$ are consistent with SM expectations.
These results are presently limited by statistics and will profit from 
the upcoming larger data sets.

\begin{table}[ht]
\begin{center}
\tbl{Published measurements and lower limits of $R=BR(t \to Wb)/BR(t \to Wq)$ and  $|V_{tb}|$ from CDF\protect\cite{cdf-run2-br} and D\O\ \protect\cite{d0-run2-br} in Run II.}
{\begin{tabular}{lcc}\toprule
& CDF ($\int \cal{L}$dt=160 $\rm pb^{-1}$) & D\O\  ($\int \cal{L}$dt=230 $\rm pb^{-1}$) \\ 
\colrule
R & $1.12^{+0.27}_{-0.23}$ & $1.03^{+0.19}_{-0.17}$\\
R (95\% C.L.) & $>$0.61 & $>$0.61\\
R (68\% C.L.) & --  & $>$0.78\\
$|V_{tb}|$ (95\% C.L.) & $>$0.78 & $>$0.78\\
$|V_{tb}|$ (68\% C.L.) & -- & $>$0.88\\
\botrule
\end{tabular}
\label{Btb-Vtb}}
\end{center}
\end{table}

\paragraph{First direct determination of $|V_{tb}|$ by \dzero\ :}
Single top quark cross section measurement provides a direct determination
of the CKM matrix element $|V_{tb}|$ since the cross section is proportional
to the square of this quantity ($\sigma \propto |V_{tb}|^2$). 
A value inconsistent with the SM expectation $|V_{tb}| \simeq$1 would be 
a signature for new physics such as a fourth quark family. The \dzero\ experiment has
performed the first direct measurement of $|V_{tb}|$ based on the 
single top quark cross section measurement derived from decision trees analysis
(discussed in section 5.2) using about 0.9 fb$^{-1}$ of data\cite{STD0evidencePRL}. This result makes
no assumptions on the unitarity of CKM matrix or the number of families,
but it does require a few assumptions. First, it is assumed that the single 
top quark production meachanism only involves interaction with a $W$ boson,
and not from one of the various beyond SM scenarios that include extra scalar
and vector bosons or FCNC interactions. The second assumption made is that
$|V_{td}|^2 + |V_{ts}|^2 \ll |V_{tb}|^2$, which is experimentally supported by
the $BR(t \to Wb)/BR(t \to Wq)$ measurements done on \ttbar\ events. This
requirement implies that $BR(t \to Wb) \simeq$100\% and that single top quark
production is completely dominated by the $tbW$ production. Finally,
it is assumed that the $tbW$ interaction is CP-conserving and of the 
$V-A$ type, but it is allowed to have an anomalous strength. 
Two measurements have been performed, one for the strength of the $V-A$
coupling $|V_{tb}f_1^L|$ in the $Wtb$ vertex, where $f_1^L$ is an arbitrary
left-handed form factor, with no requirement that it be less than one;
and one assuming $f_1^L=$1 resulting in $|V_{tb}|$ being restricted to between
0 and 1. The limits calcualted using Baysian approach are: 
$|V_{tb}f_1^L|=$1.3$\pm$0.2 and 0.68$<|V_{tb}|<$1 at the 95\% C.L.\cite{STD0evidencePRL}.

%% file: proper/charge.tex
Many of the currently measured properties of the top quark are still 
only poorly known, and the indirect constraints set by precision 
electroweak data leave plenty of room for new physics. In particular, 
its electric charge ($q_{t}$), one of the most fundamental quantities
characterizing a particle, has not been directly measured so far. 
It still remains not only to confirm that the discovered quark has a 
charge $+\frac{2}{3}$ as assigned by the SM, but also to measure
the strength of its electromagnetic (EM) coupling to rule
out anomalous contributions to its EM interactions.
In the top quark analyses of the CDF and D\O\ experiments, the correlations of 
the $b$-quarks and the $W$ bosons in 
$p \bar{p} \to t \bar{t}\to W^+W^- b \bar{b}$ 
are not uniquely determined which results in a two-fold ambiguity in the 
pairing of $W$ bosons and $b$-quarks, and, hence, in the electric
charge assignment of the top quark. In addition to the SM 
assignment $t \to W^+ b$, it is conceivable that the `t-quark' is 
an exotic quark $Q$ with charge q$=-\frac{4}{3}$ which decays 
via $Q \to W^- b$. This alternative interpretation is consistent 
with current precision electroweak data. It is possible to 
fit $Z\to \ell^+ \ell^-$ and $Z\to b \bar{b}$ data assuming a top
quark mass of $m_{t}=$270 GeV, provided that the right-handed
$b$-quark mixes with the isospin $+\frac{1}{2}$ component of an
exotic doublet of charge $-\frac{1}{3}$ and $-\frac{4}{3}$
quarks, $(Q_1~,Q_4)_{R}$\cite{exotic-top-paper}.
In such a scenario, the particle 
we have been exploring may in fact be the $-\frac{4}{3}$ charge 
quark and the top quark, with $m_{t}$=270~GeV, would have so 
far escaped detection at the Tevatron. \\

Specific $q_{t}$ measurements are therefore required to rule out or 
confirm one of the hypotheses. In order to determine $q_{t}$, 
one can study the charge of its decay products or investigate 
photon radiation in $t \bar{t}$ events. 
The latter method\cite{qtop-photon-rad} consists in cross-section 
measurements of $t \bar{t}$ pairs where a photon is radiated, 
either at the production ($p \bar{p} \to t \bar{t} \gamma$) or the 
decay ($p \bar{p} \to t \bar{t}$, $t \to Wb \gamma$); 
the radiation being dependent on the value of $q_{t}$. 
Measurement of $q_{t}$ using radiative $t \bar{t}$ processes is 
hopeless at the Tevatron\cite{qtop-photon-rad} - even with 20 $fb^{-1}$ 
of data, a charge $-\frac{4}{3}$ top quark can be excluded 
at $\approx$95\% confidence level. However, at the LHC, with 10 fb$^{-1}$, 
one can hope to determine $q_{t}$ with a precision of $\approx$10\%. 
The first method - measuring $q_{t}$ by reconstructing the charges 
of its decay products (the final-state $b-$jets and the W bosons) 
for semi-leptonic decays - seems feasible at the Tevatron. 
The $W$ boson charge in leptonic decays is accurately given by the lepton 
coming from it and the $b$-jet charge can be determined from a 
measurement of charges associated with the tracks in the jet. 
It should be noted that since both $t$ and $\bar{t}$ quarks are 
present in every event, the analysis is 
only sensitive to the absolute value of $q_{t}$. \\

The \dzero\ experiment 
has performed the first determination\cite{d0-topcharge-paper} 
of $q_{t}$ with 
$\sim$370~pb$^{-1}$ of Run II data on a double-tagged semi-leptonic 
($p \bar{p} \to t \bar{t}\to l^{\pm} \nu jj' b \bar{b}$) sample of 
$t \bar{t}$ candidate events. The first step in the measurement 
involves selecting a pure sample of $t \bar{t}$ events in data in the
$\ell+$jets channel. Events with an 
isolated high $p_T$ ($>20~GeV$) $e$ or $\mu$ accompanied by 
four or more high $p_T$ ($>15~GeV$) jets are considered for the analysis.
Events with $W$ bosons are selected by requiring \met\ $>20$ GeV. To remove the multijet background, \met\ is required to 
be acollinear with the lepton direction in the transverse plane. 
The purity of the sample is significantly enhanced by requiring at least 
two jets in the event to have a secondary vertex $b$-tag. In the resulting sample of 21 candidate events, $Wb \bar{b}$ 
production is the largest background and represents $\approx$5\% of the sample. \\

Each selected $t \bar{t}$ event has two ``legs'', 
one with a leptonically decaying $W_l$ ($t \to Wb \to l\nu b_l$) and one 
with a hadronically decaying $W_h$ ($t \to Wb \to q \bar{q'}b_h$).
The second step in the analysis consists in assigning the correct jets and 
leptons to the correct ``leg'' of the event, uniquely specifying which
$b$-jet comes from the same top (or anti-top) quark as the lepton. 
A constrained kinematic fit is performed for this purpose. The four highest
$p_T$ jets can be assigned to the set of final state quarks according to
many permutations and there are at least two ways to assign the $b-$jets
to $b_l$ and $b_h$. For each permutation, the measured four vectors
of the jets and leptons are fitted to the $t \bar{t}$ event hypothesis and 
the $b_l$ and $b_h$ jets are identified by selecting the permutation with the 
highest probability of arising from a $t \bar{t}$ event. For 16 of the 21 selected events, the kinematic fit converges leading to 
correct assignment, thus providing 32 measurements of 
$q_{t}$ since top quark charge is measured twice in each event.
In each $t \bar{t}$ event, the observable $|q_{t}|$ is computed, which is the sum 
of the lepton charge from the $W$-boson decay and the charge of $b_l$.
The $b-$jet charge is determined by taking a $p_T$-weighted sum of the 
charges of the  tracks with $p_T$ $>0.5~GeV$ contained within a cone 
of $\Delta R=$0.5 around the $b-$jet's axis and thus one can distinguish 
between the $b$ and $\bar{b}$ on a statistical basis. 
The performance of this jet charge algorithm is calibrated using 
an independent dijet data sample enriched with semileptonic $b$ quark decays. 
In the calibration sample, the charge of the $b-$jets is derived from the 
lepton charge, accounting for $c-$jet contamination and $B$ 
oscillation corrections. The performance of the algorithm is 
also calculated as a function of the $p_T$ and $\eta$ of the jet, 
so that kinematic differences between the control sample and 
the $t \bar{t}$ are accounted for. \\

\begin{figure*}[!h!tbp]
\begin{center}
\epsfig{figure=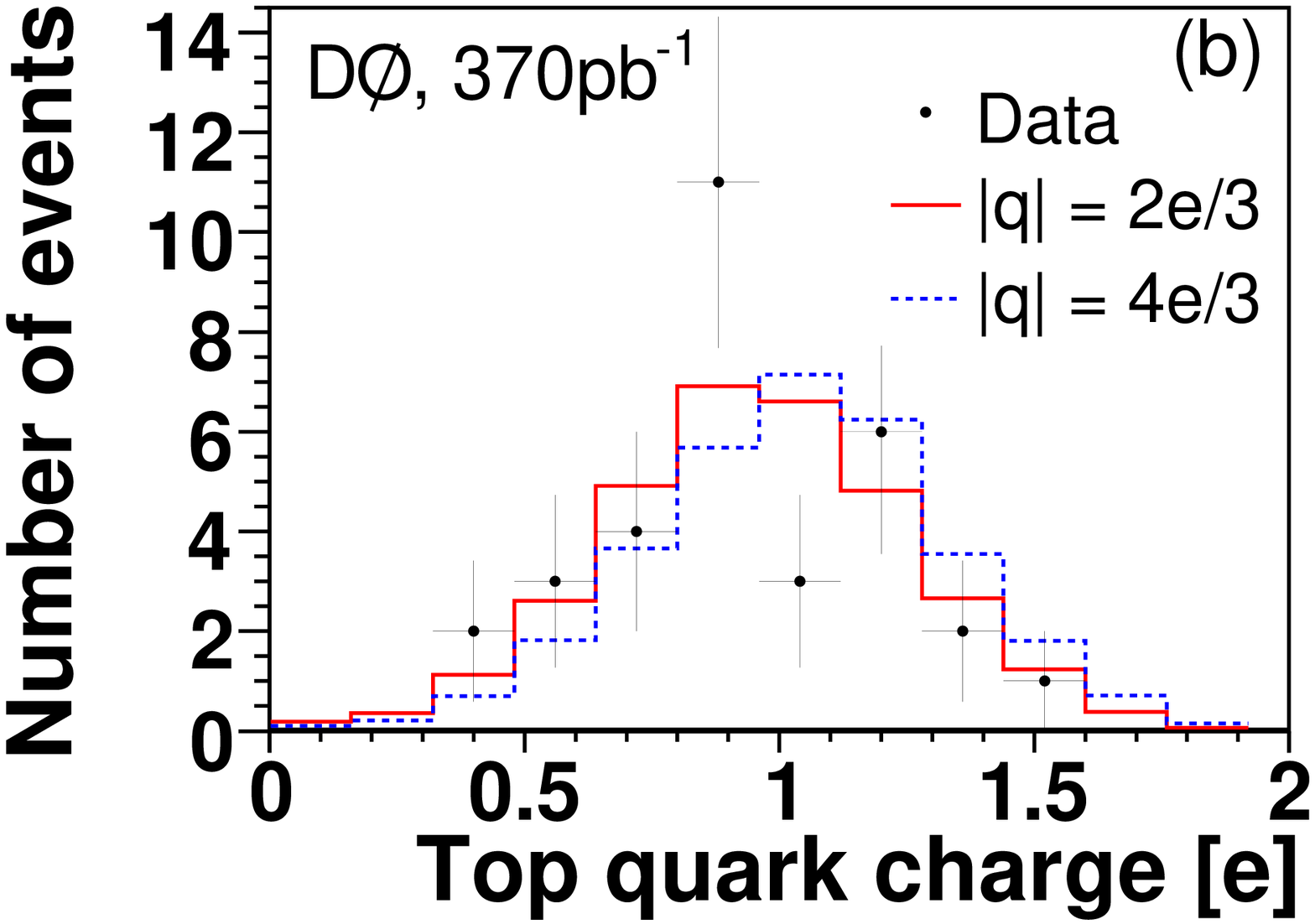,width=0.40\textwidth}
\epsfig{figure=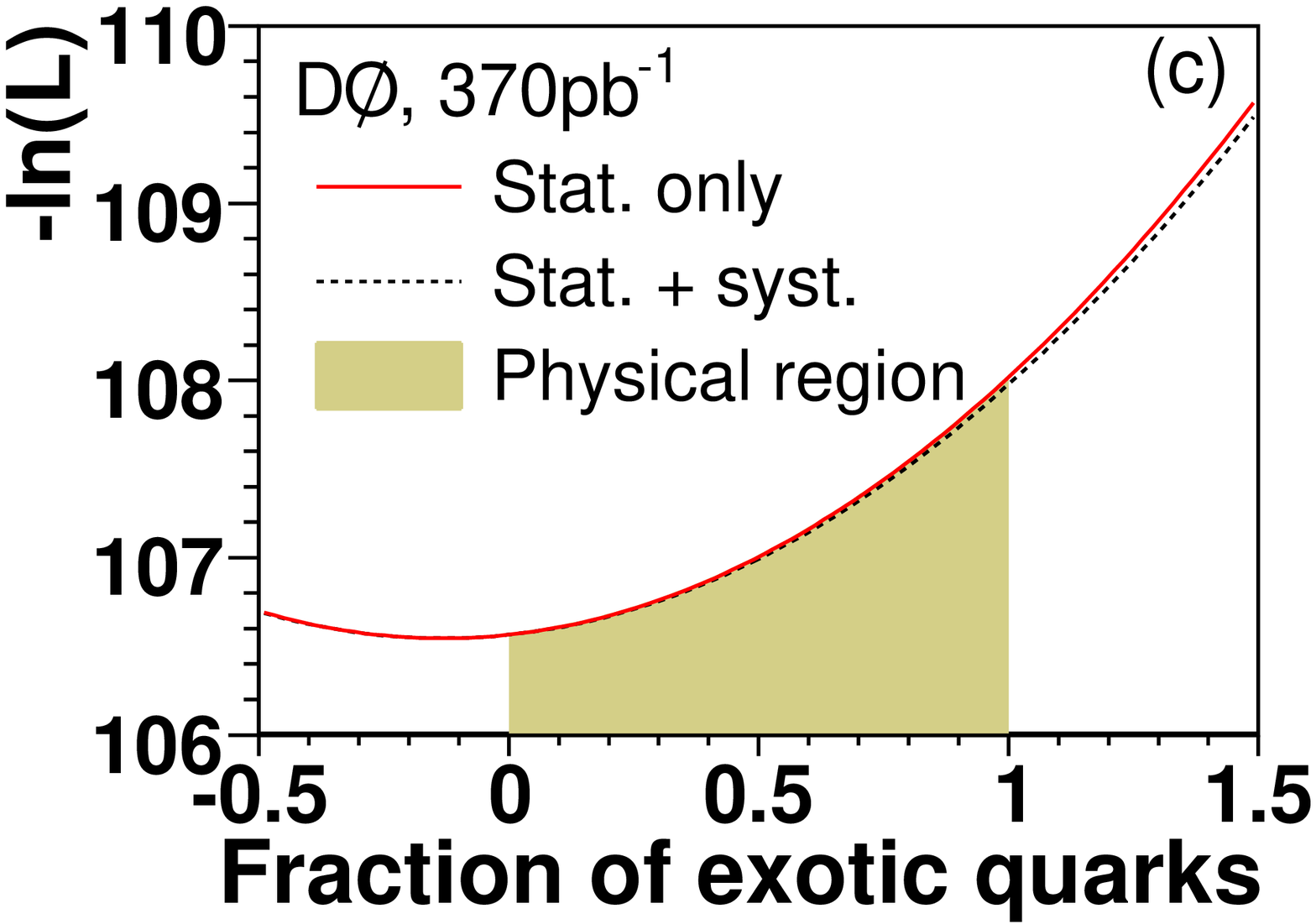,width=0.40\textwidth}
\end{center}
\vspace*{8pt}
\caption{The 32 measured values of the top quark charge 
(binned) compared to the Standard Model and 
exotic scenario templates (left)\protect\cite{d0-topcharge-paper}. Likelihood fit of the fraction of 
exotic quark pairs in the selected data sample (right)\protect\cite{d0-topcharge-paper}.
}
\label{fig:topchargedata} 
\end{figure*}

The subsequent step is to use the expected shapes of jet charge for $b$-jet 
and $\bar{b}$-jet
in data to derive the expected shape of $|q|$ for the SM ($q_{t}=+$2/3)
and the exotic ($q_{t}=-$4/3) scenario. The resulting SM and exotic
templates, normalized to unity, are used as probability density functions, 
$p^{\rm sm}$  and $p^{\rm ex}$, respectively, in the C.L. calculation.
Figure.~\ref{fig:topchargedata} compares the charge distribution of the 
top quark candidates reconstructed in data with +2/3 and -4/3 charge 
hypotheses.


A likelihood ratio defined as 
\begin{equation}
  \Lambda = \frac{\prod_i p^{\rm sm}(q_i)}{\prod_i p^{\rm ex}(q_i)}
\end{equation}
is then computed to determine the most probable model where the numerator
(denominator) measures the likelihood of the observed set of charges 
$q_i$ arising from a SM top quark (exotic quark). The subscript $i$ runs over
all 32 available measurements. The value of the $\Lambda^{\rm data}$ measured 
in data is compared with the expected $\Lambda^{\rm sm}$ and 
$\Lambda^{\rm ex}$ distributions derived by performing ensemble tests 
using the SM and exotic 
scenarios respectively. 
The observed set of charges is found to agree well with those of 
a SM top quark. For the exotic heavy quark hypothesis, only 7.8\% of the
pseudo-experiments give a higher $\Lambda$ ratio than the one measured in 
data. Therefore,
\dzero\ yields a $p-$value, corresponding to the probability of consistency
with the exotic model, of 7.8\%.
The top quark is indeed consistent with 
being the SM $|q_{t}|=$2/3 quark. Figure~\ref{fig:topchargedata}
shows the fraction of exotic quark pairs ($\rho$) in the data
obtained by performing an unbinned maximum likelihood fit to the 
observed set of $q_i$ which yields $0 \leq \rho <$0.52 at the 68\% $C.L$ and 
$0 \leq \rho <$0.80 at the 90\% $C.L$.

%% file: bsm/bsm.tex
Several alternative models predict substantially different production
and decay mechanisms for the top quark than were described in
Section~\ref{sec:prodDecay}.  For instance, the possibility
to detect a heavy resonance which decays to \ttbar\ is testable
using elements of the techniques for kinematic reconstruction.  The
final states are sensitive to alternative decay chains from the top quark,
such as thru a charged Higgs or $W$' boson.  This section will review
the experimental efforts in these directions.

{\subsection{\ttbar\ resonance search}
\label{sec:resonance}}
\input{bsm/resonance.tex}

\subsection{Charged Higgs decays}
\input{bsm/chHiggs.tex}

\subsection{$W^{\prime}$ in Top quark decays}
\input{bsm/Wprime.tex}

%% file: bsm/resonance.tex
It may be that the top quark is a special member of the fermion family.  Its
large mass has provoked suspicion that it may have a unique and significant role
to play in electroweak symmetry breaking.  
Topcolor models\cite{toptechnicolor1,toptechnicolor2} 
suggest that there are new strong interactions which 
can generate the large top quark mass while leaving other quarks light.
This interaction results in a massive \ttbar\ condensate, sometimes termed
$Z'$.  Technicolor\cite{technicolor}, on the
other hand, posits interactions at the electroweak scale which provide
electroweak symmetry breaking without the need for a scalar Higgs field.  
In the extended models, this gives masses to all
fermions.  However, the generated value of $m_t$ is too small.  
The top-color assisted technicolor\cite{toptechnicolor1,toptechnicolor2} model 
combines these approaches to predict the 
correct $m_t$.  Variants of this approach predict a $Z'$ which 
couples preferentially to third
generation fermions and has no couplings to leptons.  

\subsubsection{General Methods}

Efforts to look for a \ttbar\ resonance have employed the $\ell+$jets final
state.  In such events, a full kinematic reconstruction is performed as in the mass
analyses described in Section \ref{sec:ljmtop}, with the exception that $m_t$  
is either held fixed or within a window determined by uncertainty on the mass. 
All jet combinations  and neutrino solutions are tried. The chosen
combination is one that minimizes the $\chi^2$, 
given the known errors and measured parameters of the event.
The combinatorics of jet assignment mean that this choice is not always the
correct one, although $b-$tagging can enhance the probability.
Generally, a cut on the $\chi^2$ is applied to further reduce backgrounds.
The solution for each event permits the calculation of
the apparent \ttbar\ invariant mass, $M_{t\bar{t}}$, from that event.

A narrow resonance with $\Gamma_{Z'}=0.012 M_{Z'}$ is the standard assumption
for the signal.  Generally, this width is substantially less than the resolution
inherent in the experimental measurements, so natural widths up to a few percent
may still be accomodated.  However, this assumption does mean that any results are
not strictly applicable to some models, such as Kaluza-Klein scenarios.
The primary backgrounds to
the search are standard \ttbar\ production, $W+$jets and QCD multijet production.
Single top quark and diboson events give small but non-negligible contributions.  
Evidence of resonant production is sought
by fitting the expected $M_{\ttbar}$ resonance plus the expected \ttbar\ and backgrounds
to the observed spectrum in data.  A limit on any anomalous production cross section 
results on a limit on the mass of the resonant state.

\subsubsection{\dzero\ Searches}

In 2004, \dzero\ published a search for \ttbar\ resonance in Run I data\cite{d0r1ttreson}.
A sample of 41 $\ell+$jets events, 4 with soft $\mu-$tag selection. were used
for the analysis.
The resonance signal was modeled with {\sc Pythia} in a mass range
of 400 GeV $<M_{Z'}<$ 850 GeV.  The {\sc cteq3m} $pdf$ was used.  
The \ttbar\ background was simulated using {\sc Herwig}.  
The $W+$jets background was modeled with {\sc Vecbos}, and
{\sc Herwig} was used for the parton shower simulation.  
The instrumental background from multijet
production was derived from a data sample in which actual leptons have been 
rejected. No evidence for 
resonant production was observed.  A Bayesian fit~\cite{d0limit} was performed which 
considered three components separately: signal ($Z'\rightarrow\ttbar$), standard
\ttbar\ production, and $W+jets$ and multijet events.  The latter element had $W$ and
QCD fractions set from the template mass analysis.  
The resulting limit is $M_{Z'} > 560$ GeV$/c^2$  at 95\% C.L..

\begin{figure*}[!h!tbp]
\begin{center}
\epsfig{figure=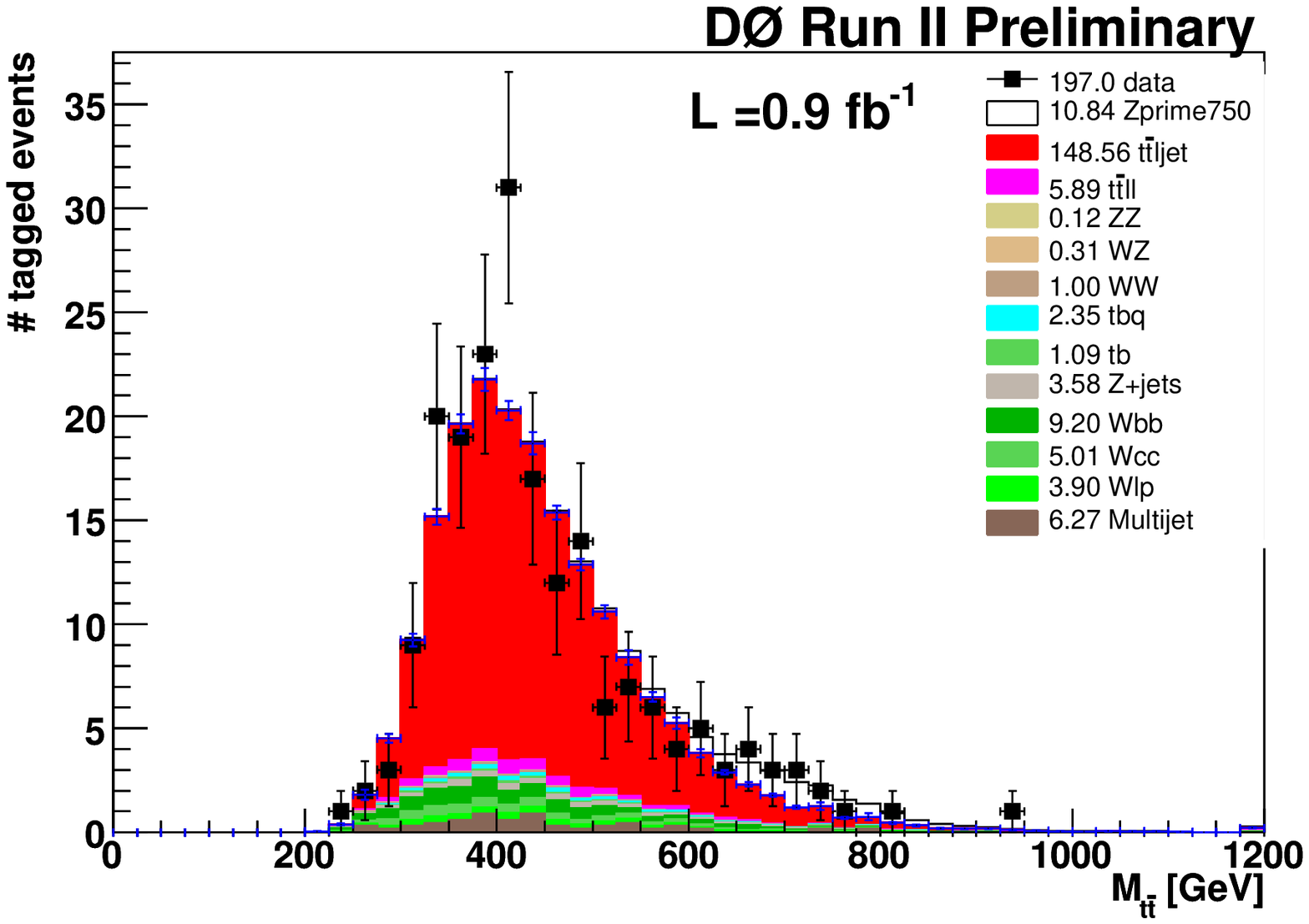,width=0.45\textwidth}
\epsfig{figure=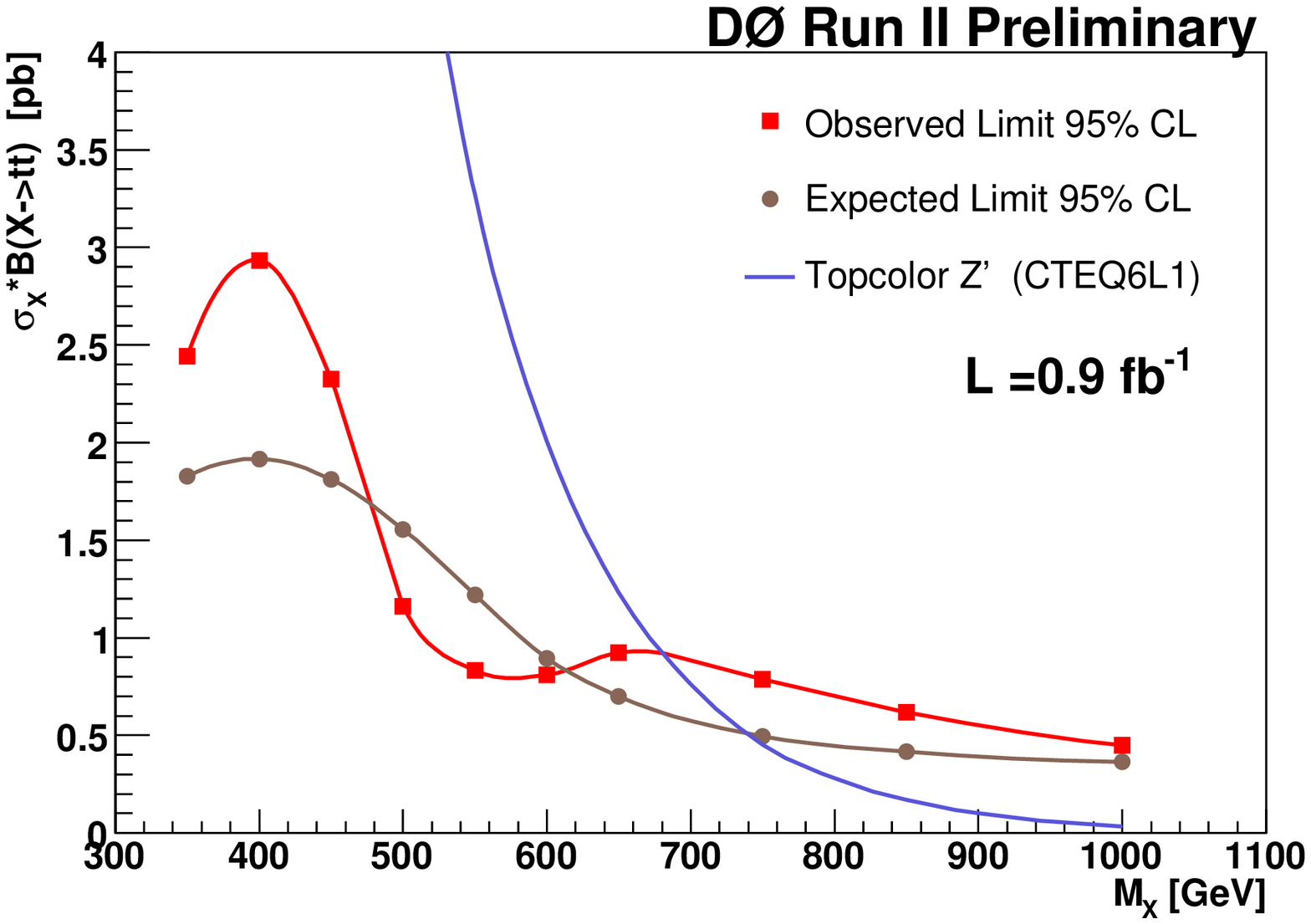,width=0.45\textwidth}
\end{center}
\vspace*{8pt}
\caption{Distribution of \ttbar\ invariant mass, $M_{\ttbar}$, from \dzero\ Run II
data (left) \protect\cite{d0res900}.  Expected and observed limits, with comparison
to the $Z'$ estimation is also shown (right).}
\label{fig:d0ttRes}
\end{figure*}

Two preliminary studies have been executed in Run II using template methods
for kinematic reconstruction.  A mass range 350 GeV$/c^2<M_{Z'}<$900 GeV$/c^2$ was considered
in both analyses.  The first search used 370 pb$^{-1}$
of $\ell+$jets events tagged with a secondary vertex tagger.  Signal Monte Carlo
samples were generated using {\sc Pythia}.  Studies indicated that the choice
of jet-to-parton configuration was correct about 65\% of the time after the 
$\chi^2$ cut.  The \ttbar\ and $W+$jets backgrounds
were generated using {\sc Alpgen} plus {\sc Pythia}.  The QCD background was
estimated via matrix method in the data.  The data yielded 108 events which
showed no evidence of a resonance.  A Bayesian approach was used to extract a
limit.  Kinematic shape uncertainties arose from jet energy calibration and efficiency
and the knowledge of the kinematic dependence of the $b-$tag and mis-tag rates.
Normalization uncertainties came from theoretical uncertainty in the \ttbar\
cross section as well as the integrated luminosity.  Correlations among
uncertainties were included in the fit.  A limit of 
$M_{Z'}<680$ GeV$/c^2$ was obtained at 95\% C.L.~\cite{d0ResR2}.
A further study has been performed with 197 $\ell+$jets events obtained from
900 pb$^{-1}$ of data.  These events were tagged with a neural network
tagger.  In this sample, 154 events were expected from standard \ttbar\
process.  Much of this analysis is similar to the 370 pb$^{-1}$ analysis.
A limit of $M_{Z'}>680$ GeV$/c^2$ was obtained~\cite{d0res900}.  Results are
shown in Fig. ~\ref{fig:d0ttRes}.

\subsubsection{CDF Searches}

An initial search for the $t\overline t$  resonance was conducted by the CDF collaboration using
Run I data~\cite{cdfr1ttreson}. 
One of the leading three jets in each event were required to be $b-$tagged with a
secondary vertex. {\sc Vecbos} was used to model $W+$jets events. {\sc Pythia} was used  
to model the signal as $Z'\rightarrow \ttbar$.  In order to improve the 
reconstructed mass resolution of the $\ttbar$ system,
the requirement on the computed top quark masses was relaxed to fall in the range 150 GeV to 
200 GeV.  This allowed a more efficient acceptance of combinations with 
correct jet-parton assignments.  Twenty events were removed out of the initial sample of 
83 because they yielded a poor minimum $\chi^2$.
No evidence of a resonance was observed in the data.
A binned likelihood fit of the data to signal and
background was  performed to extract a limit on the resonance production rate.
Masses from 400 GeV to one TeV were scanned.  The primary sources of systematic uncertainty 
were from signal and background shapes, and jet-related systematics such as energy scale 
and gluon radiation.  
Uncertainties were incorporated into the limit calculation as Gaussian uncertainties.
The 95\% C.L. upper limit for $\Gamma_{Z'} = 0.012 M_{Z'}$ is $M_{Z'} >480$ GeV.

\begin{figure*}[!h!tbp]
\begin{center}
\epsfig{figure=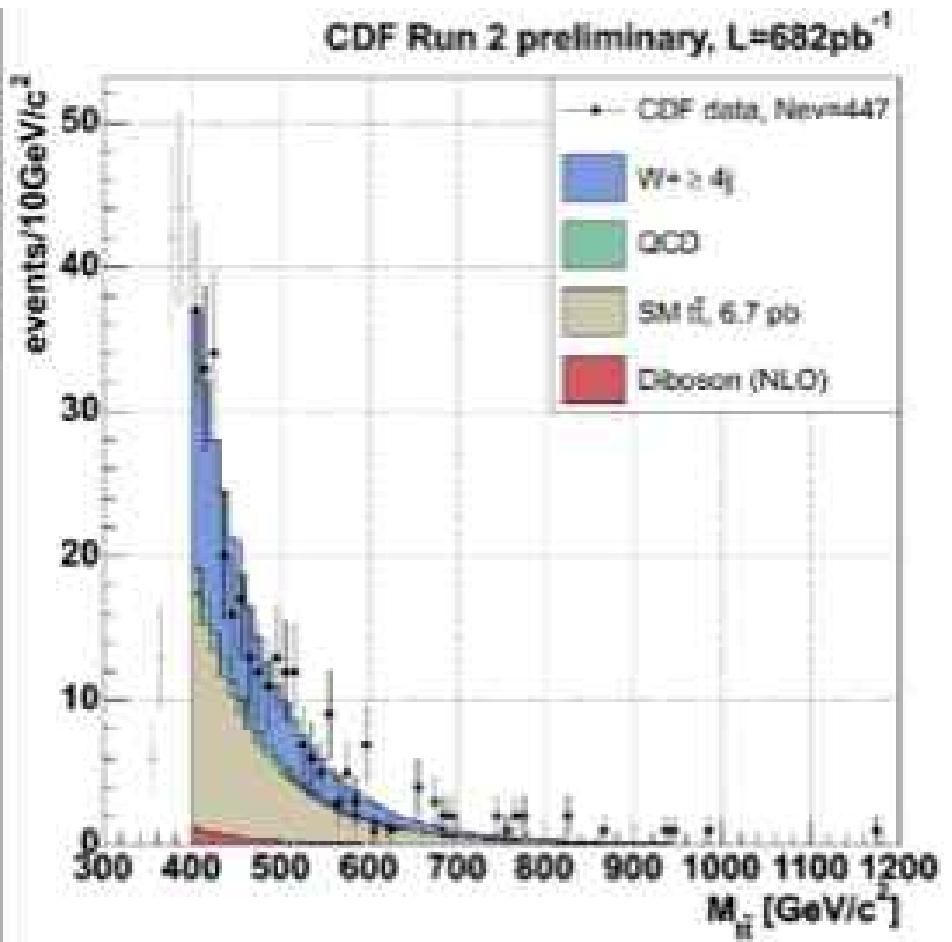,width=0.45\textwidth}
\epsfig{figure=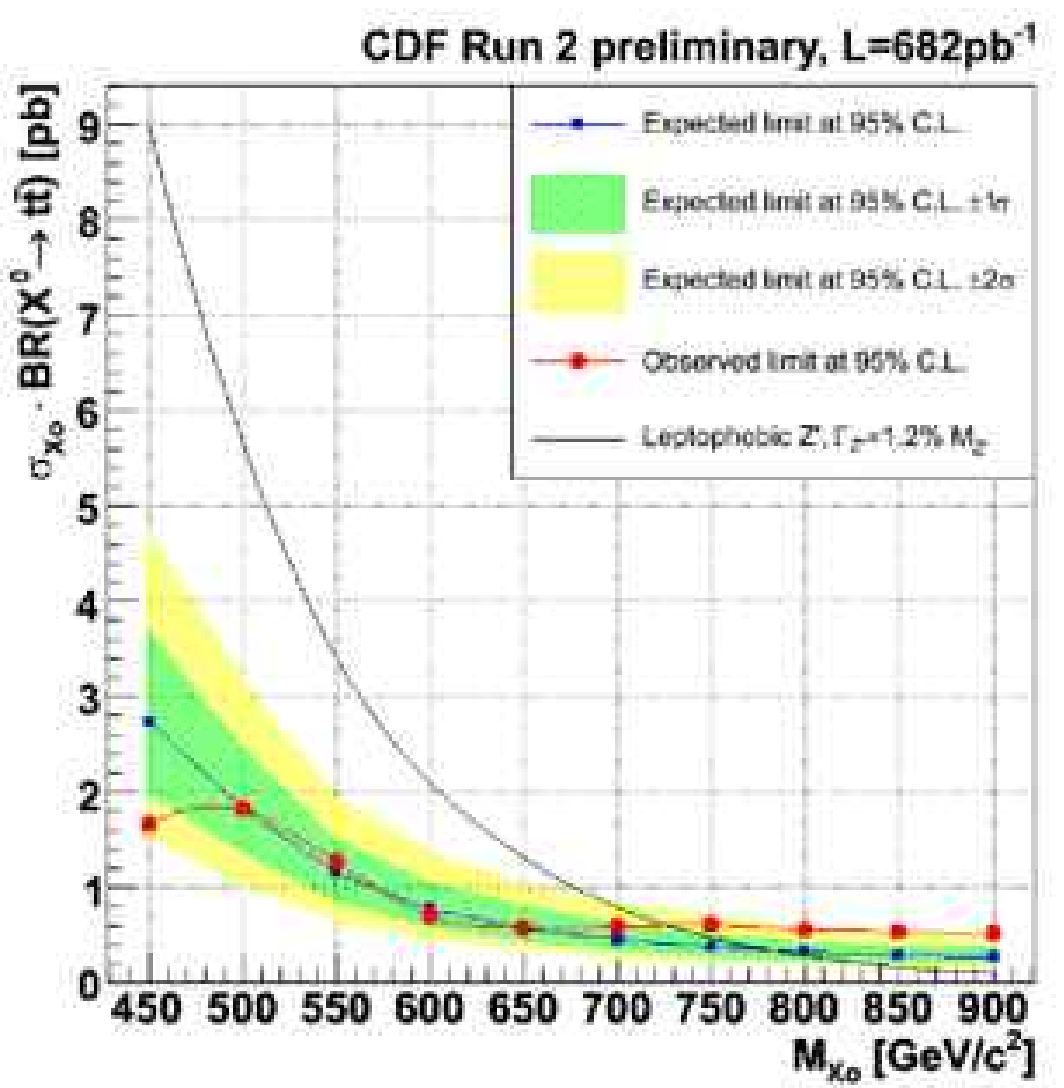,width=0.45\textwidth}
\end{center}
\vspace*{8pt}
\caption{Distribution of \ttbar\ invariant mass, $M_{\ttbar}$, from CDF Run II
data (left) \protect\cite{cdfResR2}.  Expected and observed limits, with comparison
to the $Z'$ estimation is also shown (right).}
\label{fig:cdfttRes}
\end{figure*}

CDF has improved upon this result by looking at 680 pb$^{-1}$ of Run II data~\cite{cdfResR2}.
The signal and standard \ttbar\ production were modeled with {\sc Pythia}.  Multijet background
came from data and $W+$jets events were modeled with {\sc Alpgen} plus {\sc Herwig}.  
The top quark and boson pair processes were normalized based on the theoretical cross
sections and the estimated luminosity to be 199 and 14 events, respectively.  The QCD and $W+$jets backgrounds were then taken as
the remaining contribution to the event yield observed in data: 450 events.  The QCD/$W$ ratio
was fixed at 10\% as taken from the cross section measurement~\cite{cdfljtopo194}.
Event reconstruction was done via matrix element fit using resolution functions for the jets.
All possible jet configurations were summed over in this approach, and a
probability distribution vs. $m_{\ttbar}$ was obtained.  Simulations showed that the
mean of this distribution was best correlated with the actual resonance mass.  The
mass resolution was set by the jet energy calibration and the unknown $p_z$ of the neutrino.
Incorrect jet assignments also produced a significant low mass tail.  No evidence for
an excess over standard expectations was observed.  The presence of a resonance would be
detected by a likelihood calculation based on the distributions $<m_{\ttbar}>$ for signal
of varying masses and the backgrounds.  Systematic uncertainties were
extracted by varying the jet energy scale, initial and final state radiation, and the $Q^2$
for $W+$jets production by $\pm 1\sigma$ and generating these mass distributions for use in the likelihood fit.  A limit of $M_{Z'}<725$ GeV$/c^2$ at 95\% C.L. was extracted from the data
(see Fig. ~\ref{fig:cdfttRes}).
More recently, CDF has analyzed 955 pb$^{-1}$ of data using a template fitting method.
A limit of $M_{Z'}<720$ GeV$/c^2$ was obtained ~\cite{cdfr2resol955}.

%% file: bsm/chHiggs.tex
\subsubsection{Two Higgs Doublet Models}

In the electroweak model, the generation of mass is governed by one scalar Higgs
doublet manifesting in one observable particle, the Higgs boson.  The 
peculiar problems associated with the Higgs mechanism have led to several efforts
to produce models with extended Higgs sectors.  The simplest configuration 
has two Higgs doublets and this is included in several scenarios, including supersymmetry (SUSY).  
Two types of model exist.  Type I models have only one doublet coupling to fermions.
Type II models, however, posit one doublet which couples to up-type fermions including
charged leptons, and
another doublet that couples only to down-type fermions.  In the 
MSSM~\cite{mssmModels}, the Type II case holds.
Five physical particles are predicted: two neutral scalars ($h^0$, $H^0$), a neutral
pseudoscalar ($A^0$), and two charged scalars ($H^{\pm}$).  

Evidence of the existence of the charged Higgs has been sought in $e^+e^-$ collisions\cite{LEP_Higgs}.
Indirect searches from CLEO provide the constraint $m_H > [244 + 63/(\tan\beta)^{1/3}]$ 
GeV at the 95\% C.L.\cite{cleoChHiggs}, where $\tan\beta$ is the ratio of vacuum expectation values of the two Higgs doublets.  Observed $\tau\rightarrow \nu_{\tau} K$ 
and $K\rightarrow \nu_l (\gamma)$ branching ratios provide a limit of 0.21 GeV$^{-1} 
> \tan\beta/m_H$ at 90 \% C.L.\cite{towers}.  These limits are stricter than those quoted for
the Tevatron searches below, but are more theory dependent.  The direct search 
for $H^{\pm}$ at LEP, however, provides a hard limit of $m_H > 78.6$ GeV\cite{lep-chH-Limit}. The Tevatron searches consider potential physics beyond this limit.

\subsubsection{Production and Decay}

Single charged Higgs production would have negligible cross section at the Tevatron.
Weak pair production of charged Higgses is calculated to have a cross section of
~0.1 pb.  However, a large enhancement occurs when they are generated through the production 
and decay of top quark pairs.  Then the Higgs pair production rate is potentially as high 
as the 6.7 pb \ttbar\ cross section.  Additionally,
the top quark decay produces associated $b-$jets which allows a further background reduction relative
to direct diboson production.  

\begin{figure}[thb]
\begin{center}
\epsfig{figure=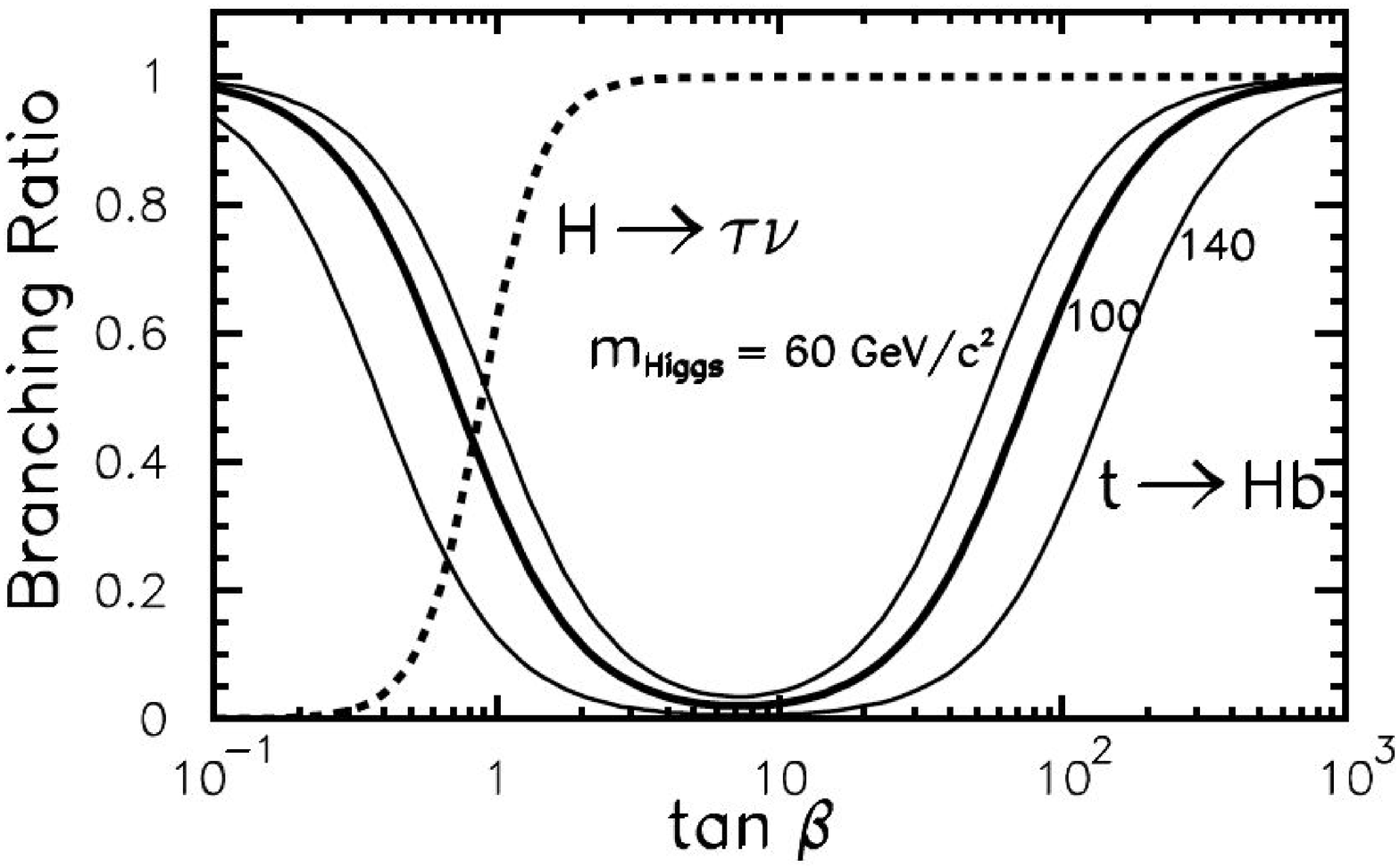,width=6.3cm}
\epsfig{figure=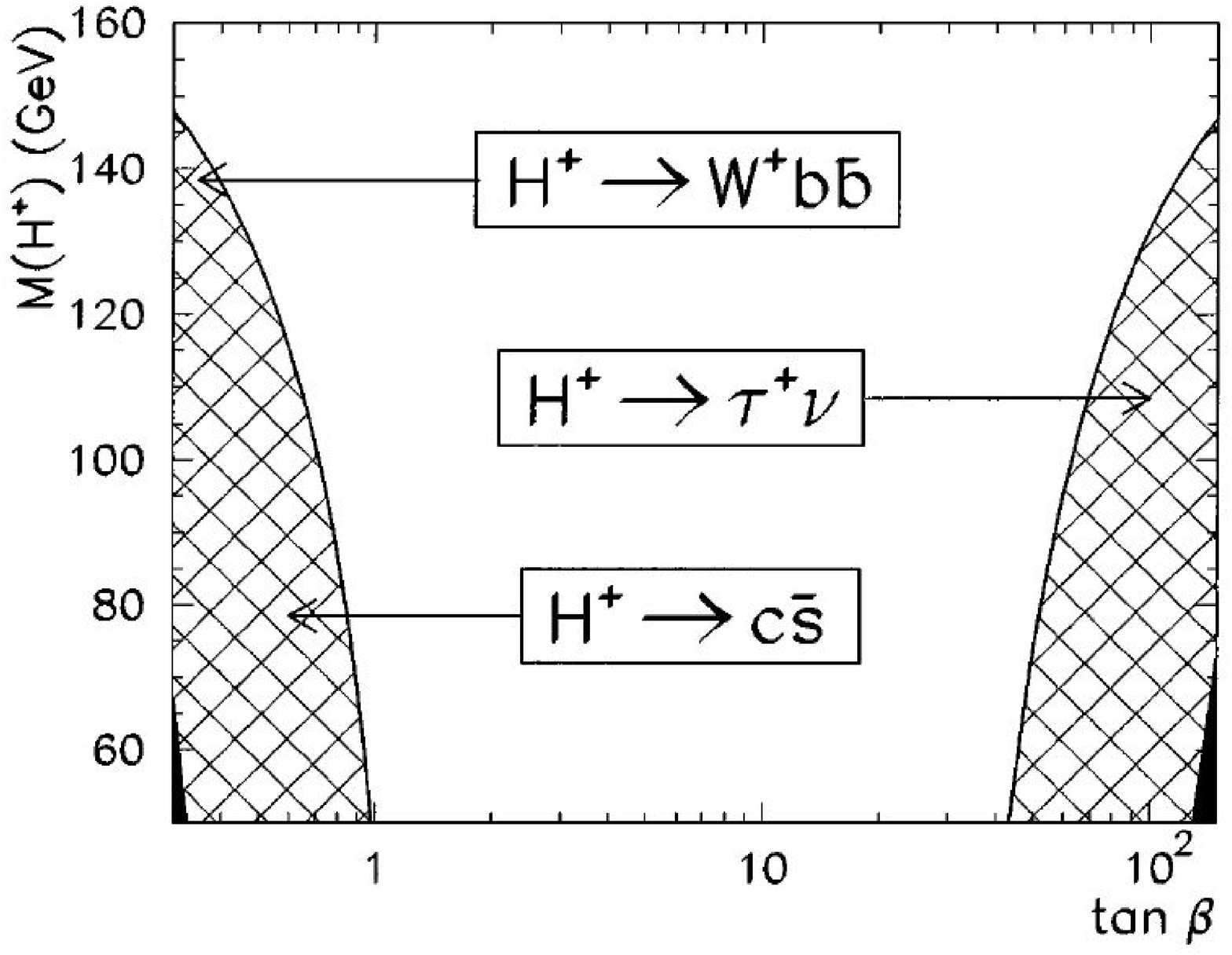,width=5cm}
\caption{At left is shown the
dependence of $BR(H\rightarrow\tau\nu)$ and $BR(t\rightarrow Hb)$ on
$\tan\beta$.  Coverage in the $m_H$ vs $\tan\beta$ plane is
shown at right when
$BR(t\rightarrow Hb) > 0.5$ (cross-hatched) and 0.9 (solid).  Regions where
specific final states dominate are indicated.}
\label{fig:tanB}
\end{center}
\end{figure}

If the $H^{\pm}$ are lighter than the top quark, then the decay $t\rightarrow Hb$ 
can compete with the standard $t\rightarrow Wb$ decay.  In particular, $BR(t\rightarrow Hb)$
$=m_t^2 \cot^2\beta + m_b^2 \tan^2 \beta + 4 m_t^2 m_b^2$
is significant when $\tan\beta$ is either very large or very small (see Fig.~\ref{fig:tanB}).  
Figure ~\ref{fig:tanB} also indicates that this branching ratio declines as
$m_H$ increases.  There are then three top quark decay modes to 
consider: $\ttbar\rightarrow HH\bbbar$, $HW\bbbar$, and $WW\bbbar$.
The final states are dictated by the $H$ and $W$ decays, and the analyses all assume
fermionic decays for the charged Higgs.  

In the Type II models, $H^{\pm}$ decays are expected to be
quite different than the normal $W$ boson decays.
Its decay branching ratios are dependent on $m_t$,
$M_H$ and $\tan\beta$.  For values of $\tan\beta>2$, the $H\rightarrow \tau\bar{\nu}$
decay dominates and grows to 100\% when $\tan\beta > 5$, as shown in Fig. \ref{fig:tanB}.  
The $\tau$ is identified explicitly only through its
hadronic decays to a narrow jet, as described in Section \ref{sec:perf}. 
For smaller $\tan\beta$ values, the $H\rightarrow c\bar{s}$ decay is primary and leads
to two quark jets in the final state.  However, when $M_H > 130$ GeV the decay chain
$H\rightarrow t b \rightarrow W\bbbar$ exceeds this mode.
The dominance of the $H\rightarrow \tau\bar{\nu}$ or $c\bar{s}$ modes 
have led to two different search strategies for low and high $\tan\beta$ regions.

\subsubsection{High $\tan\beta$ Searches}

The more direct searches
attempt to identify the presence of the $H^{\pm}$ explicitly from the presence of the
$\tau$ in a selected top quark event sample.  These searches have the benefit that they
can avoid some dependencies on theoretical calculations, since they measure
the actual rate of $\tau$ production.  In general, an assumption is made that
$BR(H\rightarrow\tau\nu) \sim$ 100\%.  CDF conducted such a search in
Run~I data~\cite{cdfchHiggs97} using \met\ triggers.
Two event topologies were explored for this analysis.  One searched for events
possessing a $\tau + l(=e,\mu,\tau) + \met +$ two jets where one of the jets had to be
tagged as coming from a $b-$quark.  The other topology exhibited two $\tau$ jets that were
not azimuthally back-to-back.  It was employed to account for the case where $M_H$ approached
$m_t$ and the $b-$jets were too soft to be reconstructed.  Instrumental backgrounds from
fake $\tau$ jets are dominant.  They are modeled from the data by folding
a $jet\rightarrow\tau$ fake rate into the multijet $+ \met$ sample.  $W$ and $Z$ boson production
was modeled using {\sc Vecbos} and normalized to the measured cross sections in data.  Standard
\ttbar\ backgrounds were also determined to be significant.  The total background
estimation was $7.4\pm2.0$ events and seven events were observed. 
To extract a limit on $H^{\pm}$ production, the polarization of the $\tau$'s should be considered because 
it affects the $\nu$ angles and produces larger \met.  {\sc Isajet} was modified to account for this
and expected signal efficiencies were extracted.  For high $\tan\beta$ (i.e. $>100$), the region 
$M_H < 147$ GeV was excluded at 95\% C.L..

Theoretical progress led to the realization that higher order radiative corrections
impact branching ratio calculations in the high $\tan\beta$ 
region\cite{coarasa,carena}.  This means that
limits in the $\tan\beta$ vs. $M_H$ plane depend on model parameters.  
CDF incorporated this into their analysis of $\ell+\tau$ events in Run I\cite{cdfchHiggs00}.  
Both tracking-centric
and calorimeter-centric $\tau$ identification schemes were used.  The explicit identification
of jets by $b-$tagging was not implemented in this analysis.  The primary backgrounds 
were $Z$ and $W$ boson production in association with jets.  \ttbar\ also provided a significant
contribution.  This was estimated by anchoring the expected event yield to CDF's measurement of
$\sigma_{\ttbar}=5.1$ pb in the 
$\ell+$jets channel.  The total background was estimated
to be $3.1\pm0.5$ events, and four events were observed.  By
considering the observed data and the estimated backgrounds and their uncertainties,
a limit was determined for $BR(t\rightarrow Hb)$.  Acceptance for $WW, WH$ and $HH$ modes 
was calculated using {\sc Pythia} with $m_t = 175$ GeV. 
{\sc Tauola} was used to provide the correct polarization of the $\tau$.  Higgs masses from 
60$/c^2$ to 160 GeV$/c^2$ were scanned, and a limit
of $BR(t\rightarrow Hb) < 0.5$ or 0.6 in the range 60 GeV $< M_H <$ 160 GeV was obtained.  In the 
MSSM at very high $\tan\beta$, this limit is not valid because the relevant Yukawa coupling
is non-perturbative.

\begin{figure}[thb]
\begin{center}
\epsfig{figure=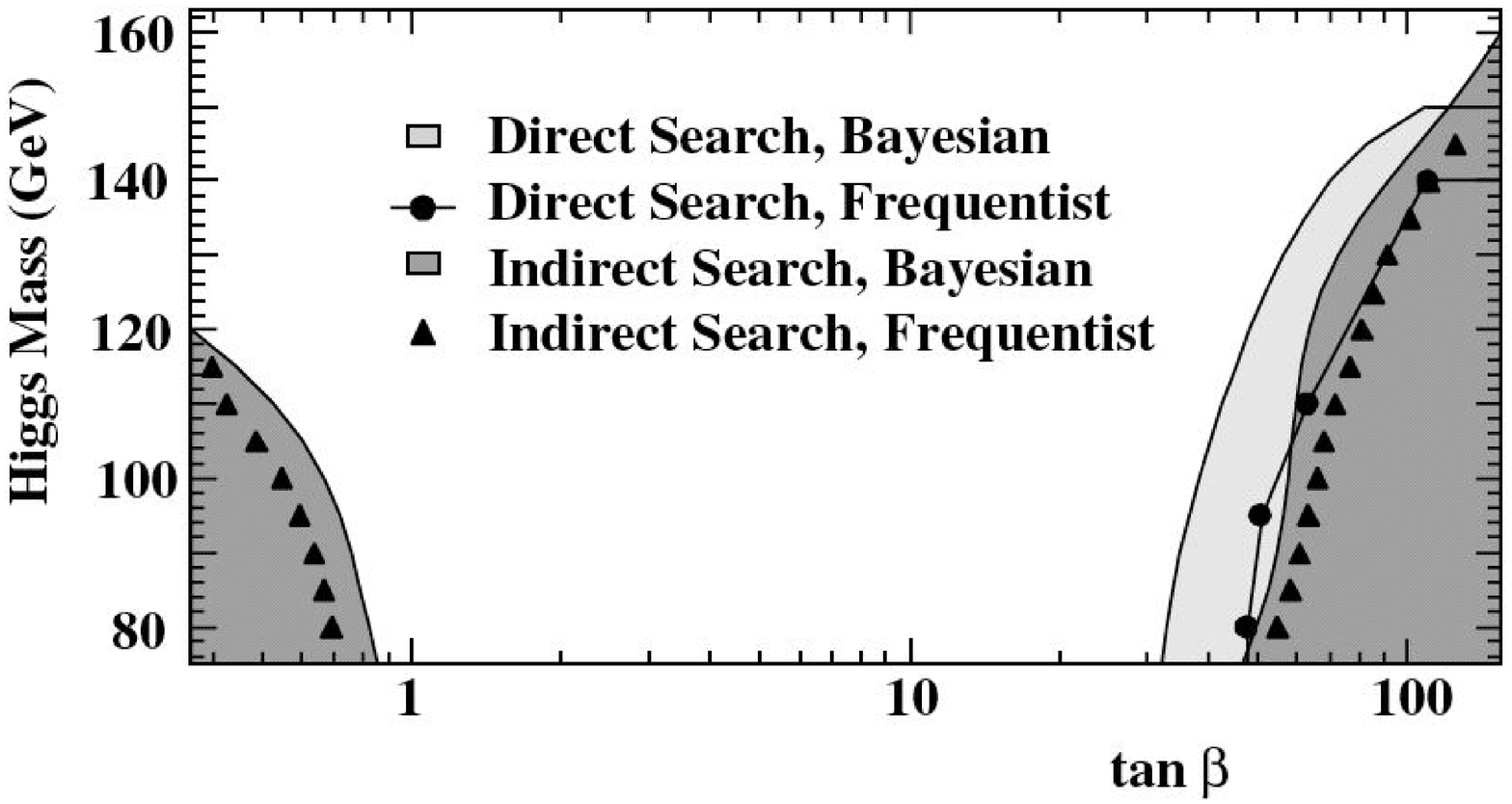,width=9.0cm}
\epsfig{figure=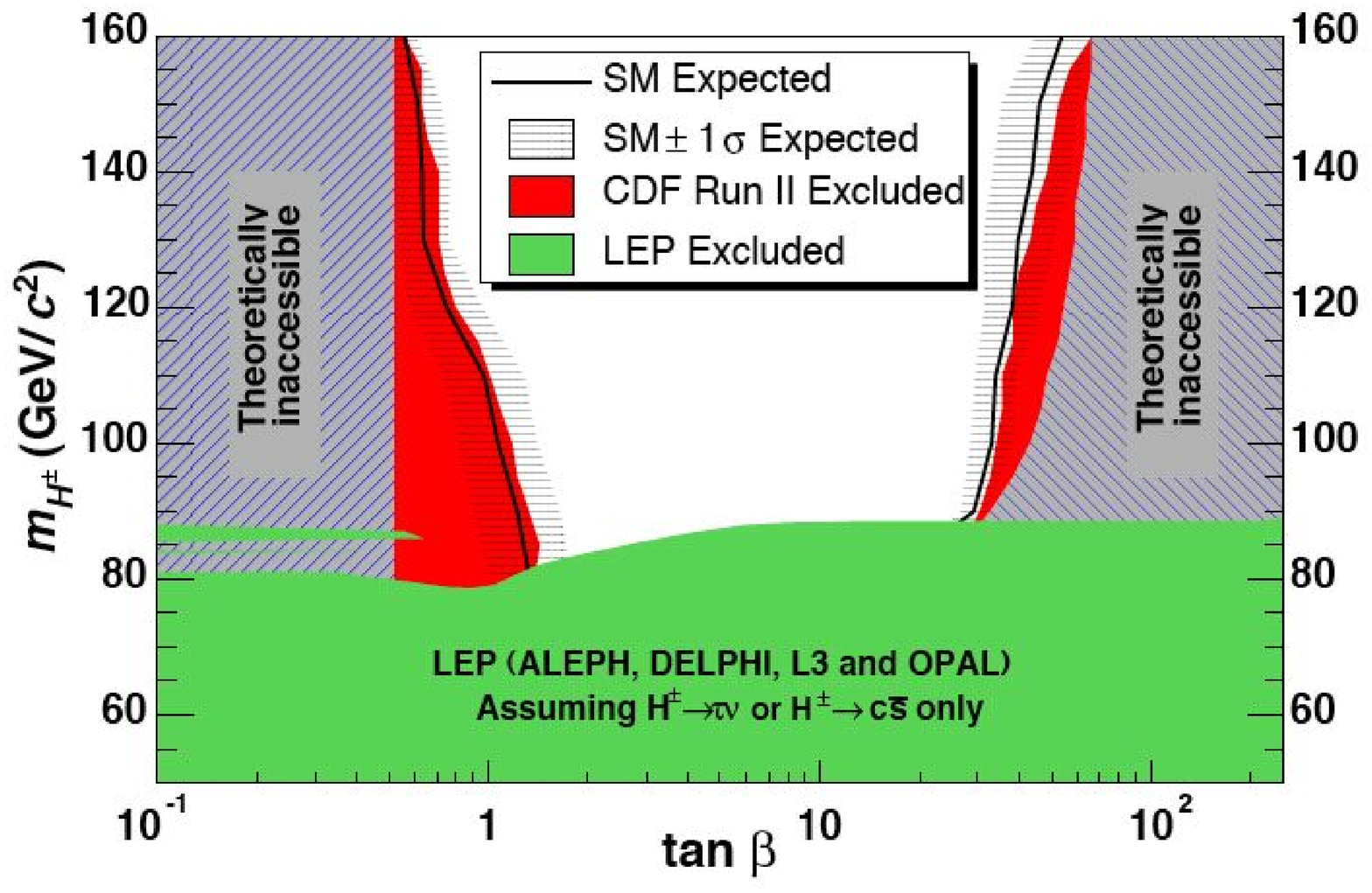,width=10.0cm}
\caption{\dzero\ Run I limits \protect\cite{d0chHiggs99} in the $M_H$ vs. $\tan\beta$ plane
incorporating direct and indirect searches (top).  CDF Run II limits
\protect\cite{cdfr2ChHiggsDecay} using radiative corrections are shown at bottom.  In the absence
of these corrections, the CDF results would be the most stringent experimental limits.}
\label{fig:chHiggsResults}
\end{center}
\end{figure}

A direct search was also pursued in 2002 using $62.2 \pm 3.1$ pb$^{-1}$ by \dzero 
\cite{d0r1ChHiggsDecay}.  Events comprised of $\tau + \met\ + jets$ and possessing a spherical distribution to the object $E_T$'s were selected.
A neural network (NN) based on JETNET\cite{jetnet} incorporated the 
\met\ and two eigenvalues of the normalized momentum tensor to discriminate signal and
background.  The NN was trained on signal generated with 
{\sc Isajet} with forced decays of $H\rightarrow\tau\nu$ and $\tau\rightarrow hadrons$.
The Higgs mass was taken as 95 GeV since it only weakly affected the NN performance.
The top quark production cross section was taken as 5.5 pb.
The instrumental background was modeled using multijet events from data.  $W+$jets 
events were modeled using {\sc Vecbos} 
plus {\sc Isajet} for soft QCD evolution.  A total of $5.2 \pm 1.6$ events were
expected from background and standard top quark processes, while three events were observed in
data.  The probability for the number of expected events to fluctuate to
the number observed was calculated for individual choices of $\tan\beta$ and 
$m_H$.  Systematic uncertainties for jet energy scale uncertainty, signal modeling
and $\tau$ identification were the main sources considered.  Values of $\tan\beta > 32.0$
were excluded at 95\% C.L. for $m_H = 75$ GeV.  The result is shown in Fig. 
\ref{fig:chHiggsResults}.

\subsubsection{Low $\tan\beta$ Region and General Searches}

As already noted, the unique signature from Higgs $\rightarrow\tau$
jets is confined to the high $\tan\beta$ region.  However, at low $\tan\beta$,
it is also possible for substantial numbers of top quarks to decay to charged
Higgses.  The Higgs decays to $c\bar{s}$ in this region are less distinctive 
than in the previous cases.  A search incorporating the low $\tan\beta$ region was 
first performed on 109.2 pb$^{-1}$ by \dzero\ in Run I.  They
employed a technique to look for the disappearance of SM top quarks\cite{d0chHiggs99}
and used the 
selections for the standard $\ell+$jets channels, both topological and soft $\mu$ tagged,
in the \ttbar\ production analysis\cite{d0r1ttcsec}.  This strategy has the
advantage of incorporating the $c\bar{s}$ decays into the search.  However, at high
$M_H >130$ and $\tan\beta<2$, the Higgs boson will preferentially decay via a virtual
top quark to $W\bbbar$.  This gives a final state that is not different
enough from standard top quarks, so the search is insensitive in this region.  Since
the presence of non-standard top quark decays would affect the measurement of
the top  quark production cross
section, the theoretical value of $5.5$ pb at 1.8 TeV 
was used and assumed to be unaffected by the presence of the
new Higgs sector.  Also, studies indicated that the non-standard decays would not 
affect the existing measured top quark mass by more than five percent, so $m_t = 175$ GeV was 
used for the analysis.  {\sc Isajet} was used to calculate signal efficiencies and was 
modified to properly account for $W\bbbar$ final states.  {\sc Pythia} was also modified 
to provide a cross-check.  Because the selection was the same as used for
the cross section measurements, those existing background estimates were valid.
The estimated background plus 
$t\rightarrow Wb$ signatures provide $30.9\pm4.0$ events compared to 30 events observed
in data.  Efficiencies were calculated for different $m_H$'s from the Monte Carlo
to extract the sensitivity at different points in the $(m_H, \tan\beta)$ plane.
Leading order theoretical calculations of branching ratios were used to calculate
the probability to observe the charged Higgs decays.  Primary 
systematic uncertainties were jet energy scale, signal model and 
particle identification.  At 95\% C.L., most of the points giving a 
$BR(t \rightarrow Hb)>0.45$ were excluded.  Combining this analysis with
the direct limit from \dzero\ gives a branching ratio limit of $BR(t\rightarrow Hb) < 0.36$
at 95\% C.L. for $0.3<\tan\beta<150$ and $m_H < 160$ GeV (see Fig. \ref{fig:chHiggsResults}).

\begin{table}[ht]
\begin{center}
\tbl{Limits by \dzero\ \protect\cite{d0chHiggs99} and CDF\protect\cite{cdfr2ChHiggsDecay} in high $\tan\beta$ region for charged Higgs branching
fraction.}
{\begin{tabular}{lcc}\toprule
Experiment & Lumi. (pb$^{-1}$) & $BR(t\rightarrow Hb)<$ @ $95\% C.L.$ \\
\colrule
\dzero\   & $66.2$ and $109.2$ & 0.36 \\
CDF       & 193                & 0.4 \\
\botrule
\end{tabular}
\label{BRchHiggs}}
\end{center}
\end{table}

Since this analysis, it was realized that
$H^{\pm}\rightarrow Wh^0$ can be significant.  
CDF has incorporated this element, along
with the developments described for their earlier analyses,
into the first search for charged Higgs bosons in Run~II\cite{cdfr2ChHiggsDecay}.  
Using 193 pb$^{-1}$,
the yield of events in four categories of final state were compared: dilepton,
$\ell+$jets with $b-$tag, $\ell+$jets with at least two
$b-$tags, and $\ell + \tau$ events.  The latter was compared to the other channels to
produce direct limits on the presence of charged Higgs bosons in top quark decay.  This analysis
was the first to use the newer Yukawa coupling and radiative corrections\cite{coarasa,carena} in the 
determination of acceptances for the sensitivity calculation.  
The analysis considered decays
of the Higgs to $\tau\nu, c\bar{s}, t^*\bar{b}$, and $Wh^0\rightarrow W\bbbar$.  {\sc Pythia}
was modified to include the $t^*\bar{b}$ decay mode and used to estimate the efficiencies
for various top quark widths, the charged Higgs mass and width, and $m_{h^0}$.
In order to properly calculate the overall acceptance for various parameter choices
in MSSM, the CPSuperH\cite{CPsuperH} generator was used. Much of the low
$\tan\beta$ region was excluded, including some of the high $m_H$ region.  
In the high $\tan\beta$ region, a limit of $BR(t\rightarrow Hb) < 0.4$ at 95\% was 
obtained, as shown in Fig. \ref{fig:chHiggsResults}.  If no assumptions are made for charged Higgs decay 
(i.e. $H\rightarrow\tau\nu$), 
the limit is relaxed to $<0.91$.  For comparison (see Table~\ref{BRchHiggs}), 
these limits would be stricter
than the previous \dzero\ limits if the radiative corrections were ignored.

%% file: bsm/Wprime.tex
The existence of new forces in nature can be revealed through the observation of
additional gauge bosons beyond those of the standard model. 
Various extensions of the SM postulate larger gauge groups\cite{topBSM,wprime-Sullivan} 
and therefore new forces associated with additional charged gauge bosons, 
which are generically called $W^{\prime}$.
For instance, the left-right symmetric model\cite{chargedHiggsModels} expands
the $SU(2)_{L} \times U(1)$ electroweak group 
to $SU(2)_{L} \times SU(2)_{R} \times U(1)$, predicting the existence
of three new gauge bosons: two charged $W^{\prime\pm}$ bosons and one neutral $Z'$ boson. 
In this model, the $W^{\prime}$ 
boson appears as a 
heavier counterpart of the left-handed $W$ and is responsible for
right-handed interactions, in the same way as the SM $W$ boson
mediates only left-handed interactions.

Previous indirect searches based on low energy phenomena such as 
$\mu$ decay, the $K_L-K_S$ mass
difference, neutrinoless atomic double beta decay, and semileptonic
branching ratio $b \to Xl\nu$ 
have resulted in stringent model-dependent limits\cite{langacker}.
Direct searches for the production of $W^{\prime}$ bosons have focused 
on its leptonic decays, $W^{\prime} \to l\nu_l$, by looking for anomalous production of 
high transverse mass $l\nu_l$ pairs. Searches using the decay mode 
$W^{\prime} \to e\nu_e$\cite{wprime-cdf-run1-e,wprime-D0-run1-e} ($W^{\prime} \to \mu\nu_{\mu}$\cite{wprime-cdf-run1-mu}) 
from Tevatron Run~I exclude a $W^{\prime}$ boson
with mass $m_{W^{\prime}}<$ 754 GeV (660 GeV) at 95\% C.L.. The combination of these 
leptonic channels yields the lower limit on $m_{W^{\prime}}$ of 
786 GeV at 95\% C.L.\cite{wprime-cdf-run1-e}.
These mass limits all assume
that the new vector boson's couplings to leptons are those given by
the SM, with the additional assumption that the neutrino
produced is much lighter than the $W^{\prime}$ boson. Searches for the $W^{\prime}$ boson in 
the quark decay channel $W^{\prime} \to q\bar{q'}$ based on resonant structure in
the dijet mass spectrum have also placed strong constraints on its mass. 
Searching in the hadronic final state avoids assumptions 
regarding the neutrino mass $m_{\nu}$, but is background limited. 
Previous direct searches in this channel have ruled out $W^{\prime}$ bosons in 
the mass range
1$<m_{W^{\prime}}<$261 GeV by UA2\cite{wprime-UA2-qq}, 300$<m_{W^{\prime}}<$420 
GeV by CDF\cite{wprime-cdf-run1-qq} and 300$<m_{W^{\prime}}<$800 
GeV by \dzero\ \cite{wprime-D0-run1-qqonly} at 95\% C.L..
 
\subsubsection{$W^{\prime}$ in Top quark sector}
\begin{figure*}[!h!tbp]
\begin{center}
\epsfig{figure=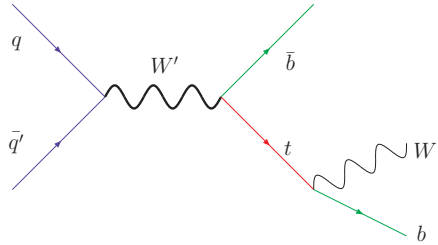,width=0.50\textwidth}
\end{center}
\vspace*{8pt}
\caption{Leading order Feynman diagram for single top quark production via a heavy $W^{\prime}$ boson. The top quark decays to a standard model $W$ boson and a $b$-quark.
}
\label{fig:wprime-feyndiag} 
\end{figure*}

The top quark sector offers great potential for the search of new gauge
bosons. The single top quark final state is especially sensitive to the 
presence of an additional heavy boson, owing to the decay chain $W^{\prime} \to t\bar{b}$,
where the top quark decays to a $b$-quark and a $W$ boson 
which subsequently decays leptonically or hadronically. 
The leading order
Feynman diagram for $W^{\prime}$ boson production resulting in single top quark events
is shown in Fig. \ref{fig:wprime-feyndiag}. 

This diagram is identical to that of the SM
$s$-channel single top quark production where the SM $W$ boson
appears as a virtual particle. The $W^{\prime}$ boson also has a $t-$channel exchange
that leads to the single top quark final state. However, the cross section 
for a $t-$channel $W^{\prime}$ process is much smaller than the SM
$t-$channel single top quark production due to the high mass of the $W^{\prime}$ boson.
A heavy $W^{\prime}$ boson signal would appear as a peak in the 
$t\bar{b}$ invariant mass distribution in these processes.
Although this search is only sensitive to $W^{\prime}$ bosons with mass above the 
$t\bar{b}$ kinematic threshold of approximately 180 GeV, it is relatively free 
of background compared to the $W^{\prime} \to q\bar{q'}$ decay mode because of the
distinct signature from the top quark decay $t \to Wb$. Furthermore, the interpretation
of the data is less sensitive to assumptions regarding the right-handed neutrino
($\nu_R$) sector or the lepton couplings of the $W^{\prime}$ boson.

\subsubsection{Run I search for $W^{\prime} \to t\bar{b}$}
CDF attempted the first search for a $W^{\prime}$ boson decaying 
to a $t\bar{b}$ final state using a 106 pb$^{-1}$ of Run I data 
sample\cite{wprime-cdf-run1-top}.
The analysis considers the $\ell+$jets final state topology of single top quark
production arising from leptonic decays of $W$ boson, $W \to e\nu,\mu\nu$. 
The event signature consists of 
a high-$p_T$ lepton, significant \met\ from the neutrino, 
and two $b$-quark jets. 
The candidate events are selected by requiring 
a high-$p_T$ electron or muon with large \met\ 
and accompanied by two or three jets, at least one of them being $b$-tagged. 
The fully simulated {\sc Pythia} Monte Carlo was used to determine the expected 
contribution from signal events in the data sample as a function of
$W^{\prime}$ mass. The $W^{\prime}$ boson is required to have a right-handed coupling to the 
$t\bar{b}$ final state since negligible signal yield differences
between right-handed and left-handed couplings are expected.
The analysis considered both the cases when $m_{W^{\prime}} < m_{\nu_{R}}$ 
and $m_{W^{\prime}} \gg m_{\nu_{R}}$, where $m_{\nu_{R}}$ is the mass
of $\nu_R$ that couples to the $W^{\prime}$. 
The analysis attempted to interpret the data for
$m_{W^{\prime}}>$225 GeV since the acceptance calculation becomes increasingly uncertain as 
one nears the $t\bar{b}$ kinematic threshold. The dominant background contributions 
to this search arises from the pair
production of top quarks, single top quark production, and the associated 
production of $W$ bosons with one or more heavy 
quarks ($Wb\bar{b}$, $Wc\bar{c}$) which
provides the largest single background contribution. 
A candidate sample of 57 events agreed reasonably with the 48$\pm$6
expected from background. Therefore, no significant evidence for 
$W^{\prime}$ boson production is seen.

\begin{figure*}[!h!tbp]
\begin{center}
\epsfig{figure=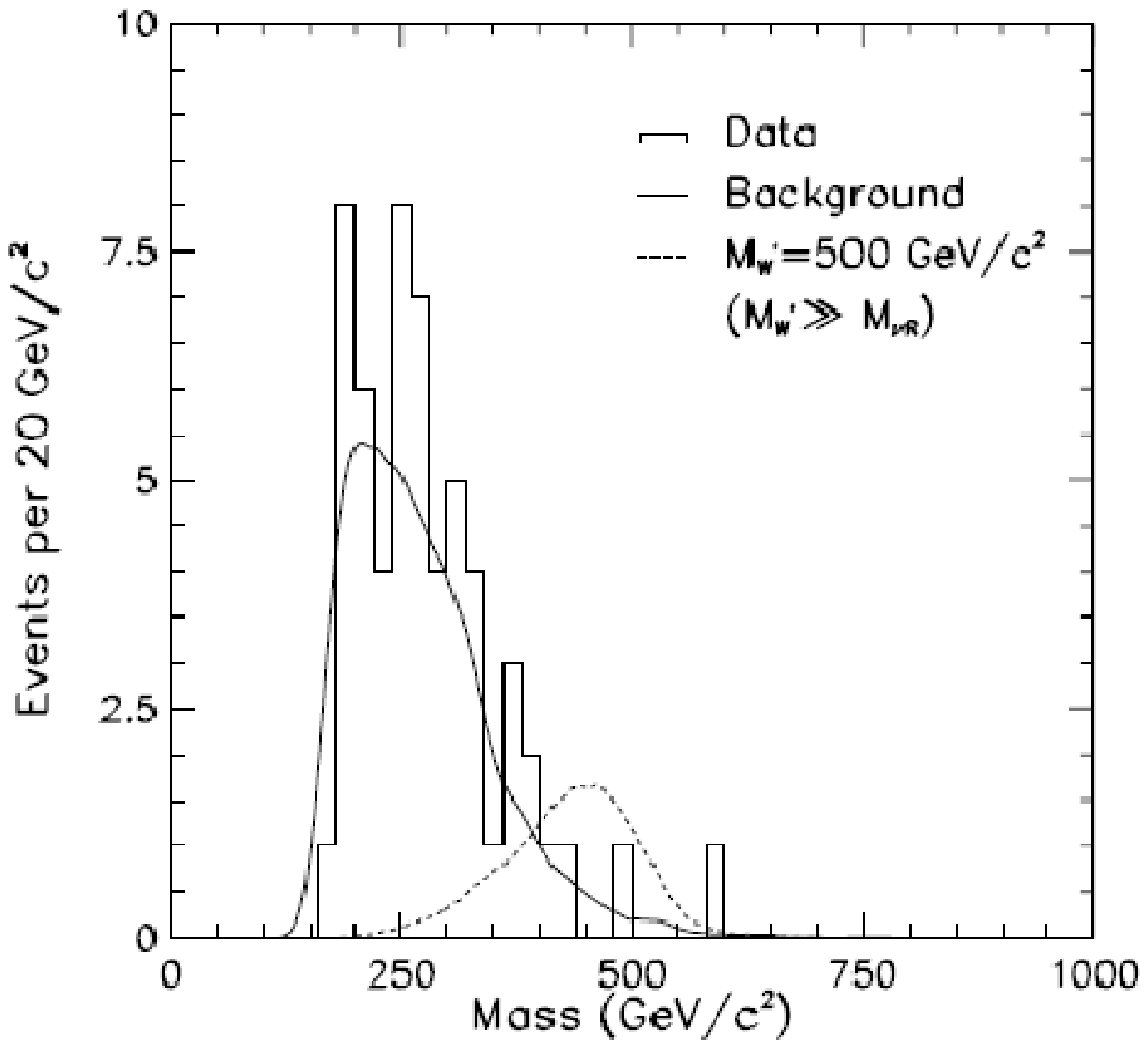,width=0.45\textwidth}
\epsfig{figure=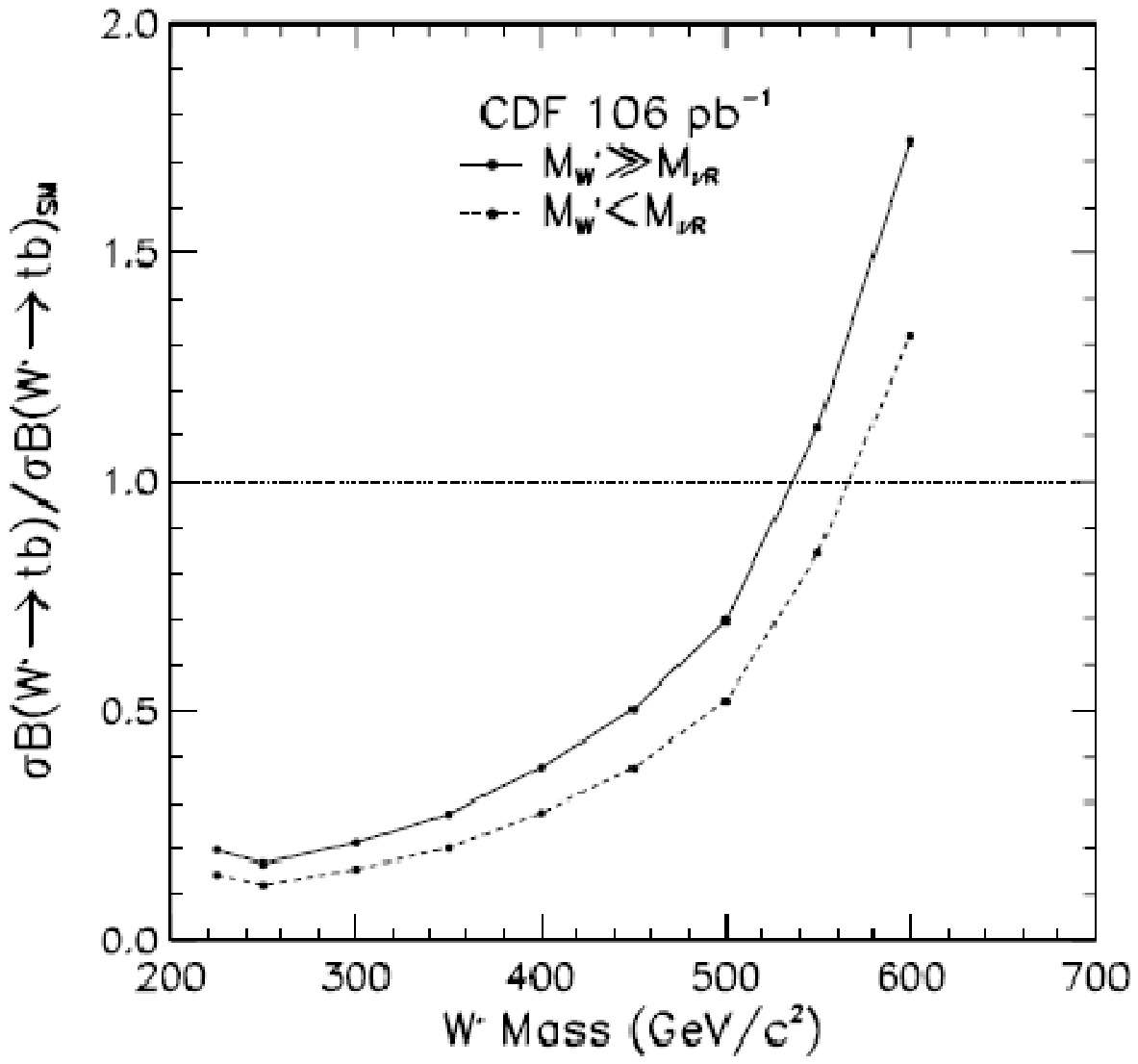,width=0.45\textwidth}
\end{center}
\vspace*{8pt}
\caption{CDF's Run I serach for $W^{\prime}$ boson in top quark decays\protect\cite{wprime-cdf-run1-top}. \textbf{Left}: The $Wb\bar{b}$ mass spectrum of the candidate events after constraining the 
lepton-neutrino invariant mass to the $W$ boson mass. The distribution expected from the 
production of a $W^{\prime}$ boson with a mass of 500 GeV is illustrated by the dashed curve. The 
distribution expected from the background processes is shown by the solid curve.
\textbf{Right}: The upper limits on the $W^{\prime}$ boson production cross section as a function of the
$W^{\prime}$ boson mass. Limits are shown for the case $m_{W^{\prime}} \gg M_{\nu_{R}}$ (solid) and $m_{W^{\prime}} < M_{\nu_{R}}$ (dashed). The intercepts at [$\sigma . BR(W^{\prime} \to t\bar{b}) /  \sigma . BR(W^{\prime} \to t\bar{b})_{SM}$]=1 correspond to the 95\% C.L. limits on the $W^{\prime}$ boson mass with SM strength couplings.}  
\label{fig:cdf-wprime} 
\end{figure*}

In such a scenario, the analysis sets a limit on the $W^{\prime}$ production cross section times
branching ratio, $\sigma \times BR(p\bar{p} \to W^{\prime} \to t\bar{b})$, employing the invariant mass distribution 
of the $Wb\bar{b}$ final state. The $Wb\bar{b}$ mass distribution for the 57 
candidate event sample is shown in Fig. \ref{fig:cdf-wprime} and is compared with the expected mass 
distribution for a $W^{\prime}$ boson with $m_{W^{\prime}}$=500 GeV and the 
sum of the background processes. 
To estimate the size of a potential signal contribution, an unbinned maximum likelihood 
fit is performed to the observed 
mass distribution, allowing for both a signal and background contribution for 
different values of $m_{W^{\prime}}$ ranging from 225 to 600 GeV.
Based on the fit results for the fraction of events arising from 
$W^{\prime}$ production, the analysis sets 95 \% C.L. upper limits using a Bayesian
technique on the relative 
contribution of a $W^{\prime}$ boson for each value of $m_{W^{\prime}}$, 
$\sigma \times BR(W^{\prime} \to t\bar{b}) /  \sigma \times BR(W^{\prime} \to t\bar{b})_{SM}$, 
where the denominator is the expected $\sigma \times B$ for a given $W^{\prime}$ boson 
mass assuming SM couplings. Figure \ref{fig:cdf-wprime} shows the 
upper limits on the  $W^{\prime}$ 
boson production cross section as a function of the $m_{W^{\prime}}$ for the two 
different assumptions on $m_{\nu_{R}}$. This analysis 
excludes a $W^{\prime}$ boson at 
95 \% C.L. with masses 225 $<m_{W^{\prime}}<$536 GeV 
for $m_{W^{\prime}} \gg m_{\nu_{R}}$ and 225 $<m_{W^{\prime}}<$566 GeV assuming $m_{W^{\prime}} < m_{\nu_{R}}$.

\subsubsection{Run II search for $W^{\prime} \to t\bar{b}$}
\begin{figure*}[!h!tbp]
\begin{center}
\epsfig{figure=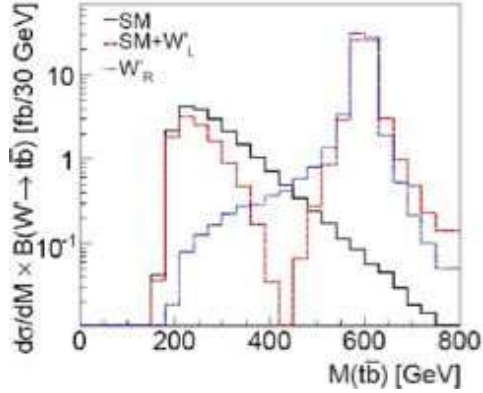,width=0.50\textwidth}
\end{center}
\vspace*{8pt}
\caption{The invariant mass of the top-bottom quark system at the parton level
for different models of $W^{\prime}$ boson production\protect\cite{wprime-D0-run2-top}. Shown are the SM s-channel distribution, the $W^{\prime}_L \to t\bar{b}$ boson distribution, including the
interference with the SM contribution, and the $W^{\prime}_R \to t\bar{b}$ boson 
contribution, for a $W^{\prime}$ boson mass of 600 GeV.}
\label{fig:D0-modeling} 
\end{figure*}

In Run II, \dzero\ has conducted a comprehensive search for the $W^{\prime}$ boson 
in the top quark decay channel by analyzing 230 pb$^{-1}$ 
of data\cite{wprime-D0-run2-top}. 
The search considers three
models of $W^{\prime}$ boson production. In each case, the CKM mixing matrix elements for 
the $W^{\prime}$ boson is set to the SM values. The first model ($W^{\prime}_L$) assumes the coupling
of the $W^{\prime}$ boson to SM fermions to be identical to that of the SM $W$ boson. Under
these assumptions, there is interference between the SM $s$-channel single top quark
process and the $W^{\prime}$ boson production process,
although this interference term is small for large 
$W^{\prime}$ boson masses. In the second and third model ($W^{\prime}_R$),
the $W^{\prime}$ boson has only right-handed interactions.  
In the second model, the $W^{\prime}_R$ boson is allowed to decay both to 
leptons and quarks, whereas in the third 
model it is only allowed to decay to quarks.
The branching fraction for the 
decay $W^{\prime} \to t\bar{b}$ is about $3/12$ when the 
decays to quarks or leptons are both included and is about $3/9$ 
when the decay to leptons is not allowed\cite{wprime-Sullivan}. 
{\sc COMPHEP} 4.4.3 matrix element 
event generator\cite{STcomphep} is used 
for the modeling of the $W^{\prime}$ boson production process. 
Fig. \ref{fig:D0-modeling} compares the 
invariant mass distribution for the $W^{\prime}$ models with
left-handed coupling (including interference) and right-handed
coupling (no interference) with the SM $s$-channel 
single top quark production, for a $W^{\prime}$ boson mass of 600 GeV. 

Table~\ref{tab:NLO-Wprime} shows the NLO cross sections\cite{wprime-Sullivan} for a $W^{\prime}$ boson $\times$ branching fraction
to $t\bar{b}$ for three different $W^{\prime}$ boson models. The systematic uncertainty on the 
cross section varies between about 12\% at a mass of 600 
GeV and 18\% at a mass of 800 GeV.
The theoretical $W^{\prime}$ boson production cross section is more than 15 pb 
for masses 
between 200 and 400 GeV for all three models\cite{wprime-Sullivan}. 
The upper limits on the single top quark production cross section in
the $s$-channel were 6.4 pb by \dzero\ based on the analysis of same 
dataset\cite{STPLBD0} and 13.6 pb by CDF\cite{RunIIcdf_result}, which do not depend much on whether 
the $W^{\prime}$ boson coupling is left-handed or right-handed. 
Thus, the $W^{\prime}$ boson production decaying to $t\bar{b}$ is excluded 
in this mass region and the analysis explores the region of even higher 
masses  for $W^{\prime}$ searches.

\begin{table}[ht]
\begin{center}
\tbl{NLO production cross section for a $W^{\prime}$ boson $\times$ branching fraction to $t\bar{b}$ for three different $W^{\prime}$ boson models\protect\cite{wprime-Sullivan}. 
 The production cross section for $W^{\prime}_L$ interactions also include the SM $s$-channel contribution as well as the interference term between the two.}
{\begin{tabular}{lcccc}\toprule
 & Cross section $\times$ $BR(W^{\prime} \to t\bar{b}$ [pb]  &  &  \\
\colrule
$W^{\prime}$ mass [GeV] &  SM+$W^{\prime}_L$ & $W^{\prime}_R(\to  l \: or \: q)$  & $W^{\prime}_R(\to q)$ only\\ \hline
600 & 2.17 & 2.10 & 2.79\\
650 & 1.43 & 1.25 & 1.65\\
700 & 1.03 & 0.74 & 0.97\\
750 & 0.76 & 0.44 & 0.57\\
800 & 0.65 & 0.26 & 0.34\\
\botrule
\end{tabular}
\label{tab:NLO-Wprime}}
\end{center}
\end{table}

This search adopts a strategy similar to that used in CDF's Run I analysis,
i.e. looks for 
events that are consistent with $W^{\prime} \to t\bar{b}$ production 
and the $W$ boson from
the top quark decaying leptonically ($W \to e\nu,\mu\nu$; 
including $W \to \tau\nu$ with $\tau \to e\nu$). 
It utilizes the same dataset, basic event selection, and background
modeling as the \dzero\ single top quark search described in  Section~\ref{secSTD0search}, and Ref.~\refcite{STPLBD0}.
It selects signal-like events and separates the data into independent analysis
sets based on final-state lepton flavor (electron or muon) with two or three jets 
and $b$-tag multiplicity, single tagged or double tagged. The independent datasets are later 
combined in the final statistical analysis.

\begin{figure*}[!h!tbp]
\begin{center}
\epsfig{figure=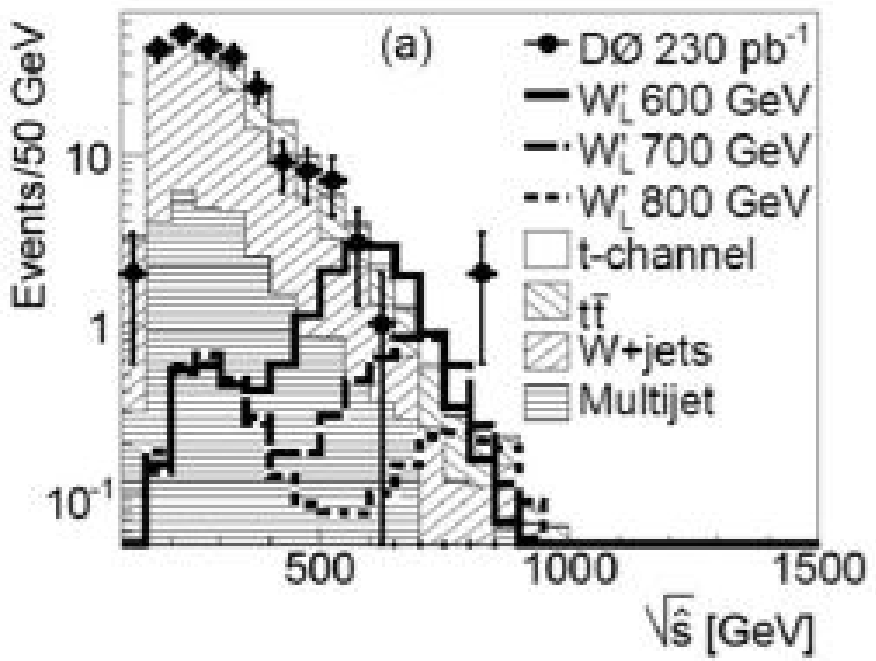,width=0.45\textwidth}
\epsfig{figure=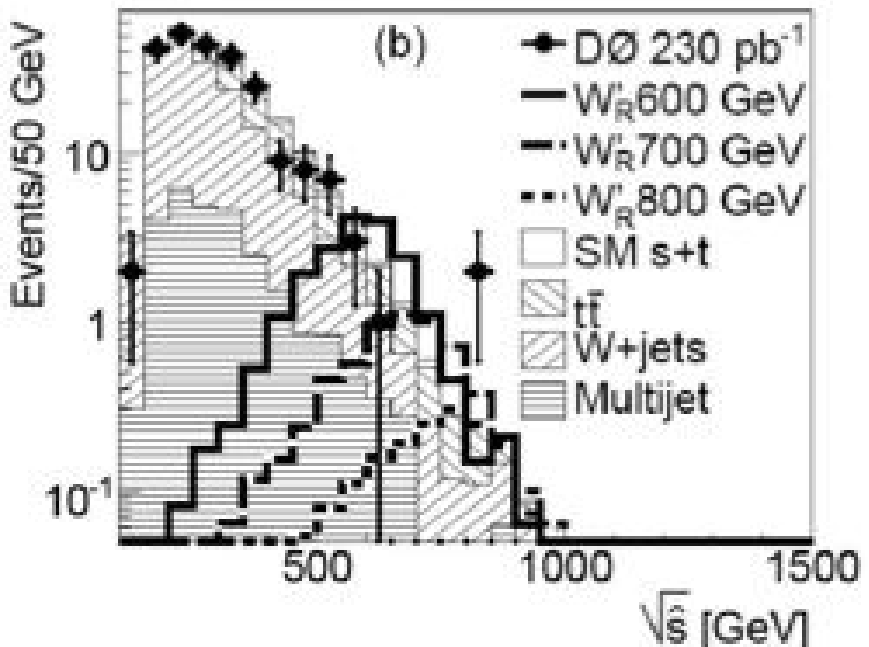,width=0.45\textwidth}
\end{center}
\vspace*{8pt}
\caption{\dzero\ 's Run II search for $W^{\prime}$ boson in top quark decays\protect\cite{wprime-D0-run2-top}. The distribution of $W^{\prime}$ boson invariant mass for several values of
$W^{\prime}$ boson mass and background processes for (a) left-handed $W^{\prime}$
boson couplings, and (b) right-handed couplings when only the decay to quarks
is allowed. Electron, muon, single-tagged and double-tagged events are combined.}
\label{fig:D0-run2-xsection} 
\end{figure*}

The large mass of the $W^{\prime}$ boson is expected to set it apart from
all background processes, hence a complete kinematic reconstruction of 
the invariant mass of the $W^{\prime}$ boson is performed by adding the four-vectors of 
all reconstructed final state objects: the jets, the lepton, and the neutrino.
Figure \ref{fig:D0-run2-xsection} shows the distribution of effective mass 
of $\ell+$ \met\ $+$ jets when the $\ell\nu$ invariant mass is
constrained to the $W$ boson mass. The data, sum of background, and
expected $W^{\prime}$ boson contributions are shown for different couplings
and masses. The observed event yield is
consistent with the background predictions within uncertainties.
The dominant sources of systematic uncertainty on the signal and
background acceptances arises from the modeling of $b$ hadrons
in the simulation, the jet energy scale, object identification 
and trigger efficiencies, and modeling of jet fragmentation.

\begin{figure*}[!h!tbp]
\begin{center}
\epsfig{figure=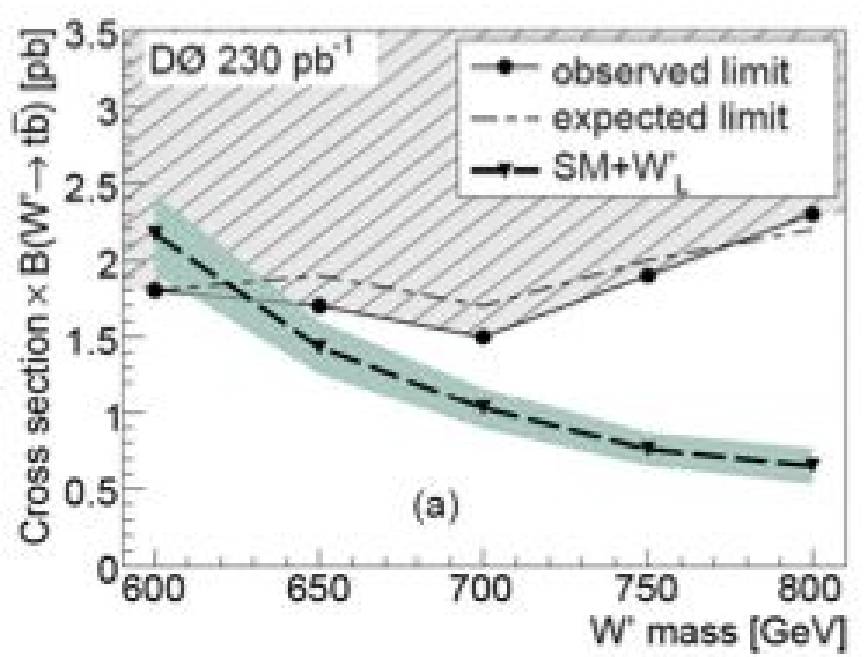,width=0.45\textwidth}
\epsfig{figure=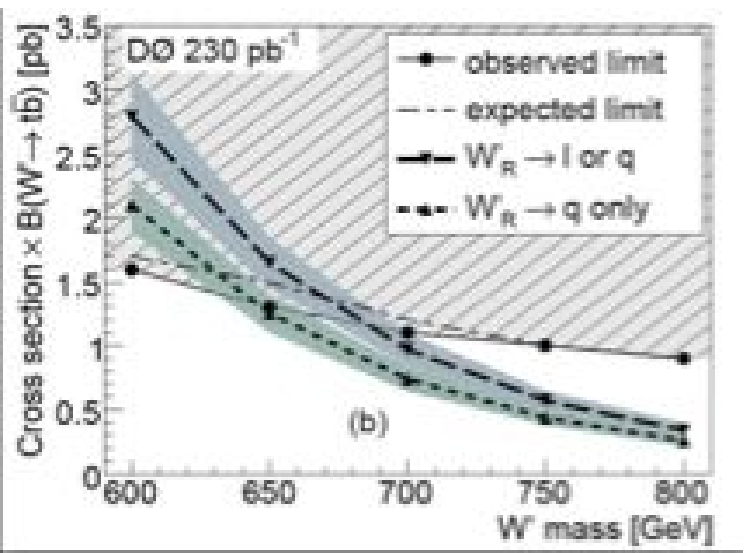,width=0.45\textwidth}
\end{center}
\vspace*{8pt}
\caption{\dzero\ 's Run II search for $W^{\prime}$ boson in top quark decays\protect\cite{wprime-D0-run2-top}. Cross section times branching fraction limits at the 95\% C.L. 
versus the mass of the $W^{\prime}$ boson 
with (a) left-handed couplings and (b) right-handed couplings. Also shown are
the NLO cross sections and the expected limits. The shaded regions above the 
circles are excluded by the measurement.}
\label{fig:D0-run2-limit}
\end{figure*}

In the absence of a significant excess over the background predictions,
a binned likelihood analysis is performed on the observed invariant mass distribution
to obtain upper limits on 
the $\sigma \times BR(p\bar{p} \to W^{\prime} \to t\bar{b})$ 
for discrete $W^{\prime}$ mass points in each model. Figure \ref{fig:D0-run2-limit} 
shows the $\sigma \times BR(p\bar{p} \to W^{\prime} \to t\bar{b})$ limits together 
with the NLO cross sections and the expected limits, along with their uncertainties. 
At the 95\% C.L., the shaded areas above the solid lines 
are excluded in this analysis. The intersection of the solid line with the lower 
edge of the uncertainty band on the predicted cross section defines the 95\% C.L. 
lower mass limit for each model. Together with the limit from the SM $s$-channel 
single top quark search, the presence of a $W^{\prime}$ boson with 
SM-like left-handed coupling is excluded if it has a mass between 200 and 610 GeV. 
In addition, the presence of a $W^{\prime}$ boson with right-handed couplings 
that is allowed to decay to leptons and quarks (only quarks) is excluded if it has 
mass between 200 and 630 GeV (670 GeV). This is the first direct search
limit for the $W^{\prime}$ boson which takes into account 
interference with the SM properly. It is also the most stringent limit 
on the presence of a $W^{\prime}$ boson in top quark decays.
 
More recently, CDF has also produced a preliminary result\cite{wprime-cdf-run2} on the search 
for resonant $W^{\prime}$ boson which decays via 
$W^{\prime} \to t\bar{b}$ using a sample of 
approximately 1 fb$^{-1}$. 
As in the previous analyses, the search looks for unexpected structure
in the spectrum of the invariant mass distribution of the reconstructed
$W$ boson and two leading jets ($M_{Wjj}$). Expected contributions from SM
processes are derived from selections and background studies of the single 
top quark analysis. Resonant $t\bar{b}$ production is modeled as simple
$W^{\prime}$ with SM-like couplings to fermions. 
The analysis finds no evidence for resonant $W^{\prime}$ 
production and excludes a $W^{\prime}$
boson for $m_{W^{\prime}}<$760 GeV in the case 
of $m_{W^{\prime}} \gg m_{\nu_{R}}$ and $m_{W^{\prime}}<$790 GeV in the case
of $m_{W^{\prime}} < m_{\nu_{R}}$ at 95\% C.L..

%% file: prospects/conclu.tex
\subsection{Tevatron synopsis}

The Tevatron program in Run II is currently undergoing rapid progress and
measurements of the top quark are greatly improving.
We have seen a flowering of techniques for the isolation and
study of the top quark. 
Some approaches pursued in Run I, such as the topological selection of
$\ell+$ jets events or the secondary vertex tagging 
have been adopted by both experiments as useful strategies.
New selection mechanisms such as $\ell+$ track in the
dilepton case, and $b$-jet tagging  tagging based on neural network
discriminant in general have produced competitive
results and significantly expand the fraction of top quark
events that are analyzable.
Attempts to maximize sensitivity for various measurements have generally meant that 
selections permit significant background yield.  This necessitated
improvements 
to background modeling that included increased
ability to determine important properties of background in data.  An example
involves the study of
the flavor content of the $b-$tagged backgrounds.  Usage of Monte Carlos
has been improved by efforts to more correctly describe jet production.

Substantial increases in integrated luminosity have so far brought
the measurement of the \ttbar\ cross section from 30\% to $< 15\%$ 
uncertainties.  These measurements are in general agreement
with the theoretically predicted value.
Many different channel definitions have been tried
within the dilepton, $\ell+$ jets and all-jets categories.  So far, no
serious discrepancies have been observed in the relative 
rates of these channels.
Extensive kinematic tests of the various channels also reveal no clear
difference from standard model expectations: production and decay seem
well-modeled.  The Run II measurements will surpass the 10\% level of
precision. They are currently systematically dominated and of comparable
precision as the theoretical uncertainties.
This will hopefully motivate further theoretical work.  It will also
provide an interesting test when it comes to comparing LHC and Tevatron production.

The major physical significance of $m_t$ has motivated both Tevatron experiments
to devote considerable effort to its measurement.
Already, Run II has surpassed initial expectations.
Improved techniques have been a significant part of this effort.  Both experiments
have attempted a variety of fitting techniques to exploit the data 
to the fullest extent.  A critical contribution in this regard 
has been the matrix element approach, which is now being used on all channels.
Soon measurements in all channels  will be dominated by systematic
uncertainties.  In Run I,
it was clear from the outset that the calibration of jet energies
would be a major concern.  This still remains the case, and the
measurement of top quark properties, particularly the mass, has driven
major progress in this area.  One of the results of this effort is
that we now see the use of the top quark sample as a means to anchor
this calibration via the use of jets from $W\rightarrow jj$ decays 
$in$ $situ$.  
This is a critical transition that will have significant
impact on future Tevatron and LHC physics.
Currently, the cumulative uncertainty in $m_t$ is 1\%, with each experiment
achieving 1.5\% in their best measurements.  This result has been 
achieved  with only a 
fraction of the expected total Run II luminosity.  As energy scale and
other uncertainties are further reduced, 
the measurement will likely become even more precise. 
Such a measurement will remain competitive with the LHC 
for some time to come.

An exciting aspect of the field is the rising experimental capability to
test electroweak physics through the properties of this particle. 
With the advent of a signal for single top quark production, a new direct
probe of electroweak interactions is available.  The measurement of
$|V_{tb}|$ is an important element of this effort.  Run II should
yield a 10\% measurement of this quantity.  Strong
constraints have been laid on potential $V+A$ couplings at the $Wtb$ vertex.
The top quark charge is already determined to be like its $c$ and $u$ brethren.

Direct searches of new physics in the production and decay of the top quark
have yielded significant constraints.  Techniques to look for 
charged Higgs Bosons in top quark decay have been refined as the calculational
picture has improved.  Limits are occupying an increasing fraction of the
experimentally available parameter space.  With the increased statistics,
better limits of $t\overline t$ resonance production are being set, 
as well as on the
presence of potential extra $W'$ bosons in single top quark production.  

{\subsection{Looking to the Future}
\label{sec:future}}

Many of the basic techniques described in
this review can be used at the LHC, and there will be many new challenges.
Instantaneous luminosities of $10\times$ relative to the Tevatron, 
and \ttbar\ cross sections $100\times$ larger 
will provide a thousand-fold increase in statistical
samples.  This will allow, for instance, very large event samples 
with both $b-$jets identified.
Early studies indicate that 1\% precision in $m_t$ measurement
is accessible (e.g. Ref. \refcite{atlasTDR}).  To achieve or surpass
this level will require large, well-constrained samples.
The top quark 
pair production and decay will be measurable across a wide array of
channels.  The measurements that will improve most dramatically are the
other top quark properties, such as $V_{tb}$ or $W$ boson helicity.  The
searches for new physics will also gain considerable sensitivity from the
ability to make precise analysis of the rates and kinematics of final state
particles in top quark events.

\begin{figure*}[!h!tbp]
\begin{center}
\epsfig{figure=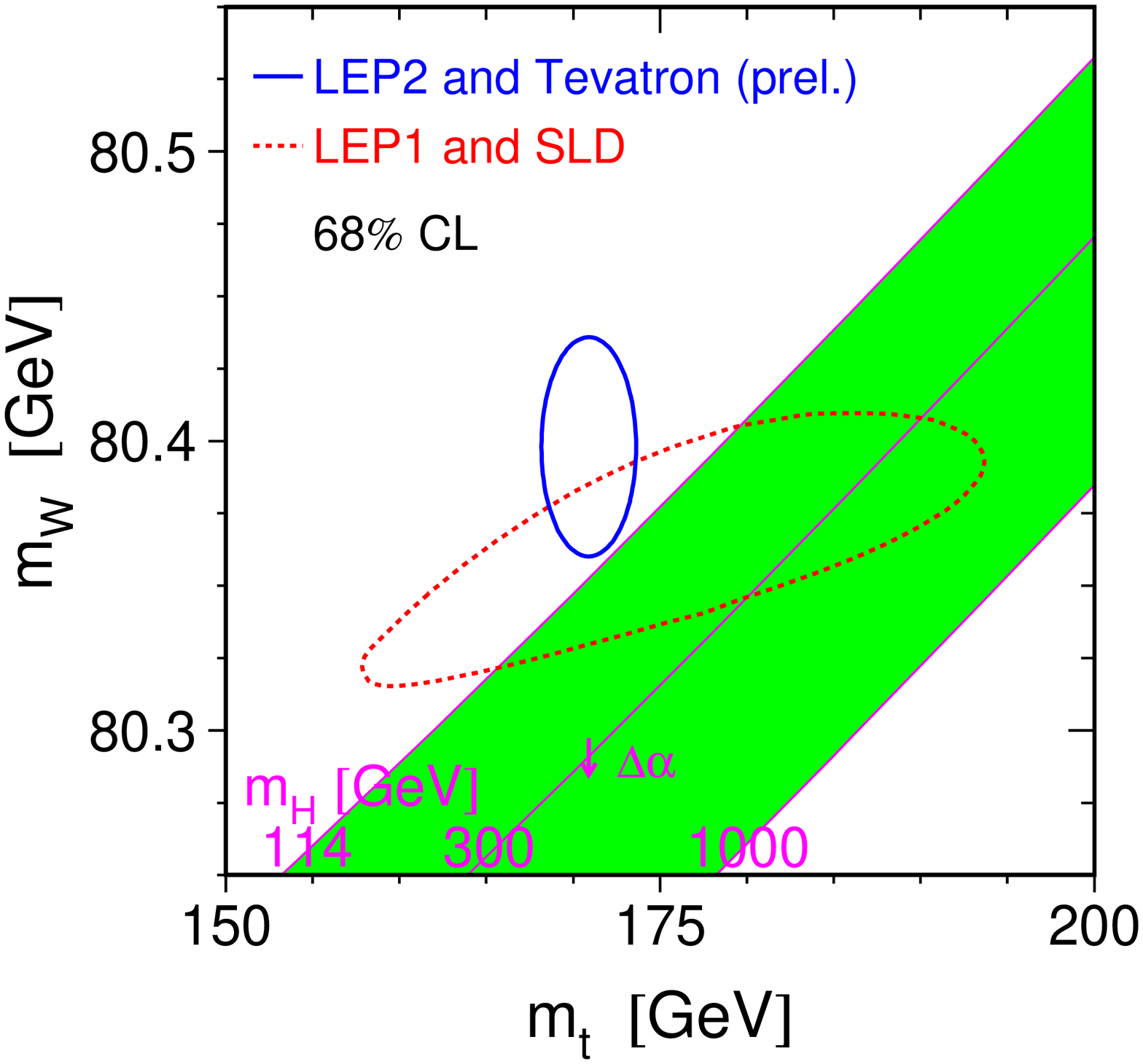,width=6cm}
\epsfig{figure=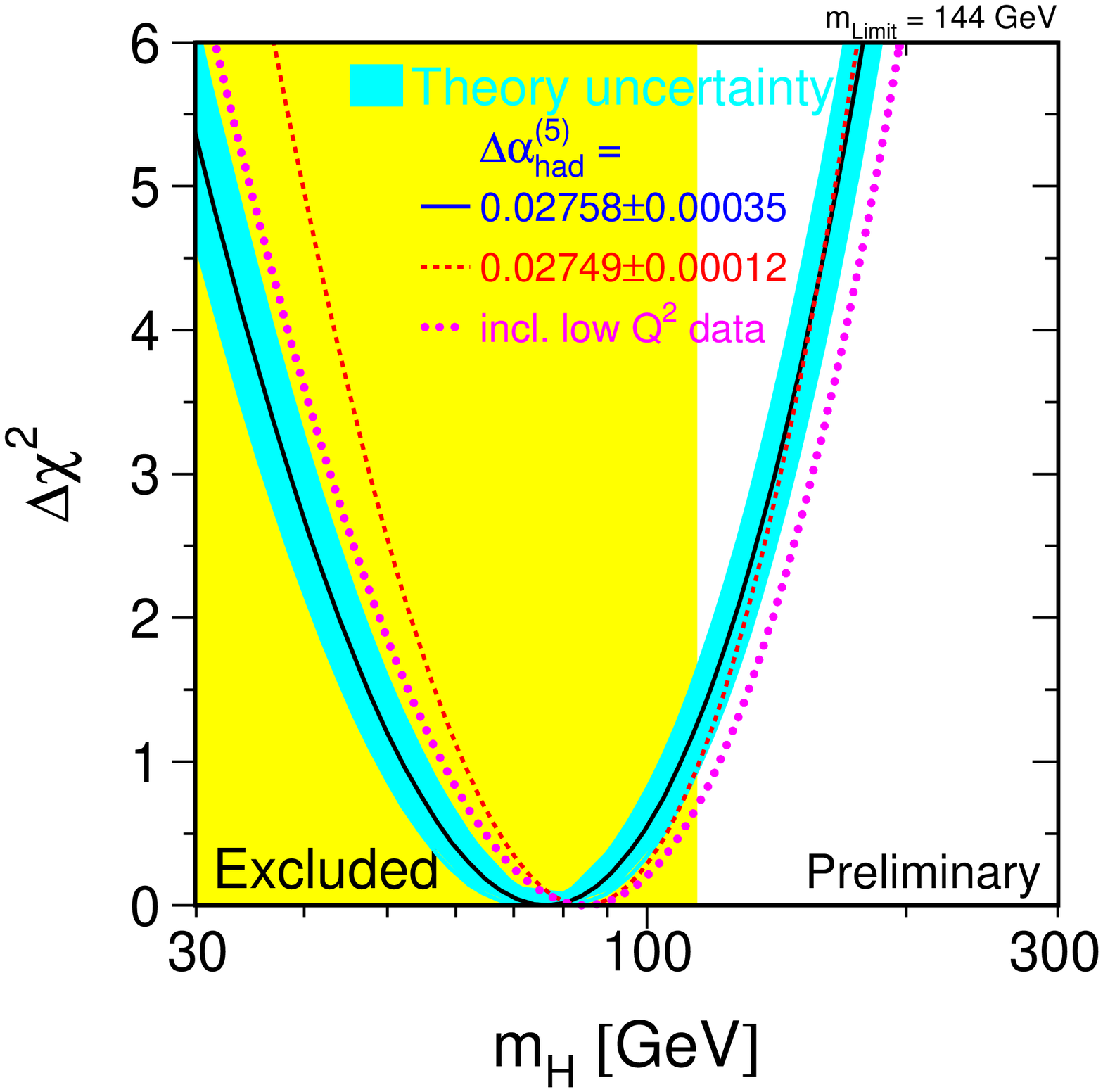,width=6cm}
\caption{
The expected region of Higgs mass in the $m_W$ vs. $m_t$ plane (left).
The combined Tevatron measurement of $m_t$ (and $m_W$) are shown.
Also shown is the fit to all electroweak data (right), in which
the value of $m_t$ provides substantial weight \protect\cite{tewg}.}
\label{fig:mHiggs}
\end{center}
\end{figure*}

\begin{figure*}[!h!tbp]
\begin{center}
\epsfig{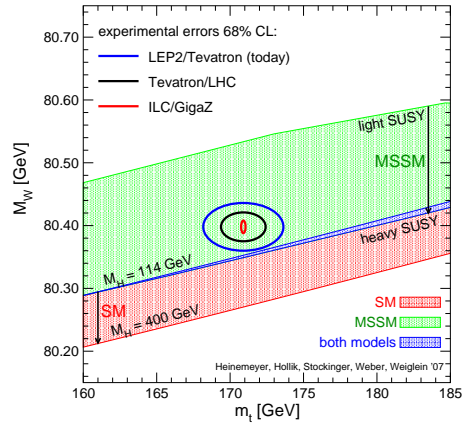}
\caption{
Regions in $m_W$ vs. $m_t$ plane favored by MSSMs and by the standard
model \protect\cite{mssmHeyn}.  Anticipated sensitivities from future
measurements are also shown.}
\label{fig:mssm}
\end{center}
\end{figure*}

Finally, the study of the top quark is inextricably linked with the
question of electroweak symmetry breaking and the Higgs mechanism.  
By including the measured value of $m_t$ in the global
electroweak fit one obtains the most stringent constraint on the mass of the
Higgs boson. The preferred value for $m_H$ is $76^{+33}_{-24}$  GeV\cite{tewg}, 
which is not favored by the $m_H > 114.7$ GeV limit from 
direct searches at LEP-II\cite{LEP_Higgs}.  This result is shown in
Fig. \ref{fig:mHiggs}.  The 95\%
confidence level upper limit on the Higgs boson mass set by this analysis is 
$m_H  <144$
 GeV. 
It is interesting to consider the measurements in relation to potential new physics.
For example, Figure \ref{fig:mssm} indicates the regions in the $m_W$ vs. $m_t$ plane
favored by MSSMs ~\cite{mssmHeyn,twoLoop} and by the 
standard model.  
If current central values of $m_W$ and $m_t$ remain the same as more
precise measurements are made, a signature of new physics may arise.
In the standard model, improving the precision with which the top quark mass is measured will
be a crucial ingredient in further tightening the constraints on the Higgs 
boson mass.  

The most favored region provides for a complex
set of Higgs boson production and decay processes.  In this sense, the study
of the Higgs boson has much in common with what confronted experimentalists
for the top quark discovery.  
Final state channels including most of the objects 
identifiable at a hadron collider detector: $e, \gamma, \mu, \nu, jets, \tau$, 
and $b-$jets.  This implies that the full range of experimental
capabilities at the experiments will need to be mastered to maximize
sensitivity.  Multiple channels will be used to gain
significance before any one channel sees a pristine signal. 
Decay chains include gauge bosons, and there are 
concurrent $W/Z+X$ backgrounds that go with this.
Some of the modeling questions and background estimation
approaches that have been important in study of the top quark will also be 
important for the Higgs boson search.

One thing is clear, the race is on in earnest for the Higgs boson.  Will the 
Tevatron have a shot at a signal?
Will the LHC effort turn on rapidly and overwhelm the search quickly?  Is
there a Higgs mechanism at all, or are there new strong dynamics in its place
which elevate the top quark to a more fundamental status. The study of
the top quark lies at an important cross-roads in the pursuit of answers
to these questions.
We look forward with fascination to the chance to work with our colleagues
in the coming years' study.

\section{Acknowledgments}

We sincerely thank the staff of Fermilab for providing a wonderful 
vehicle with which to study this new and fascinating particle.  With 
gratitude, we thank our colleagues on CDF and \dzero\ for the 
effort and experience they have shared in making the measurements described
in this paper.  We have benefited from the
input of several colleagues in the drafting of this review.  Special
thanks go to Ulrich Heintz, Boaz Klima, Avto Kharchilava, 
Matteo Cacciari, Nikolaos
Kidonakis, Lisa Shabalina, Florencia Canelli and Kirsten Tollefson 
for their thoughtful feedback on various aspects of this paper.  Compilation
of this review was funded in part by DE-FG02-O4ER41299 with Southern Methodist
University, by DE-FG02-91ER40688 with Brown University and by NSF-0555632 with
SUNY Buffalo.

%% file: toprev.bbl
\begin{thebibliography}{20}
\bibitem{d0Obs}D\O\ Collab. (S. Abachi et al.), \textit{Phys. Rev. Lett.} {\bf 74}, 2632 (1995).
\bibitem{cdfObs}CDF Collab. (F. Abe et al.), \textit{Phys. Rev. Lett.} {\bf 74}, 2626 (1995).
\bibitem{bhatRev} P. Bhat, H. Prosper and S. Snyder, ``Top Quark Physics at the Tevatron'', \textit{\textit{Int. J. Mod. Phys.}} A {\bf 13}, 5113 (1998).
\bibitem{dhimanRev} D. Chakraborty, J. Konigsberg and D. Rainwater, ``Top Quark Physics'', \textit{\textit{Ann. Rev. Nucl. Part. Sci.}} {\bf 53}, 301 (2003).
\bibitem{dawsonRev} S. Dawson, ``The Top Quark, QCD and New Physics'', Lect. given at TASI 2002 Summer School, U. of Colorado, Boulder, CO, June 2-28 (2002); {\bf hep-ph/0303191}; {\bf BNL-71205-2003-CP}.
\bibitem{wagnerRev} W. Wagner, ``Top Quark Physics in Hadron Collisions'', \textit{\textit{Rep. Prog. Phys.}} {\bf 68}, 2409 (2005).
\bibitem{salam-weinberg}A. Salam, \textit{Elementary Particle Physics} (Almqvist and Wiksells,
Stockholm, 1968); S. Weinberg, \textit{\textit{Phys. Rev. Lett.}} {\bf 19}, 1264 (1967).
\bibitem{higgs}P. W. Higgs,  \textit{Phys. Lett.} {\bf 12}, 132 (1964); \textit{Phy. Rev.} {\bf 145}, 1156 (1966). F. Englert and R. Brout, \textit{Phys. Rev. Lett.} {\bf 13}, 321 (1964).
\bibitem{qcd}C. Quigg, \textit{Gauge Theories of the Strong, Weak, and Electromagnetic Interactions} (Addison-Wesley, 1983).
\bibitem{quarkmod}M. Gell-Mann, \textit{Phys. Lett.} {\bf 8}, 214 (1964); G. Zweig,\textit{ CERN Report}, 8182/Th.401, (1964).
\bibitem{fieldFeyn-stringFrag}R. D. Field and R. P. Feynman, \textit{Nucl. Phys.} B {\bf 136}, 1 (1978); B. Andersson et al., \textit{Phys. Rep.} {\bf 97}, 31 (1983). 
\bibitem{CKM}M. Kobayashi and T. Maskawa, \textit{Prog. Theor. Phys.} {\bf 49}, 652 (1973).
\bibitem{GIM}S. Glashow, J. Iliopolous and L. Maiani, \textit{Phys. Rev.} D {\bf 2}, 1285 (1970).
\bibitem{tauDiscover}M. L. Perl et al., \textit{Phys. Rev. Lett.} {\bf 35}, 1489 (1975).
\bibitem{bDiscover}S. W. Herb et al., \textit{Phys. Rev. Lett.} {\bf 39}, 252 (1977).
\bibitem{georgi}H. Geori and S. Glashow, \textit{Nucl. Phys.} B {\bf 167}, 173 (1980)
\bibitem{kane}G. Kane and M. Peskin, \textit{Nucl. Phys.} B {\bf 195}, 29 (1982)
\bibitem{georgi2}H. Georgi and A. Pais, \textit{Phys. Rev.} D {\bf 19}, 2746 (1979)
\bibitem{bjorken}J. Bjorken and K. Lane, in \textit{Proc. of Int. Neutrino Conf.}, Baksan, 1977, ed. by M. Markov et al., (Nauka, Moscow, 1978).
\bibitem{gursey}F. Gursey, P. Raymond and P. Sikivie, \textit{Phys. Lett.} B {\bf 60}, 177 (1976).
\bibitem{fcnclimits}UA1 Collab. (A. Bean et al.), \textit{Phys. Rev.} D {\bf 35}, 3533 (1987).
\bibitem{BmesonMix}ARGUS Collab. (H. Albrecht et al.), \textit{Phys. Lett.} B {\bf 192}, 245 (1987); CLEO Collab.( J. Bartelt et al.), \textit{Phys. Rev. Lett.} B {\bf 71}, 1680 (1993).
\bibitem{topless}E. Ma,\textit{ Phys. Rev. Lett.} {\bf 57}, 535 (1986).
\bibitem{Zwidth} Particle Data Group (K. Hagiwara et al.), \textit{Phys. Rev.} D {\bf 66}, 010001 (2002).
\bibitem{T3b}JADE Collab. (E. Elsen et al.), \textit{ Z. Phys.} C {\bf 46}, 349 (1990); CELLO
Collab. (H. Behrend et al.), \textit{Z. Phys.} C {\bf 47}, 333 (1990); TOPAZ Collab.
(A. Shimonaka et al.), \textit{Phys. Lett.} B {\bf 268}, 457 (1991). 
\bibitem{LEPT3b}The LEP Electroweak Working Group, the SLD Electroweak and
Heavy Flavor Groups, \textit{Phys. Rep.} {\bf 427}, 257 (2006).
\bibitem{petraLim} Review of Particle Properties, Particle Data Group,\textit{Phys. Lett.} B {\bf 239}, VII 167 (1990).
\bibitem{ua1Lim} UA1 Collab. (C. Albajar et al.), \textit{Z. Phys.} C {\bf 37}, 505 (1988);
\textit{Z. Phys.} C {\bf 48}, 1 (1990).
\bibitem{ua2Lim} UA2 Collab. (T. Akessan et al.), \textit{Z. Phys.} C {\bf 46}, 179 (1990).
\bibitem{cdfLim} CDF Collab. (F. Abe et al.), \textit{Phys. Rev. Lett.} {\bf 68}, 447 (1992);
\textit{Phys. Rev.} D {\bf 45}, 3921 (1992).
\bibitem{d0limit131}D\O\ Collab. (S. Abachi et al.), \textit{Phys. Rev. Lett.} {\bf 724}, 2138 (1994).
\bibitem{topBSM}T. M. P. Tait and C.-P. Yuan, \textit{Phys. Rev.} D {\bf 63}, 014018 (2000).
\bibitem{factoriz}J. C. Collins, D. E. Soper and G. Sterman, \textit{Nucl. Phys.}
B {\bf 263}, 37(1986).
\bibitem{LO}H. M. Georgi et al., \textit{ Ann. Phys.} {\bf 114}, 273 (1978); L. M. Jones and H. W. Wyld, \textit{Phys. Rev.} D {\bf 17}, 1782 (1978); M. Gluck, J. F. Owens and E. Reya, \textit{Phys. Rev.} D {\bf 17}, 2324 (1978); J. Babcock, D. Sivers and S. Wolfram, \textit{Phys. Rev.} D {\bf 18}, 162 (1978); K. Hagiwara and T. Yoshino, \textit{Phys. Lett.} B {\bf 80}, 282 (1979); B. L. Combridge, \textit{Nucl. Phys.} B {\bf 151}, 429 (1979); W. Beenakker et al., \textit{Nucl. Phys.} B {\bf 411}, 343 (1994).
\bibitem{NLO}P. Nason, S. Dawson and R. Ellis, \textit{Nucl. Phys.} B {\bf 303}, 607 (1988); W. Beenakkeret al., \textit{Phys. Rev.} D {\bf 40}, 54 (1989).
\bibitem{NLOb}W. Beenakker, et al., \textit{Nucl. Phys.} B {\bf 351}, 507 (1991).
\bibitem{sterman}G. Sterman, \textit{Nucl. Phys.} B {\bf 281}, 310 (1987).
\bibitem{catani}S. Catani and L. Trentadue, \textit{Nucl. Phys.} B {\bf 327}, 323 (1989); S. Catani and L. Trentadue, \textit{Nucl. Phys.} B {\bf 353}, 183 (1991).
\bibitem{HQresum}N. Kidonakis and G. Sterman, \textit{Phys. Lett.} B {\bf 387}, 867 (1996); N. Kidonakis and G. Sterman, \textit{Nucl. Phys.} B {\bf 505}, 321 (1997).
\bibitem{BCMN}R. Bonciani, S. Catani, M. Mangano and P. Nason, \textit{Nucl. Phys.} B {\bf 529}, 424 (1998).
\bibitem{cacciari}M. Cacciari, et al., \textit{J. High Energy Phys.} {\bf 04}, 068 (2004).
\bibitem{kinchoice}N. Kidonakis, F. Laenen, S. Moch, and R. Vogt, \textit{Phys. Rev.} D {\bf 64}, 114001 (2001); N. Kidonakis, \textit{Phys. Rev.} D {\bf 64}, 014009 (2001).
\bibitem{kidonakis}N. Kidonakis and R. Vogt,\textit{Phys. Rev.} D {\bf 68}, 114014 (2003).
\bibitem{toptechnicolor2}K. Lane and E. Eichten, \textit{ Phys. Lett.} B {\bf 352}, 382 (1995);
 \textit{Phys. Lett.} B {\bf 433}, 96 (1998).
\bibitem{ST-xsec-sullivan}Z. ~Sullivan, \textit{Phys. Rev.} D {\bf 70}, 114012 (2004). 
\bibitem{ST-xsec-kidonakis}N.~Kidonakis,\textit{Phys. Rev.} D {\bf 74}, 114012 (2006); N.~Kidonakis, \textit{Phys. Rev.} D {\bf 75}, 071501 (2007).
\bibitem{ST-xsec-a}T. M. P. Tait,\textit{ Phys. Rev.} D {\bf 61}, 034001 (2000).
\bibitem{ckm_measurement} S.Eidelman et al., \textit{Phys. Lett.} B {\bf  592}, 1 (2004).
\bibitem{alpgen}M. Mangano et al., \textit{J. High Energy Phys.} {\bf 07}, 001 (2003)
\bibitem{pythia}T. Sjostrand et al.,\textit{Comput. Phys. Commun.} {\bf 135}, 238 (2001).
\bibitem{herwig}G. Marchesini et al., \textit{Comput. Phys. Commun.} {\bf 67}, 465 (1992); G. Corcella et al., \textit{J. High Energy Phys.} {\bf 01}, 010 (2001).
\bibitem{matching1}M. Mangano,\textit{ http://cepa.fnal.gov/patriot/mc4run2/MCTuning/061104/mlm.pdf} (2004).
\bibitem{matching2}S. Mrenna and P. Richardson, \textit{J. High Energy Phys.} {\bf 05}, 040 (2004).
\bibitem{matchingCatani}S. Catani et al., \textit{J. High Energy Phys.} {\bf 0111}, 063 (2001); CERN-TH-2000-367 (2000).
\bibitem{tauola}S. Jadach et al., \textit{Comp. Phys. Commun.} {\bf 76}, 361 (1993).
\bibitem{QQgen}P. Avery, K. Read and G. Trahern, \textit{CLEO Report} {\bf CSN-212}, 1985 (unpublished).
\bibitem{tevatron}H. T. Edwards, \textit{Ann. Rev. Nucl. Part. Sci.} {\bf 35}, 605 (1985).
\bibitem{d0run1}D\O\ Collab. (S. Abachi, et al.), \textit{ Nucl. Instrum. Meth.} A {\bf 338}, 185 (1994). 
\bibitem{cdfrun1}CDF. Collab. ( J. Antos et al.), \textit{ Nucl. Instrum. Meth.} A {\bf 360}, 118 (1995). 
\bibitem{cdfrun2}CDF Collab. ( C. Newman-Holmes et al.), The CDF Upgrade, \textit{Fermilab-CONF-} {\bf 96-218-E}, 1996 (unpublished); R. Blair et al., Technical Design Report, \textit{Fermilab-PUB-} {\bf 96-390-E},1996 (unpublished).
\bibitem{d0run2}D\O\ Collab. (V. Abazov et al.), The Upgraded D0 Detector, \textit{Nucl. Instrum. Meth.} A {\bf  565}, 463 (2006).
\bibitem{d0muon}D\O\ Collab. (V. Abazov, et al.), \textit{FERMILAB-PUB-} {\bf 05-034-E} (2005).
\bibitem{ConeAlg} CDF Collab.( F. Abe et al.), \textit{Phys. Rev.} D {\bf 45}, 1448 (1992);
G. C. Blazey et al., RunII Jet Physics, in \textit{ Proc. Workshop: QCD and Weak Boson Physics in RunII}, eds. U. Bauer, R. K. Ellis, D. Zappenfeld (FERMILAB-PUB-{\bf 00-297}, 2000).
\bibitem{D0runIJES}D\O\ Collab.( B. Abbott et al.), \textit{Phys. Rev. Lett.} {\bf  86}, 1707 (2001); \textit{Nucl. Instrum. Methods Phys. Res. Sect.} A {\bf  424}, 352 (1999).
\bibitem{evtgen}D. Lange, \textit{ Nucl. Instrum. Meth. Phys. Res.} A {\bf 462}, 152 (2001).
\bibitem{MCgenRev}S. R. Slabopitsky, Event generators for top quark production and decays, in \textit{ Proc. Int. Workshop Top Quark Physics} (Coimbra, Portugal, 2006), Proc. of Sci. TOP2006:019, 2006, {\bf hep-ph/0603124}.
\bibitem{CDFMCgen} M. Dobbs et al., Les Houches Guidebook to Monte Carlo Generators for Hadron Collider Physics, {\bf hep-ph/0403045}.
\bibitem{STstmc}E.E. Boos et al., \textit{Phys. Atom. Nucl.} {\bf 69}, 1317 (2006).
\bibitem{STcomphep}  {C\rm{omp}\sc{hep}}~Collab. (E.~Boos {\sl et al.}),
\textit{ Nucl. Instrum. Methods Phys. Res.} A {\bf 534}, 250 (2004).
\bibitem{STmadevent}T. Stelzer and W.F. Long, \textit{Comput. Phys. Commun.} {\bf 81},
  337 (1994); F. Maltoni and T. Stelzer, \textit{J. High Energy Phys.} {\bf 02}, 027 (2003).
\bibitem{geant3} R.~Brun {\sl et al.}, GEANT Detector Description and Simulation Tool, \textit{ CERN Program Library} {\bf W5013} (1994).
\bibitem{sphericity}V. Barger, J.Ohnemus and R. J.N. Phillips, 
 \textit{ Phys. Rev.} D {\bf 48}, 3953 (1993).
\bibitem{d0r2llcsec}D\O\ Collab. (V. Abazov et al.), \textit{Phys. Lett.} B {\bf 626}, 55 (2005).
\bibitem{d0r1ttcsec}D\O\ Collab.(V. Abazov, et al.), \textit{Phys. Rev.} D {\bf 67}, 012004 (2003); \textit{Phys. Rev. Lett.} {\bf 79}, 1203 (1997).
\bibitem{LEPZmass} LEP Electroweak Working Group:\textit{ http://lepewwg.web.cern.ch/LEPEWWG/plots/summer2005}.
\bibitem{llkin}CDF Collab. (D. Acosta et al.), \textit{Phys. Rev. Lett.} {\bf 95}, 022001 (2005).
\bibitem{d0425llcsec}D\O\ Collab. (V. Abazov, et al.), \textit{Phys. Rev.} D 76, 052006 (2007).
\bibitem{d01fbllcsec}
D\O\ Collab. (V. Abazov, et al.), \textit{D0-CONF-5477} (2007); D\O\ Collab. (V. Abazov, et al.), \textit{D\O\ -CONF-5465} (2007); D\O\ Collab. (V. Abazov, et al.), \textit{D\O\ -CONF-5371} (2007).
\bibitem{cdfr2llcsec}CDF Collab. (D. Acosta et al.), \textit{Phys. Rev. Lett.} {\bf 93}, 142001 (2004).
\bibitem{cdfr1llcsec}CDF Collab. (F. Abe et al.), \textit{Phys. Rev. Lett.} {\bf 80}, 2773 (1998).
\bibitem{cdf360llcsec}CDF Collab. (D. Acosta et al.), subm. to \textit{Phys. Rev.} D Rapid. Comm. (2006).
\bibitem{cdf1fbllcsec}CDF Collab. (D. Acosta et al.), in Proc. of Lepton-Photon Conf. (2007); \textit{CDF-CONF-8802,8770}.
\bibitem{vecbos}F. A. Berends et al., \textit{Phys. Lett.} B {\bf 357}, 32 (1991).
\bibitem{tauCsec} D\O\ Collab. (V. Abazov, et al.), \textit{D\O\ -CONF-5234} (2006).
\bibitem{chargedHiggsModels}J. C. Pati and A. Salam, \textit{Phys. Rev.} D {\bf 10}, 275 (1974); R. N. Mohapatra and J. C. Pati, \textit{Phys. Rev.} D {\bf 11}, 566 (1975); G. Senjanovic and R. N. Mohapatra, \textit{Phys. Rev.} D {\bf 12}, 1502 (1975); T. G. Rizzo, \textit{Phys. Rev.} D {\bf 25}, 1355 (1982);D {\bf 27}, 657(A) (1983).
\bibitem{cdfljtopo194}CDF Collab., (D. Acosta, et al.), \textit{Phys. Rev.} D {\bf 52}, 052003 (2005).
\bibitem{ljtopo230}D\O\ Collab. (V. Abazov, et al.), \textit{Phys. Lett.} B {\bf 626}, 45 (2005).
\bibitem{d0lj425topo}D\O\ Collab. (V. Abazov, et al.), \textit{Phys. Rev.} D {\bf 76}, 092007 (2007).
\bibitem{d0lj900topo}D\O\ Collab. (V. Abazov, et al.), \textit{D\O\ -CONF-5262} (2007).
\bibitem{berends}F. Berends et al., \textit{Nucl. Phys.} B {\bf 357}, 32 (1991).
\bibitem{cdfr1ljsltcsec}CDF Collab. (T. Affolder, et al.), \textit{Phys. Rev.} D {\bf 64}, 032002 (2001).
\bibitem{cdfr2ljmutagcsec}CDF Collab. (D. Acosta, et al.), \textit{Phys. Rev.} D {\bf 72}, 032002 (2005).
\bibitem{cdf760mutagcsec}CDF Collab. (D. Acosta, et al.), \textit{CDF-CONF-8565} (2006).
\bibitem{d0r2ljsmt}D\O\ Collab. (R. Harrington), in \textit{Proceedings of Lake Louise Winter Institute 2007} (Alberta, Canada, Feb 2007) World Scientific Publ. Co., Singapore, (in press).
\bibitem{cdfr2btagkinfitcsec}CDF Collab. (D. Acosta, et al.), \textit{Phys. Rev.} D {\bf 71}, 072005 (2005).
\bibitem{cdfr2ljbtagtopocsec}CDF Collab. (D. Acosta, et al.), \textit{Phys. Rev.} D {\bf 71}, 052003 (2005).
\bibitem{cdfr2btag162csec}CDF Collab. (D. Acosta, et al.),\textit{ Phys. Rev.} D {\bf 71}, 052003 (2005).
\bibitem{cdfr2ljbtagcsec}CDF Collab. (D. Acosta, et al.), \textit{Phys. Rev.} D {\bf 71}, 052003 (2005).
\bibitem{cdf1fbbtagcsec}CDF Collab. (D. Acosta, et al.), \textit{CDF-CONF-8795}, (2007).
\bibitem{ljbtag230}D\O\ Collab. (V. Abazov, et al.), \textit{Phys. Lett.} B {\bf 626}, 35 (2005).
\bibitem{cdfr2btag318csec}CDF Collab. (A. Abulencia, et al.), \textit{Phys. Rev. Lett.} {\bf 97}, 082004 (2006). 
\bibitem{cdfr2jetprobcsec}CDF Collab. (A. Abulencia, et al.), \textit{Phys. Rev.} D {\bf 74}, 072006 (2006).
\bibitem{ljbtag425}D\O\ Collab. (V. Abazov et al.), \textit{Phys. Rev.} D {\bf 74}, 112004 (2006).
\bibitem{d0900ljbtag}D\O\ Collab. (V. Abazov et al.), \textit{D\O\ -CONF-5355} (2007).
\bibitem{d0r1alljetscsec}D\O\ Collab. (B. Abbott, et al.), \textit{Phys. Rev. Lett.} {\bf 83}, 1908 (1999); \textit{Phys. Rev.} D {\bf 60}, 012001 (1999).
\bibitem{cdfr1alljetscsec}CDF Collab. (F. Abe et al.), \textit{Phys. Rev. Lett.} {\bf 79}, 1992 (1997).
\bibitem{cdfr2alljetscsec}CDF Collab. (D. Acosta et al.), \textit{Phys. Rev.} D {\bf 74}, 072005 (2006).
\bibitem{d0r2alljetscsec}D\O\ Collab. (V. Abazov et al.), \textit{ Phys.\ Rev.\ } D {\bf 76}, 072007 (2007).
\bibitem{ROOT}R. Brun and F. Rademakers, \textit{Nucl. Inst. Meth. in Phys. Res.} A {\bf 389}, 81 (1997).
\bibitem{STD0evidencePRL} D\O\ Collab. (V. Abazov et al.),{\bf  hep-ex/0612052}, acc. to \textit{Phys. Rev. Lett.}.
\bibitem{STD0prdone}D\O\ Collab. (V. Abazov et al.),{\bf hep-ex/0604020}, subm. to \textit{Phys. Rev.} D. 
\bibitem{d0runI} D\O\ Collab. (B. Abbott, et al.), \textit{Phys. Rev. D Rapid Comm.} {\bf 63}, 031101 (2001); D\O\ Collab. (V. Abazov et al.), \textit{Phys. Lett.} B {\bf 517}, 282 (2001).
\bibitem{STAcosta} CDF Collab. (D. Acosta et al.), 
Phys. Rev. D {\bf 65}, 091102 (2002); \textit{ Phys.\ Rev.\ } D {\bf 69}, 052003 (2004).
\bibitem{RunIIcdf_result} CDF Collab. (D. Acosta et al.),
\textit{Phys. Rev.} D {\bf 71}, 012005 (2005).
\bibitem{STPLBD0} D\O\ Collab. (V. Abazov et al.), \textit{Phys. Lett.} B {\bf 622}, 265 (2005).
\bibitem{STvariables} E.~Boos and L.~Dudko, \textit{Nucl. Instrum. Methods} A {\bf 502}, 486 (2003).
\bibitem{STMahlon:1995zn} G.~Mahlon and S.~Parke,\textit{ Phys. Rev.} D {\bf 53}, 4886 (1996); S.~Parke and Y.~Shadmi, \textit{Phys. Lett.} B {\bf 387}, 199 (1996);
G.~Mahlon and S.~J.~Parke, \textit{Phys.\ Rev.\ } D {\bf 55}, 7249 (1997).
\bibitem{STmlpfit} J.~Schwindling,
\\ {\verb+http://schwind.home.cern.ch/schwind/MLPfit.html+}.
\bibitem{STNNBayes} M.~Feindt, e-Print Archive physics/0402093 (2004).
\bibitem{STdt-breiman} L.~Breiman {\it et al.}, {\it Classification and Regression Trees} (Wadsworth, Stamford, 1984).
\bibitem{STDTboost}Y.~Freund and R.E.~Schapire, in {\it Proceedings of the Thirteenth   International Conference on Machine Learning}, ed. L.~Saitta (Morgan Kaufmann, San Francisco, 1996), p.~148.
\bibitem{STFNALtalk}  D\O\ , {\it Evidence for Single Top Quark
 Production and First Measurement of |$V_{tb}$|} presented on Dec 8th,
 2006 at the Fermilab Joint Theory and Experimental Seminar.
\bibitem{STbayesianNN} R.M.~Neal, {\it Bayesian Learning of Neural Networks}
(Springer-Verlag, New York, 1996); Used ``Software for Flexible Bayesian
Modeling'' package, {\texttt
http://www.cs.toronto.edu/$\sim$radford/fbm.software.html}.    
\bibitem{CDF_top_SUSY07}
CDF Collab. (S.~Leone),
\textit{Published Proceedings 15th International Conference on Supersymmetry 
and the Unification of Fundamental Interactions (SUSY07)}, 
(Karlsruhe, Germany, July 26-August 1, 2007), Librix Publishers, Budapest (in press), FERMILAB-CONF-07-581-E. 
\bibitem{tewg}D\O\ and CDF Collab. Top Quark Mass Combination, \textit{The Tevatron
Electroweak Working Group}, {\bf hep-ex/0703034}; FERMILAB-TM-2380-E (2007).
\bibitem{mssmModels}S. Dimoupoulos and H. Georgi, \textit{Nucl. Phys.} B {\bf 193}, 10 (1981).
\bibitem{llmtopMlb} CDF Collab. (F. Abe et al.), \textit{Phys. Rev. Lett.} {\bf 80}, 2779 (1998). 
\bibitem{d0r1llmtop}D\O\ Collab. (B. Abbott, et al.), \textit{Phys. Rev.} D {\bf 60}, 052001 (1999); \textit{Phys. Rev. Lett.} {\bf 80}, 2063 (1998).
\bibitem{nuWTcdf1}CDF Collab. (F. Abe et al.), \textit{Phys. Rev. Lett.} {\bf 82}, 271 (1999).
\bibitem{nuWTcdf2}CDF Collab. (A. Abulencia et al.), \textit{Phys. Rev.} D {\bf 73}, 112006 (2006).
\bibitem{dalitz}R. Dalitz and G. Goldstein, \textit{Phys. Rev.} D {\bf 45}, 1531 (1992).
\bibitem{kondo}K. Kondo, \textit{J. Phys. Soc. Jpn.}  {\bf 57}, 4126 (1988); \textit{J. Phys. Soc. Jpn.}  {\bf 60}, 836 (1991).
\bibitem{llMEcdf2}CDF Collab. (A. Abulencia et al.), \textit{Phys. Rev. Lett.} {\bf 96}, 152002 (2006); 
\textit{Phys. Rev.} D {\bf 74}, 032009 (2006).
\bibitem{MEmtop}D\O\ Collab.(V. Abazov et al.),\textit{ Nature} {\bf 429}, 638 (2004).
\bibitem{d0r2llmtPLB} D\O\ Collab.(V. Abazov, et al.), \textit{Phys. Lett.} B, {\bf 655}, 7 (2007).
\bibitem{d0r2nuWT1fb} D\O\ Collab. (V. Abazov, et al.), \textit{D\O\ -CONF-5347} (2007).
\bibitem{d0r2nuWTmoments} D\O\ Collab. (V. Abazov, et al.), \textit{D\O\ -CONF-5171} (2006).
\bibitem{d0r2MWT1fb} D\O\ Collab. (V. Abazov, et al.), \textit{D\O\ -CONF-5463} (2007).
\bibitem{PDEref} L. Holmstrom, R. Sain and H. E. Miettinen, \textit{Comput. Phys. Commun.} {\bf 88}, 195 (1995).
\bibitem{cdfR2lltemp} CDF Collab. (A. Abulencia et al.), \textit{Phys. Rev.} D {\bf 73}, 112006 (2006).
\bibitem{cdfllME1fb} CDF Collab. (A. Abulencia, et al.), \textit{Phys. Rev.} D {\bf 75}, 031105 (2007).
\bibitem{MEref}G.~Mahlon and S.~J.~Parke, \textit{Phys.\ Lett.\ } B {\bf 411}, 173 (1997).
\bibitem{KB_thesis} 
``A Precision Measurement of the Top Quark Mass,'' K.~M.~Black, Ph.D. thesis,
Boston University, Boston, USA, 2005.
\bibitem{CDF_template} CDF Collab. (A.~Abulencia {\sl et al.}), 
 \textit{Phys. Rev. Lett.} {\bf 96}, 022004 (2006).
\bibitem{CDF_template_2} CDF and \dzero\ Collabs. (M. Zielinski),
in \textit{Proceedings of Physics at LHC} (Cracow, Poland, 2006), Acta Physica Polonica (2007), 
{\bf hep-ex/0610017}.
\bibitem{D0_ljmass_RunI}   D\O\ Collab. (S. Abachi, et al.),
 \textit{ Phys.\ Rev.\ Lett.\ } {\bf 79}, 1197 (1997);\\
  D\O\ Collab. (B. Abbott, et al.), \textit{ Phys.\ Rev.\ } D {\bf 58}, 052001 (1998).
\bibitem{D0_ljmass_RunII}  
 D\O\ Collab. (V. Abazov et al.), \textit{Phys. Rev.} D {\bf 74}, 092005 (2006).
\bibitem{D0_ME_W07} 
 D\O\ Collab. (M.~Wang),\textit{Proceedings of 42nd Rencontres de Moriond 
on QCD and High Energy Hadronic Interactions}, 
(La Thuile, Aosta Valley, Italy, March 2007), The' Gioi Publishers (2007)..
\bibitem{CDF_ME} CDF Collab. (E.~Brubaker),
in \textit{Proceedings 41st Rencontres de Moriond on 
Electroweak Interactions and Unified Theories} ( La Thuile, Aosta Valley, 
Italy, March 11-18, 2006), The' Gioi Publishers (2006), FERMILAB-CONF-06-124-E..
\bibitem{Kondo} K.~Kondo, Waseda University, \textit{RISE Technical Report} {\bf No.05-01} (2005), {\bf hep-ex/0508035}.
\bibitem{QDLM} CDF Collab. (A. Abulencia, et al.), \textit{Phys. Rev.} D {\bf 73}, 092002 (2006). 
\bibitem{D0_ideogram_RunII} 
D\O\ Collab. (V. Abazov et al.), \textit{Phys. Rev.} D {\bf 75}, 
092001 (2007); FERMILAB-PUB-07/039-E.
\bibitem{W_Delphi} DELPHI Collab. (P.~Abreu et al.),
\textit{Eur. Phys. J.} C {\bf 2}, 581 (1998); 
\textit{Phys. Lett.} B {\bf 462}, 410 (1999);
\textit{Phys. Lett.} B {\bf 511}, 159 (2001); 
M.~Mulders, \textit{Ph.D.~thesis} (FOM, Amsterdam \&  Amsterdam U.), Sep 2001, 226pp.
\bibitem{cdf_lxy}
CDF Collab. (A. Abulencia et al.), \textit{Phys. Rev.} D {\bf 75}, 
071102 (2007);
\bibitem{cdfr2mt6jtemp} CDF Collab. (A. Abulencia, et al.), \textit{Phys. Rev.} D 76, 072009 (2007).
\bibitem{d0r1alljetsmtop} D\O\ Collab. (V. Abazov et al.), \textit{Phys. Lett.} B {\bf 606}, 25 (2005).
\bibitem{cdfr2mt6jIdeo} CDF Collab. (A. Abulencia, et al.), \textit{Phys. Rev. Let.} 98, 142001 (2007).
\bibitem{cdfR2alljetsMass}CDF Collab. (A. Abulencia, et al.), \textit{CDF-CONF-8709} (2007).
\bibitem{EWWG}\textit{The LEP Electroweak Working Group}, LEPEWWG/2006-01, http://lepewwg.web.cern.ch/LEPEWWG/stanmod/summer2006/s06.ew.ps.gz.
\bibitem{LEP_Higgs}{The LEP Working Group for Higgs Boson Searches}\textit{Phys. Lett.} B {\bf 565}, 61 (2003).

\bibitem{tWb} G. Kane, C.-P. Yuan, and G. Ladinsky, \textit{Phys. Rev.} D {\bf 45},
  124, (1992). 
\bibitem{cdfWhelR1}CDF Collab. (D. Acosta et al.), \textit{Phys. Rev.} D {\bf 71}, 031101(R) (2005)
\bibitem{cdfWpolR1}CDF Collab. (T. Affolder et al.), \textit{Phys. Rev. Lett.} {\bf 84}, 216 (2000)
\bibitem{cdfWhelR2a}CDF Collab. (A. Abulencia et al.), \textit{Phys. Rev.} D {\bf 73}, 111103(R) (2006)
\bibitem{cdfWhel700}CDF Collab. (A. Abulencia et al.), \textit{Phys. Rev. Lett.} {\bf 98}, 072001 (2007)
\bibitem{Whel06Dzero} 
D\O\ Collab.(V. Abazov et al.), \textit{Phys. Rev.} D {\bf 75}, 031102(R) (2007).
\bibitem{Whel05Dzero} 
D\O\ Collab.(V. Abazov et al.), \textit{Phys. Rev.} D {\bf 72}, 011104(R) (2005).
\bibitem{Whel07Dzero} 
D\O\ Collab.(V. Abazov et al.), subm. to \textit{Phys. Rev. Lett.}(2007); FERMILAB-PUB-07/588-E.
\bibitem{fcnc} G.Eilam, J.L.Hewett, A.Soni, \textit{Phys. Rev.} D {\bf 44}, 1473 (1991).
\bibitem{cdf-run1-br}CDF Collab. (T. Affolder et al.), \textit{Phys. Rev. Lett.} {\bf 86}, 3233 (2001).
\bibitem{cdf-run2-br} CDF Collab.(D. Acosta et al.) , \textit{Phys. Rev. Lett.} 95, 102002 (2005).
\bibitem{d0-run2-br}  D\O\ Collab. (V. Abazov et al.), \textit{Phys. Lett.} B {\bf 639}, 616 (2006).
\bibitem{900pbRmeas} \dzero\ Collab. (V. Abazov, et al.), subm. to \textit{Phys. Rev. Lett.}; FERMILAB-PUB-08-010-E, hep-ex/0801.1326.
\bibitem{exotic-top-paper} D. Chang, W. Chang, and E. Ma, \textit{Phy. Rev.} D {\bf 59}, 091503 (1999); {\bf 61}, 037301 (2000); D. Choudhary, T. M. Tait and C. E. Wagner, \textit{Phy. Rev.} D {\bf 65}, 053002 (2005).
\bibitem{qtop-photon-rad} U. Baur, M. Buice and L. H. Orr, \textit{Phy. Rev.} D {\bf 64}, 094019 (2001).
\bibitem{d0-topcharge-paper} D\O\ Collab. (V. Abazov et al.), \textit{Phys. Rev. Lett.} {\bf 98}, 041801 (2007).
\bibitem{toptechnicolor1}C. T. Hill, \textit{ Phys. Lett.} B {\bf 345}, 483 (1995).
\bibitem{technicolor}S. Weinberg, \textit{ Phys. Rev.} D {\bf 13}, 974 (1976); L. Susskind, \textit{ Phys. Rev.} D {\bf 20}, 2619 (1979); S. Dimopoulos and L. Susskind,\textit{ Nucl. Phys.} B {\bf 155}, 237 (1979); E. Eichten and K. Lane, \textit{Phys. Lett.} B {\bf 90}, 125 (1980).
\bibitem{cdfr1ttreson}CDF Collab. (T. Affolder et al.), \textit{Phys. Rev. Lett.} {\bf 85}, 2062 (2000).
\bibitem{d0r1ttreson}D\O\ Collab. (V. Abazov et al.), \textit{Phys. Rev. Lett.} {\bf 92}, 221801 (2004).
\bibitem{d0limit}I. Bertram {\sl et al.}, \textit{FERMILAB-TM-} {\bf 2104} (2000).
\bibitem{d0ResR2} D\O\ Collab. (C. Schwanenberger), in \textit{Proceedings of International Europhysics Conference on High Energy Physics} (Lisboa, Portugal, 2005), Proc. of Sci. (2006), hep-ex/0602048.
\bibitem{d0res900}\dzero\ Collab. (V. Abazov, et al.), D0-CONF-5443 (2007).
\bibitem{cdfResR2}CDF Collab. (A. Abulencia, et al.), subm. to \textit{Phys. Rev. Lett.}, {\bf hep-ex/0709.0705} (2007).
\bibitem{cdfr2resol955}CDF Collab. (A. Abulencia, et al.), subm. to \textit{Phys. Rev. D}, {\bf hep-ex/0710.5335} (2007).
\bibitem{cleoChHiggs}M. S. Alam et al., \textit{Phys. Rev. Lett.} {\bf 74}, 2885 (1995).
\bibitem{towers}S. Towers, submitted to \textit{Phys. Lett.} B, {\bf hep-ex/0004022}.
\bibitem{lep-chH-Limit}LEP Higgs Working Group, \textit{LHWG Note} {\bf 2001-05}.
\bibitem{cdfchHiggs97}CDF Collab. (F. Abe, et al.), \textit{Phys. Rev. Lett.} {\bf 79}, 357 (1997).
\bibitem{coarasa}J. Coarasa, J. Guasch and J. Sola, \textit{Preprint UAB-FT-451}, (1998); talk
presented at IVth Intl. Symp. on Rad. Corr. (RADCOR), Barcelona, Sep. 8-12, 1998; {\bf hep-ph/9903212}.
\bibitem{carena}M. Carena, D. Garcia, U. Nierste and C. Wagner, \textit{Nucl. Phys.} B {\bf 577}, 88 (2000).
\bibitem{cdfchHiggs00}CDF Collab. (T. Affolder, et al.), \textit{Phy. Rev.} D {\bf 62}, 012004 (2000).
\bibitem{d0r1ChHiggsDecay}D\O\ Collab. (V. Abazov et al.), \textit{Phys. Rev. Lett.} {\bf 88}, 151803 (2002).
\bibitem{jetnet}JETNET package, http://www.thep.lu.se/public$_{-}$html/jetnet$_{-}$30$_{-}$manual/jetnet$_{-}$30$_{-}$manual.html.
\bibitem{d0chHiggs99} D\O\ Collab. (B. Abbott, et al.), \textit{Phys. Rev. Lett.} {\bf 82}, 4975 (1999).
\bibitem{cdfr2ChHiggsDecay}CDF Collab. (A. Abulencia et al.), \textit{Phys. Rev. Lett.} {\bf 96}, 042003 (2006).
\bibitem{CPsuperH}J. Lee et al.,\textit{Comput. Phys. Commun.} {\bf 156}, 283 (2004).
\bibitem{wprime-Sullivan}Z. Sullivan, \textit{Phys. Rev.} D {\bf 66}, 075011 (2002).
\bibitem{langacker}P. Langacker and S.U. Shankar, \textit{Phys. Rev.} D {\bf 40}, 1569 (1989) and references therein.

\bibitem{wprime-cdf-run1-e} CDF Collab. (T. Affolder et al.), \textit{Phys. Rev. Lett.} {\bf 87}, 231803 (2001).
\bibitem{wprime-D0-run1-e} D\O\ Collab. (S. Abachi et al.), \textit{Phys. Rev. Lett.} {\bf 76}, 3271 (1996).
\bibitem{wprime-cdf-run1-mu} CDF Collab. (F. Abe et al.), \textit{Phys. Rev. Lett.} {\bf 84}, 5716 (2000).
\bibitem{wprime-UA2-qq}UA2 Collab. (J. Alitti et al.), \textit{Nucl. Phys.} B {\bf 400}, 3 (1993).
\bibitem{wprime-cdf-run1-qq} CDF Collab. (F. Abe et al.), \textit{Phys. Rev.} D {\bf 55}, 5263 (1997).
\bibitem{wprime-D0-run1-qqonly}D\O\ Collab. (V. Abazov et al.), \textit{Phys. Rev.} D {\bf 69}, 111101 (2004).
\bibitem{wprime-cdf-run1-top} CDF Collab. (D. Acosta et al.), \textit{Phys. Rev. Lett.} {\bf 90}, 081802 (2003).
\bibitem{wprime-D0-run2-top} D\O\ Collab. (V. Abazov et al.), \textit{Phys. Lett.} B {\bf 641}, 423 (2006).
\bibitem{wprime-cdf-run2}CDF and D\O\ collab. (M. Datta), in \textit{Proceedings of Hadron Collider Physics Symposium 2007} (La Biodola, Isola d'Elba, Italy, May 2007) Elsevier, Nucl. Phys. B: Proc. Suppl. (in press).
\bibitem{atlasTDR}ATLAS Collab., ATLAS-TDR-15, CERN/LHCC/99-15 (1999).
\bibitem{mssmHeyn}S. Heinemeyer, W. Hollik, D. Stockinger, A.M. Weber and G. Weiglein, 
Journ. of High Energy Phys. 0608:052, (2006); S. Heinemeyer, W. Hollik and G. Weiglein, Phys. Rept. 425, 265 (2006).
\bibitem{twoLoop}A. Djouadi, P. Gambino, S. Heinemeyer, W. Hollik, C. Jünger and G. Weiglein, Phys. Rev. Lett. 78, 3626 (1997); Phys. Rev. D 57, 4179 (1998); S. Heinemeyer, G. Weiglein, JHEP 10, 072 (2002); J. Haestier, S. Heinemeyer, D. Stockinger and G. Weiglein, J. High Energy Phys. 12, 027 (2005).
\end{thebibliography}
